\documentclass[12pt,preprint]{aastex}
\newcommand{\muv}{\ensuremath{\mu_v }}
\newcommand{\mubr}{\ensuremath{\mu_{br} }}
\newcommand{\xiso}{\ensuremath{X_{\mbox{\scriptsize iso}}}}
\newcommand{\eiso}{\ensuremath{E_{\mbox{\scriptsize iso}}}}
\newcommand{\xgt}{\ensuremath{X_\gamma^{\mbox{\scriptsize (T)}}}}
\newcommand{\egt}{\ensuremath{E_\gamma^{\mbox{\scriptsize (T)}}}}
\newcommand{\xgi}{\ensuremath{X_\gamma^{\mbox{\scriptsize (I)}}}}
\newcommand{\xbgi}{\ensuremath{\bar{X}_\gamma^{\mbox{\scriptsize (I)}}}}
\newcommand{\egi}{\ensuremath{E_\gamma^{\mbox{\scriptsize (I)}}}}
\newcommand{\epk}{\ensuremath{E_{\mbox{\scriptsize peak}}}}
\newcommand{\epkr}
       {\ensuremath{E_{\mbox{\scriptsize peak}}^{\mbox{\scriptsize (rest)}}}}
\newcommand{\davg}{\ensuremath{\langle D\rangle}}

\shorttitle{GRB Jet Profiles and Signatures}
\shortauthors{Graziani, Lamb, and Donaghy}

\begin{document}

\title{Gamma-Ray Burst Jet Profiles And Their Signatures}
\author{C. Graziani, D. Q. Lamb, and T. Q. Donaghy}
\email{carlo@oddjob.uchicago.edu, quinn@oddjob.uchicago.edu,
lamb@oddjob.uchicago.edu}
\affil{Department of Astronomy and Astrophysics, University of
Chicago, 5640 S. Ellis Avenue, Chicago, IL 60637}

\begin{abstract} 
HETE-II and BeppoSAX have produced a sample of GRBs and XRFs with known
redshifts and $E_{pk}$.  This sample provides four important empirical
constraints on the nature of the source jets:  Log $E_{iso}$ is
approximately uniformly distributed over several orders of magnitude;  the
inferred prompt energy Log $E_{\gamma}$ is narrowly distributed; the
Amati relation holds between $E_{iso}$ and $E_{pk}$; and the Ghirlanda
relation holds between $E_{\gamma}$ and $E_{pk}$.
 
We explore the implications of these constraints for GRB jet structure
during the prompt emission phase. We infer the underlying angular profiles
from the first two of the above constraints assuming all jets have the
same profile and total energy, and show that such ``universal jet'' models
cannot satisfy both constraints.

We introduce a general and efficient method for calculating relativistic
emission distributions and $E_{pk}$ distributions from jets with arbitrary
(smooth) angular jet profiles.  We also exhibit explicit analytical
formulas for emission from top-hat jets (which are not smooth).  We use
these methods to exhibit $E_{pk}$ and $E_{iso}$ as a function of viewing
angle, for several interesting families of GRB jet profiles.  We use the
same methods to calculate expected frequency distributions of $E_{iso}$ and
$E_{\gamma}$ for the same families of models.

We then proceed to explore the behavior of universal jet models under a
range of profile shapes and parameters, to map the extent to which these
models can conform to the above four empirical constraints.

\end{abstract}


%%%%%%%%%%%%%%%%%%%%%%%%%%%%%%%%%%%%%%%%%%%%%%%%%%%%%%%%%%%%%%%%%%%%%%%%%%
%%%                                                                    %%%
%%%  \S1: Introduction                                                 %%%
%%%                                                                    %%%
%%%%%%%%%%%%%%%%%%%%%%%%%%%%%%%%%%%%%%%%%%%%%%%%%%%%%%%%%%%%%%%%%%%%%%%%%%
\section{Introduction}

Since the recognition that gamma-ray burst (GRB) sources are highly
relativistic collimated jets, the study of the structure of the jets and
of the distribution of jet parameters among burst sources has been an area
of active research.  An early observational success of the jet picture of
GRBs was the discovery by \citet{f01} of a striking correlation between
GRB energy fluence and the break times of GRB afterglow light curves.
Using the simple picture of a conical, uniform jet, \citet{f01}, and later
\cite{bloom2003} inferred jet opening angles from the break times using
results on the dynamics of afterglows of \citet{sari1999} for a sample of
GRBs with known redshifts.  After correcting the isotropic-equivalent
gamma-ray energy of each GRB by the inferred solid angle subtended by its
conical jet, \citet{f01} found that the extremely broad (3-4 decades)
distribution of isotropic-equivalent energy was mapped to a remarkably
narrow ($\sim 1$ decade) distribution of inferred total energy.

The debate over the meaning of this result has given rise to two main
alternative interpretations.  The first interpretation is an elaboration
of that of \citet{f01}, in which GRB sources all have a common amount of
energy available for gamma-ray emission (they are ``standard bombs''), and
differ from one another by the opening angle of their conical jet
\citep{lamb04}. Implicit in this view is that little or no emission is
received by an observer whose line-of-sight is not included in the jet
cone.

The alternative view is that GRB jets might have structure that is richer
than a simple uniformly-emitting cone.  The attractive feature of this
picture is that if the jet structure is sufficiently rich, it might be
possible to ascribe the same jet structure to all GRB jets, and to explain
the distribution of GRB energies using only the variation of observer
lines-of-sight with respect to the jet axis \citep{rossi2002,zm02}.

These two viewpoints are sometimes presented juxtaposed as the ``uniform
jet model'' and the ``structured jet model''.  This terminology is
unfortunate, since it is clear that the real distinction under debate is
not really whether GRB jets have ``structure'', but rather whether the
observed energetics of GRBs can be explained in terms of a single parameter
(the angle between the observer's line-of-sight and the jet axis), or
whether additional parameters relating to jet structure are required.  Thus
the juxtaposition would be better thought of as the ``universal jet model''
versus the ``variable-geometry jet model,'' where a sort of middle ground
is occupied by the ``quasi-universal model'' \citep{lr2004}.

One element that has not featured prominently in discussions of jet
structure is the role of relativistic kinematics in prompt GRB emission. 
Relativistic kinematics is a central feature of the ``off-axis beaming''
models of Yamazaki and collaborators
\citep{yamazaki2002,yamazaki2003,yamazaki2004,toma2005}, which those
authors used to attempt to unify the phenomenon of GRBs with that of
``X-Ray Flashes'' or XRFs \citep{heise2001,kippen2001}. However, their
models invoke a uniform conical ``top-hat'' jet structure.  By and large,
calculations of prompt emission expected from ``structured'' jets have
essentially bypassed the role of relativistic beaming and Doppler shift,
assuming a ``what you see is what you get'' relation of the jet emission
profile to the jet radiation pattern that tacitly implies a Lorentz factor
satisfying $\gamma^{-1}\rightarrow 0$ \citep{zm02,lr2004,zhang2004}.

This is a gap that is worth plugging.  It is clear that the observed
radiation pattern from a ``structured'' GRB jet can be substantially
modified by relativistic kinematics, both spectrally and in terms of total
radiated energy.  Depending on the Lorentz factor and viewing angle,
relativistic kinematics can turn a steep jet profile into a shallow
radiation pattern, or turn a hard source spectrum into a soft observed
spectrum (or, as we shall see, vice versa).  It is certainly not safe to
assume that it is even possible to produce a given radiation pattern
without a stringent --- and possibly unphysical --- constraints on the
Lorentz factor.

\begin{figure}[t]
\begin{minipage}[t]{3.2in}
\includegraphics[width=3.2in,clip=]{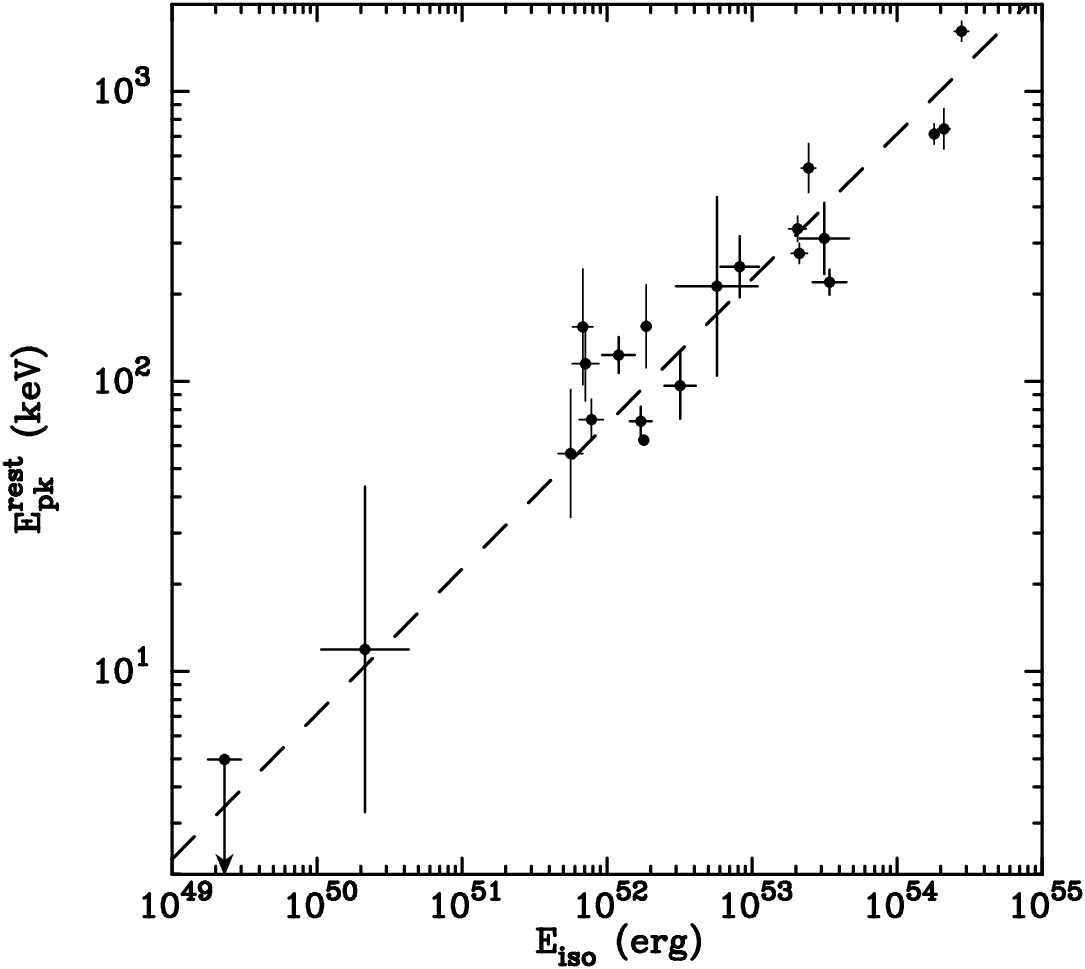}\\
\includegraphics[width=3.2in,clip=]{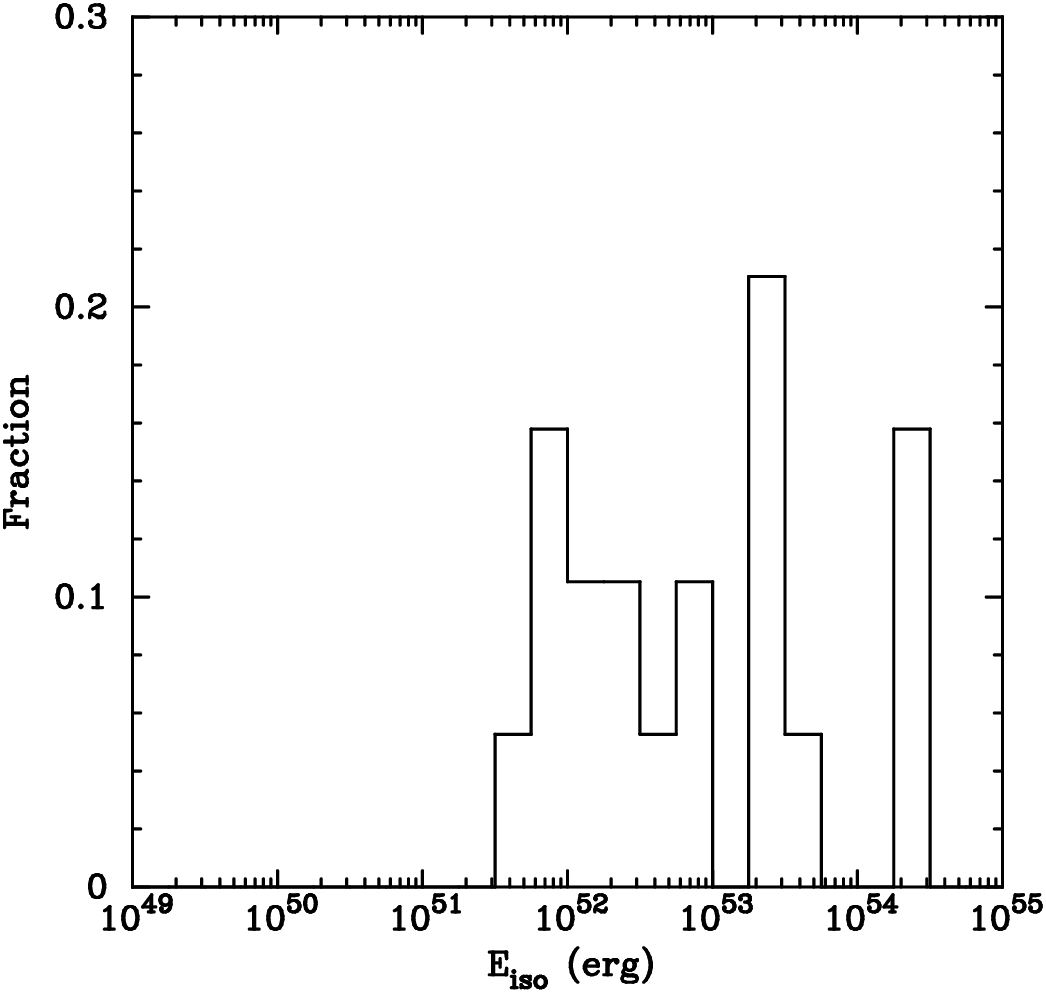}
\end{minipage}
\begin{minipage}[t]{3.2in}
\includegraphics[width=3.2in,clip=]{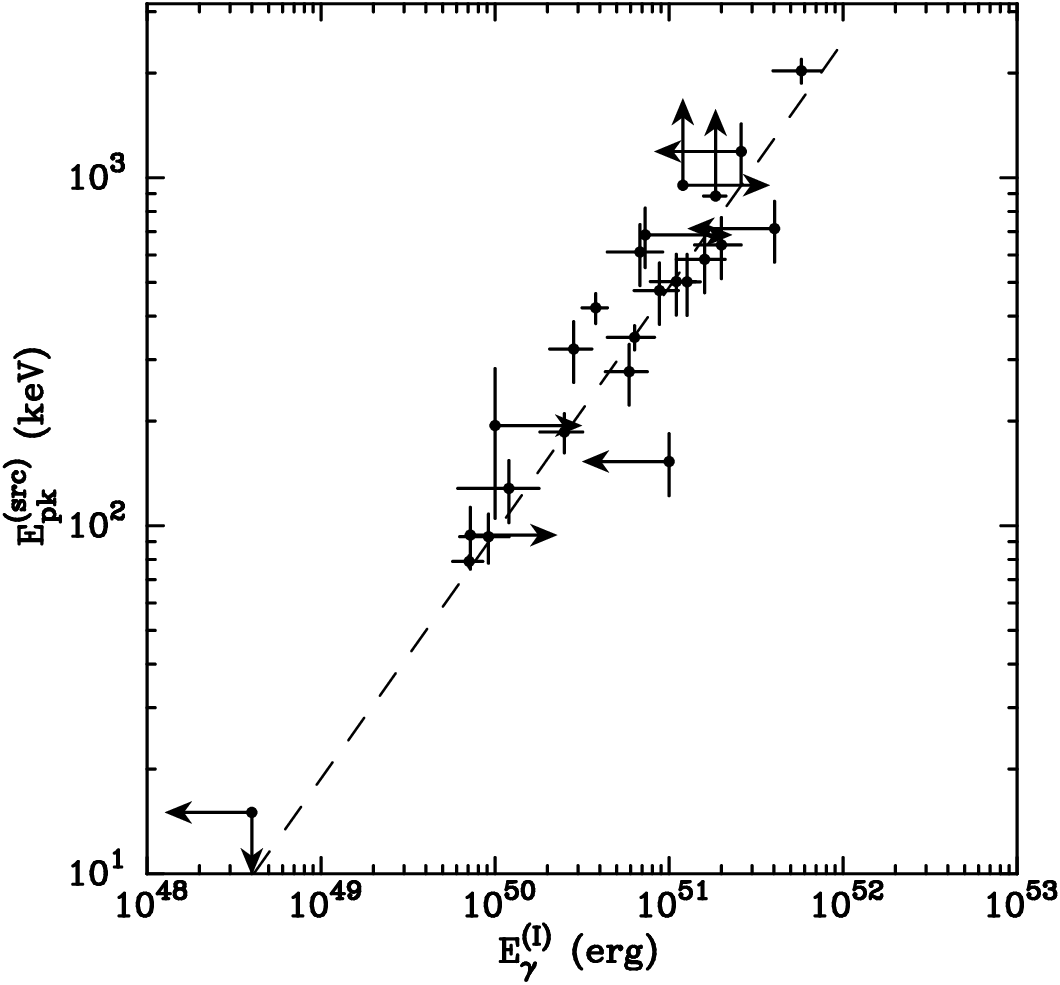}\\
\includegraphics[width=3.2in,clip=]{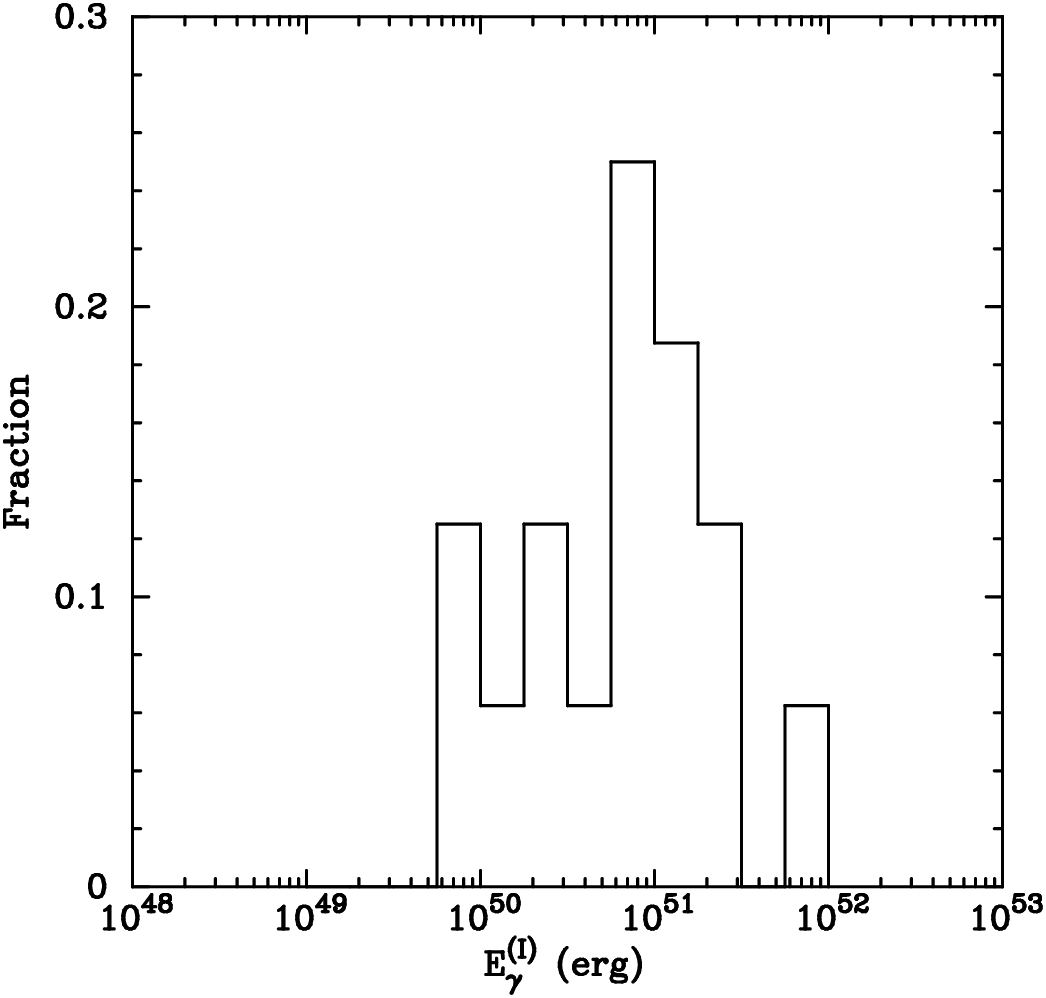}
\end{minipage}
\caption{Empirical properties of \eiso\ and \egi.  Top left:  Amati
relation.  Top right: Ghirlanda relation.  Bottom left: Histogram of
\eiso\ values, excluding XRFs.  Bottom right: Histogram of \egi\ values
for events with well-determined \egi.  The data used to make the plots in
the left-hand panels is from \citet{lamb03}.  The data used to make the
plots in the right-hand panels is from \citet{ggl2004}.}
\label{empirical_properties_figure}
\end{figure}

Certain constraints on universal jet shapes and distributions may be
inferred from observations. The empirical constraints on jet models that
we consider here are illustrated in
Fig.~\ref{empirical_properties_figure}.  They are:

\begin{itemize}

\item The distribution of GRB isotropic-equivalent energy \eiso\ 
appears to be quite broad, spanning at least four orders of magnitude, and
appears approximately uniform in $\log\eiso$, that is,
\begin{equation}
dN/d\log\eiso\approx A, 
\label{eiso_is_uniform}
\end{equation}
where $A$ is a constant \citep{amati2002,lamb03}.

\item The distribution of GRB total gamma-ray energy inferred using the
\citet{f01} procedure --- \egi\ --- appears to be quite narrow, with
most values clustered within a decade around a geometric mean whose
(model-dependent) value is somewhere in the range $5\times
10^{50}$--$3\times 10^{51}$~erg \citep{f01}.

\item Empirically, GRB isotropic energies \eiso\ and peak energies \epk\
of GRB $\nu F_\nu$ spectra appear to be satisfy the ``Amati Relation'', a
very tight correlation according to which $\epk\sim\eiso^{1/2}$
\citep{amati2002,lamb03}.

\item Empirically, GRB inferred total energies \egi\ and peak energies
\epk\ of GRB $\nu F_\nu$ spectra appear to be satisfy the ``Ghirlanda
Relation'', a very tight correlation according to which
$\epk\sim{\egi}^{0.7}$ \citep{ggl2004}.

\end{itemize}

In this paper we consider the extent to which universal jet models with
various underlying emission profiles can satisfy these observational
constraints on GRBs with afterglows and redshifts.  The plan of the paper
is as follows:

In \S\ref{distribution} we consider the effect of the choice of functional form
of the emission profile on the frequency distributions of \eiso\ and \egi. 
We consider power-law profiles, and introduce the ``Fisher'' (exponential)
profile, which is designed to satisfy Eq.~(\ref{eiso_is_uniform}) exactly.
We also inquire into the distributional properties of top-hat jet profiles
in universal jet models.

The treatment in \S\ref{distribution} uses ``bare'' emission, that is,
uncorrected by relativistic kinematics. In \S\ref{emission} we introduce a
general and efficient method of computing fully relativistic prompt
emission from jets with arbitrary (smooth) emission profiles.  We
introduce the notion of ``photon-number-weighted average Doppler shift'',
a quantity that serves as a proxy for observed \epk.  We also exhibit the
method for calculating how the frequency distributions of
\S\ref{distribution} are modified by relativistic kinematics.  Finally, we
give compact analytic expressions for relativistically-correct emission
and average Doppler factor in the special case of the top-hat jet profile.

In \S\ref{a_bunch_of_plots}, we use these ideas to explore the
characteristic angular emission profiles, \eiso-\epk\ (``Amati''), and
\egi-\epk\ (``Ghirlanda'') relations that would be observed from Universal
GRB models with Fisher, power-law, and top-hat emission profiles.  We also
show explicitly how the frequency distributions of \S\ref{distribution}
are modified by relativistic kinematics.

Our conclusions concerning the ability of universal jets based on the
various emission profiles to satisfy the observational constraints are
discussed in \S\ref{discussion}

%%%%%%%%%%%%%%%%%%%%%%%%%%%%%%%%%%%%%%%%%%%%%%%%%%%%%%%%%%%%%%%%%%%%%%%%%%
%%%                                                                    %%%
%%%  \S2:  Profiles and frequency distributions                        %%%
%%%                                                                    %%%
%%%%%%%%%%%%%%%%%%%%%%%%%%%%%%%%%%%%%%%%%%%%%%%%%%%%%%%%%%%%%%%%%%%%%%%%%%
\section{Universal Jet Profiles and Their Frequency Distributions in Energy
\label{distribution}}

For the sake of clarity of the discussion that follows, it is useful to
draw a terminological distinction between jet profiles and jet
models.

A jet {\it profile} is a distribution of emission across the surface of
the emitting shock.  Popular choices include the ``top hat'' jet, and
$\theta^{-\delta}$ power-laws of \citet{rossi2002} and \citet{zm02} (also
sometimes referred to as the ``structured jet''), where $\theta$ denotes
angular location on the jet.

A jet {\it model} of GRB emission includes a choice of profile, but also a
choice of distribution of profile parameters.  For example, the model of
\citet{zm02} selects a power-law with a fixed index, and a fixed total
energy.  The ``Variable Opening Angle'' model of \citet{lamb04}, on the
other hand, selects a top-hat profile and assumes a fixed total energy,
but a distribution of jet opening angles.

Models such as that of \citet{rossi2002} and \citet{zm02} that adopt a fixed
profile shape and total prompt energy --- and thus rely only on the
variation of observer viewing angle to produce all the distributions of
GRB energetics --- are properly called ``universal'' models.  We will
refer to models that allow for variations in profile shape as
``variable-geometry'' models.  Note that neither class of models
necessarily requires that its underlying profile  should belong to a
particular family --- one can in principle ``mix-and-match'' models and
profiles quite freely.

It is apparent that universal-jet models of GRBs are simpler than
variable-geometry models, in the sense that they require fewer parameters.
Clearly, an important question that must be addressed in this connection
is: do universal jet models have enough freedom to satisfy known
observational constraints on GRBs?  This question is the main concern of
this paper.

As a notational preliminary, let $\xiso\equiv\ln\eiso$, and
$\muv\equiv(1-\cos\theta_v)$, where $\theta_v$ is the angle between the
line-of-sight and the center of the jet.  We will also use
$\xgt\equiv\ln\egt$ and $\xgi\equiv\ln\egi$, where ``(T)'' refers to
``True'' (i.e. the actual total gamma-ray energy emitted), and ``(I)''
refers to ``Inferred'' --- the gamma-ray energy inferred using the
\citet{f01} procedure, in which the jet energy is corrected from its
``isotropic'' value using the solid angle subtended by the jet opening
angle inferred from the break time of the jet.  In general, $\egi\neq\egt$.

In this section, we assume symmetric ``2-lobed'' jets, so that $0<\muv<1$. 

%%%%%%%%%%%%%%%%%%%%%%%%%%%%%%%%%%%%%%%%%%%%%%%%%%%
%%%%%%%%%%%%%%%%%%%%%%%%%%%%%%%%%%%%%%%%%%%%%%%%%%%
\subsection{The Fisher Jet Profile}
\label{fisherdist}

Suppose we describe a Universal Jet by a function $f(\muv)$, that is
\begin{equation}
\xiso=f(\muv).
\end{equation}

How do we choose $f(\muv)$ so as to satisfy the constraint of
Eq.~(\ref{eiso_is_uniform})?

By construction, Universal jet models assume all properties of GRB
energetics are produced by variation of the viewing angle.  Naturally, this
variation is assumed uniform in $\mu_v$ for $0<\muv<1$.  In this range of
$\mu_v$, we therefore have
\begin{eqnarray}
\frac{dN}{d\xiso}&=&\frac{dN/d\muv}{d\xiso/d\muv}\nonumber\\
&=&B \left.[f^\prime(\muv)]^{-1}\right|_{f(\muv)=\xiso},
\end{eqnarray}
where $B$ is an uninteresting normalization constant.  This kind of
expression corresponds closely to expressions of \citet{guetta2005}
concerning universal jet models, the obvious difference being that here we
concern ourselves with modeling the \eiso\ distribution, and not the
observed fluence distribution.

In consequence of the limited allowed range of $\muv$, $dN/d\xiso$ can only
be constant in a limited range of $\xiso$.  Within that range, we have
$f^\prime(\muv)=\mbox{const.}$, so that

\begin{equation}
\xiso = f(\muv) = c_1\muv + c_0,
\end{equation}
and
\begin{equation}
\frac{dN}{d\xiso}=\frac{1}{c_1}\Theta(\xiso-c_0)\Theta(c_0+c_1-\xiso).
\label{unixisodist}
\end{equation}

The shape of the jet in linear energy space is
\begin{equation}
\eiso=\exp(c_1\muv + c_0),
\end{equation}
which peaks at $\muv=1$ for $c_1>0$.  The larger $c_1$, the sharper the
peak.  This functional form is called the ``Fisher Distribution'', and is
well known from the theory of statistical distributions on spheres
\citep[p.228]{m72}.  In the small-angle approximation, when
$\muv\approx(1-{\theta_v}^2/2)$, it approaches a symmetric 2-dimensional
Gaussian with standard deviation $\theta_0\equiv {c_1}^{-1/2}$, a form
considered by \citet{zm02}, \citet{zhang2004}, \citet{lr2004}, and
\citet{dz05}.  The Fisher distribution is more convenient for our
purposes, since its geometrically-natural form makes it useful for
analytical work for widths and viewing angles well outside the small-angle
approximation.

Note that the domain restriction in Eq. (\ref{unixisodist}) says that
\xiso\  varies over a domain whose breadth is $c_1=1/{\theta_0}^2$.  For a
jet with an opening angle of 5.7$^\circ$, $\theta_0=0.1$, and the
resulting range of \eiso\ is 100 e-foldings, corresponding to about 43
decades.  Naturally, instrument thresholds prevent us from probing such a
range of \eiso.

It is convenient to adjust the normalization of this expression so that it
reflects the true emitted energy, $\egt$.  Since
$\eiso=4\pi d\egt/d^2\Omega_v$, we have

\begin{eqnarray}
\egt&=&2\,\int_{0}^1 \frac{1}{2}d\muv\,e^{\muv/{\theta_0}^2+c_0}\nonumber\\
&=&e^{c_0}{\theta_0}^2\left(\exp({{\theta_0}^{-2}}) - 1\right),
\end{eqnarray}
where the factor of 2 in the first line accounts for the energy in both
lobes of the symmetric jet.

We thus have
\begin{equation}
\eiso=\frac{\egt}
{{\theta_0}^2\left(\exp({{\theta_0}^{-2}}) - 1\right)}\, 
e^{\muv/{\theta_0}^2},
\end{equation}
and
\begin{eqnarray}
\xiso&=&\xgt + {\theta_0}^{-2}\muv -
  \ln\left[{\theta_0}^2\left(\exp({{\theta_0}^{-2}}) - 1\right)\right]
\nonumber\\
&\approx&\xgt - {\theta_0}^{-2}(1-\muv) + \ln{\theta_0}^{-2},
\end{eqnarray}
where in the final line we have made the approximation ${\theta_0}^2\ll 1$.

%%%%%%%%%%%%%%%%%%%%%%%%%%%%%%%%%%%%%%%%%%%%%%%%%%%
%%%%%%%%%%%%%%%%%%%%%%%%%%%%%%%%%%%%%%%%%%%%%%%%%%%
\subsection{The \xgi\ Distribution of Universal Fisher Jets}

The quantity \eiso\ is directly related to a physical property of the GRB
--- the energy emitted per unit solid angle along the observer's
line-of-sight.  The quantity \egi, on the other hand, has a slightly
different status.  It is the result of a data-analysis procedure --- the
\citet{f01} procedure --- which multiplies \eiso\ by a solid angle
interior to a jet opening angle inferred from the light-curve break time.
This procedure is particularly adapted to top-hat profile jets, so the
physical meaning of \egi\ in the context of other profiles can be somewhat
contrived.  Nonetheless, the empirical distribution of \egi\ appears to
supply important constraints on jet models, as noted above, so it is
important to understand jet model predictions of these distributions.

In the case of top-hat jet profiles, the \citet{f01} procedure infers a
jet opening angle from the afterglow light-curve break time.  In the case
of a non-uniform universal jet, the break time does not correspond to a
jet edge.  Rather, it occurs when the bulk of the emission from the jet is
accessible to the observer, since after that time, the evolution of the
light curve is controlled by the evolution of $\gamma$ only, which is
expected to be a power-law in time.

As pointed out by \citet{kumar2003} and \citet{dz05}, the light-curve break
indicates one of two situations:  (1) the viewing angle $\theta_v$ is
larger than $\theta_0$, and the beaming angle $1/\gamma$ has widened to
the point that the {\it peak} of the jet has become visible, so that the
emission can no longer increase in brightness at the rate that it did when
the beaming cone was still climbing up the peak. Or (2) the viewing angle
$\theta_v$ is smaller than $\theta_0$, and the beaming angle $1/\gamma$
has widened until it is equal to $\theta_0$, so that most of the jet
emission is now visible, and further increase in $1/\gamma$ produces a
diminishing return in flux.  

Following \citet{kumar2003} and \citet{dz05}, we therefore assume that the
inferred break angle $\theta_{br}=\max(\theta_v,\theta_0)$, and that \egi\
is given by
\begin{equation}
\egi = \eiso \times (1 - \mubr),
\label{egfrail}
\end{equation}
where $\mubr\equiv\cos\theta_{br}$

\begin{figure}[t]
\includegraphics[width=5truein]{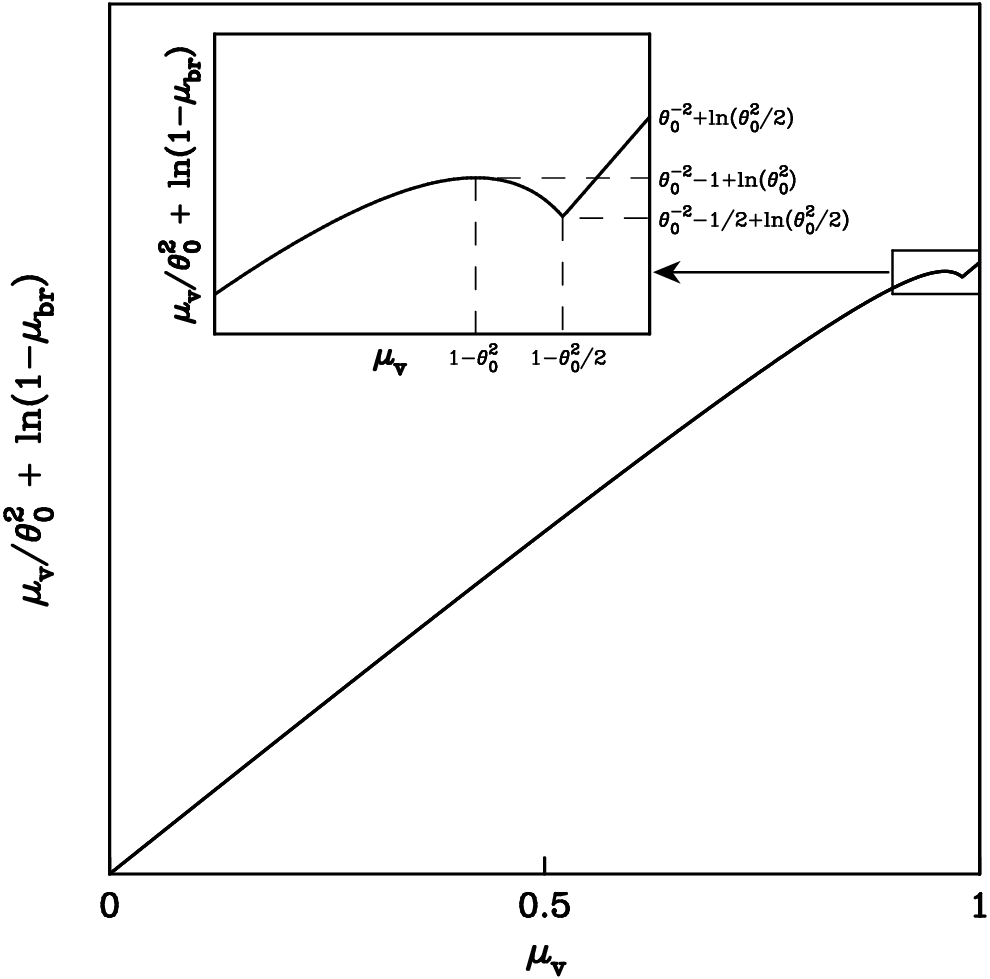}
\caption{Graphical analysis of the \egi\ distribution of Universal Fisher jets.}
\label{graphanal}
\end{figure}

We may now inquire as to the distribution $dN/d\xgi$ of Universal Fisher
jets.  \xgi\ is given by
\begin{equation}
\xgi=\xiso+\ln(1-\mubr).
\label{xgi}
\end{equation}
The shape of this function is shown in Fig.~\ref{graphanal}.  From the
figure, it is apparent that the range of \egi\ is about the same as the
range of \eiso, about $\exp({\theta_0}^{-2})$.  It also appears that the
correspondence from \xgi\ to \muv\ is single-valued in almost all of this
range, except for an small annulus of inner radius $\theta_0$ and outer
radius $\sqrt 2\theta_0$, where it is three-valued.

We may therefore write
\begin{eqnarray}
\frac{dN}{d\xgi}&=&\sum_{\mbox{\scriptsize Roots}}\frac{dN/d\muv}{d\xgi/d\muv}\nonumber\\
&=&B\times\sum_{\mbox{\scriptsize Roots}}\left.\left|{\theta_0}^{-2} - \frac{1}{1-\muv}\right|^{-1}
\right|_{\xgi=\xiso(\muv)+\ln(1-\mubr)},
\label{dndxgi}
\end{eqnarray}
where the sum is over the roots of Eq.~(\ref{xgi}), of which there is only
one outside the aforementioned annulus.

Assuming a typical $\theta_0\lesssim 0.1$, it is apparent that in most of
the range $0<\muv<1$, the first term in the summand of Eq.~(\ref{dndxgi})
dominates the second term, so that $dN/d\xgi$ is approximately also
uniform in most of its range.  It does spike (integrably) when
$1-\muv={\theta_0}^2$ (as $\theta_v$ crosses the inner radius of the
annulus), but this behavior is not enough to make the distribution a
narrow one.  Most of the probability, and therefore most of the events,
occur at a range of angles considerably larger than the core annulus,
which subtends a small solid angle.  For example, the 95\% point of the
$dN/d\xgi$ distribution occurs at a value of \xgi\  corresponding to
$\muv=0.05$. With ${\theta_0}^{-2}=100$, this value of \xgi\ is lower than
the value near the jet core by about 90 e-foldings, equivalent to about 39
decades.  Examples of the behavior of Eq.~(\ref{dndxgi}) are shown by the
``bare'' cuves in the four panels of Fig.~\ref{fisher_dndxgi}, for a range
of values of $\theta_0$.

We see that Eq.~(\ref{dndxgi}) represents an extremely broad distribution,
far broader than can actually be probed by real instruments -- obviously,
HETE, BeppoSAX, BATSE and SWIFT didn't and won't probe 40 decades of \egi
or \eiso. The key point, however, is that in the observationally
accessible region, the distribution is broad, not narrow, and is only cut
off by instrumental effects.

%%%%%%%%%%%%%%%%%%%%%%%%%%%%%%%%%%%%%%%%%%%%%%%%%%%
%%%%%%%%%%%%%%%%%%%%%%%%%%%%%%%%%%%%%%%%%%%%%%%%%%%
\subsection{The \xiso\ Distribution of ``Universal Structured'' Jets}

The universal ``structured'' jet model of \citet{rossi2002} and
\citet{zm02} does not attempt to satisfy constraint (1).  Instead, it
attempts to preserve the appearance of the \citet{f01} result --- that is,
to satisfy constraint
\begin{equation}
dN/d\xgi = N_0\times \delta(\xgi -\xbgi).
\label{egi_is_narrow}
\end{equation}
by imposing the following shape:
\begin{equation}
\eiso=K/(1-\muv),
\end{equation}
or, equivalently,
\begin{equation}
\xiso=Q - \ln(1-\muv),
\end{equation}
where $K$ and $Q$ are constants, and where we have replaced the small-angle
form with the general, spherically-correct form.

Since this form does not attempt to constrain $dN/d\xiso$, it is of
interest to calculate what this distribution might be.  This is easily
done:

\begin{eqnarray}
\frac{dN}{d\xiso}&=&\frac{dN/d\muv}{d\xiso/d\muv}\times\Theta(\xiso-Q)
\nonumber\\
&=&B\times(1-\muv)\times\Theta(\xiso-Q)\nonumber\\
&=&(BK/\eiso)\times\Theta(\xiso-Q)\nonumber\\
&=&BKe^{-\xiso}\times\Theta(\xiso-Q).
\label{zmxisodist}
\end{eqnarray}

In other words, the distribution is exponential in \xiso, with a width
scale of one e-folding (0.43 decades) and a cutoff at some low value of
\xiso.  Needless to say, the empirical distribution
(Fig.~\ref{empirical_properties_figure}) looks nothing like this.

%%%%%%%%%%%%%%%%%%%%%%%%%%%%%%%%%%%%%%%%%%%%%%%%%%%
%%%%%%%%%%%%%%%%%%%%%%%%%%%%%%%%%%%%%%%%%%%%%%%%%%%
\subsection{Jet Distributions, Broad and Narrow}
\label{broad_vs_narrow}

It appears from the foregoing that universal jet models have a serious
problem:  if you try to make a jet that has a broad \eiso\ distribution, you
find that the jet also has a broad \egi\ distribution, while forcing a
narrow \egi\ distribution results in a narrow \eiso\ distribution.

To illustrate the point, suppose we let the functional relation between
\eiso\ and 1-\muv\ be a general power-law, so that
\begin{equation}
\xiso=Q - \delta\ln(1-\muv).
\label{pljet}
\end{equation}
The universal ``structured'' jet model of \citet{rossi2002} and
\citet{zm02} corresponds to $\delta=1$.  Those authors also considered
general power-law jet profiles as well, although those profiles were
power-laws in $\theta_v$, rather than in the solid angle interior to
$\theta_v$ as here.

It then follows, by the same argument that led to Eq.~(\ref{zmxisodist}),
that the energy distributions are:
\begin{equation}
\frac{dN}{d\xiso}=C\times \delta^{-1}e^{-\xiso/\delta}\times\Theta(\xiso-Q),
\label{plxiso}
\end{equation}
and
\begin{equation}
\frac{dN}{d\xgi}=C\times (\delta-1)^{-1}e^{-\xgi/(\delta-1)}
\times\Theta(\xgi-Q).
\label{plxgi}
\end{equation}

Note that for the \citet{zm02} universal jet profile, $\delta\rightarrow
1$, and the right-hand side of Eq.~(\ref{plxgi}) tends to a
$\delta$-function.

It appears from Eqs.~(\ref{plxiso}) and (\ref{plxgi}) that the family of
models satisfying Eq.~(\ref{pljet}) is capable of producing energy
distributions of any breadth, simply by varying the value of $\delta$. The
larger the value of $\delta$, the broader the distribution.  This property
is in contrast with the properties of Fisher jets, which (by design) can
only represent broad \eiso\ distributions.  We may therefore use the
power-law jet family of models to explore the connection between the
breadth of the \xiso\ distribution and that of the \xgi\ distribution.

From Eqs.~(\ref{plxiso}) and (\ref{plxgi}) we may observe that within
this family of models, the widths of the two distributions are closely
linked: if we try to push out the width of the \xiso\ distribution by
increasing $\delta$, we will also automatically increase the width of the
\xgi\ distribution to very nearly the same width.  If, on the other hand,
we attempt to restrict the range of \xgi\ by making $\delta$ close to 1, we
thereby also restrict the range of \xiso.

It appears, then, that Universal jet models {\it force us to choose
between two observational constraints on GRBs}.  Within the context of a
universal jet model, it is not possible to satisfy both  the requirement
that $dN/d\xiso$ should be broad and the requirement that  $dN/d\xgi$
should be narrow.

%%%%%%%%%%%%%%%%%%%%%%%%%%%%%%%%%%%%%%%%%%%%%%%%%%%%%%%%%%%%%%%%%%%%%%%%%%
%%%                                                                    %%%
%%% \S3: Emission from relativistic jets                               %%%
%%%                                                                    %%%
%%%%%%%%%%%%%%%%%%%%%%%%%%%%%%%%%%%%%%%%%%%%%%%%%%%%%%%%%%%%%%%%%%%%%%%%%%
\section{Computing Relativistic Emission From Jet Profiles}
\label{emission}

In this section we present formulas and methods of computing various
observational quantities given certain jet angular emission profiles. 
Naturally, underlying emission profiles are not observed directly, but
rather they are modulated by relativistic effects due to their
relativistic bulk motion.

In this work, we are concerned with jet emission profiles that are due to
variations in emissivity across the jet, rather than due to variations in
the Lorentz factor $\gamma$.  We consider that during the prompt emission
phase of the GRB, $\gamma$ is constant across the jet and has no
appreciable time variation \citep{rees1994}.

We consider prompt energies and observed fluences, rather than prompt peak
luminosities and observed peak fluxes.  Given the assumed invariance of
$\gamma$, this approach allows us to ignore time variations in the
emission, and to represent prompt emission properties of GRBs simply in
terms of weighted integrals of jet angular profiles.

We also assume that the observer's detection process is approximately
bolometric --- that is, we ignore the spectral effects of relativistic
kinematics, so we do not concern ourselves with the kinematic
transformations of detector bandpasses, or with the integration of emission
spectra over those bandpasses.

Note that in this section, unlike in \S\ref{distribution}, we assume
``single-lobed'' jets, for the sake of simplicity of our expressions. 
A symmetric ``two-lobed'' jet may be constructed from oppositely directed
single-lobed jets, ascribing half of the total energy to each lobe.

%%%%%%%%%%%%%%%%%%%%%%%%%%%%%%%%%%%%%%%%%%%%%%%%%%%
%%%%%%%%%%%%%%%%%%%%%%%%%%%%%%%%%%%%%%%%%%%%%%%%%%%
\subsection{Observables\label{sec_obs}}

\subsubsection{\eiso\ and \egi}

It is obviously of interest to calculate the observed fluence as a
function of viewing angle, from which \eiso\ follows directly. 

In what follows, we refer to the inertial frame at rest with respect to an
element of the jet as the ``rest frame'' (of that element), and to the
inertial frame at rest with respect to the central engine as the ``source
frame.''

Suppose a jet with bulk Lorentz factor $\gamma$ radiates a source-frame
energy \egt.  Let each differential element of the jet $d^2\vec{n}$ be
labeled by the unit vector $\vec{n}$, so that $\vec{n}$ furnishes
coordinates on the jet surface.  For concreteness, we will assume
$\vec{n}$ is a source-frame (as opposed to rest-frame) vector.  Define a
source-frame unit vector $\vec{l}$ that points in the direction of the
line-of-sight to the observer.

The element of the jet at $\vec{n}$ radiates isotropically in its own rest
frame an amount of rest-frame energy per unit rest-frame solid angle
\begin{equation}
\frac{dE^{(R)}}{d^2\vec{l}^{(R)}}\,d^2\vec{n} =
\frac{\gamma^{-1}\egt}{4\pi}
\times\epsilon(\vec{n}\cdot\vec{b})\,d^2\vec{n},
\label{dedomega}
\end{equation}
Here $\vec{l}^{(R)}$ is a unit direction vector pointing in the direction
of the line-of-sight to the observer in the the rest frame of the jet
element at $\vec{n}$. The profile function
$\epsilon(\vec{n}\cdot\vec{b})$, which is azimuthally symmetric about the
direction of the jet-axis unit vector $\vec{b}$, is normalized so that
\begin{equation}
\oint\epsilon(\vec{n}\cdot\vec{b})\,d^2\vec{n}=
2\pi\int_{-1}^1dx\,\epsilon(x)=1.
\end{equation}

The usual relativistic Doppler shift may be applied to yield the
source-frame energy per unit source-frame solid angle:
\begin{eqnarray} \frac{dE}{d^2\vec{l}} &=&\oint
\frac{dE^{(R)}}{d^2\vec{l}^{(R)}}\times
\gamma^{-3}(1-\beta\vec{n}\cdot\vec{l})^{-3}\,d^2\vec{n} \nonumber\\
&=&
\frac{\egt}{4\pi}\oint \epsilon(\vec{n}\cdot\vec{b})\times
\gamma^{-4}(1-\beta\vec{n}\cdot\vec{l})^{-3}\,d^2\vec{n}.
\label{conv1}
\end{eqnarray}
Since this is an energy, rather than a luminosity, it should be converted
to an observed fluence by means of a multiplication by $(1+z)d_L(z)^{-2}$,
where $d_L(z)$ is the usual luminosity distance.

The isotropic-equivalent energy is then given by
\begin{eqnarray}
\eiso&=&4\pi\times\frac{dE}{d^2\vec{l}}\nonumber\\
&=&\egt\oint\epsilon(\vec{n}\cdot\vec{b})\times
\gamma^{-4}(1-\beta\vec{n}\cdot\vec{l})^{-3}\,d^2\vec{n}.
\label{eiso}
\end{eqnarray}

We will also be interested in the inferred total energy, \egi, which is
given by
\begin{eqnarray}
\egi&=&2\pi(1-\cos\theta_{br})\times\frac{dE}{d^2\vec{l}}\nonumber\\
&=&\frac{1}{2}(1-\cos\theta_{br})\times\eiso,
\label{eginferred}
\end{eqnarray}
where $\theta_{br}=\max(\theta_v,\theta_0)$, and $\theta_v$ is the viewing
angle --- the angular distance between the observer and the axis of the
jet.  Again, this assumes all the inferred energy is ascribed to a single
jet lobe.

\subsubsection{Peak Energy}
\label{peake}

Another interesting calculation is the effective \epk\ observed along the
viewing direction, assuming a common rest-frame \epk\ for the entire jet.
For universal jet models, this quantity can be placed in relation to
\eiso\ and \egi, in order to examine whether the empirical \epk-\eiso\
relation of \citet{amati2002}, or the empirical \epk-\egi\ relation of
\citet{ggl2004} can be reproduced.

Technically, \epk\ is the peak of the $\nu F_\nu$ spectrum obtained by
superposing the appropriately Doppler-shifted and suitably
Doppler-weighted spectra received from various parts of the jet. 
Performing this superposition explicitly presents difficulties, such as
choosing a detailed shape of the underlying emission spectrum in such a
way as to yield a Band (GRB) function \citep{band1993} spectrum for the
superposition.  This is the approach chosen by
\citet{yamazaki2002,yamazaki2003,yamazaki2004}.

Our approach is to calculate the ``photon-number averaged \epk.'' We
ascribe a common rest-frame \epkr\ to the entire jet, and calculate the
average Doppler-shifted \epk\ in the viewing direction, where the average
is weighted by photon number fluence.  This is tantamount to computing the
average of the Doppler shift
$D\equiv\gamma^{-1}(1-\beta\vec{n}\cdot\vec{l})^{-1}$, according to the
weighting function
$\epsilon(\vec{n}\cdot\vec{b})\times\gamma^{-2}(1-\beta\vec{n}\cdot\vec{l})^{-2}$
and applying it to the rest-frame \epk.

In other words, we want to be able to calculate
\begin{equation}
\epk=\davg\times\epkr,
\label{epk}
\end{equation}
where \epkr\ is the rest-frame peak energy, and
\begin{equation}
\davg=\frac
{\oint d^2\vec{n}\,\epsilon(\vec{n}\cdot\vec{b})\times
       \gamma^{-3}(1-\beta\vec{n}\cdot\vec{l})^{-3}}
{\oint d^2\vec{n}\,\epsilon(\vec{n}\cdot\vec{b})\times
       \gamma^{-2}(1-\beta\vec{n}\cdot\vec{l})^{-2}}.
\label{dbar}
\end{equation}
The accuracy of Eq.~(\ref{epk}) as an estimator of the peak of the
effective spectrum depends on the nature of the underlying rest-frame
spectrum.  In general, there will certainly be systematic uncertainties
and biases in \davg\epkr\ as an estimator of observed peak energy.
Nonetheless, it seems reasonable to expect that in its functional
dependence on $\theta_v$, in its correlations with \eiso\ and \egi, and in
its frequency distribution in universal jet models, the behavior of
\davg\epkr\  should not be too dissimilar from the behavior of the peak
energy of the true composite spectrum.  Since these are the properties
that concern us, we adopt \davg\ as a proxy for \epk\ in what follows.

One useful special case of \davg\ is $\epsilon(\vec{n}\cdot\vec{b})=1$ ---
isotropic outflow.  In this case, it is straightforward to show that
$\davg=\gamma$.  The reason this is useful is that this is also the limit
to which \davg\ should tend when the profile $\epsilon$ varies weakly on
angular scales comparable to $\gamma^{-1}$ on the part of the jet that is
moving parallel to the line-of-sight.  It thus provides a valuable sanity
check in practical computations.

\subsubsection{Relativistic Corrections to Frequency
Distributions of \eiso\ and \egi.\label{sec_edist}}

It is interesting and important to understand the extent to which our
\S\ref{distribution} results on \eiso\ and \egi\ distributions of universal jet
models are affected by relativistic corrections.

What is required is the form of $dN/d\eiso$ and $dN/d\egi$, where instead
of being given directly in terms of the emission profile as we assumed
above, $\eiso$ and $\egi$ are given by Eqs.~(\ref{eiso}) and
(\ref{eginferred}).

Since in universal jet models the distribution of \eiso\ is entirely due to
a uniform distribution in $\mu_v\equiv\cos\theta_v$, we may write
\begin{equation}
\frac{dN}{d\eiso}=\frac{dN/d\mu_v}{|d\eiso/d\mu_v|}=
\frac{\mbox{const.}}{|d\eiso/d\mu_v|}.
\label{dndeiso}
\end{equation}

We therefore need to calculate $d\eiso/d\mu_v$ starting from
Eq.~(\ref{eiso}).  This is accomplished by differentiating the Doppler
factor in the integrand with respect to $\mu_v$.  By adopting
spherical-polar coordinates $(\theta_v, \phi_v)$ for $\vec{l}$ with the
$z$-axis aligned with $\vec{b}$, and differentiating the quantity
$\vec{n}\cdot\vec{l}$ with respect to $\mu_v$ while holding
the $\phi_v$ and $\vec{n}$ fixed, it may be shown that
\begin{equation}
\frac{d(\vec{n}\cdot\vec{l}\,)}{d\mu_v}=
\frac{\vec{n}\cdot(\vec{b}-\mu_v\vec{l})}{1-{\mu_v}^2},
\label{dndotldmu}
\end{equation}

It follows then that if we differentiate Eq.~(\ref{eiso}) with respect to
$\mu_v$ we obtain
\begin{equation}
\frac{d\eiso}{d\mu_v}=\frac{3\beta\egt}{\gamma^4(1-{\mu_v}^2)}
\oint d^2\vec{n}\,
\frac{\epsilon(\vec{n}\cdot\vec{b})}
     {(1-\beta\vec{n}\cdot\vec{l})^4}
(\vec{n}\cdot\vec{b}-\mu_v\vec{n}\cdot\vec{l}).
\label{deisodmuv}
\end{equation}

We also require
\begin{equation}
\frac{dN}{d\egi}=\frac{dN/d\mu_v}{|d\egi/d\mu_v|}=
\frac{\mbox{const.}}{|d\egi/d\mu_v|}.
\label{dndegi}
\end{equation}

In Eq.~(\ref{dndegi}) the factor $d\egi/d\mu_v$ may be related to
$d\eiso/d\mu_v$ using Eq.~(\ref{eginferred}):
\begin{equation}
\frac{d\egi}{d\mu_v}=\frac{1}{2}(1-\mu_v)\frac{d\eiso}{d\mu_v}-\frac{1}{2}\eiso.
\label{dgidmuv}
\end{equation}

A special limiting case is the ``needle'' jet limit,  where a narrow jet
is observed off-axis \citep{yamazaki2002,yamazaki2003,yamazaki2004}.  In
this case, $\epsilon(\vec{n}\cdot\vec{b})$ is so narrow that it acts as a
$\delta$-function. By Eq.~(\ref{eiso}), the radiation pattern is then
proportional to $(1-\beta\mu_v)^{-3}$.  It is straightforward to show that
in this limit, we have
\begin{equation}
\lim \frac{dN}{d\log\eiso} = {\eiso}^{-1/3}.
\label{eiso_limit}
\end{equation}
Furthermore, using Eq.~(\ref{eginferred}), and using $\beta\approx 1$, we
can also show that
\begin{equation}
\lim \frac{dN}{d\log\egi} = {\egi}^{-1/2}.
\label{egi_limit}
\end{equation}
These limits furnish useful sanity checks, and are interesting in their
own right in the context of the off-axis beaming model
\citep{yamazaki2002,yamazaki2003,yamazaki2004}.

%%%%%%%%%%%%%%%%%%%%%%%%%%%%%%%%%%%%%%%%%%%%%%%%%%%
%%%%%%%%%%%%%%%%%%%%%%%%%%%%%%%%%%%%%%%%%%%%%%%%%%%
\subsection{Spherical Convolution}

We present in this section an efficient and general technique for
computing the energy emission distributions of GRB jets with arbitrary
(smooth) emission profiles.

\subsubsection{Method}

The quantities of interest described in the previous section are all
expressed in terms of convolutions on the 2-sphere of two functions that
are azimuthally symmetric but not mutually concentric.  That is, they are
of the form
\begin{equation}
\oint d^2\vec{n}\,f^{(1)}(\vec{n}\cdot\vec{q\,}^{(1)})\times
                  f^{(2)}(\vec{n}\cdot\vec{q\,}^{(2)}).
\label{conv0}
\end{equation}

Now suppose we represent each of the two functions $f^{(1)}$ and $f^{(2)}$
on the two-sphere by their expansion in spherical harmonics.  Since they
are azimuthally symmetric, only the $m=0$ spherical harmonics appear, that
is
\begin{equation}
f^{(i)}(\vec{n}\cdot\vec{q\,}^{(i)})
=\sum_{l=0}^\infty f^{(i)}_l \frac{2l+1}{2}
P_l(\vec{n}\cdot\vec{q\,}^{(i)})
\label{expand_f}\\
\end{equation}
where the $P_l(x)$ are the usual Legendre polynomials, normalized so that
$\int_{-1}^1 P_l(x)^2\,dx = 2/(2l+1)$.  By the orthogonality of the
$P_l(x)$, it follows immediately that
\begin{equation}
f^{(i)}_l=\int_{-1}^1 dx\,P_l(x)\,\times f^{(i)}(x).
\label{coef_f}
\end{equation} 

If we let the $z$-direction be along $\vec{q\,}^{(1)}$ and define
spherical polar coordinates ($\theta$, $\phi$) for $\vec{n}$, so that
$\vec{n}\cdot\vec{q\,}^{(1)}=\cos\theta$, then in Eq.~(\ref{expand_f}) we
may replace $P_l(\vec{n}\cdot\vec{q\,}^{(1)})$ by
$\sqrt{4\pi/(2l+1)}Y_l^0(\theta, \phi)$ in the expression for $f^{(1)}$.

If we introduce spherical polar coordinates ($\theta_2$, $\phi_2$) for the
direction vector $\vec{q\,}^{(2)}$ (so that
$\cos\theta_2=\vec{q\,}^{(1)}\cdot\vec{q\,}^{(2)})$, we may also expand
$P_l(\vec{n}\cdot\vec{q\,}^{(2)})$ in the expression for $f^{(2)}$ in
Eq.~(\ref{expand_f}), using the addition theorem of spherical harmonics: %
\begin{equation}
P_l(\vec{n}\cdot\vec{q\,}^{(2)})=\frac{4\pi}{2l+1}\sum_{m=-l}^{l}
Y_l^m(\theta_2,\phi_2)\,Y_l^m(\theta,\phi)^*.
\label{add_thm}
\end{equation}

Substituting Eqs.~(\ref{expand_f}) and
(\ref{add_thm}) into Eq.~(\ref{conv0}), and using the orthogonality of
spherical harmonics, we obtain
\begin{equation}
\oint d^2\vec{n}\,f^{(1)}(\vec{n}\cdot\vec{q\,}^{(1)})\times
                  f^{(2)}(\vec{n}\cdot\vec{q\,}^{(2)}) =
\sum_{l=0}^\infty (2l+1)\,\pi\,P_l(\vec{q\,}^{(1)}\cdot\vec{q\,}^{(2)})
           \,f^{(1)}_l\,f^{(2)}_l.
\label{conv2}
\end{equation}

Now, if either $f^{(1)}$ or $f^{(2)}$ is approximately band-limited to
Legendre polynomials with $l<l_0$, then the sum in Eq.~(\ref{conv2}) is
over a finite number of terms.  We shall assume the approximate
band-limitedness of the jet profile $\epsilon(\vec{n}\cdot\vec{b})$ in
what follows.  For jets characterized by an opening half-angle $\theta_0$,
the band limit $l_0$ will be in excess of $\pi/\theta_0$.

The usefulness of Eq.~(\ref{conv2}) hinges upon our ability to calculate
the {\it Legendre transform} integrals of Eq.~(\ref{coef_f}), and to do so
accurately and efficiently.  We now consider how this is to be
accomplished.

It is in principle possible to use the well-known recursion relations for
the Legendre polynomials to obtain expressions for the higher-$l$
transforms in terms of lower-$l$ transforms, at least for the transforms
of the Doppler factor and for the Fisher jet profile.  Such a recursion
scheme is unquestionably efficient.  Unfortunately, it turns out to be
inaccurate --- the resulting recursion relations are unstable, and for
physically interesting values of $l_0$ the transforms at the largest
values of $l$ can be dominated by amplified numerical noise.

The ``Fast Legendre Transform'' (FLT) algorithm of \citet{dh94} is
well-suited to our purposes.  It is the spherical analogue of the
celebrated FFT of \citet{ct65}, and has already become somewhat well-known
in astrophysical circles, among analysts of CMB data \citep{osh98,wg01}

\citet{dh94} showed that for band-limited functions $f^{(i)}(x)$ with band
limit $l_0$, the Legendre transform of Eq.~(\ref{coef_f}) is exactly
given by the quadrature formula
\begin{equation}
f^{(i)}_l=\sum_{j=0}^{2l_0-1}a_j^{(l_0)}\,f^{(i)}(\cos\theta_j)\,
                             P_l(\cos\theta_j),
\label{quadrature}
\end{equation}
where
\begin{eqnarray}
\theta_j&\equiv&\frac{\pi j}{2l_0}\label{q_absc}\\
a_j^{(l_0)}&\equiv&\frac{2}{l_0}\sin\left(\frac{\pi j}{2l_0}\right)
\sum_{l=0}^{l_0-1}(2l+1)^{-1}\sin\left[\frac{(2l+1)\pi j}{2l_0}\right].
\label{q_weights}
\end{eqnarray}
In Eq.~(\ref{q_weights}) we have corrected the normalization in the
expression for the weights of \citet{dh94}, which was missing a factor
$\sqrt 2$.

Eq.~(\ref{quadrature}) is the {\it Discrete Legendre Transform} (DLT),
which is entirely analogous to the Discrete Fourier Transform.  The
band-limit $l_0$ plays a role analogous to that the Nyquist frequency --
any power in the Legendre spectrum of $f^{(i)}(x)$ above $l_0$ is aliased
by the DLT down to harmonics below $l_0$.  For this reason, computation by
means of spherical expansion is suitable for smooth jet profiles, such as
the Fisher profile or the power-law profile.  Profiles with sharp edges,
such as the top-hat jet profile, can be expected to have substantial power
at very high harmonics, and are therefore less well-suited to computation
by this method.  We will exhibit compact analytic emission formulas for
the top-hat jet profile in \S\ref{sec_uniform} below.

Naive evaluation of the DLT evidently results in an expensive algorithm,
one that is $\mathcal{O}({l_0}^2)$.  Physically interesting GRB jets are
believed to be characterized by beaming angles
$\gamma^{-1}\sim 10^{-2}$---$10^{-3}$~radians, and by jet opening angles
$\gtrsim 10^{-2}$~radians, implying values of $l_0$ in the several hundreds,
or even above 1000.  Particularly if many integrals are to be evaluated (as
in a population-synthesis simulation, for example), the naive algorithm
exacts a prohibitive cost.

Fortunately, \citet{dh94} also exhibited the FLT:  a fast ---
$\mathcal{O}(l_0\log(l_0)^2)$ --- algorithm modeled after the ``Divide and
Conquer'' strategy of the FFT \citep{ct65}.  We have implemented this
algorithm as a C library, and find it quite satisfactory for our present
purposes.  All computations of emission from Fisher and power-law profile
jets exhibited in the next section were performed using this method.

\subsubsection{Savings}

In \S\ref{sec_obs} we saw that the spherical convolutions in which we are
interested combine $\epsilon(\vec{n}\cdot\vec{b})$ or
$(\vec{n}\cdot\vec{b})\epsilon(\vec{n}\cdot\vec{b})$ with
$(1-\beta\vec{n}\cdot\vec{l})^{-n}$ ($n=2,3,4$), or with
$(\vec{n}\cdot\vec{l})(1-\beta\vec{n}\cdot\vec{l})^{-4}$. A DLT (that is,
a list of $l_0$ numbers) is required for each of the above convolvees.  If
each such DLT were computed as a separate FLT, the computational burden
would rise to near-irritating levels.

Fortunately, it is not necessary to compute an FLT for each of these DLTs.
The well-known recursion relation
$(2l+1)xP_l(x)=(l+1)P_{l+1}(x)+lP_{l-1}(x)$ among the Legendre
Polynomials, together with the relation
$(1-\beta x)^{-n+1}=(1-\beta x)\times(1-\beta x)^{-n}$ allow us to relate
the DLT of $(1-\beta x)^{-n}$ ($n=2,3$), and the DLT of
$x(1-\beta x)^{-4}$ to the DLT of $(1-\beta x)^{-4}$, and also to relate
the DLT of
$x\times\epsilon(x)$ to the DLT of $\epsilon(x)$.

It is thus only necessary, for the purposes of the present work, to
calculate the DLTs of $\epsilon(x)$ and of $(1-\beta x)^{-4}$.  The other
DLTs are then computable essentially for free.  If it were not of interest
to calculate the distributions of \S\ref{sec_edist}, it would only be
necessary to compute the DLTs of $\epsilon(x)$ and of $(1-\beta x)^{-3}$,
in order to compute \eiso\ and \epk.

Note also that since the two required DLTs are independent of the viewing
angle $\theta_v$, the processing expense of calculating those DLTs is
incurred once, after which the stored DLTs may be reused for many values
of the viewing angle $\theta_v$.

\subsubsection{Numerical Limitations}

Eq.~(\ref{conv2}) has the form of an inner product with a weighting
function proportional to $P_l(\vec{q\,}^{(1)}\cdot\vec{q\,}^{(2)})$.  In
the cases we consider here, the argument of the Legendre polynomial is the
cosine of the viewing angle, $\cos\theta_v$.  When $\theta_v$ is close to
zero, the Legendre polynomial is nearly 1, and the terms in the inner
product add constructively.  This behavior is necessary so that maximum
emission is obtained when the observing direction is aligned with the jet
axis.

As the observing direction moves away from the jet axis, terms in the
inner product must begin to engage in some mutual cancellation, in order
to produce the expected decrease in fluence.  If the underlying profile
spans many orders of magnitude (as the Fisher profile may easily do), then
it may come to pass that the cancellations required to produce the
necessary suppression of fluence at large $\theta_v$ cannot occur in
practical computations, as they are drowned out by numerical noise due to
the finite machine precision.

The most commonly encountered form of floating-point double precision is
constructed from 8-byte words, which afford a relative spacing between 
adjacent representable floating point numbers of about $10^{-14}$.  This
is the smallest relative magnitude that can be attained, using such
arithmetic, in calculations that depend on delicate cancellation.

If we demand a minimum accuracy of 1\%, we may therefore expect that the
computable dynamic range of \eiso\ (say) is no more than about 12 decades,
using 8-byte reals.  This expectation is in fact borne out by experience. 
Of course this is plenty of dynamic range for the present purposes. 
Should greater dynamic range be nonetheless required, some additional
decades may be secured by adopting extended precision (typically 16-byte
floating point words), and possibly computing with machine architectures
endowed with a larger natural word size.

\subsubsection{``Smooth'' Power-Law Profiles}

Power-law profiles diverge at the jet axis.  It is customary to deal with
this deplorable behavior by truncating the power-law at some small angle
\citep{zm02}.  This practice represents a problem for our
spherical-expansion approach, since the sharp edge at the cutoff injects
unwanted spectral power into the DLT of $\epsilon_{\mbox{\scriptsize
PL}}$ at high values of $l$. 

Consequently, the normalized power-law profile that we actually use in
this work is
\begin{equation}
\epsilon_{\mbox{\scriptsize PL}}(\cos\theta)=
\kappa(1-\cos\theta+{\theta_0}^2/2)^{-\delta},
\label{pl_profile_norm}
\end{equation}
where we have introduced a smooth cut-off $\theta_0$, rather than truncate
the power-law.  The resulting profile has a smooth, cusp-free peak in
$\theta$ at $\theta=0$.

The (single-lobe) normalization factor $\kappa$ is given by
\begin{equation}
\kappa=\left\{
\begin{array}{c@{\quad:\quad}l}
\frac{1-\delta}{2\pi\left[(2+{\theta_0}^2/2)^{1-\delta}-
                            ({\theta_0}^2/2)^{1-\delta}\right]}
&\delta\neq 1\\
\frac{1}{2\pi\log(4{\theta_0}^{-2}+1)}
&\delta=1
\end{array}
\right.
\end{equation}

The effective angular width of this profile is a function of both
$\theta_0$ and $\delta$.  A simple estimator of this width may be
constructed by making a Gaussian approximation to the jet near the peak. 
The width of this Gaussian is then approximately
\begin{eqnarray}
\theta_w&\equiv&\left|
\frac{d^2\log\epsilon_{\mbox{\scriptsize PL}}}{d\theta^2}
\right|^{-1/2}\nonumber\\
&=&\theta_0/\sqrt{2\delta}.
\label{pl_width}
\end{eqnarray}
Thus for a reasonable range of $\delta$ near $\delta=1$, $\theta_0$ is
itself not a bad estimator of the jet width.  For large $\delta$, however,
it is advisable to regard $\theta_w$, rather than $\theta_0$, as the
effective angular width of the profile.

In particular, when we calculate $\egi=\eiso\times(1-\cos\theta_{br})$ for
power-law profiles, we always use a break angle
$\theta_{br}=\max(\theta_v,\theta_w)$, rather than the ``naive''
prescription $\theta=\max(\theta_v, \theta_0)$.

\subsubsection{Emission From Afterglow Jets}

The main focus of the current work is prompt GRB emission, so we are not
really concerned with afterglow emission.  Nonetheless, it is worthwhile
to digress briefly on the applicability of our spherical convolution
strategy to afterglows.

In prompt emission, it is quite reasonable to assume that $\gamma$ is
constant throughout the duration of the burst \citep{rees1994}.  This
assumption underlies our expressions for the various emission formulas
derived above.

Afterglow emission is another story.  The external shock model features a
deceleration of the shock front, due to interaction with the ISM
\citep{piran1999}.  This means that photons received simultaneously by the
observer from different parts of the shock front were actually emitted at
epochs characterized by different values of $\gamma$ --- there is a
relativistic ``look-back'' effect that must be taken properly into account
\citep{woods1999,granot1999}.  It may seem that this effect compromises
the usefulness of the methods introduced in this section, by making the
main emission formulas take forms that cannot be cast as convolutions on
the 2-sphere.

In fact, subject to the assumption that the {\it shape} of the underlying
emission profile is invariant (so that only the emission amplitude varies
with time), it is still possible to write the emission formulas as 2-sphere
convolutions, so that the present approach using decompositions in Legendre
polynomials may still be valuable for afterglow studies.

This is the case because at each observer time $t_o$, a photon received
from the element of the jet at $\vec{n}$ was emitted at a time
$t(t_o,\vec{n}\cdot\vec{l}\,)$ that is a function of $\vec{n}\cdot\vec{l}$,
where $\vec{l}$ is the viewing direction.  This function is computable in
terms of the jet velocity history, $\beta(t)$.  If the jet emission profile
(now a function of time) has by assumption the form
\begin{equation}
E(t,\vec{n}) = A(t)\epsilon(\vec{n}\cdot\vec{b}),
\label{decompose_profile}
\end{equation}
then the instantaneous flux is proportional to a convolution like that of
Eq.~(\ref{conv0}), where
\begin{eqnarray}
f^{(1)}(\vec{n}\cdot\vec{b}\,) &=& \epsilon(\vec{n}\cdot\vec{b}),\\
f^{(2)}(\vec{n}\cdot\vec{l}\,) &=&
A\left(t(t_o,\vec{n}\cdot\vec{l})\right) \times
\left[ 1 - \beta\left(t(t_o,\vec{n}\cdot\vec{l}\,)\right) \vec{n}\cdot\vec{l}
\,\right]^{-3}.
\end{eqnarray}

It follows that under a moderately restrictive assumption about the time
development of the emission profile, afterglow emission may also be
treated by decomposition into Legendre polynomials, as described above. A
possible generalization of this approach that relaxes the above
restriction somewhat is to consider profiles constituted of linear
superpositions of terms like the one in Eq.~(\ref{decompose_profile}).

Note another assumption made here, however:  we still require that
$\gamma$ is uniform across the jet.  In fact, afterglow jet models with
spatial variation in $\gamma$ are considered in the literature
\citep{rossi2002,granot2003}. If the uniform-$\gamma$ assumption is
removed, the required factorization of the integrand into
$f^{(1)}(\vec{n}\cdot\vec{b}\,)$ and $f^{(2)}(\vec{n}\cdot\vec{l}\,)$
fails, and direct expansion in Legendre polynomials is no longer an
available strategy.  If the variation of $\gamma$ across the jet is not
too strong, however, it is still possible to adopt the spherical expansion
method after expanding the Doppler factor in powers of the deviation of
$\gamma$ from its value on the jet axis,
$\gamma(\vec{n})-\gamma(\vec{b})$.

%%%%%%%%%%%%%%%%%%%%%%%%%%%%%%%%%%%%%%%%%%%%%%%%%%%
%%%%%%%%%%%%%%%%%%%%%%%%%%%%%%%%%%%%%%%%%%%%%%%%%%%
\subsection{Emission From Top-Hat Jet Profiles\label{sec_uniform}}

As mentioned above, the sharp edge of a top-hat jet profile forces unwanted
power onto the higher harmonics of its DLT.  Fortunately, there are
closed-form analytical expressions for the emission from a top-hat jet
profile, which we present here.

\subsubsection{Emission Formulas}

The emission profile of a top-hat jet of opening half-angle $\theta_0$ is
\begin{equation}
\epsilon_{TH}(\cos\theta)=\left[2\pi(1-\cos\theta_0)\right]^{-1}
\Theta(\cos\theta-\cos\theta_0),
\end{equation}
where $\Theta(x)$ is the Heaviside step function.  From Eq.~(\ref{eiso})
we then have
\begin{eqnarray}
\eiso&=&\frac{\egt}{2\pi\gamma^4(1-\cos\theta_0)}
\oint d^2\vec{n}\,\frac{\Theta(\vec{n}\cdot\vec{b}-\cos\theta_0)}
                       {(1-\beta\vec{n}\cdot\vec{l})^3}\nonumber\\
&=&\frac{\egt}{2\pi\gamma^4(1-\cos\theta_0)}
\int_{\cos\theta_0}^1d\cos\theta\,\int_{-\pi}^\pi\frac{d\phi}
{\left[1-\beta(\cos\theta_v\cos\theta+\sin\theta_v\sin\theta\cos\phi)\right]^3},
\label{eiso_uniform_int}
\end{eqnarray}
where again $\theta_v$ is the angle between $\vec{b}$ and $\vec{l}$.

The integral over $\phi$ may be performed using formula 3.661.4 of
\citet[p.383]{gr65}.  The remaining integral over $\cos\theta$ may then be
performed by means of the successive substitutions
$y=\beta\cos\theta-\cos\theta_v$, $\tan q=\gamma y/\sin\theta_v$.  The result
is
\begin{equation}
\eiso=\frac{\egt}{2\beta\gamma^4(1-\cos\theta_0)}
\left[f(\beta-\cos\theta_v) - f(\beta\cos\theta_0-\cos\theta_v)\right],
\label{eiso_uniform}
\end{equation}
where
\begin{equation}
f(z)\equiv
\frac{\gamma^2(2\gamma^2-1)z^3+(3\gamma^2\sin^2\theta_v-1)z+2\cos\theta_v\sin^2\theta_v}
{(z^2 + \gamma^{-2}\sin^2\theta_v)^{3/2}}.
\label{f_function}
\end{equation}

From Eq.~(\ref{dbar}), the average Doppler factor $\langle D\rangle$ is
expressed as a ratio of two integrals, one of which is the same as the one
in Eq.(\ref{eiso_uniform_int}), while the other is quite similar and may be
calculated using the same technique.  The result is
\begin{equation}
\langle D\rangle=
\gamma^{-1}\,
\frac
{f(\beta-\cos\theta_v) - f(\beta\cos\theta_0-\cos\theta_v)}
{g(\beta-\cos\theta_v) - g(\beta\cos\theta_0-\cos\theta_v)},
\label{dbar_uniform}
\end{equation}
where $f(z)$ is as in Eq.~(\ref{f_function}), while the function $g(z)$ is
given by
\begin{equation}
g(z)\equiv
\frac
{2\gamma^2z+2\cos\theta_v}
{(z^2 + \gamma^{-2}\sin^2\theta_v)^{1/2}}.
\label{g_function}
\end{equation}

The frequency distribution of \eiso\ is generally given by
Eq.~(\ref{dndeiso}).  This formula may be applied with
Eq.~(\ref{eiso_uniform}) to yield an expression for $dN/d\eiso$ that is
suitable for computation.  This expression is algebraically obvious, but
burdensome, and will be omitted here.

%%%%%%%%%%%%%%%%%%%%%%%%%%%%%%%%%%%%%%%%%%%%%%%%%%%%%%%%%%%%%%%%%%%%%%%%%%
%%%                                                                    %%%
%%% \S4 A million plots                                                %%%
%%%                                                                    %%%
%%%%%%%%%%%%%%%%%%%%%%%%%%%%%%%%%%%%%%%%%%%%%%%%%%%%%%%%%%%%%%%%%%%%%%%%%%
\section{Jet Profiles And Their Signatures}
\label{a_bunch_of_plots}

We now turn directly to the observable signatures of universal jet models
based on Fisher, power-law, and top-hat jet profiles.  

Our approach is to explore the parameter space of universal jet models
based on these families of profiles, assessing features of the landscape
for their resemblance or dissimilarity to the empirically-known properties
of GRBs discussed earlier.  We do not intend here to make quantitative
judgments of how well the models ``fit'' the data.  Rather, we wish to
construct a map of the parameter space of universal-jet GRB models, so as
to clarify the strengths and weaknesses of each profile choice.  This
allows one to choose which aspects of the data can be modeled successfully
using a particular family of universal-jet models, and which must be
``evaded'' by invoking new model features.  It also allows easier
interpretation of the results of population-synthesis simulations that
attempt to reproduce the distributional and correlational properties of
\eiso, \egi, and \epk, irrespective of whether the models adopted are of
the universal or variable-geometry variety.

This map is also a useful tool with respect to models that make more
realistic observable choices than ours.  Our variables are admittedly
simplistic, insofar as comparison with data from real GRB detectors are
concerned.  \eiso\ is a bolometric quantity, which makes no reference to
detector bandpass, and $\davg\times\epkr$ is a somewhat crude
representation of the true spectral peak.  More realistic modeling
ascribes an actual rest-frame spectrum to the jet, possibly even allowing
for spectral variation over the surface of the shock.  Even more realistic
would be the introduction of time variability, so as to allow more
detailed modeling of detector thresholds.

What such modeling gains in fidelity, however, it loses in performance. 
It is an extremely burdensome task to compute emission, correlation, and
distributional properties from models that must integrate numerically
across the shock surface, the photon frequency domain, and the time
domain.  Such investigations cannot currently be carried out over wide
swaths of parameter space.  This is where our map can be valuable, by
indicating qualitatively the kinds of models and relatively narrow ranges
of parameters that are worth a closer look, and where builders of
complicated models might consider unlimbering their heavy machinery.

In this section we consider the following families of universal GRB jet
models:

\begin{itemize}
\item Fisher profile universal jets

\item Power-law profile universal jets, with power-law indices $\delta$=1,
2, and 8.

\item Top-hat profile universal jets. 

\end{itemize}

For each family of models, we exhibit six kinds of plots:  \eiso\ versus
$\theta_v$, \davg\ versus $\theta_v$, \eiso\ versus \davg\ (``Amati'' plot),
\egi\ versus \davg\ (``Ghirlanda'' plot), $dN/d\log\eiso$ frequency plot, and
$dN/d\log\egi$ frequency plot.

In each case we allow $\gamma$ to take the values 10, 33.3, 100, and 333.

Each family of profiles has a parameter $\theta_0$, which characterizes
the angular width of the profile.  Its interpretation is specific to the
functional form of the profile. In almost all cases, we allow $\theta_0$
to take the values 0.1, 0.3, 0.01, and 0.003.  The one exception is the
power-law profile with $\delta=8$.  In this case we actually select
$\theta_0$ = 0.4, 0.12, 0.04, and 0.012, so as to make the effective width
estimator $\theta_w$ of Eq.~(\ref{pl_width}) take the values $\theta_w$
=0.1, 0.3, 0.01, and 0.003.

The region of parameter space covered by these choices of parameter values
includes a subregion where $\gamma^{-1}$ is larger than the characteristic
width of the underlying jet profile ($\theta_0$ or $\theta_w$, as the case
may be).  This subregion of parameter space may be considered unphysical
on hydrodynamic grounds \citep{rhoads1999,sari1999}, since a hydrodynamic
jet expands sideways rather than expanding ballistically with a constant
shape in this regime.  Furthermore, general arguments of hydrodynamic causality
during the acceleration phase suggest that the angular scale of the jet
should be quite comparable to $\gamma^{-1}$ \citep[Lazzati 2005, private
communication]{stern2003}.
On the other hand, there is evidence that MHD jets
can sustain degrees of collimation that easily violate this constraint
\citep{lapenta2005}.  We do not intend to discuss the details of these
physical pictures here.  We simply consider this region in
our study for the sake of completeness.

%%%%%%%%%%%%%%%%%%%%%%%%%%%%%%%%%%%%%%%%%%%%%%%%%%%
%%%%%%%%%%%%%%%%%%%%%%%%%%%%%%%%%%%%%%%%%%%%%%%%%%%
\subsection{Emission From Fisher Jet Profiles}

In what follows, we use the single-lobe normalized Fisher jet profile:
\begin{equation}
\epsilon_{\mbox{\scriptsize Fisher}}(\cos\theta)=
\frac{e^{\cos\theta/{\theta_0}^2}}{4\pi{\theta_0}^2\sinh{\theta_0}^{-2}}.
\label{fisher_profile_norm}
\end{equation}

\subsubsection{Emission}
\label{sec_fisher_emission}

\begin{figure}[t]
\begin{minipage}[t]{3.2in}
\includegraphics[width=3.2in,clip=]{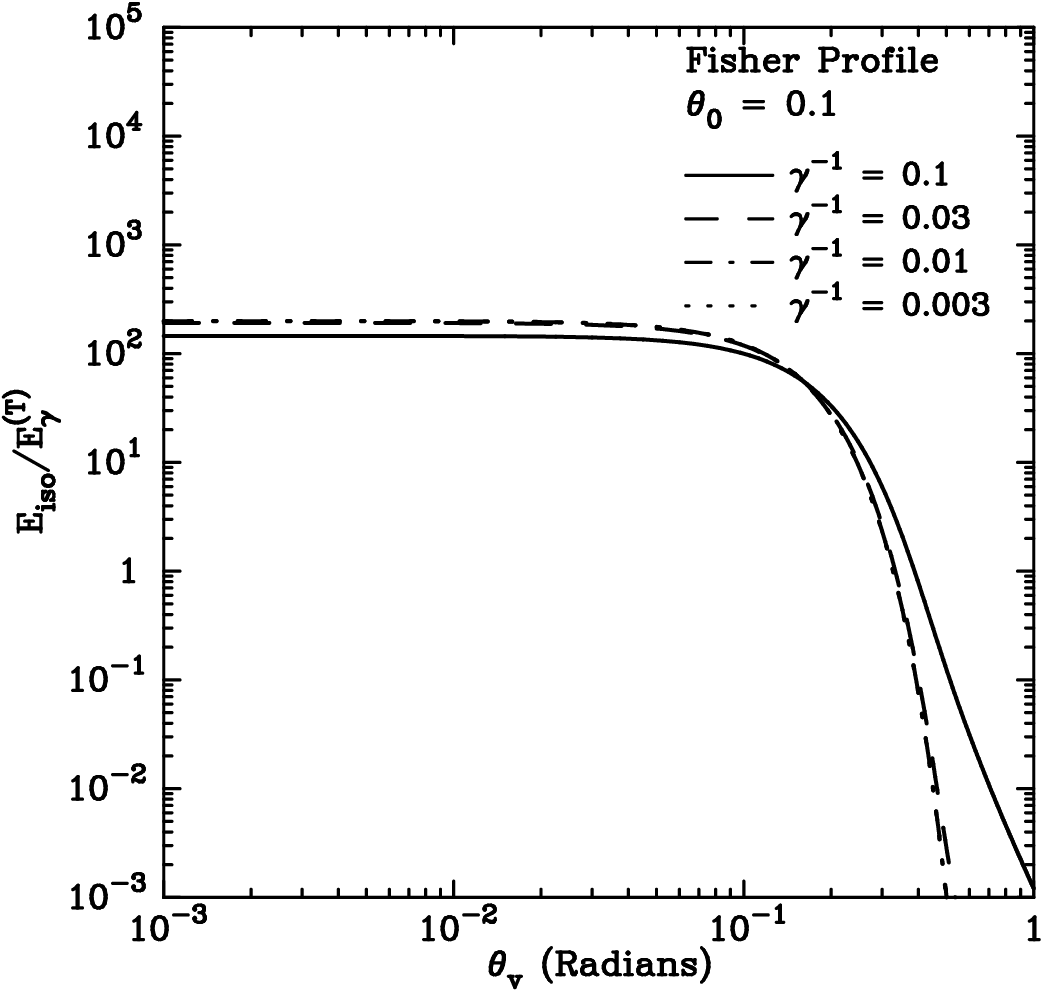}\\
\includegraphics[width=3.2in,clip=]{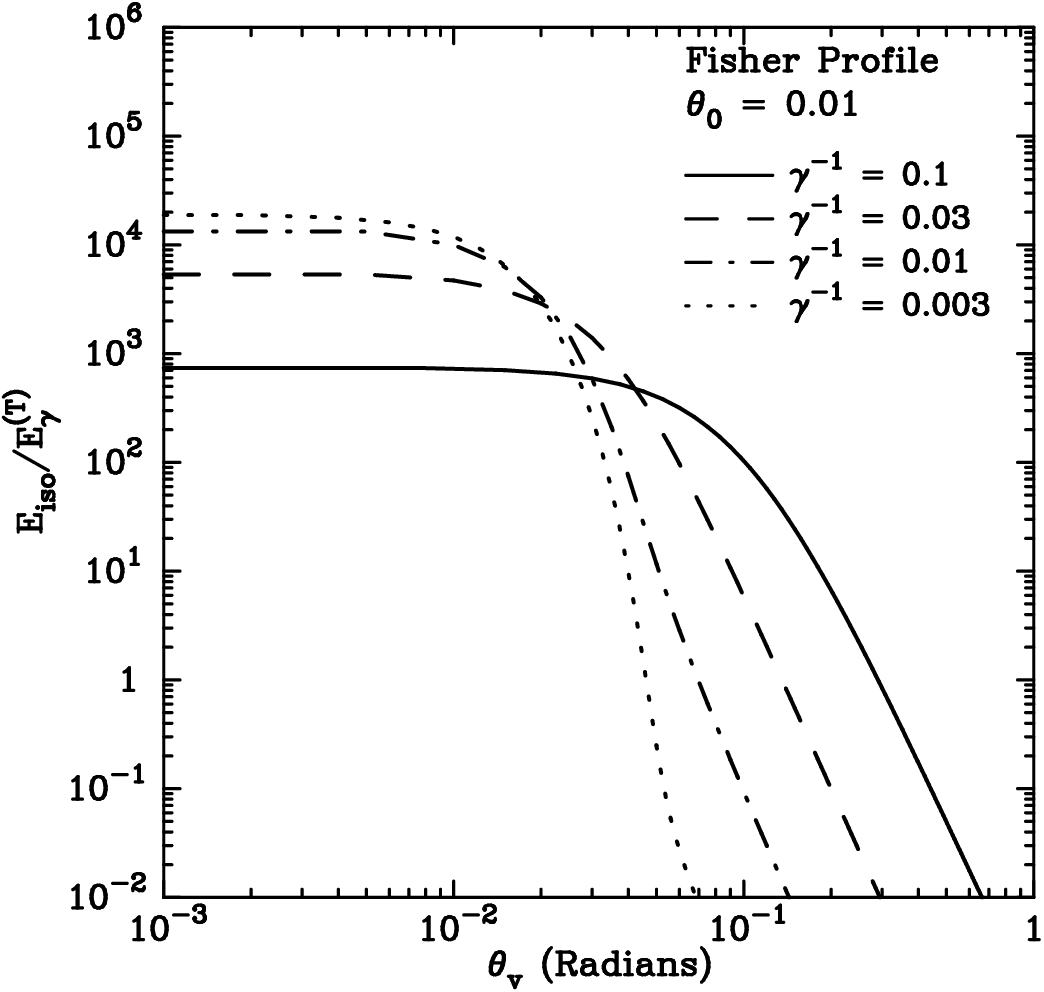}
\end{minipage}
\begin{minipage}[t]{3.2in}
\includegraphics[width=3.2in,clip=]{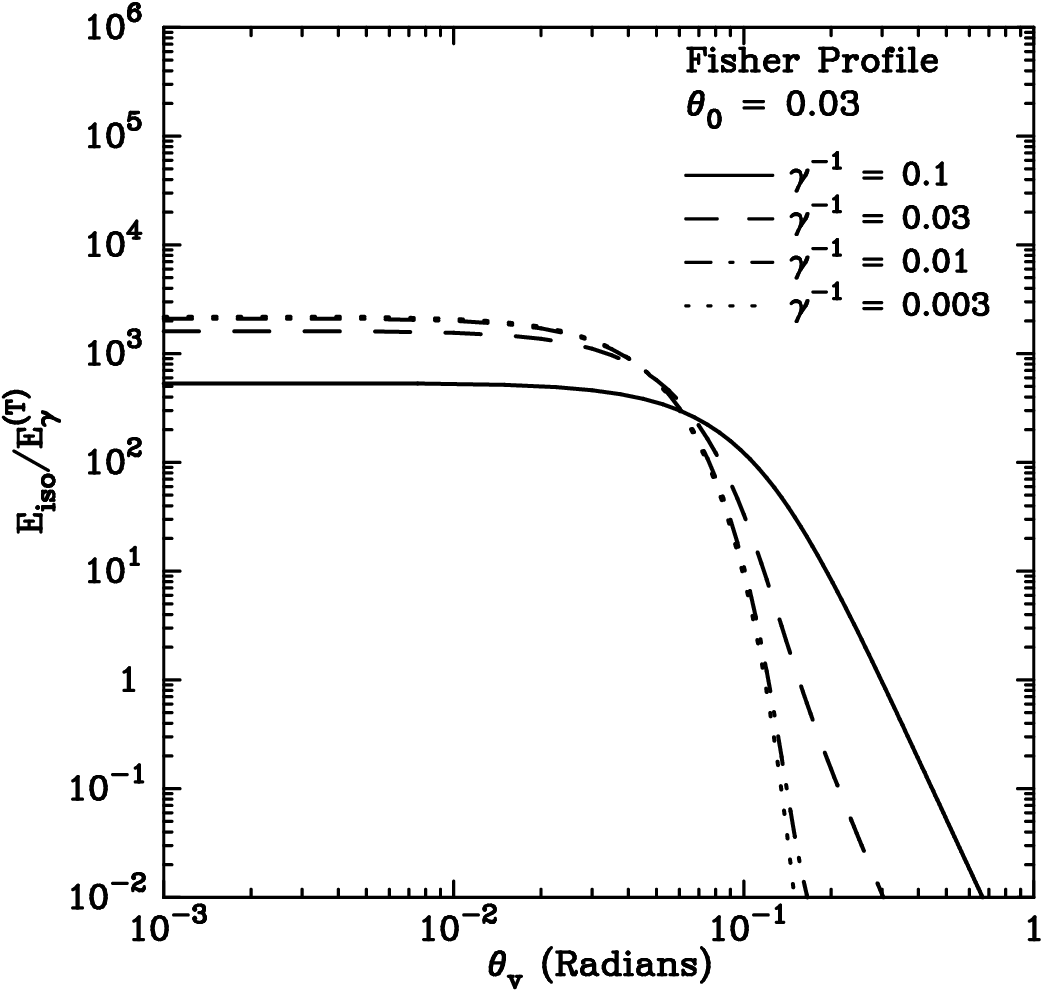}\\
\includegraphics[width=3.2in,clip=]{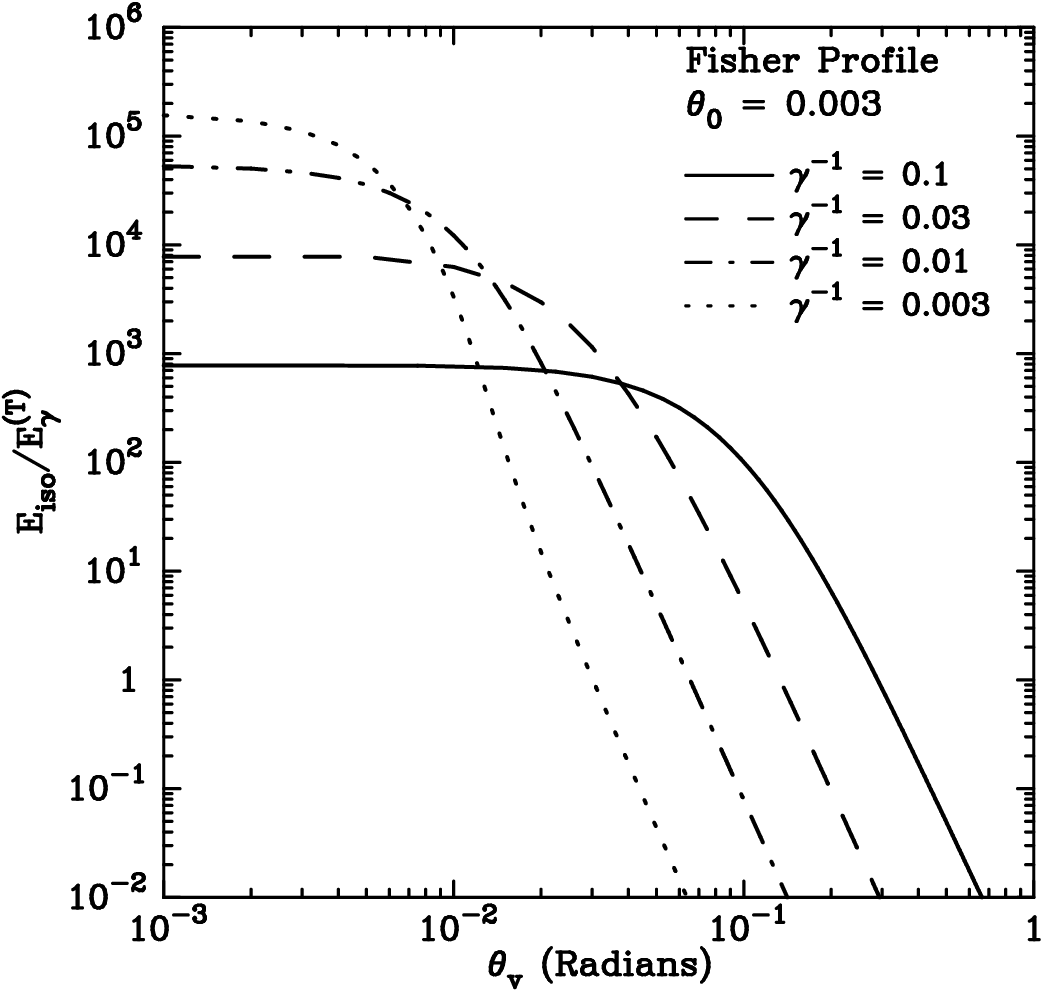}
\end{minipage}
\caption{\eiso\ as a function of viewing angle for emission from
relativistic jets with Fisher emission profiles.  Each panel shows the
result for $\gamma^{-1}$ = 0.1, 0.03, 0.01, and 0.003.  Top left: 
$\theta_0=0.1$.  Top right: $\theta_0=0.03$.  Bottom left:
$\theta_0=0.01$.  Bottom right: $\theta_0=0.003$.}
\label{fisher_emission}
\end{figure}

The four panels of Fig.~\ref{fisher_emission} show the behavior of \eiso\ 
as a function of viewing angle $\theta_v$.  The noteworthy feature of
these plots is that the value of $\gamma^{-1}$ sets a lower limit to the
effective angular size of the emission pattern.  The characteristic
angular width of radiation patterns from jets with $\theta_0<\gamma^{-1}$
is set by $\gamma^{-1}$, not $\theta_0$, since in this limit the jet
effectively resembles  ``needle jet,'' whose emission pattern is
essentially set by the Doppler function.  This effect is particularly
conspicuous in the lower-right hand panel of Fig.~\ref{fisher_emission},
in which the intrinsic profile width is 0.003 radians,  while all of the
emission patterns are in fact considerably broader than this.

This jet-broadening effect is a ubiquitous feature of relativistic
kinematics, that recurs in all the other profiles as well.

\begin{figure}[t]
\begin{minipage}[t]{3.2in}
\includegraphics[width=3.2in,clip=]{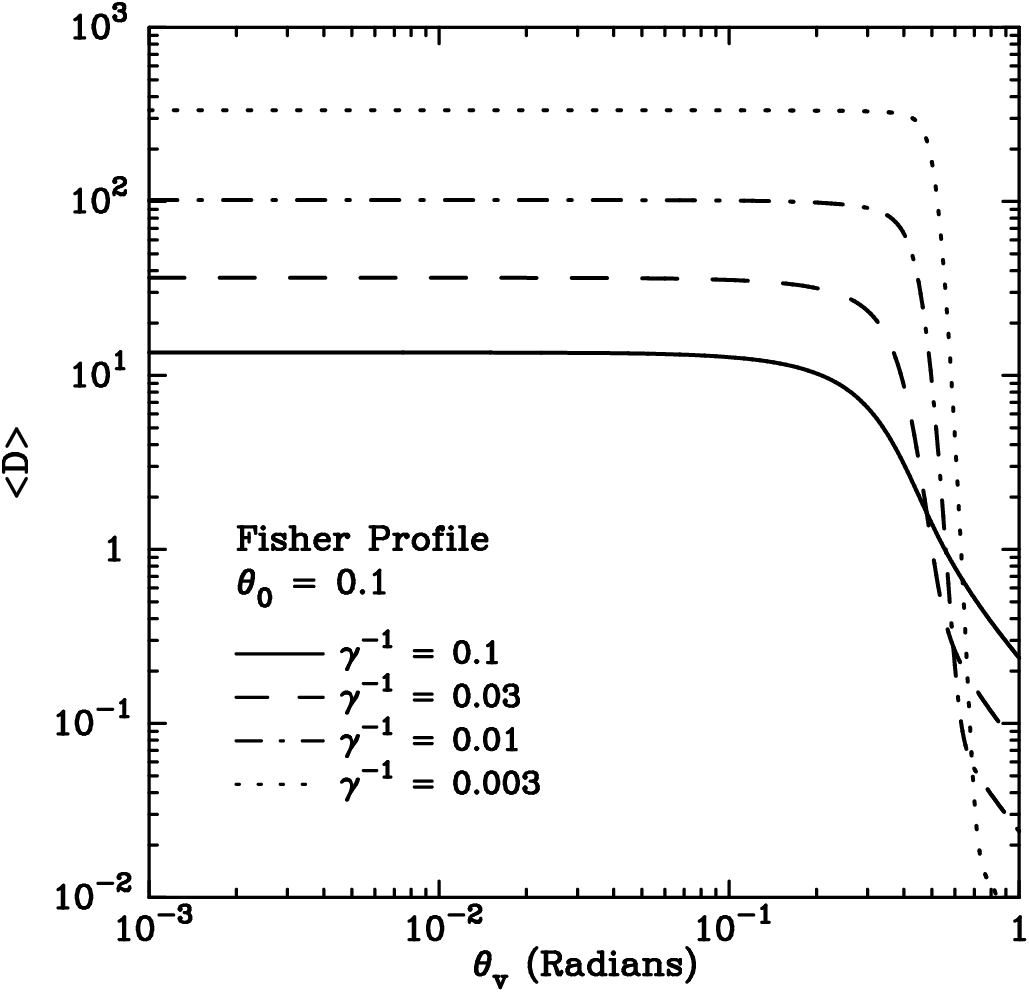}\\
\includegraphics[width=3.2in,clip=]{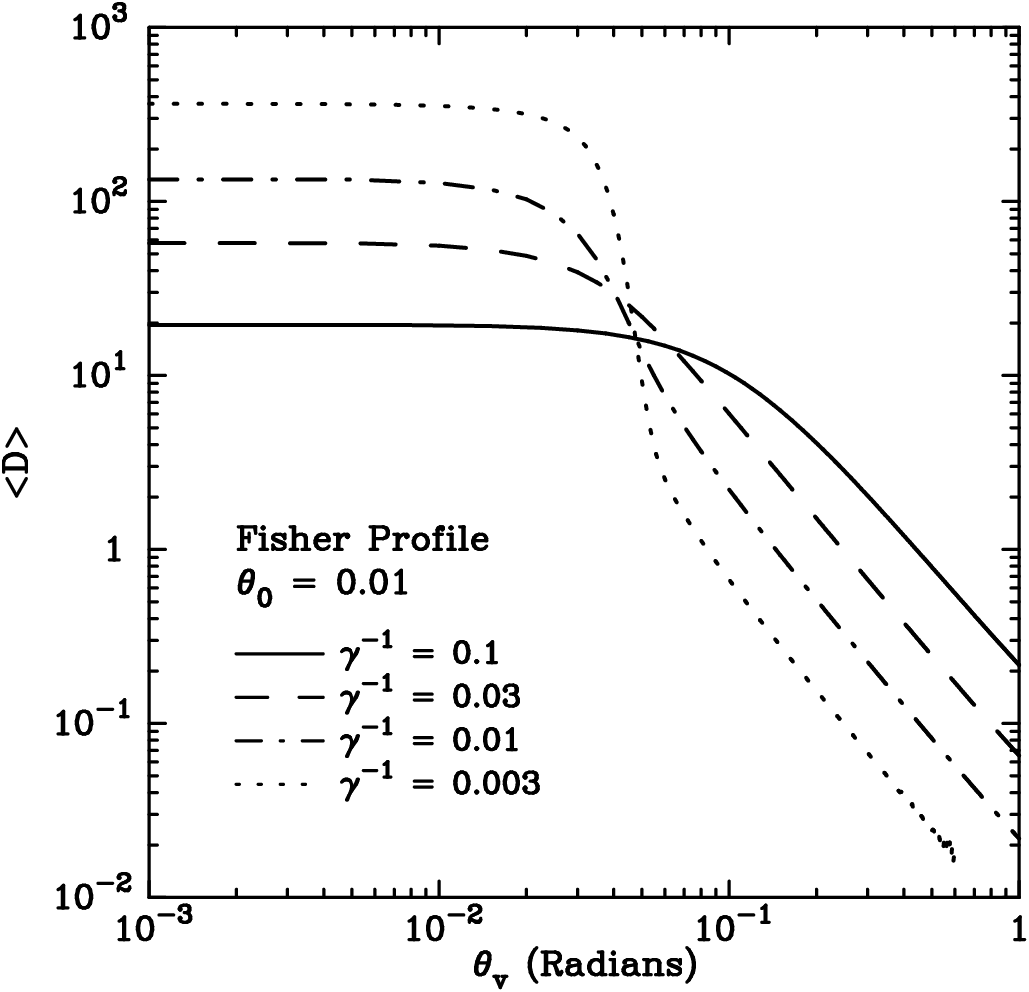}
\end{minipage}
\begin{minipage}[t]{3.2in}
\includegraphics[width=3.2in,clip=]{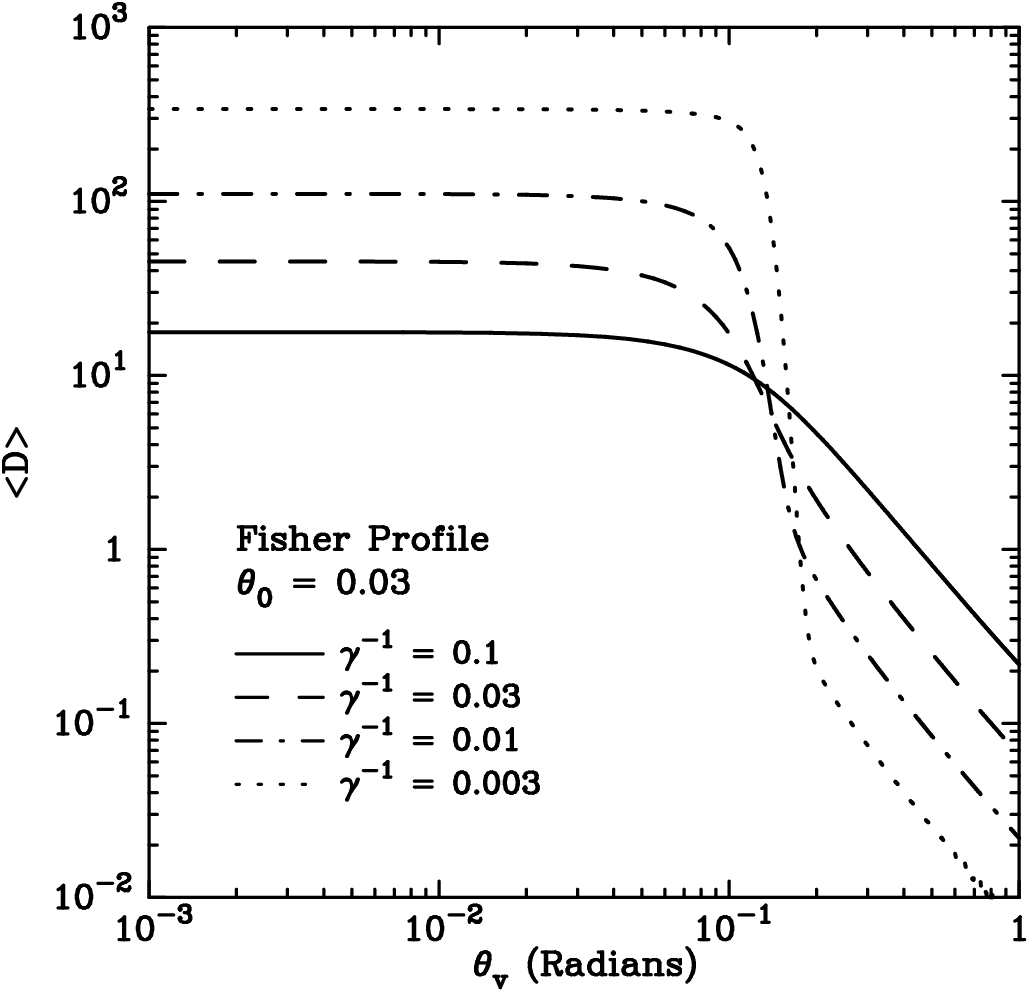}\\
\includegraphics[width=3.2in,clip=]{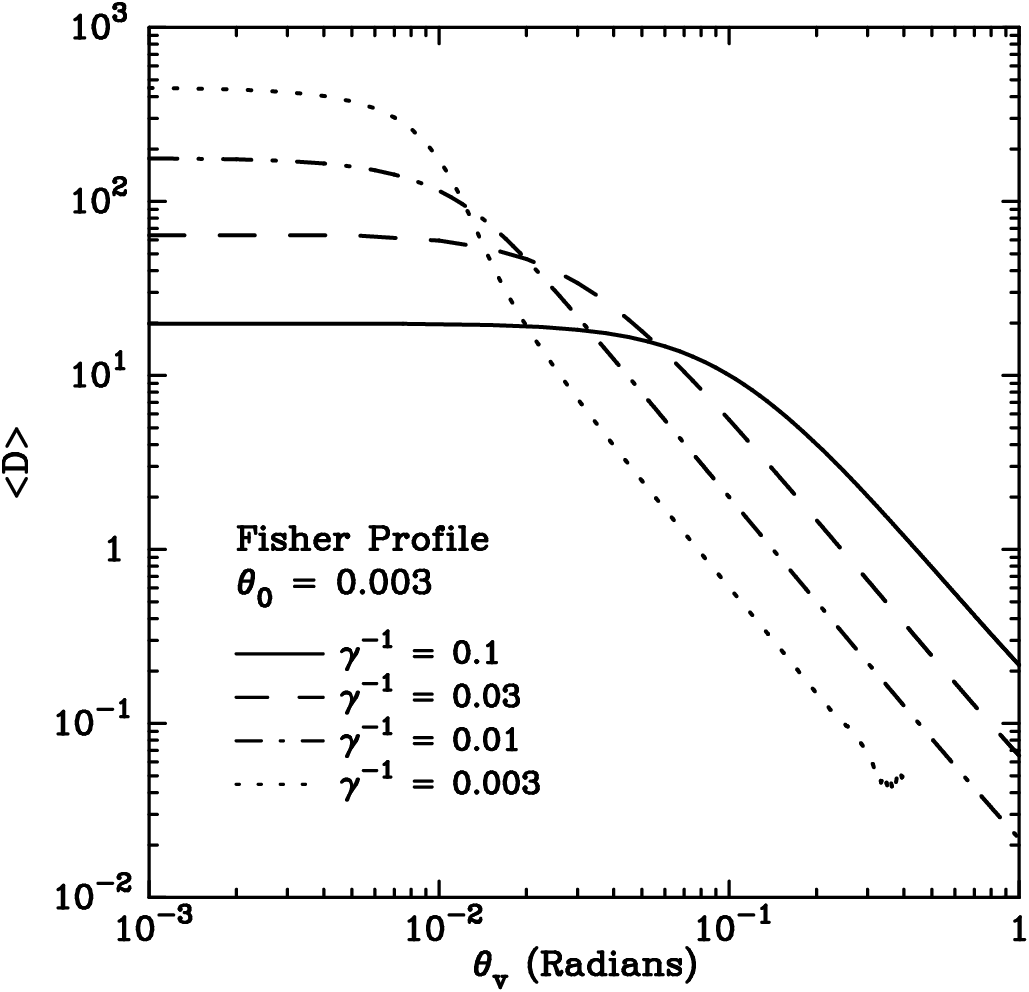}
\end{minipage}
\caption{Photon-number-averaged Doppler factor \davg\ as a function of
viewing angle for emission from relativistic jets with top-hat emission
profiles. Each panel shows the result for $\gamma^{-1}$ = 0.1, 0.03, 0.01,
and 0.003.  Top left:  $\theta_0=0.1$.  Top right: $\theta_0=0.03$. 
Bottom left: $\theta_0=0.01$.  Bottom right: $\theta_0=0.003$.  Note the
curious ``shelf'' behavior for $\gamma^{-1}\ll\theta_0$.}
\label{fisher_epk}
\end{figure}

The four panels of Fig.~\ref{fisher_epk} show the behavior of \davg\ as a
function of viewing angle $\theta_v$.  Here a curious feature requires
explanation:  there is a ``shelf'' that appears in the \davg\ function
near the jet axis, which for large $\gamma$ extends considerably beyond
the value of $\theta_0$.  For $\theta_0=0.1$, the $\gamma=333$ curve has a
shelf that extends out to $\theta_v=0.5$, taking up about 12\% of a
hemisphere.

\begin{figure}[t]
\begin{minipage}[t]{3.2in}
\includegraphics[width=3.2in,clip=]{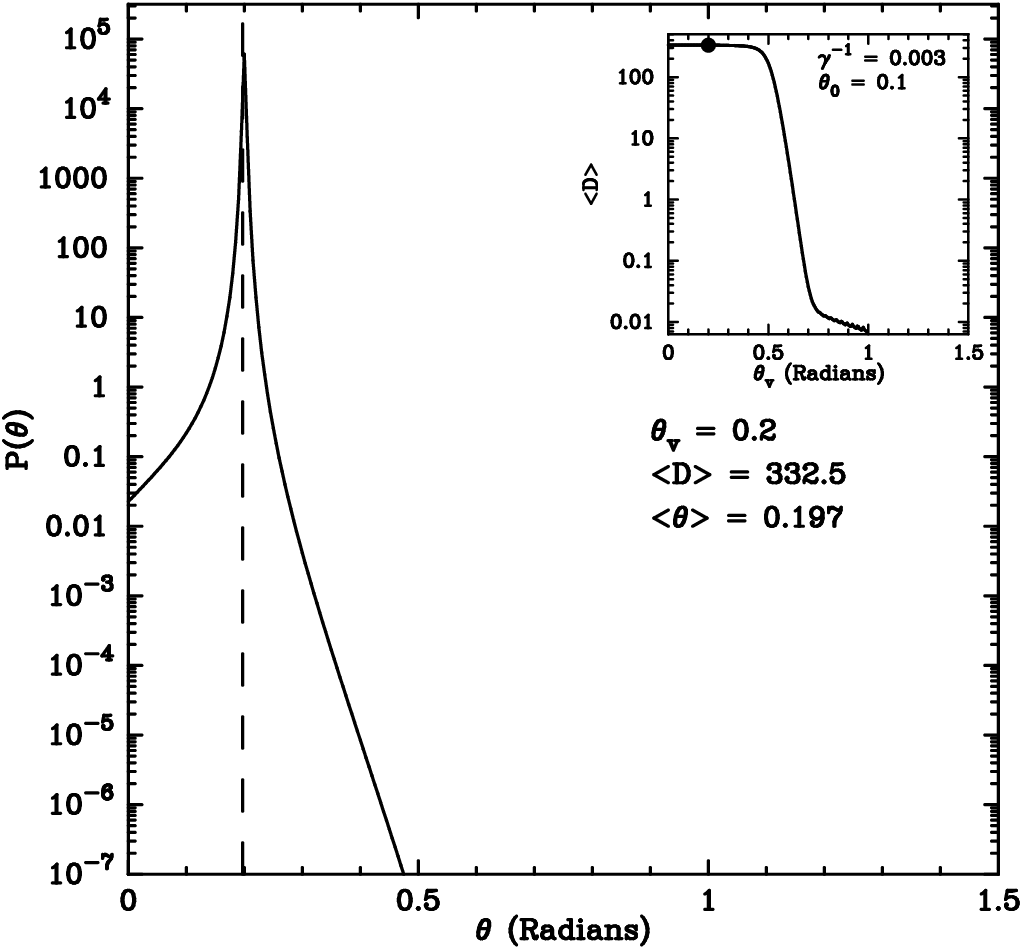}\\
\includegraphics[width=3.2in,clip=]{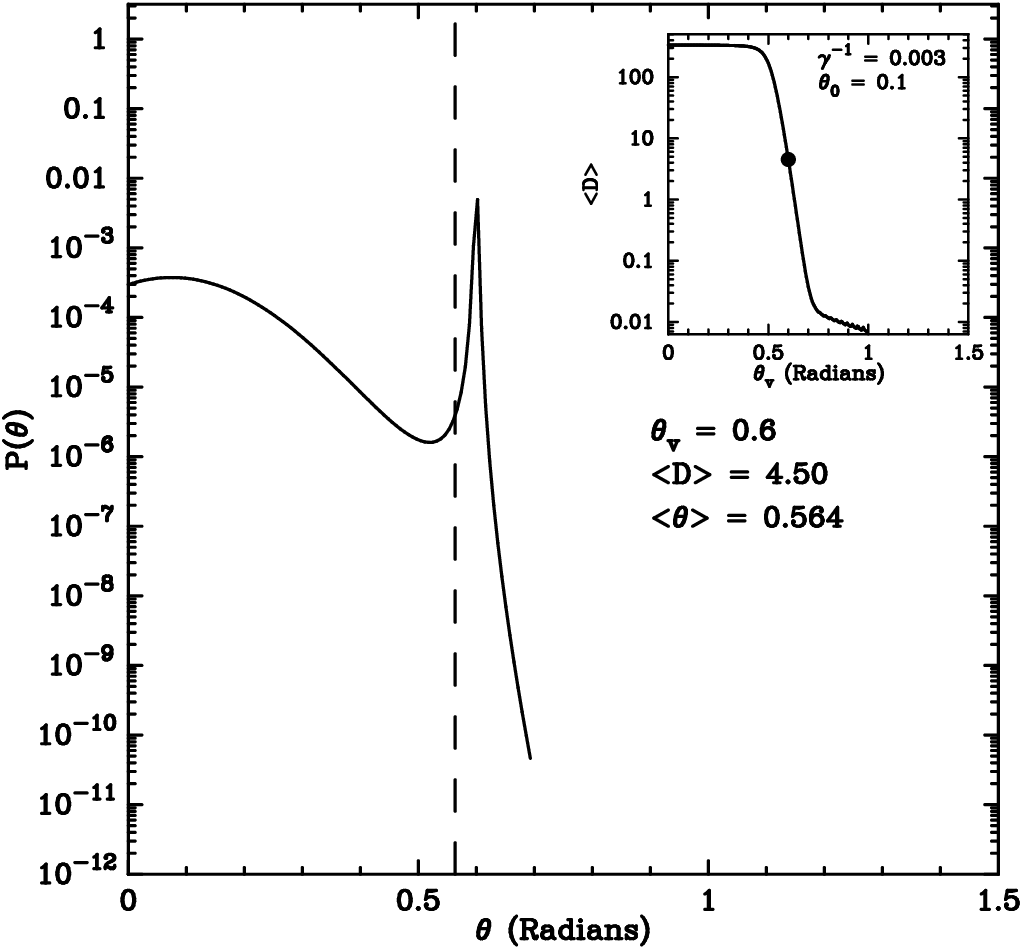}
\end{minipage}
\begin{minipage}[t]{3.2in}
\includegraphics[width=3.2in,clip=]{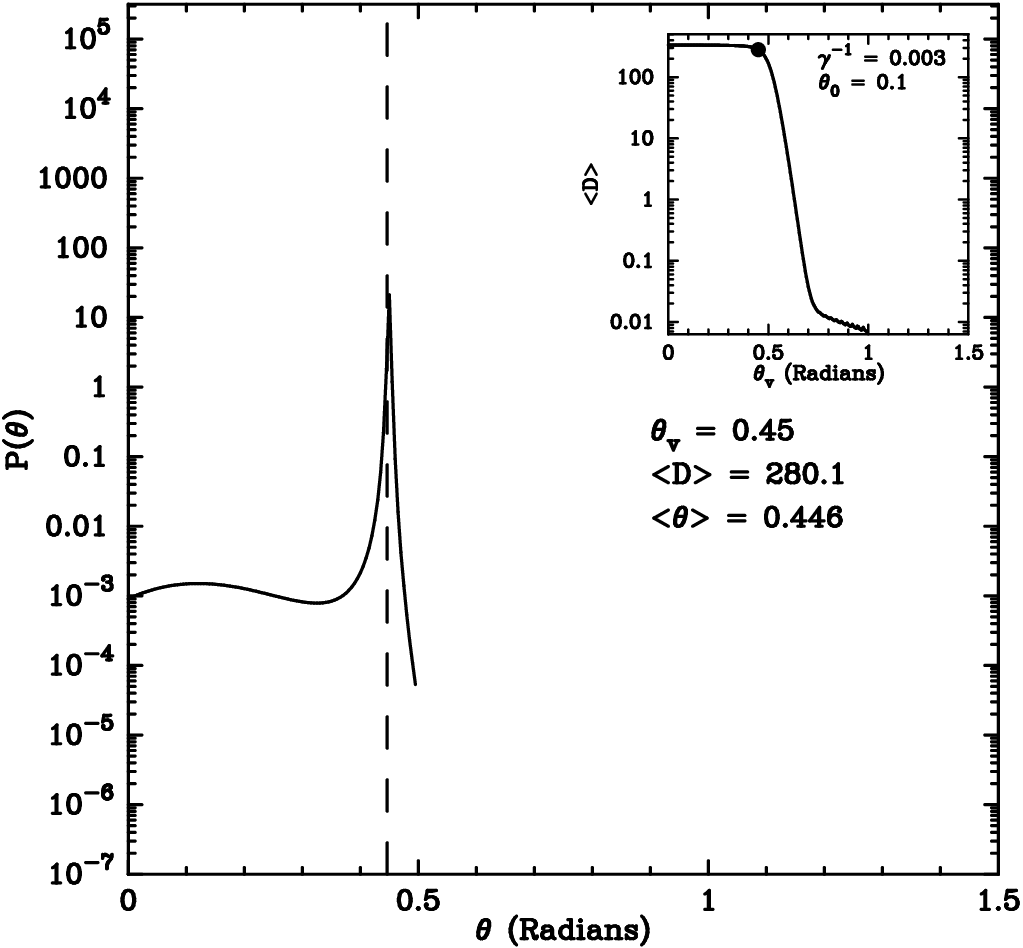}\\
\includegraphics[width=3.2in,clip=]{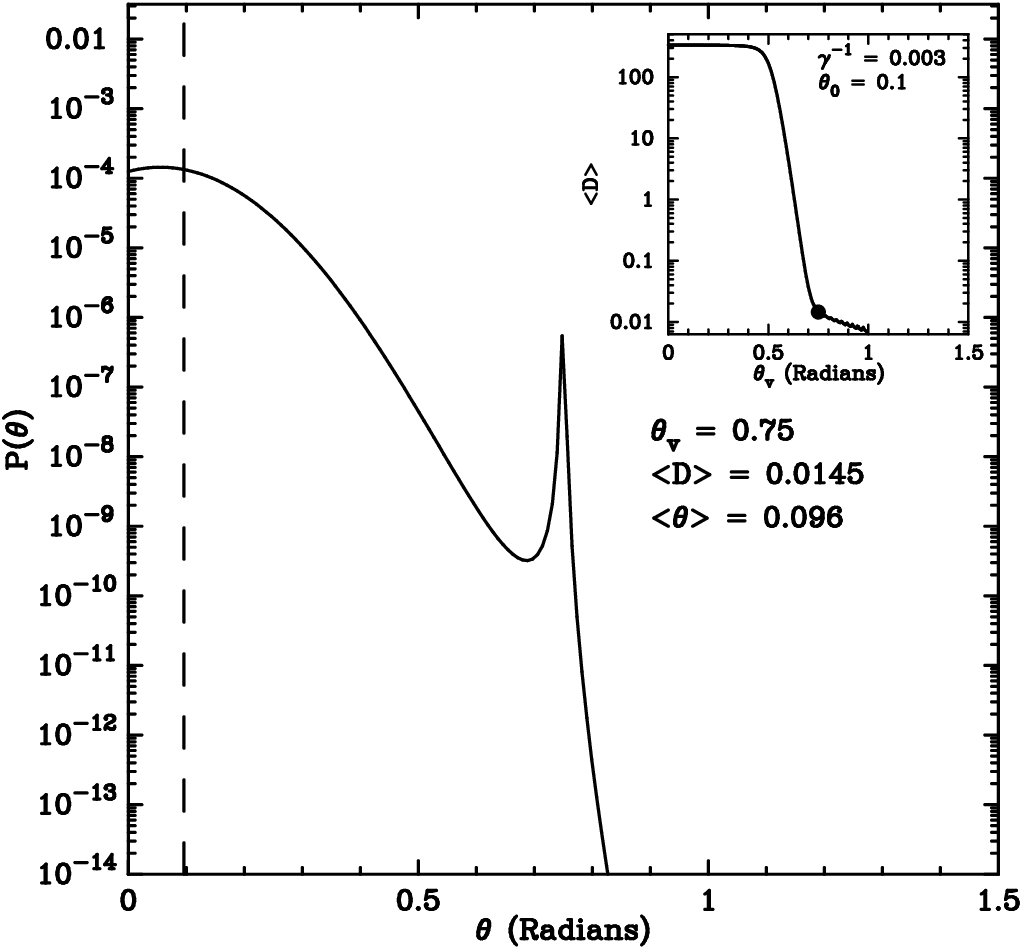}
\end{minipage}
\caption{Photon provenance distributions for a Fisher jet with
$\gamma^{-1}=0.003$, $\theta_0=0.1$.  These plots illustrate the origin of
the ``shelf'' behavior in the profile of \davg, as seen in
Fig.~\ref{fisher_epk}.  Top left: $\theta_v=0.2$.  Top right:
$\theta_v=0.45$.  Bottom left: $\theta_v=0.6$.  Bottom right:
$\theta_v=0.75$. In each panel, the $x$-axis is angular location on the
jet, while the $y$-axis is the (un-normalized) probability that a photon
along $\theta_v$ should have been emitted from that part of the jet.  The
inset shows the current $\theta_v$ and \davg, represented as a dot on the
\davg\  profile (now shown with a linear angular scale).  The vertical
dashed line shows the ``mean'' $\langle\theta\rangle$ obtained from the
current value of \davg.}
\label{fisher_epk_why}
\end{figure}

The explanation of the shelf is seen in the four panels of
Fig.~\ref{fisher_epk_why}.  These plots show the photon ``provenance''
distribution for an observer viewing a Fisher jet with parameters
$\gamma=333$, $\theta_0=0.1$, at viewing angle $\theta_v$.  The curve in
each panel is a slice through the plane containing $\vec{l}$ and $\vec{b}$
of the function $\epsilon_{\mbox{\scriptsize
Fisher}}(\vec{b}\cdot\vec{n})\times(1-\beta\vec{l}\cdot\vec{n})^{-2}$,
plotted as a function of angle $\theta$ from $\vec{b}$.  Ignoring the
azimuthal direction for the sake of simplicity, this is the
(un-normalized) probability that a photon directed along $\vec{l}$ should
have been emitted from a location on the shock surface that is at an angle
$\theta$ from the jet axis.  The four panels show this curve for four
different viewing angles --- $\theta_v$ = 0.2, 0.45, 0.6, and 0.75.  The
corresponding point on the corresponding $\davg-\theta_v$ plot of
Fig.~\ref{fisher_epk} is shown in the inset.  Angular scales are
represented linearly in these  plots, the better to bring out the detail.

The dashed line shows the position of the ``Doppler-mean'' viewing angle
$\langle\theta\rangle$ --- that is the angle such that
$\gamma^{-1}[1-\beta\cos(\theta_v-\langle\theta\rangle)]^{-1}=\davg$.  It
represents the provenance of the photons most responsible for giving \davg\
its current value.

It can be seen from these plots that $\langle\theta\rangle$ is simply
tracking the mass of the distribution, whose location and extent is
determined by the interplay between the Fisher function and the Doppler
photon distribution.  Near the axis of the jet, the peak of the Doppler
factor dominates, and almost all the received photons originate from
locations on the jet that are moving towards the observer (top left panel).

When the viewing angle gets large enough, the Doppler peak starts to peel
away from the jet axis.  The provenance distribution begins to
develop a bimodal character (with the Doppler mode still dominant), so the
mean photon provenance shifts slightly away from the part of the jet moving
towards the observer, towards the axis of the jet.  This is the beginning
of the shelf drop-off (top right panel).

As the viewing angle continues to grow, the jet axis peak of the
distribution becomes more and more competitive with the Doppler peak,
pulling $\langle\theta\rangle$ substantially away from the part of the jet
moving towards the observer (bottom left panel).

Finally, at large viewing angles, the Doppler peak is eclipsed by the jet
axis peak.  The value of $\langle\theta\rangle$ approaches $\theta_0$, and
changes very little thereafter.  From here on out, the jet appears to the
observer as a needle, with no angular structure as far as \davg\ is
concerned (bottom right panel).

The presence of this shelf at the bright end of universal Fisher jet models
with $\gamma^{-1}\ll\theta_0$ is potentially very constraining on such
models.  We have already noted that in the case of $\gamma=333$,
$\theta_0=0.1$, the shelf takes up 12\% of a hemisphere.  In the context of
a universal model, this means that for these parameters, the brightest 12\% 
of GRBs should belong to a sub-population with constant \davg\, and hence
with constant \epk.  Such a concentration might very well be conspicuous. 
As we will see, it also strongly distorts the \eiso--\davg\ and
\egi--\davg\ relations away from their empirical forms.

\subsubsection{\eiso\ -- \davg\ and \egi\ -- \davg\ Correlations}
\label{fisher_corr}

\begin{figure}[t]
\begin{minipage}[t]{3.2in}
\includegraphics[width=3.2in,clip=]{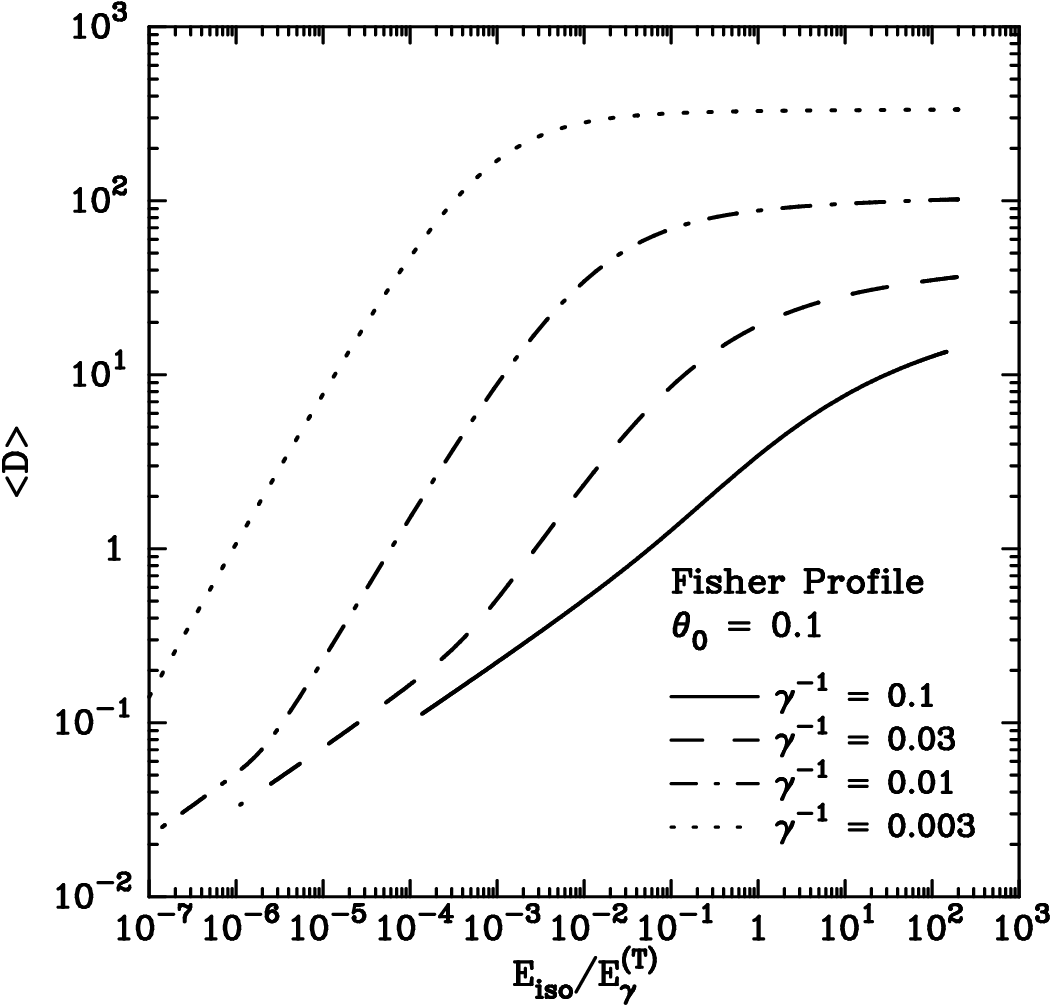}\\
\includegraphics[width=3.2in,clip=]{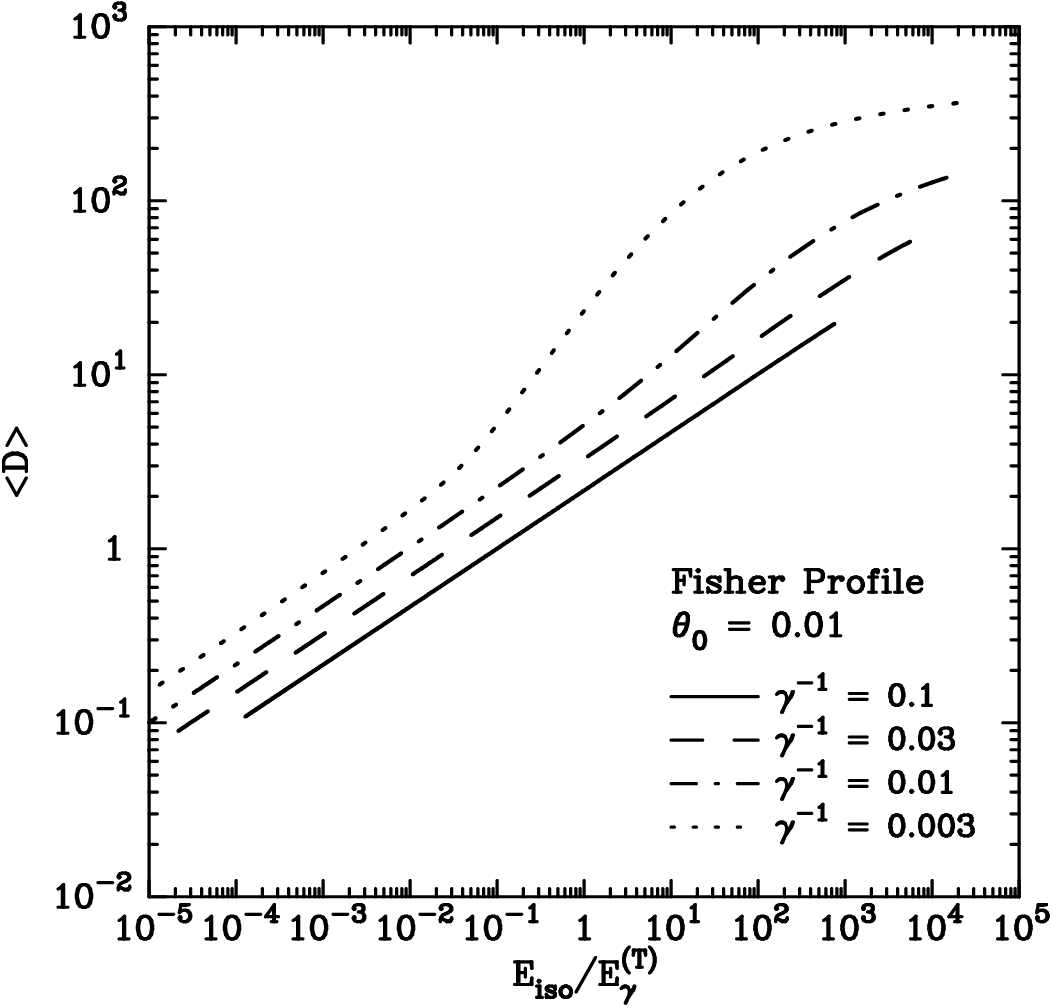}
\end{minipage}
\begin{minipage}[t]{3.2in}
\includegraphics[width=3.2in,clip=]{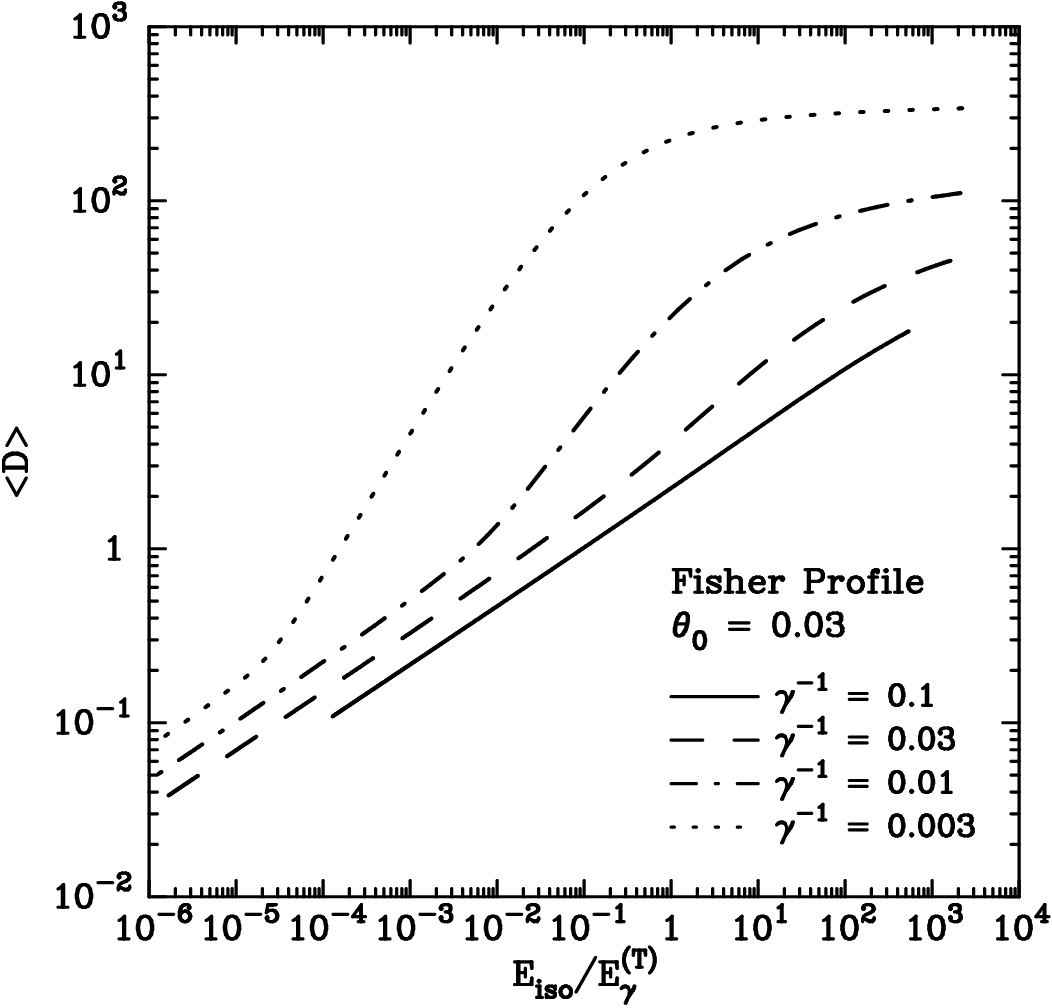}\\
\includegraphics[width=3.2in,clip=]{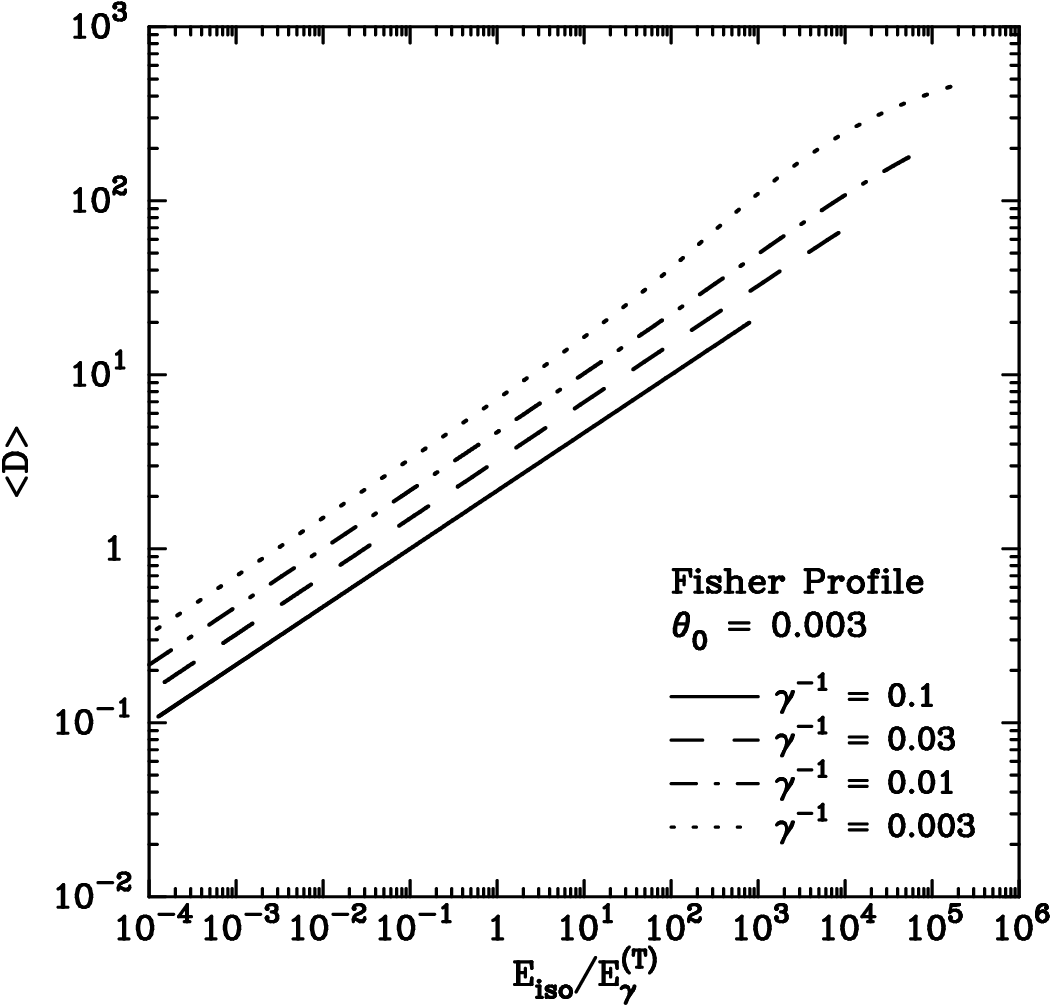}
\end{minipage}
\caption{\eiso--\davg\ (``Amati'') plots for universal jet models based on
Fisher profiles.  Each panel shows the result for $\gamma^{-1}$ =
0.1, 0.03, 0.01, and 0.003.  Top left:  $\theta_0=0.1$.  Top right:
$\theta_0=0.03$.  Bottom left: $\theta_0=0.01$.  Bottom right:
$\theta_0=0.003$.}
\label{fisher_amati}
\end{figure}

The four panels of Fig.~\ref{fisher_amati} show the behavior of the
\eiso--\davg\ (``Amati'') correlation in universal Fisher models with our
standard parameter choices.  The axis scales are chosen so that a slope of
0.5 appears as a 45$^\circ$ line, in order to ease the search for an
``Amati-like'' slope.

The 1/3 slope characteristic of off-axis emission from ``needle'' jets is
seen to hold for $\theta_0\ll\gamma^{-1}$.  This part of parameter space
is unpromising, since the empirical slope is known to be 0.5
\citep{amati2002,lamb03}.

For $\gamma^{-1}\ll\theta_0$, there does appear a section of the
\eiso--\davg\ curve that seems to have the empirically-observed slope. 
This part of the curve does not, however, include the brightest --- and
most easily observable --- events.  In fact, the ``Amati-like'' part of
the curve corresponds to the narrow ``cliff'' of Fig.~\ref{fisher_epk}.
For $\gamma=333$, $\theta_0=0.1$ (lower left panel), for example, it
comprises only viewing angles $\theta_v$ between 0.5 and 0.7, subtending
only 11\% of a hemisphere, compared to the 12\% subtended by the shelf. 
The case of $\gamma=333$, $\theta_0=0.03$ (upper right panel) seems more
promising, since now the shelf sets in at about $\theta_v=0.1$, subtending
a much smaller solid angle.

It appears, then, that a fit of a universal Fisher jet to ``Amati''-type
data could be extremely constraining on the model parameters.  The
necessity of producing the Amati slope, while at the same time not
inflicting on the model predictions a sub-population of events from the
shelf with zero slope that is brighter than, and comparably numerous to,
the ``Amati-like'' sub-population, would in all likelihood result in a
narrow acceptable region of $\gamma$-$\theta_0$ parameter space.

\begin{figure}[t]
\begin{minipage}[t]{3.2in}
\includegraphics[width=3.2in,clip=]{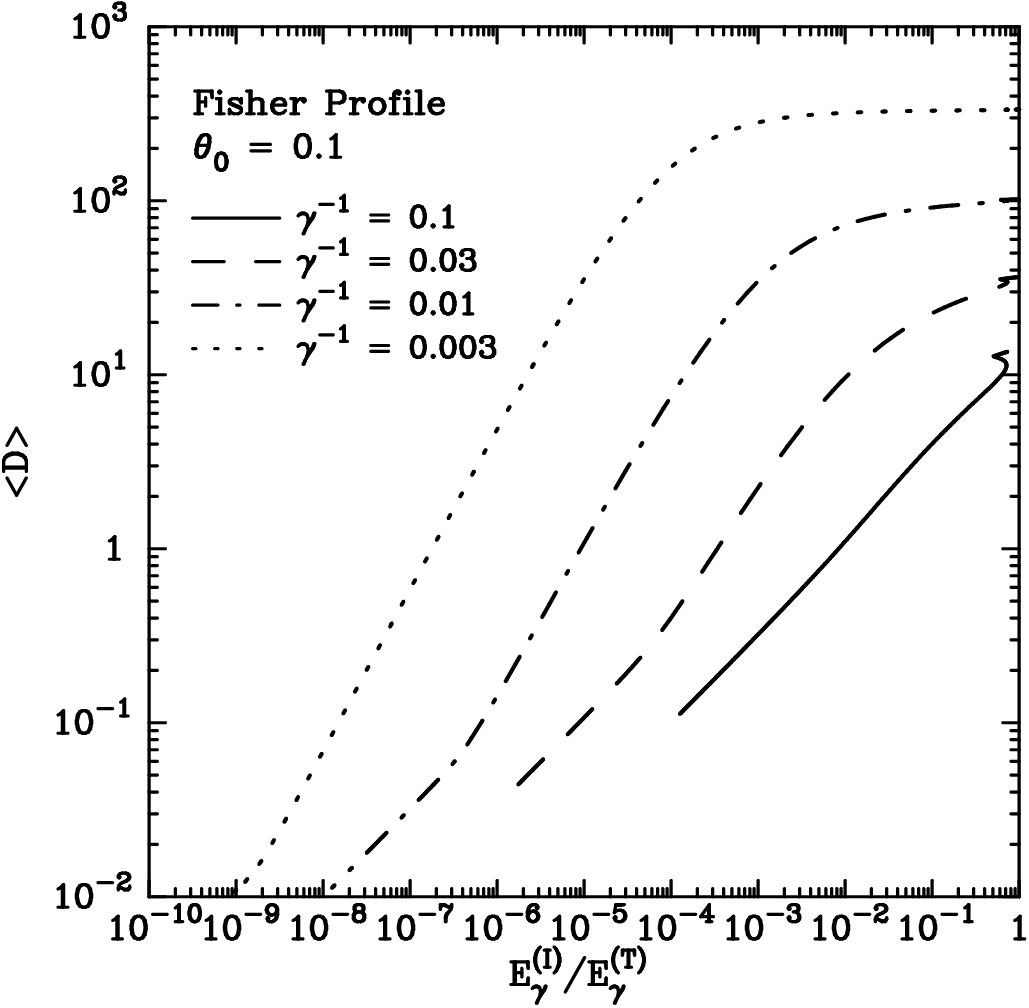}\\
\includegraphics[width=3.2in,clip=]{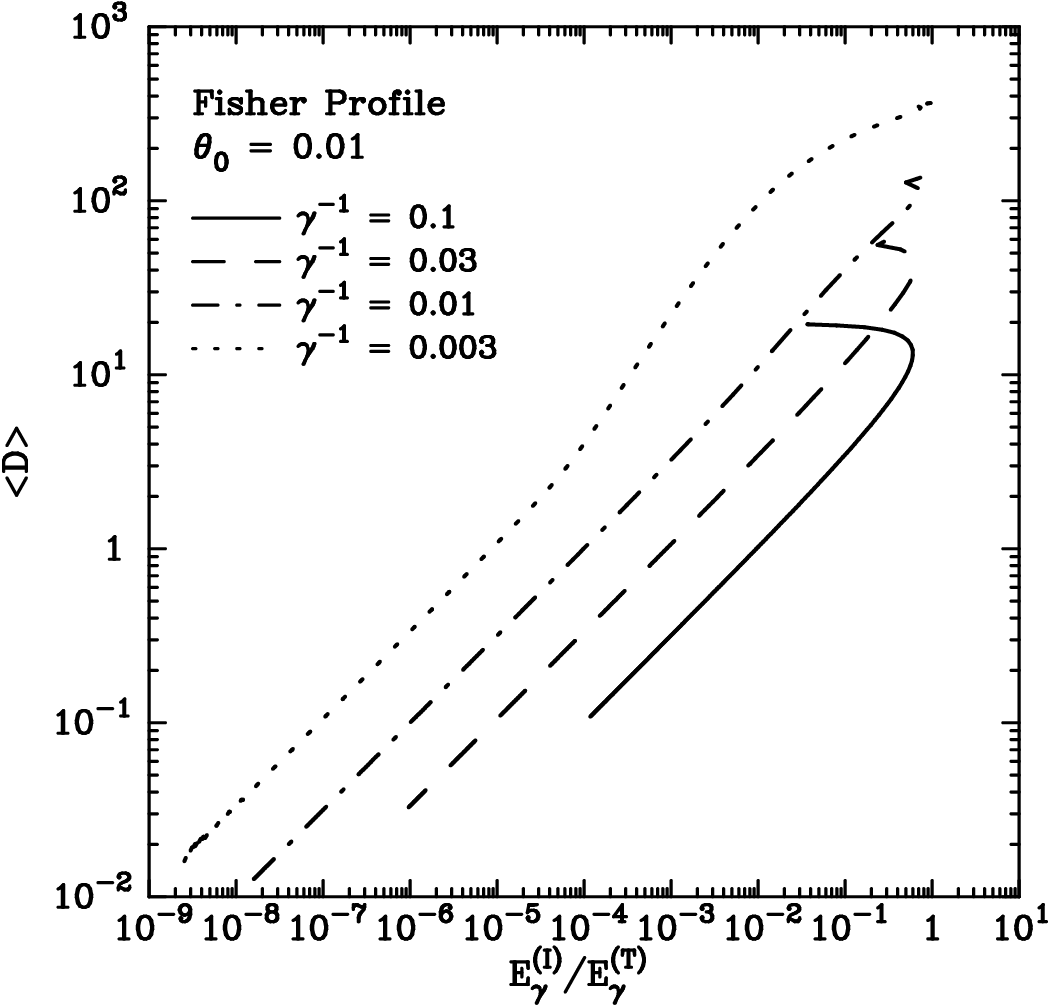}
\end{minipage}
\begin{minipage}[t]{3.2in}
\includegraphics[width=3.2in,clip=]{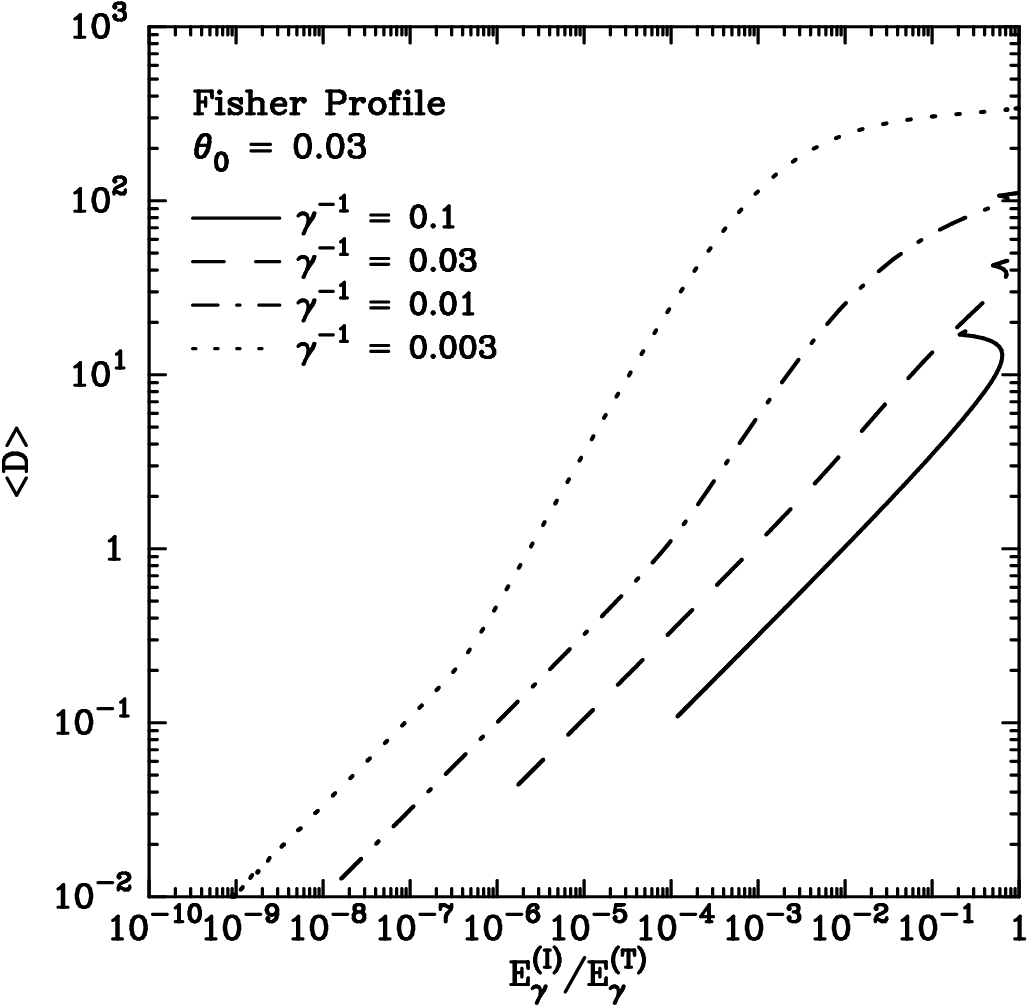}\\
\includegraphics[width=3.2in,clip=]{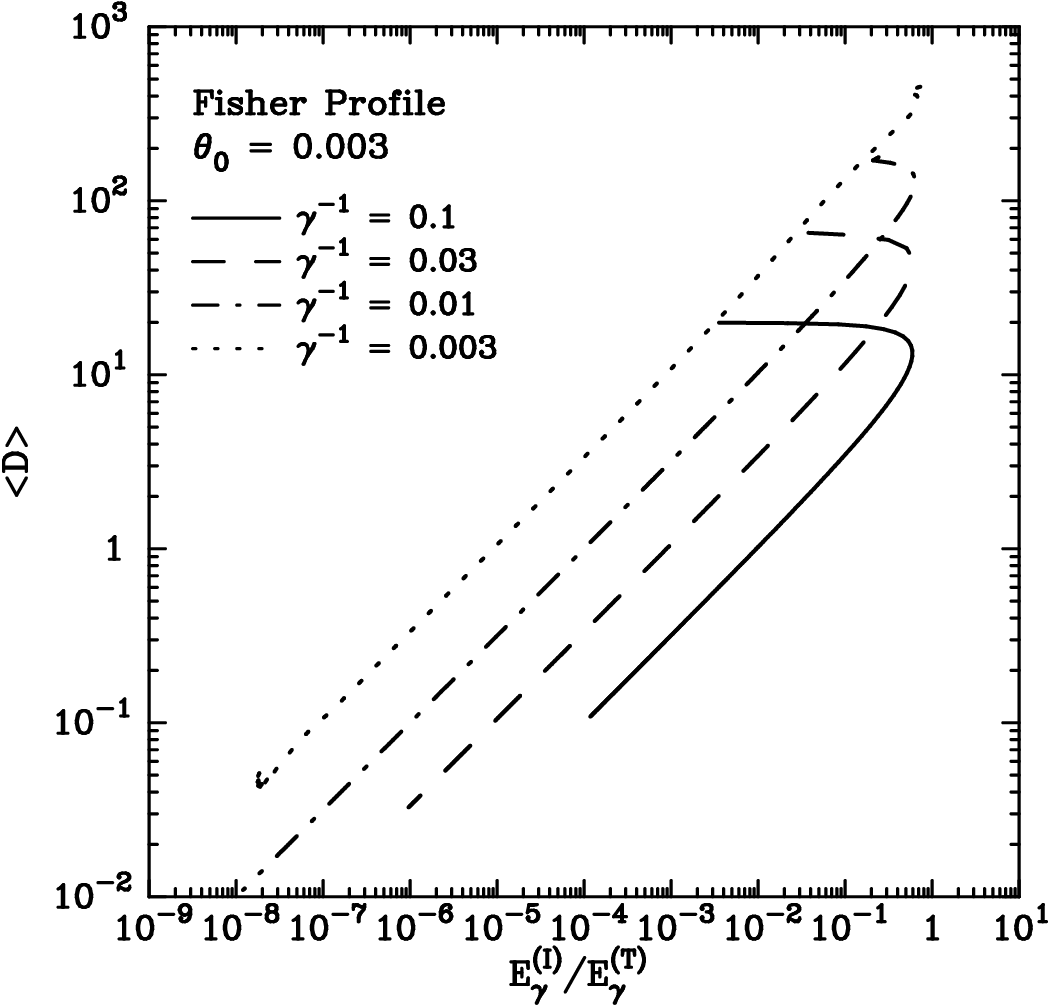}
\end{minipage}
\caption{\egi--\davg\ (``Ghirlanda'') plots for universal jet models based on
Fisher profiles.  Each panel shows the result for $\gamma^{-1}$ =
0.1, 0.03, 0.01, and 0.003.  Top left:  $\theta_0=0.1$.  Top right:
$\theta_0=0.03$.  Bottom left: $\theta_0=0.01$.  Bottom right:
$\theta_0=0.003$.}
\label{fisher_ghirlanda}
\end{figure}

The four panels of Fig.~\ref{fisher_ghirlanda} show the behavior of the
\egi--\davg\ (``Ghirlanda'') correlation in universal Fisher models with
our standard parameter choices.

The $\theta_0<\gamma^{-1}$ limit appears to feature a slope of 0.5, as
opposed to the empirically-determined value of 0.7 \citep{ggl2004}, so
again this part of parameter space is not promising territory for this
family of models.

For $\gamma^{-1}\ll\theta_0$, there is a steeper slope, corresponding again
to the drop-off from the shelf (as may be determined by comparison with
Fig.~\ref{fisher_epk}).  For $\theta_0=0.1$, $\gamma=333$ (upper left
panel), this slope is 1. It drops down to zero at high values of \egi. 
Similar behavior is seen for $\theta_0=0.03$, $\gamma=333$ (upper right
panel).  The difference between the two curves is that the flattening-out
takes place over a narrower range in $\theta_v$ when $\theta_0=0.03$ than
when $\theta_0=0.1$, reducing the anomalous zero-slope sub-population.  So
just as for the Amati case, it appears that a fit of a universal Fisher jet
to ``Ghirlanda''-type data could produce very strict constraints on
$\gamma$ and $\theta_0$.

One feature of Fig.~\ref{fisher_ghirlanda} requiring an explanation is the
hook-like structure observed for $\gamma^{-1}>\theta_0$, in which \egi\
appears to fold back upon itself.

This feature is another side-effect of the jet-broadening effect noted
above.  The relativistic kinematics is cutting off the jet profile when
$\theta_v\sim\gamma^{-1}$, long before $\theta_v$ approaches $\theta_0$. 
Therefore, the factor $1-\muv$ in the definition of \egi\ 
(Eq.~\ref{egfrail}) begins to depress the value of \egi\ earlier and more
deeply than it could the ``bare'' fisher profile.  With reference to the
schematic Fig.~\ref{graphanal}, the maximum of the curve is occurring at
lower values of \muv, and the curve plunges to much lower values before it
``recovers'' at $\theta_v=\theta_0$.

What this indicates is that when relativistic kinematics are taken into
account, it would be best to apply the \citet{f01} procedure using a break
angle $\theta_{br}=\max(\theta_v,\theta_0,\gamma^{-1})$, in order to avoid
such artifacts.

\subsubsection{\eiso\ and \egi\ Frequency Distributions}

\begin{figure}[t]
\begin{minipage}[t]{3.2in}
\includegraphics[width=3.2in,clip=]{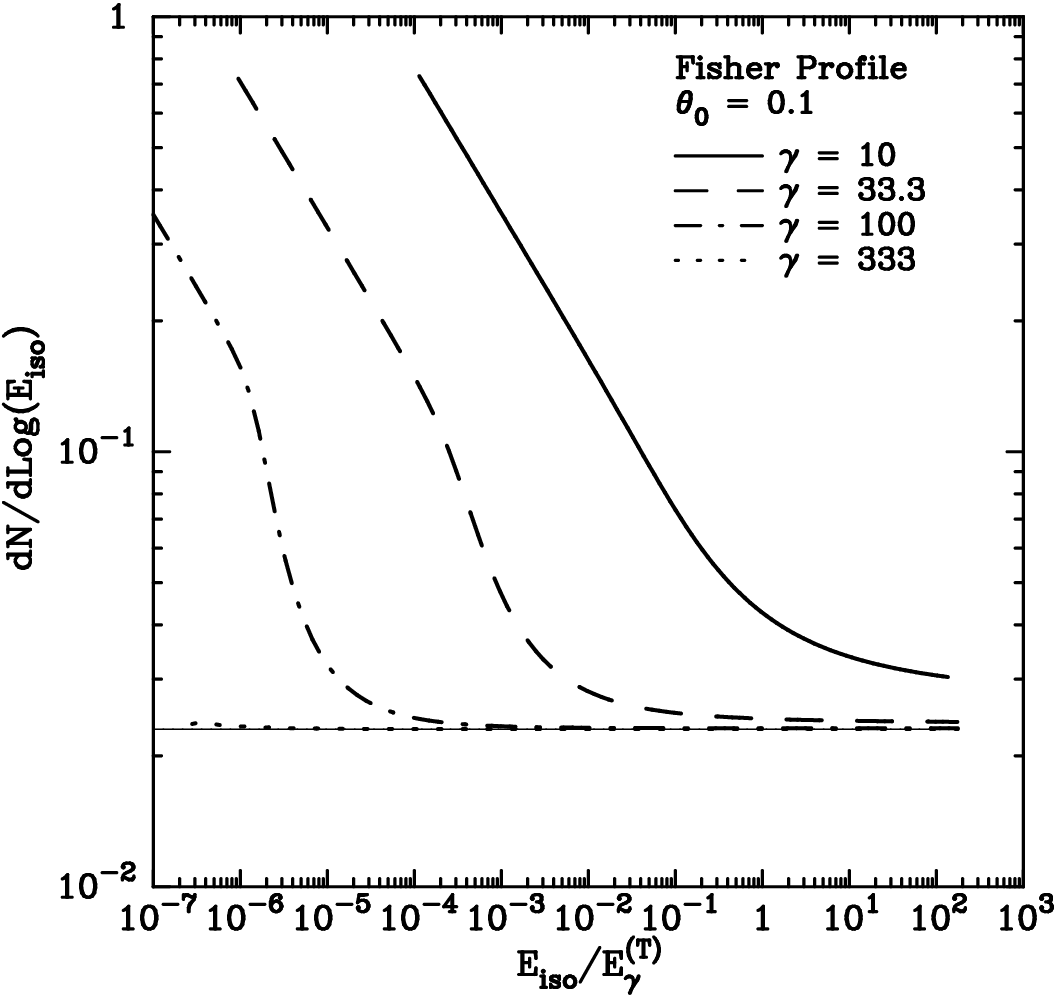}\\
\includegraphics[width=3.2in,clip=]{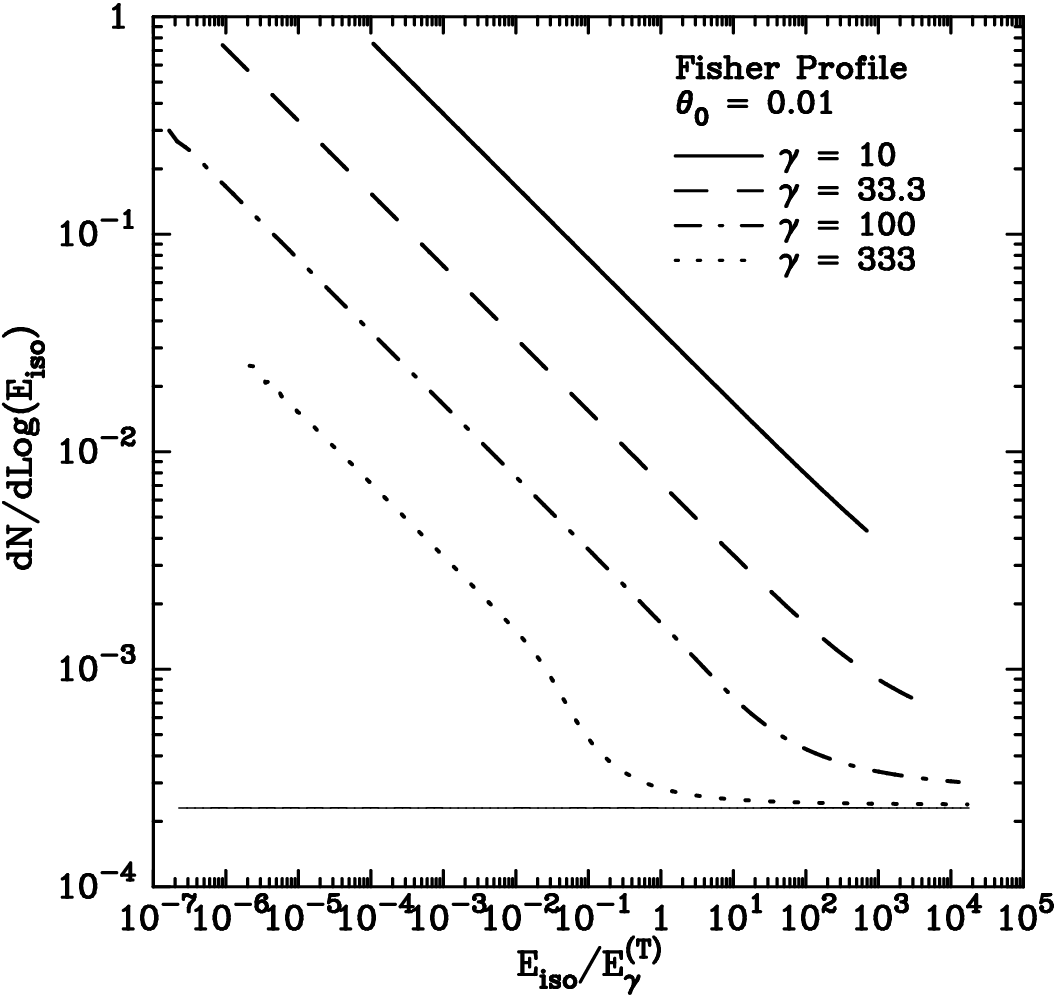}
\end{minipage}
\begin{minipage}[t]{3.2in}
\includegraphics[width=3.2in,clip=]{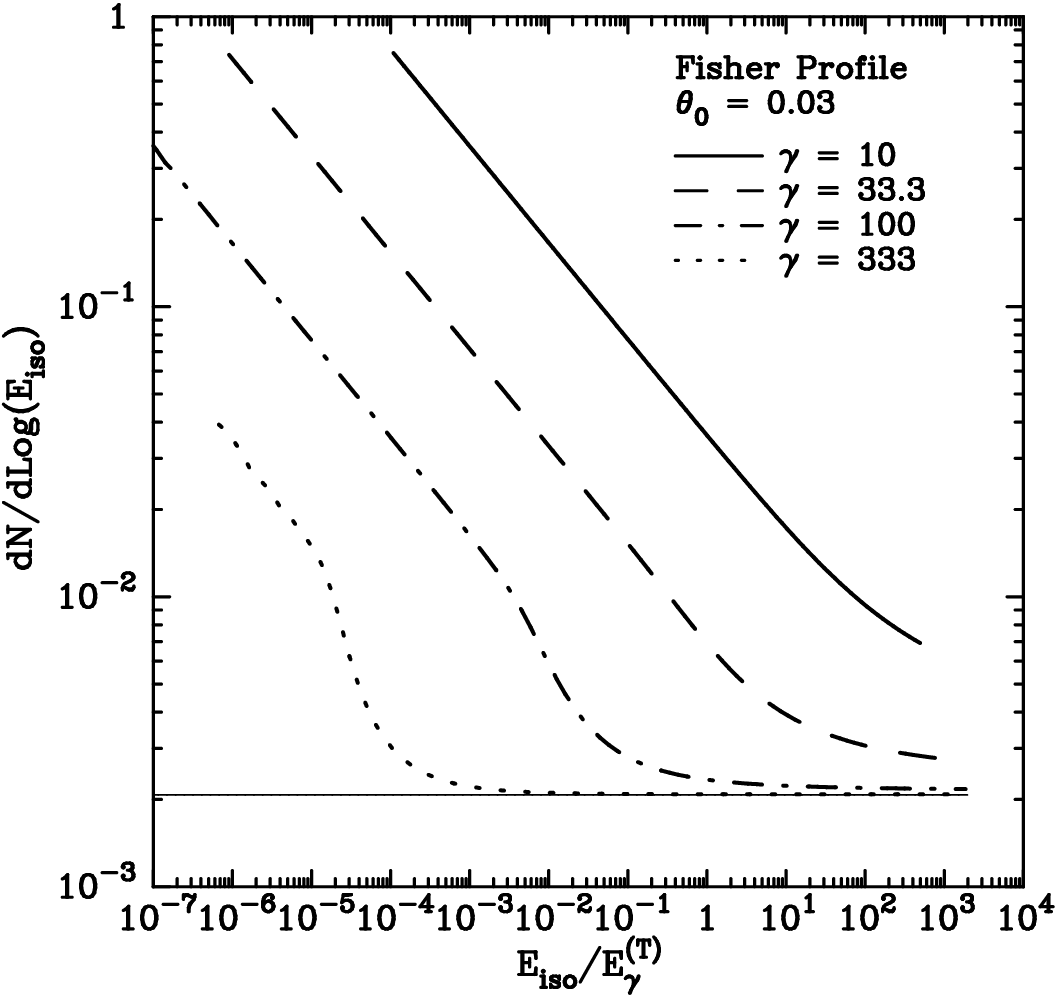}\\
\includegraphics[width=3.2in,clip=]{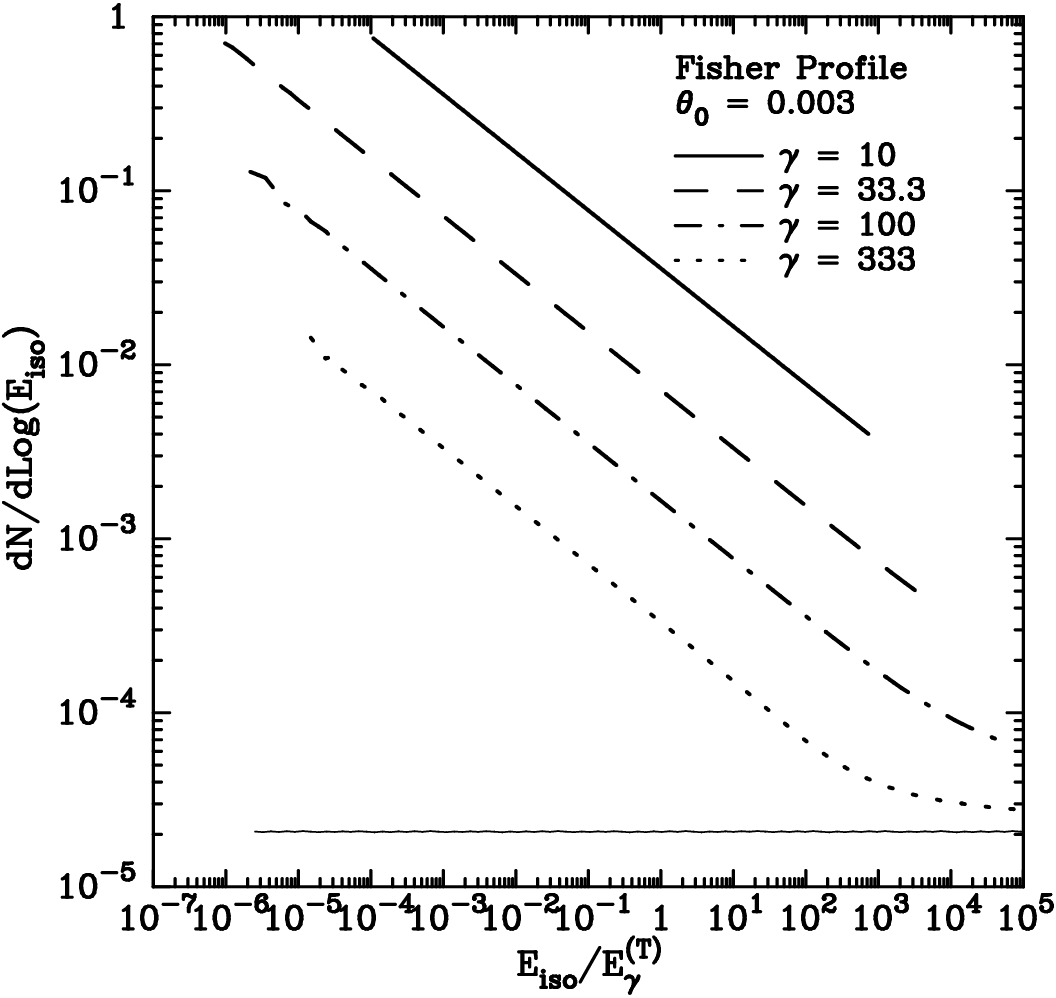}
\end{minipage}
\caption{$dN/d\log(\eiso)$ distributions for universal jet models based on
Fisher profiles.  Each panel shows the result for $\gamma$ = 10, 33.3,
100, and 333.  Top left:  $\theta_0=0.1$.  Top right:
$\theta_0=0.03$.  Bottom left: $\theta_0=0.01$.  Bottom right:
$\theta_0=0.003$.  The vertical scale is arbitrary.  The horizontal
light solid line shows the ``bare'' distribution due to the underlying
profile --- essentially the $\gamma\rightarrow\infty$ limit.}
\label{fisher_dndxiso}
\end{figure}

\begin{figure}[t]
\begin{minipage}[t]{3.2in}
\includegraphics[width=3.2in,clip=]{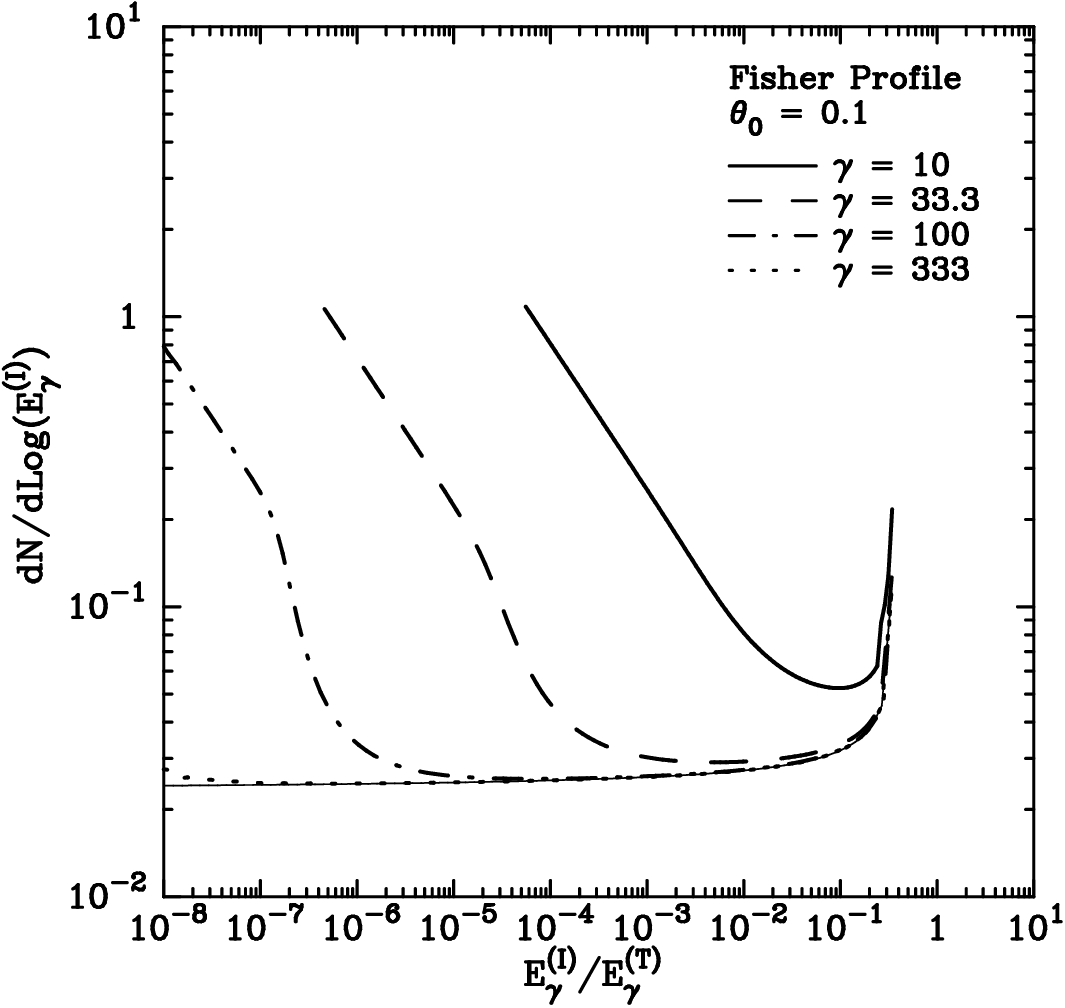}\\
\includegraphics[width=3.2in,clip=]{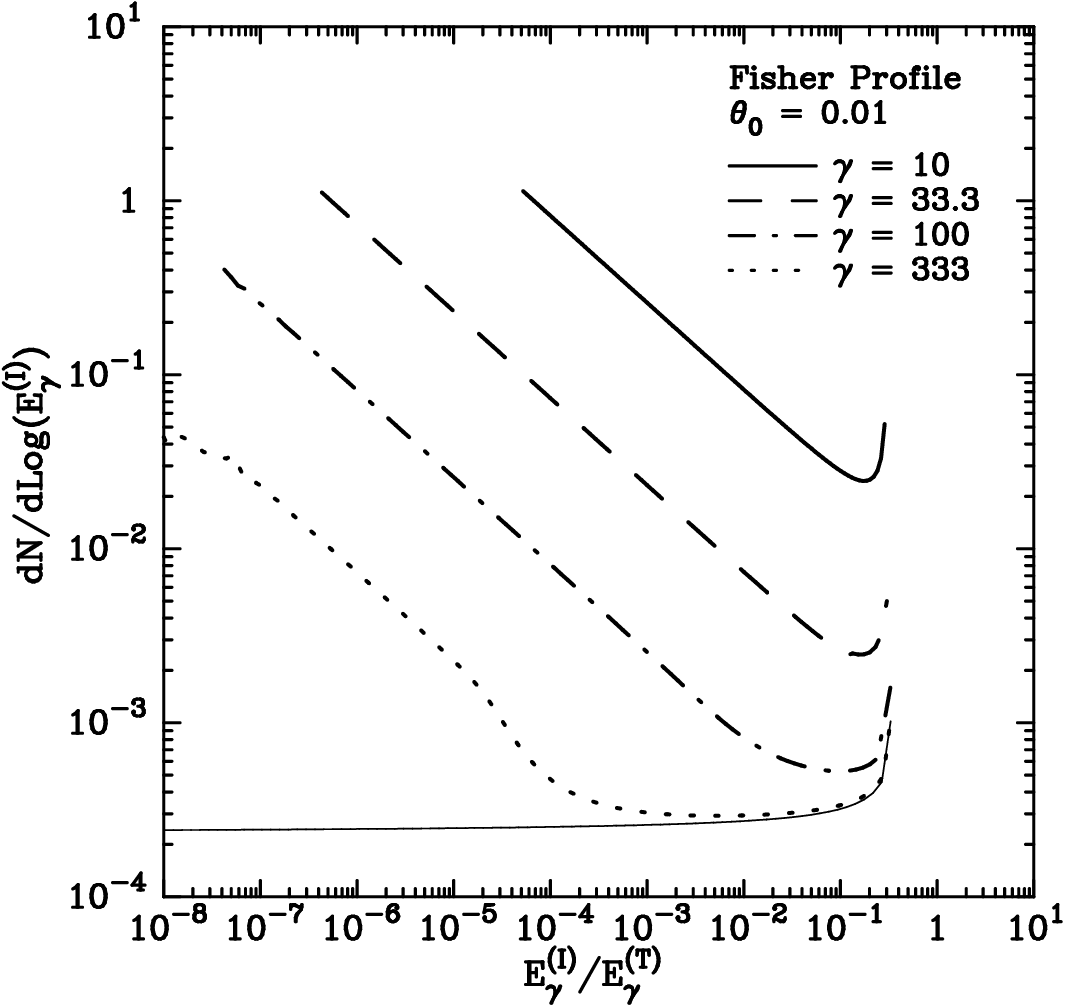}
\end{minipage}
\begin{minipage}[t]{3.2in}
\includegraphics[width=3.2in,clip=]{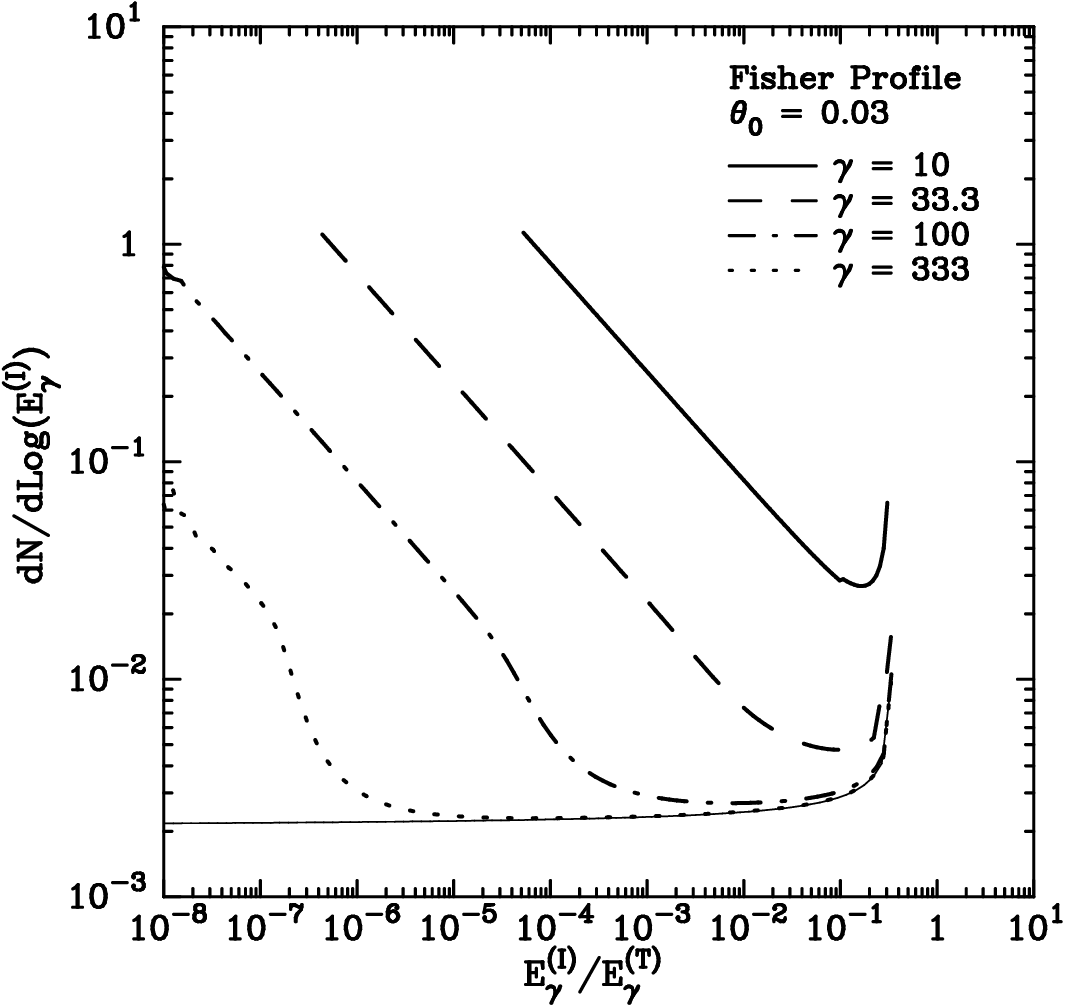}\\
\includegraphics[width=3.2in,clip=]{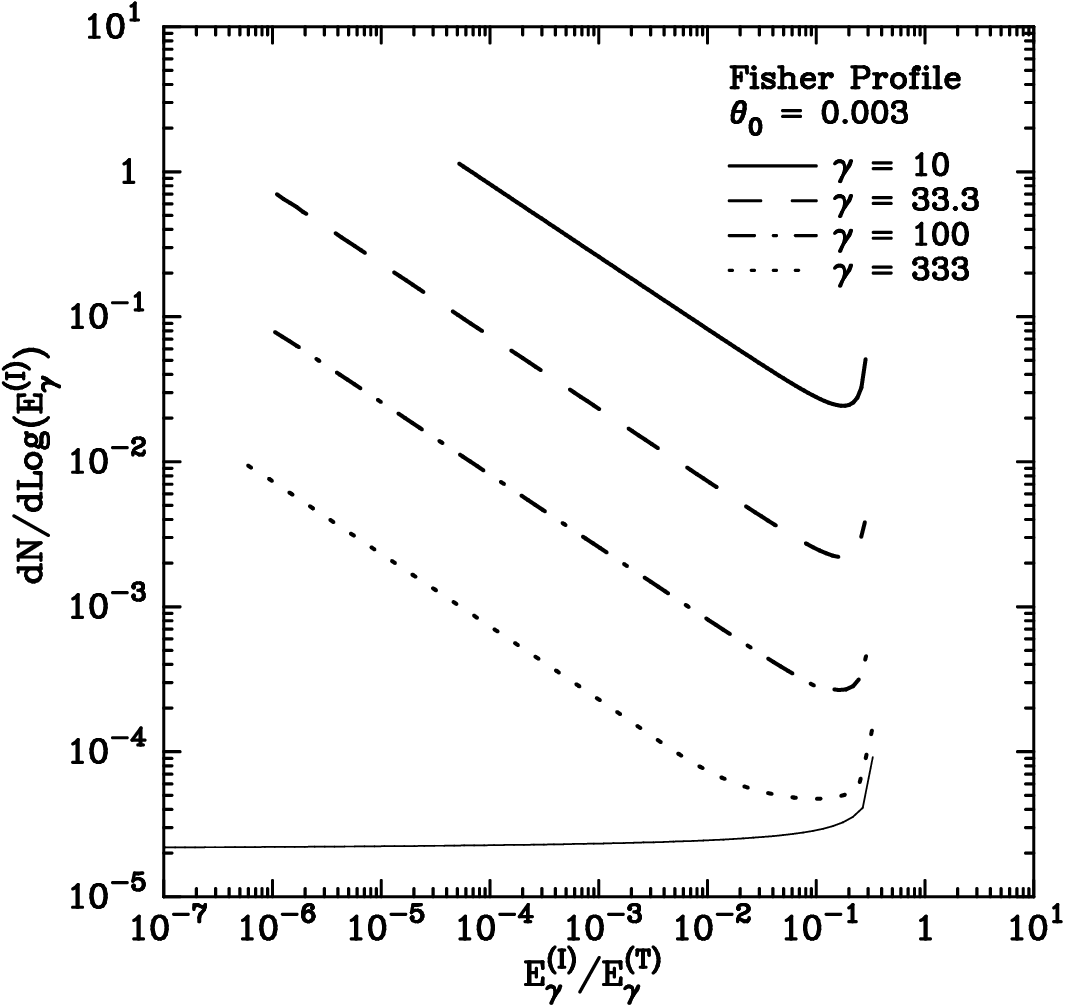}
\end{minipage}
\caption{$dN/d\log(\egi)$ distributions for universal jet models based on
Fisher profiles.  Each panel shows the result for $\gamma$ = 10, 33.3,
100, and 333.  Top left:  $\theta_0=0.1$.  Top right:
$\theta_0=0.03$.  Bottom left: $\theta_0=0.01$.  Bottom right:
$\theta_0=0.003$.  The vertical scale is arbitrary.  The light solid
line shows the ``bare'' distribution due to the underlying profile ---
essentially the $\gamma\rightarrow\infty$ limit.}
\label{fisher_dndxgi}
\end{figure}

The four panels of Fig.~\ref{fisher_dndxiso} show the effect of
relativistic kinematics on the uniform \eiso\ frequency distributions of
universal Fisher jets that were derived in \S\ref{fisherdist}.  In each
panel, the light solid line shows the ``bare'' value of the distribution,
corresponding to the (uniform) shape derived just from the profile. 
Essentially, this is the $\gamma\rightarrow\infty$ limit.

We see here that with $\gamma^{-1}\ll\theta_0$, the uniform distribution
that was built into the Fisher profile by construction is largely preserved
for bright events.  As $\theta_0$ decreases, the jet emission behaves more
and more like off-axis emission from a needle jet, which by
Eq.~(\ref{eiso_limit}) is a power-law with a slope of -1/3.  Accordingly,
the frequency distributions depart more and more from the uniform shape
that was due to the Fisher profile.

The four panels of Fig.~\ref{fisher_dndxgi} show the effect of relativistic
kinematics on the broad \egi\ distributions of universal Fisher jets that
were derived in \S\ref{fisherdist}.  The ``bare'' distribution is again
shown by the light solid line.  Just as for the \eiso\ frequency plots,
these distributions remain broad, in resemblance to the ``bare'' version,
until $\theta_0$ becomes so small that the emission is characteristic of
an off-axis needle jet, at which point they assume the shape of a
power-law with a slope of -1/2, as expected from Eq.~(\ref{egi_limit}).

%%%%%%%%%%%%%%%%%%%%%%%%%%%%%%%%%%%%%%%%%%%%%%%%%%%
%%%%%%%%%%%%%%%%%%%%%%%%%%%%%%%%%%%%%%%%%%%%%%%%%%%
\subsection{$\delta=1$ Power-Law Jet Profiles}
\label{sec_pl_1.0}

\subsubsection{Emission}

\begin{figure}[t]
\begin{minipage}[t]{3.2in}
\includegraphics[width=3.2in,clip=]{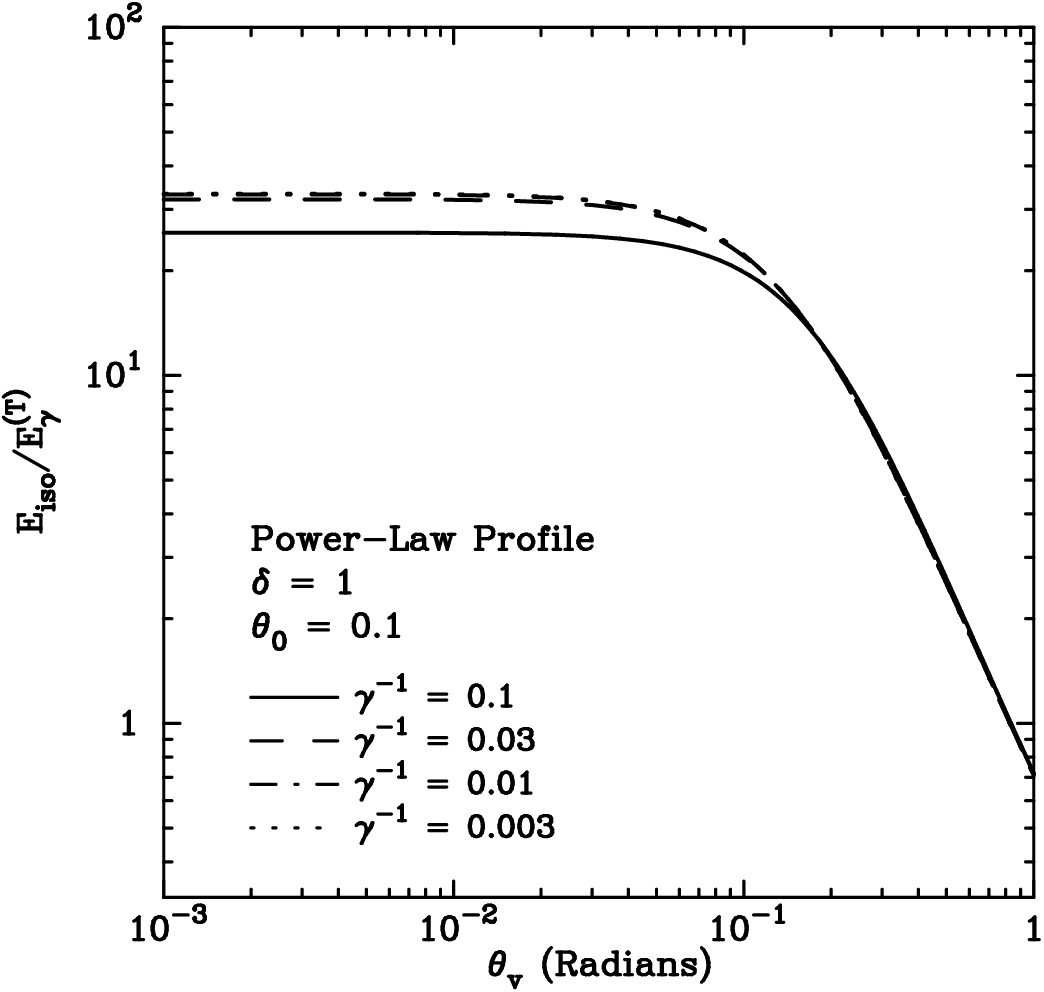}\\
\includegraphics[width=3.2in,clip=]{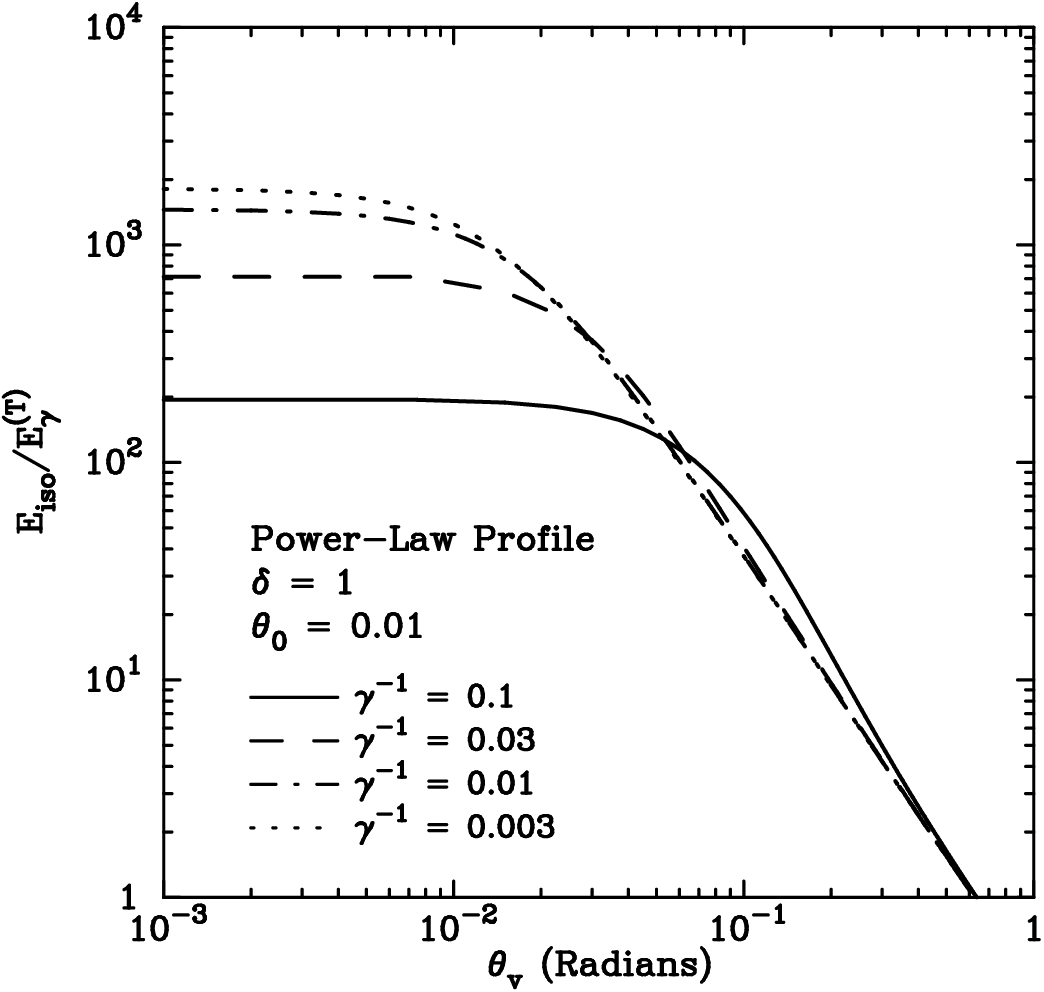}
\end{minipage}
\begin{minipage}[t]{3.2in}
\includegraphics[width=3.2in,clip=]{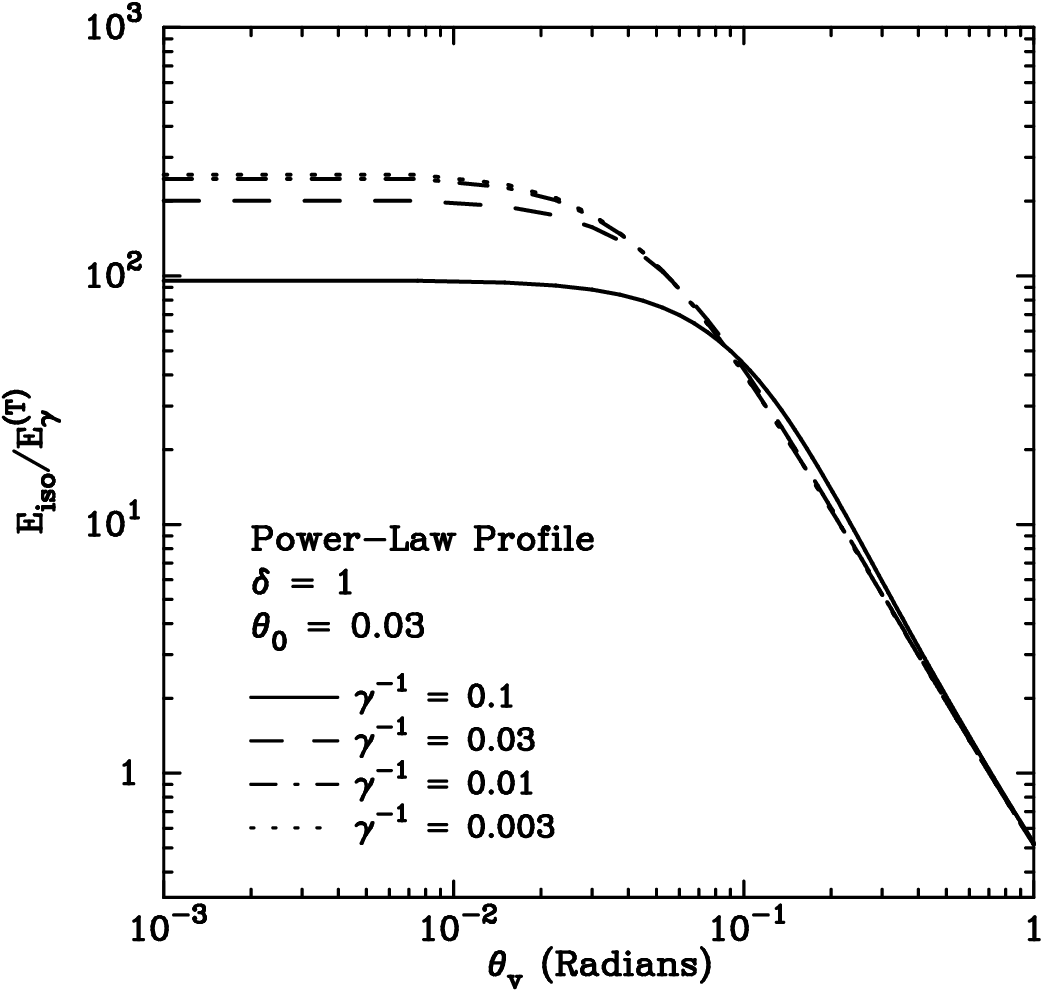}\\
\includegraphics[width=3.2in,clip=]{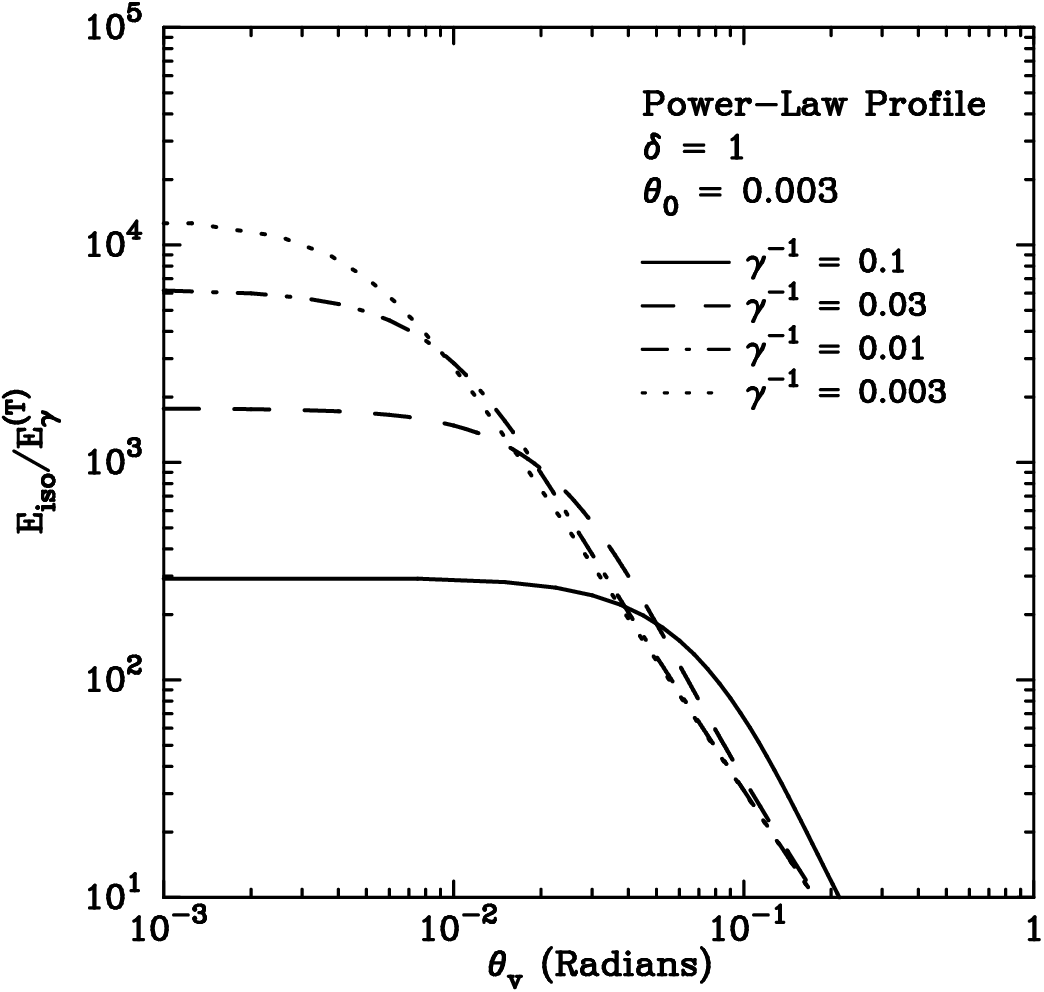}
\end{minipage}
\caption{\eiso\ as a function of viewing angle for emission from
relativistic jets with $\delta=1$ power-law emission profiles.  These
figures help illustrate the fact that $\gamma^{-1}$ sets a lower limit on
the effective angular size of the jet.  Each panel shows the result for
$\gamma^{-1}$ = 0.1, 0.03, 0.01, and 0.003.  Top left:  $\theta_0=0.1$. 
Top right: $\theta_0=0.03$.  Bottom left: $\theta_0=0.01$.  Bottom right:
$\theta_0=0.003$.}
\label{pl_1.0_emission}
\end{figure}

The four panels of Fig.~\ref{pl_1.0_emission} show the behavior of \eiso\ 
as a function of viewing angle $\theta_v$ for the power-law profile with
index $\delta=1$.  Once again, we see the ``jet-broadening'' effect, in
which the effective angular size of the jet radiation pattern is set by
whichever is larger of $\theta_0$ and $\gamma^{-1}$.  These radiation
patterns have broad tails in every case, reflecting the fact that this
profile falls off more gradually than the Doppler photon distribution
function $\gamma^{-2}(1-\beta\mu)^{-2}$.

\begin{figure}[t]
\begin{minipage}[t]{3.2in}
\includegraphics[width=3.2in,clip=]{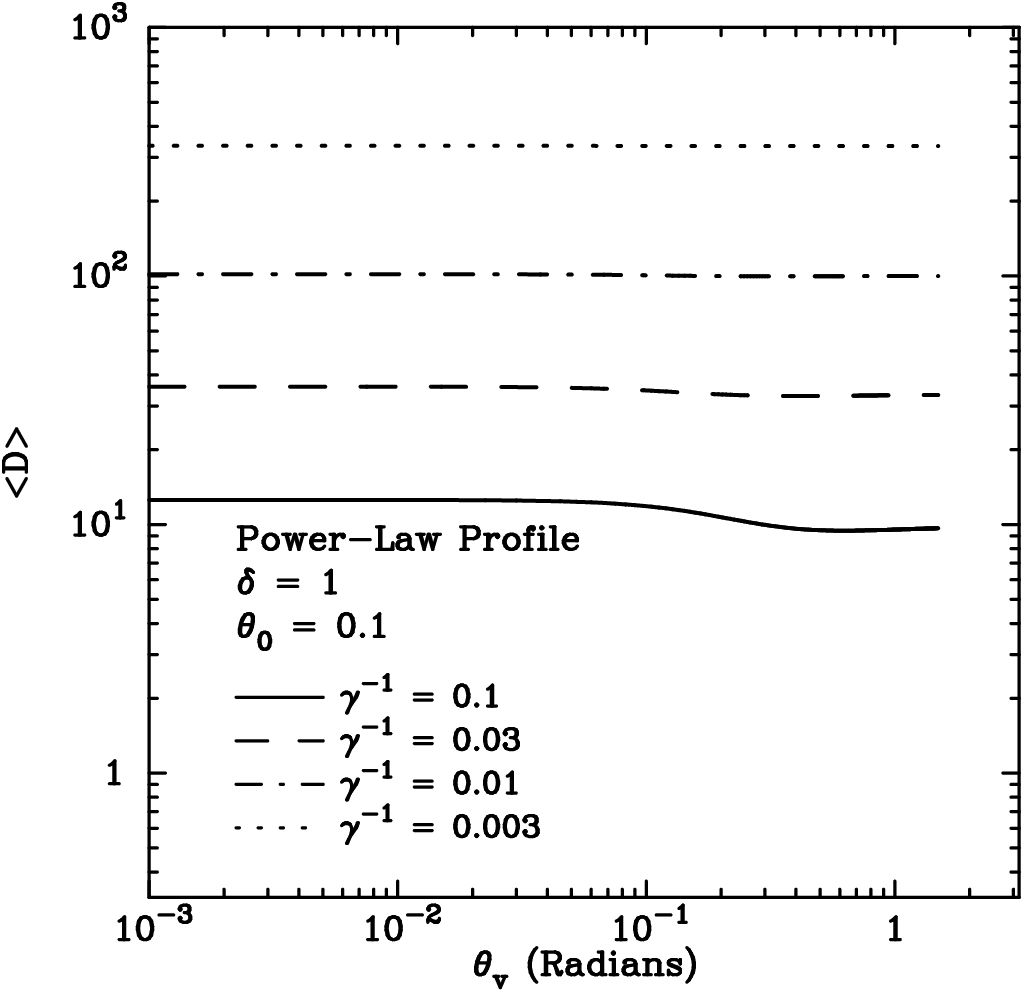}\\
\includegraphics[width=3.2in,clip=]{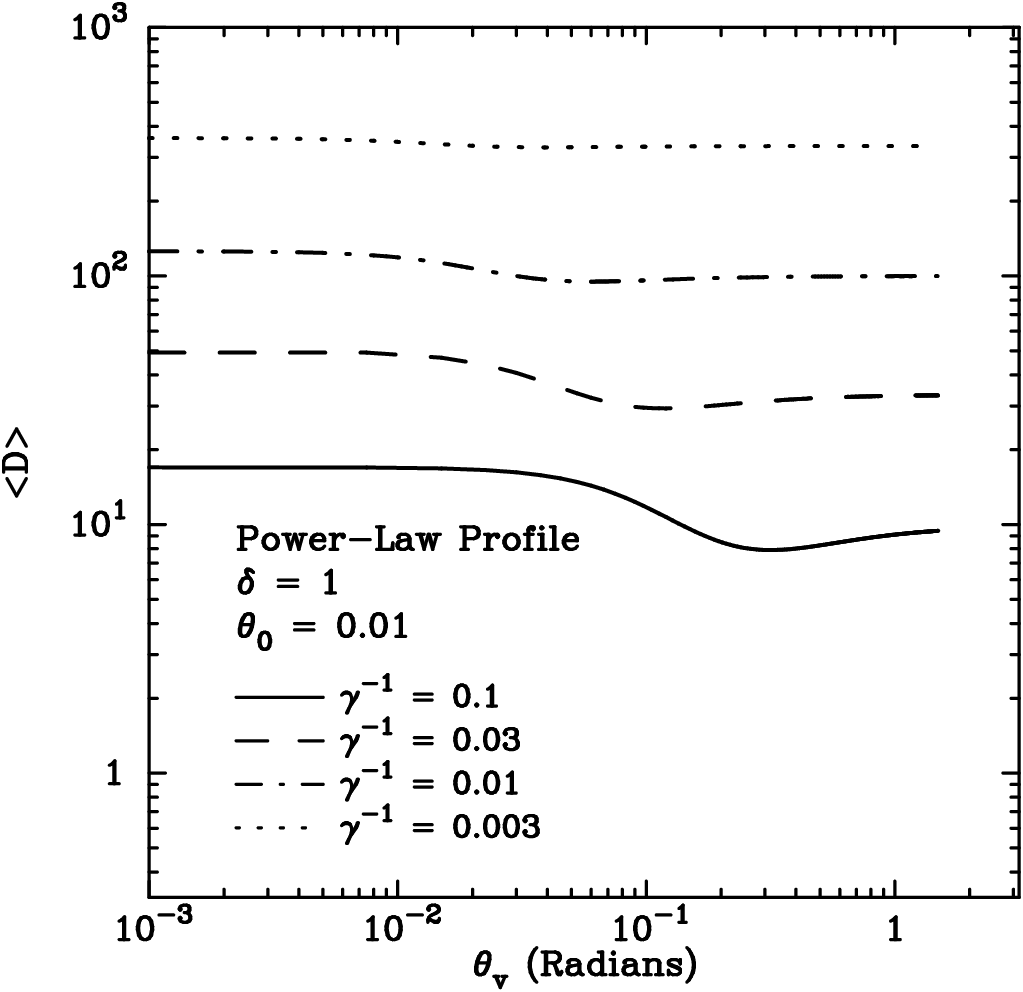}
\end{minipage}
\begin{minipage}[t]{3.2in}
\includegraphics[width=3.2in,clip=]{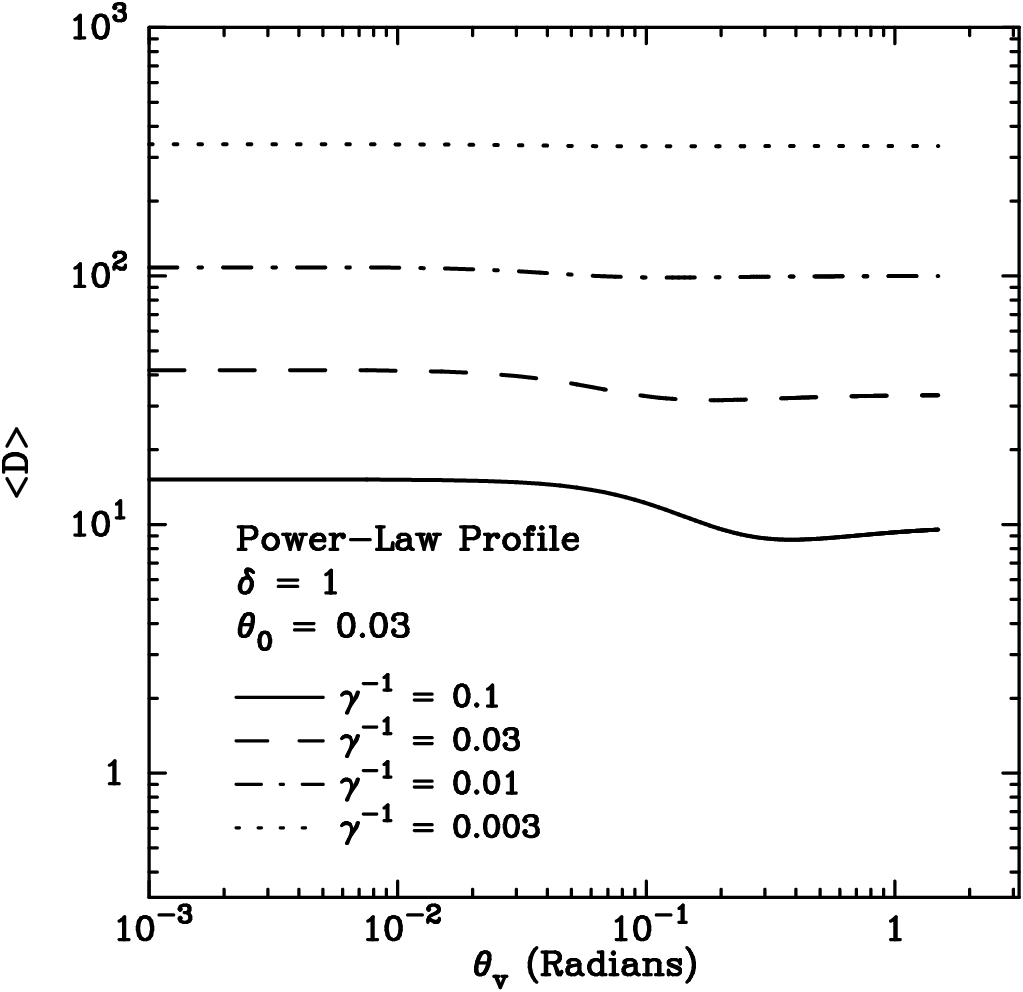}\\
\includegraphics[width=3.2in,clip=]{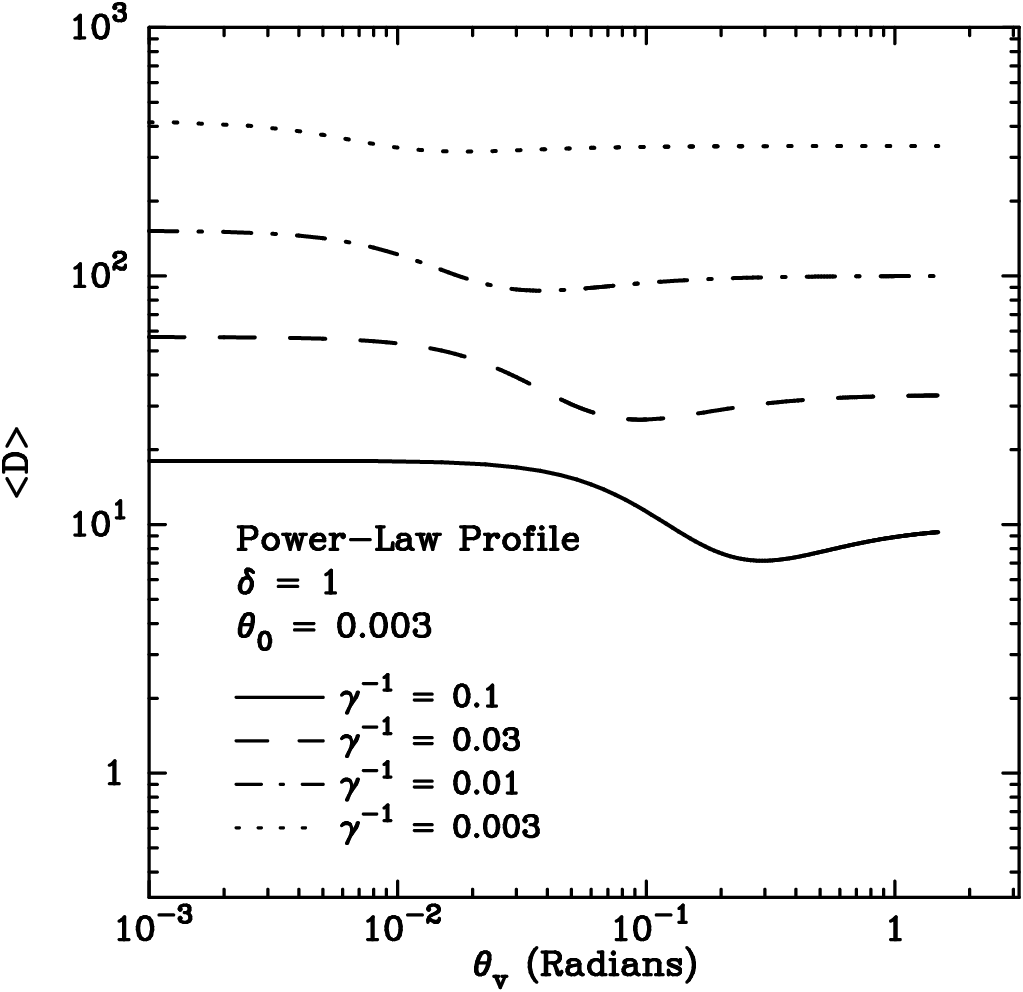}
\end{minipage}
\caption{Photon-number-averaged Doppler factor \davg\ as a function of
viewing angle for emission from relativistic jets with $\delta=1$
power-law emission profiles. Each panel shows the result for $\gamma^{-1}$
= 0.1, 0.03, 0.01, and 0.003.  Top left:  $\theta_0=0.1$.  Top right:
$\theta_0=0.03$.  Bottom left: $\theta_0=0.01$.  Bottom right:
$\theta_0=0.003$.  At large $\theta_v$, $\davg\rightarrow\gamma$.  It is
noteworthy that the dynamic range of \davg\ is small compared to the other
profiles we consider.}
\label{pl_1.0_epk}
\end{figure}

The four panels of Fig.~\ref{pl_1.0_epk} show the behavior of \davg\ as a
function of viewing angle $\theta_v$.  The striking feature of these plots is
the extremely restricted dynamic range of \davg.

The large-$\theta_v$ limit in each case is $\davg=\gamma$, as is expected
by the fact that far from the jet axis, this profile falls so gradually
that it looks nearly uniform (recall from \S\ref{peake} that $\gamma$ is
the ``uniform outflow'' limit of \davg).  This reason for this
large-$\theta_v$ limit of \davg\ also helps explain why the dynamic range
of \davg\ is so limited --- this profile varies quite gently almost
everywhere, so its \davg\ deviates only moderately from $\gamma$ almost
everywhere.  The only real chance that \davg\ has to run away is near
$\theta_v=0$, but even here it is eventually reined in by the cutoff.  The
smaller $\theta_0$, the wider the available range of \davg, but it is
clear that uncomfortably small values of $\theta_0$ will be required in
order to allow this profile to produce the kind of range of \epk\ --- at
least two orders of magnitude --- that is observed in GRBs.

\subsubsection{\eiso\ -- \davg\ and \egi\ -- \davg\ Correlations}

\begin{figure}[t]
\begin{minipage}[t]{3.2in}
\includegraphics[width=3.2in,clip=]{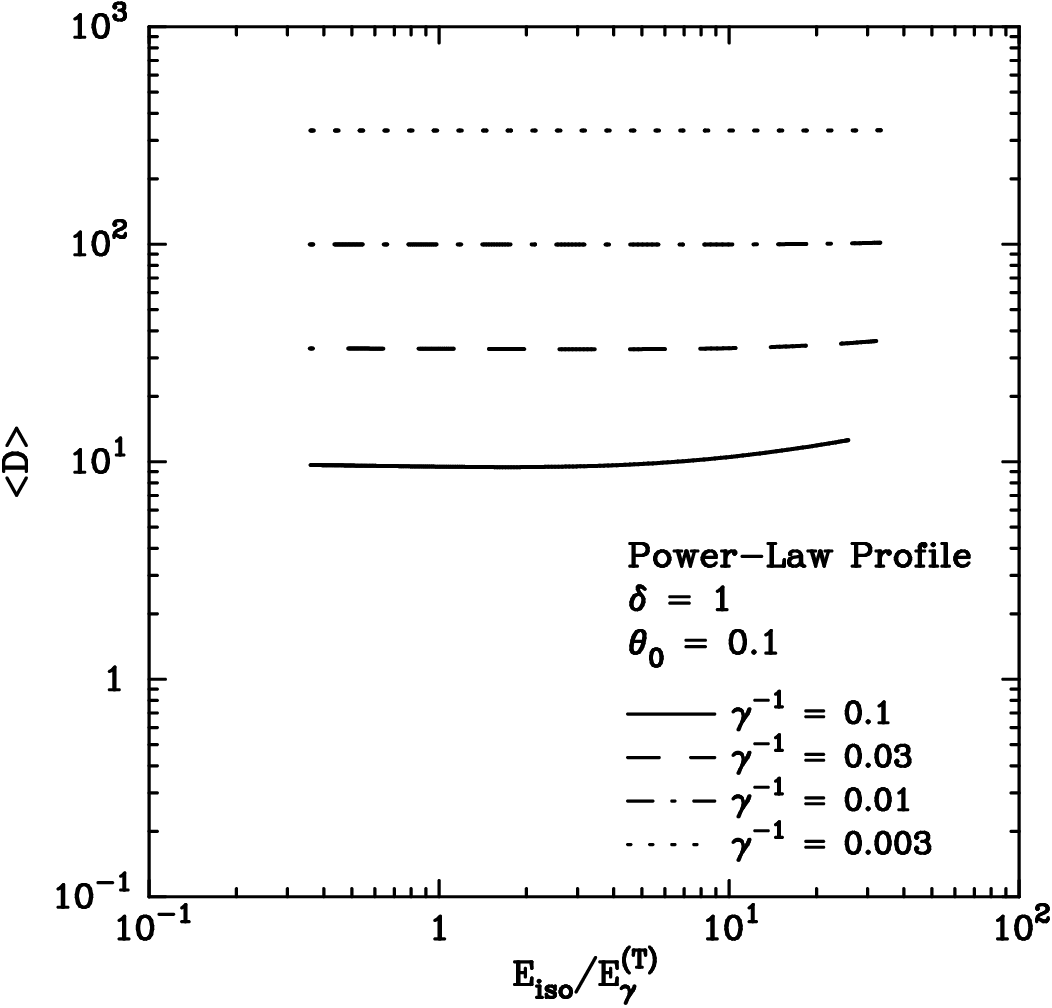}\\
\includegraphics[width=3.2in,clip=]{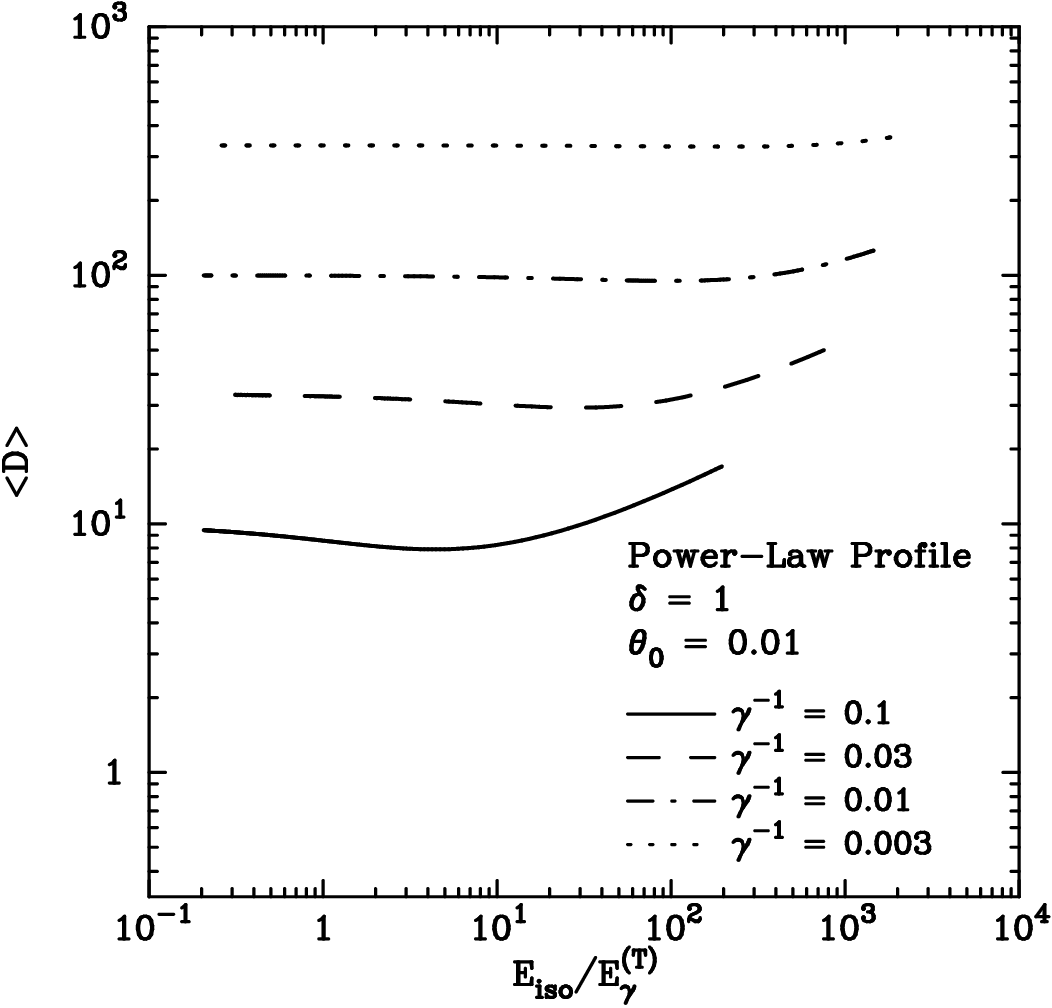}
\end{minipage}
\begin{minipage}[t]{3.2in}
\includegraphics[width=3.2in,clip=]{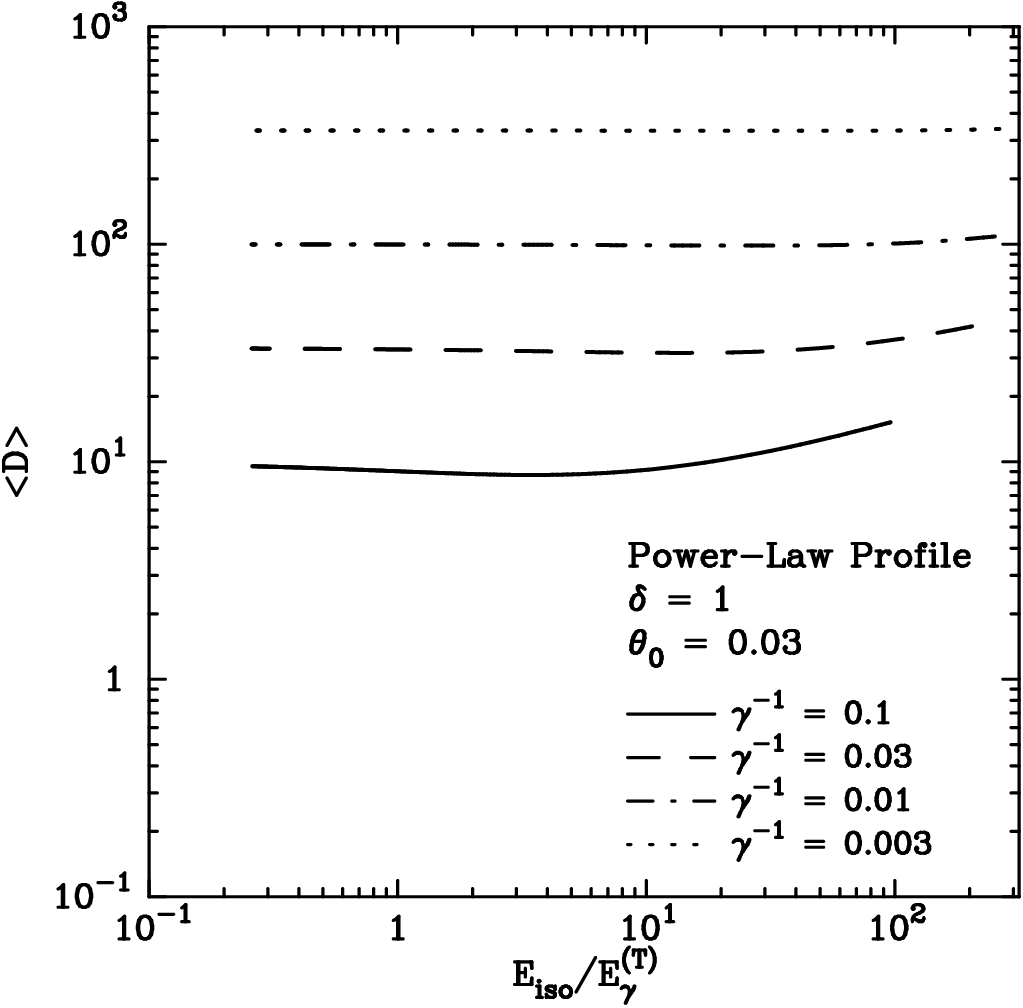}\\
\includegraphics[width=3.2in,clip=]{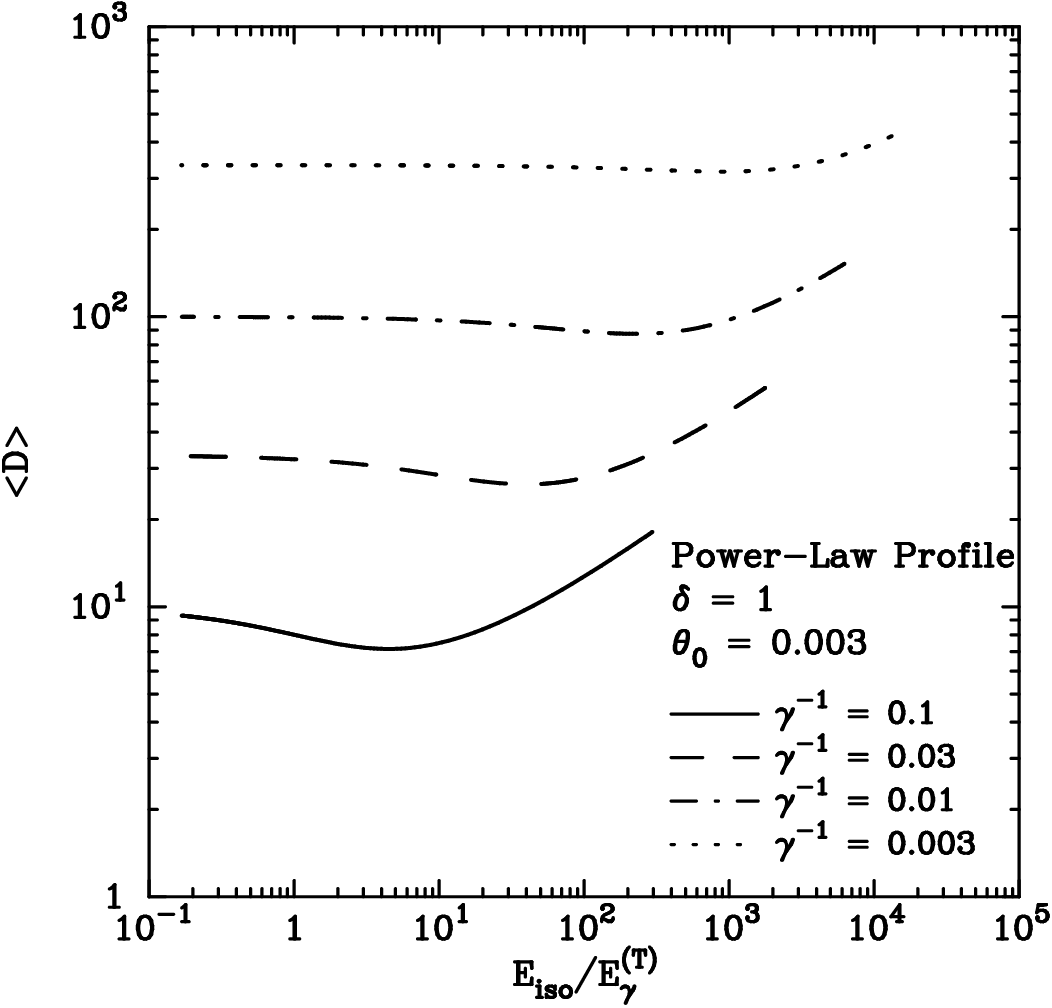}
\end{minipage}
\caption{\eiso--\davg\ (``Amati'') plots for universal jet models based on
$\delta=1$ power-law profiles.  Each panel shows the result for $\gamma^{-1}$ =
0.1, 0.03, 0.01, and 0.003.  Top left:  $\theta_0=0.1$.  Top right:
$\theta_0=0.03$.  Bottom left: $\theta_0=0.01$.  Bottom right:
$\theta_0=0.003$.}
\label{pl_1.0_amati}
\end{figure}

The four panels of Fig.~\ref{pl_1.0_amati} show the behavior of the
\eiso--\davg\ (``Amati'') correlation in universal $\delta=1$ power-law
models with our standard parameter choices.

Here we see the some of the trouble created by the limited dynamical range
of \davg.  The curves stay more-or-less flat for much of the dynamical
range of \eiso.  The high-$\gamma$ curves do suddenly rear up at high
\eiso, but not nearly enough to make a convincing Amati plot, and in any
event, comparison with Figs.~\ref{pl_1.0_epk} and \ref{pl_1.0_emission}
show that this is going on over a very small region of $\theta_v$,
corresponding to a tiny fraction of observed bursts in a universal jet
model.

\begin{figure}[t]
\begin{minipage}[t]{3.2in}
\includegraphics[width=3.2in,clip=]{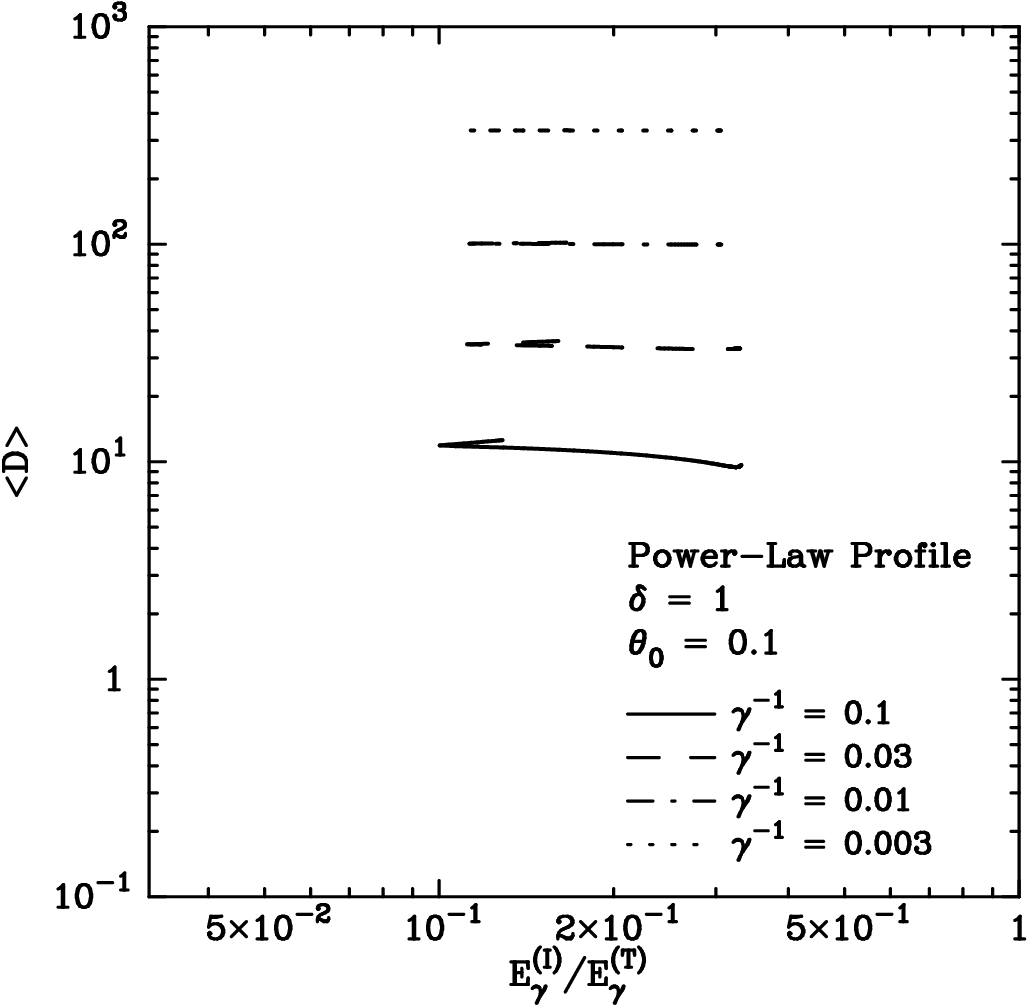}\\
\includegraphics[width=3.2in,clip=]{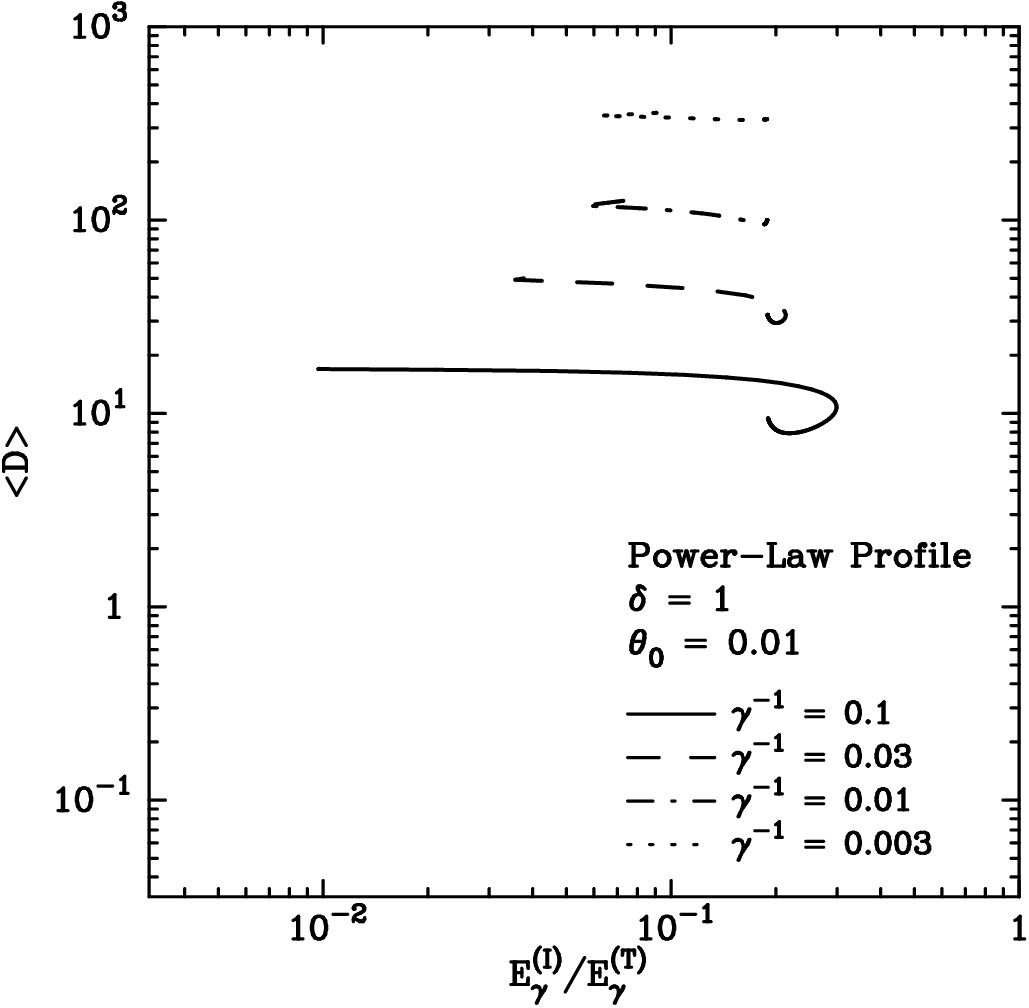}
\end{minipage}
\begin{minipage}[t]{3.2in}
\includegraphics[width=3.2in,clip=]{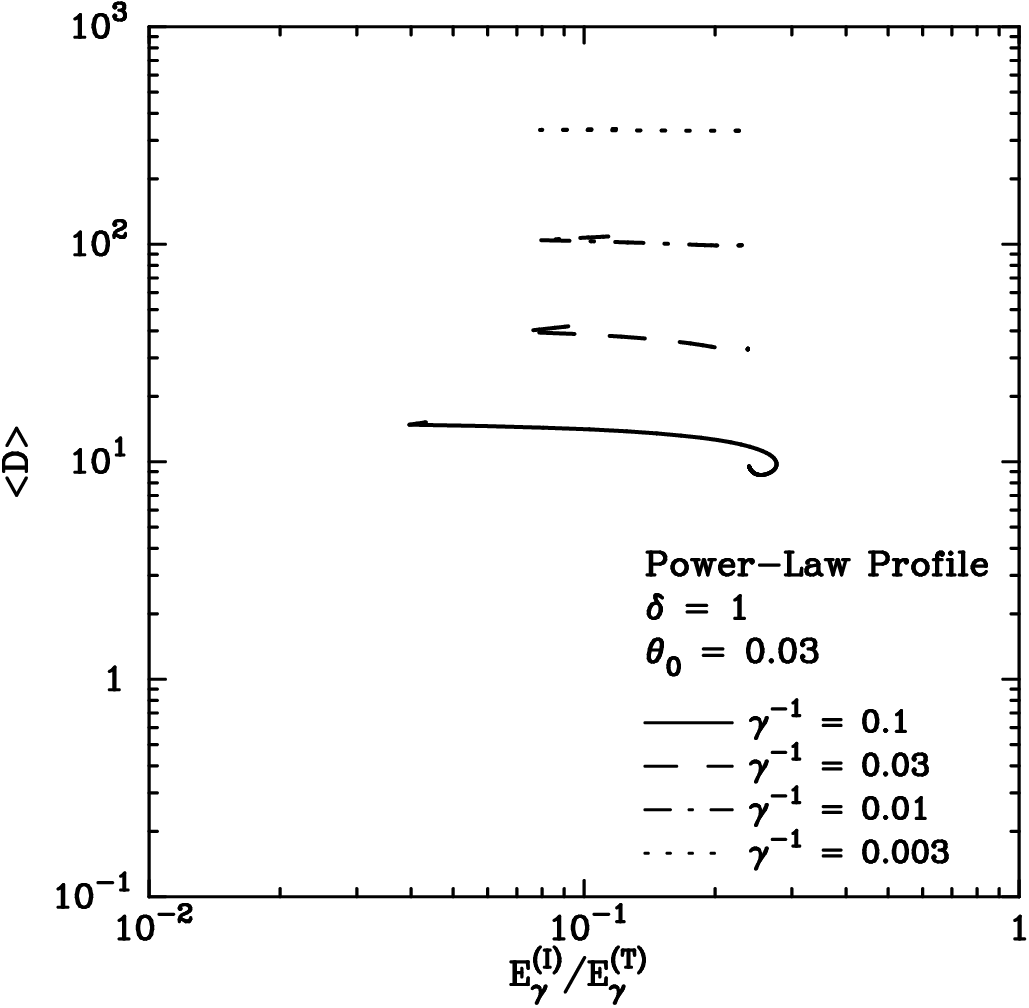}\\
\includegraphics[width=3.2in,clip=]{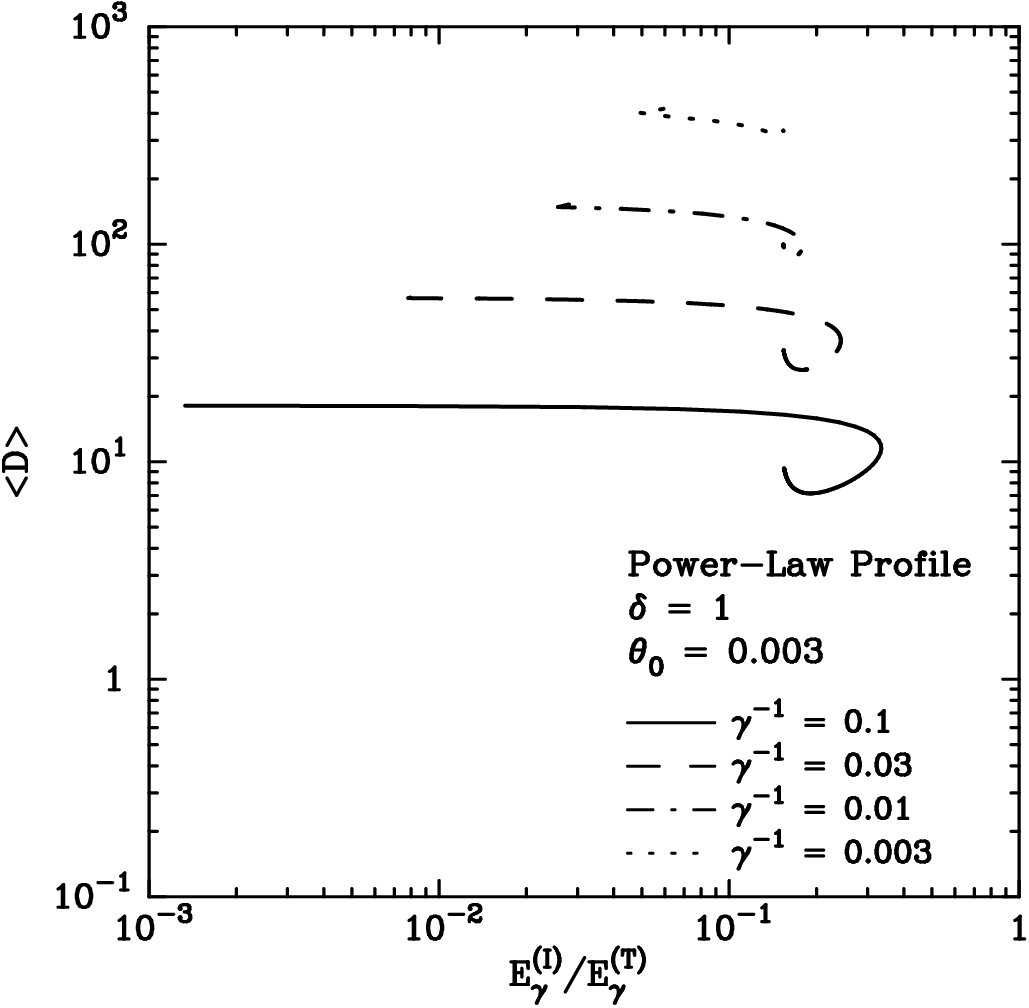}
\end{minipage}
\caption{\egi--\davg\ (``Ghirlanda'') plots for universal jet models based
on $\delta=1$ power-law profiles.  Each panel shows the result for $\gamma^{-1}$ =
0.1, 0.03, 0.01, and 0.003.  Top left:  $\theta_0=0.1$.  Top right:
$\theta_0=0.03$.  Bottom left: $\theta_0=0.01$.  Bottom right:
$\theta_0=0.003$.}
\label{pl_1.0_ghirlanda}
\end{figure}

The four panels of Fig.~\ref{pl_1.0_ghirlanda} show the behavior of the
\egi--\davg\ (``Ghirlanda'') correlation in universal $\delta=1$ power-law
models with our standard parameter choices.

There seems to be no hope for a real ``Ghirlanda'' relation with this
profile.  Due to the limited dynamic range of \davg, the trajectory is
essentially flat, with a hook-like structure at the very end due to the
same side-effect of relativistic jet-broadening that was remarked upon in
\S\ref{fisher_corr}.

\subsubsection{\eiso\ and \egi\ Frequency Distributions}

\begin{figure}[t]
\begin{minipage}[t]{3.2in}
\includegraphics[width=3.2in,clip=]{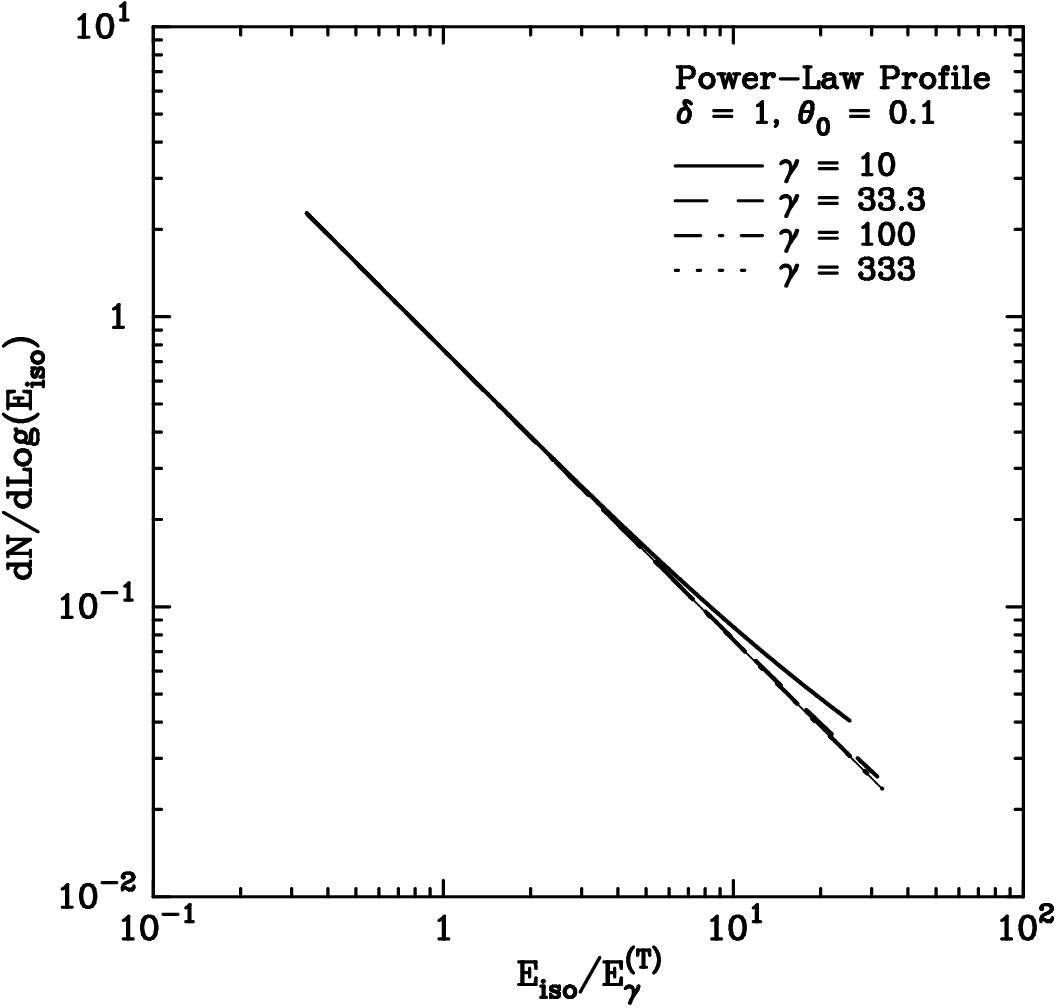}\\
\includegraphics[width=3.2in,clip=]{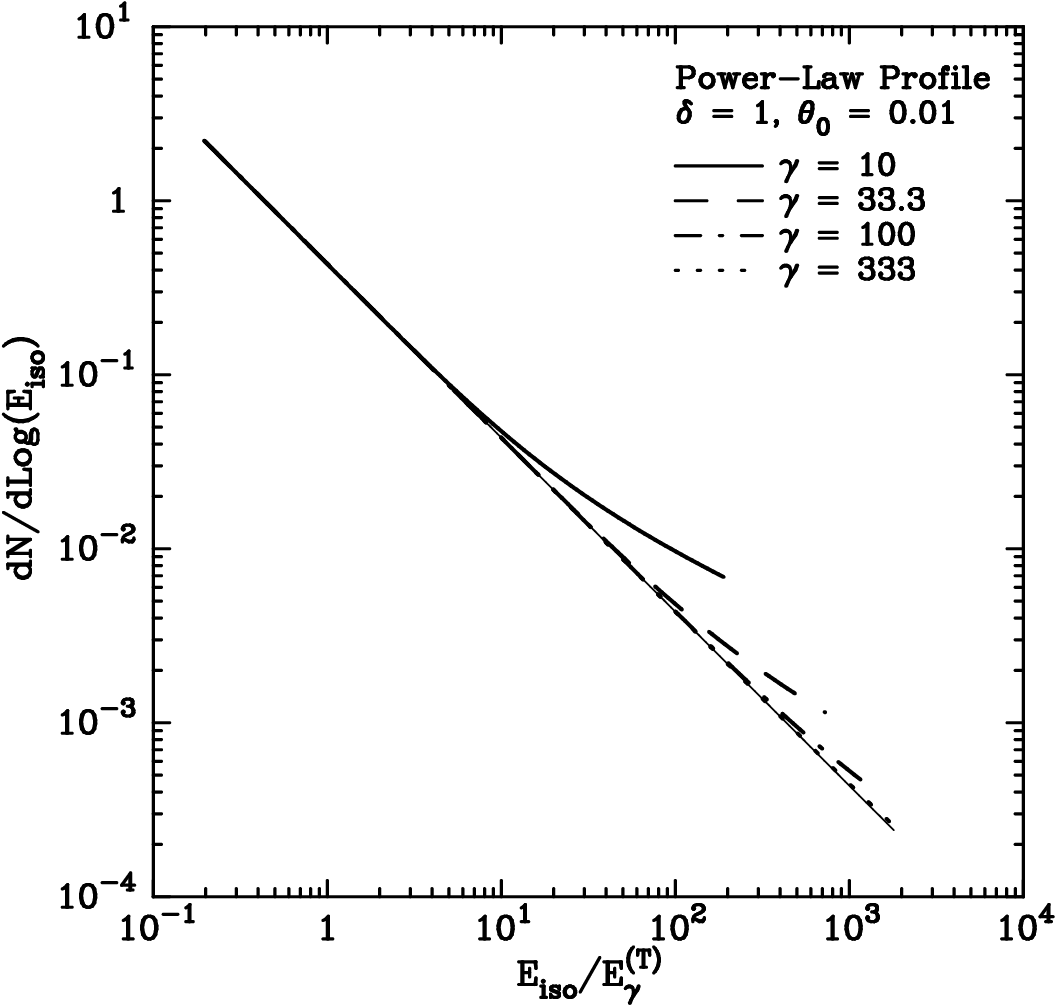}
\end{minipage}
\begin{minipage}[t]{3.2in}
\includegraphics[width=3.2in,clip=]{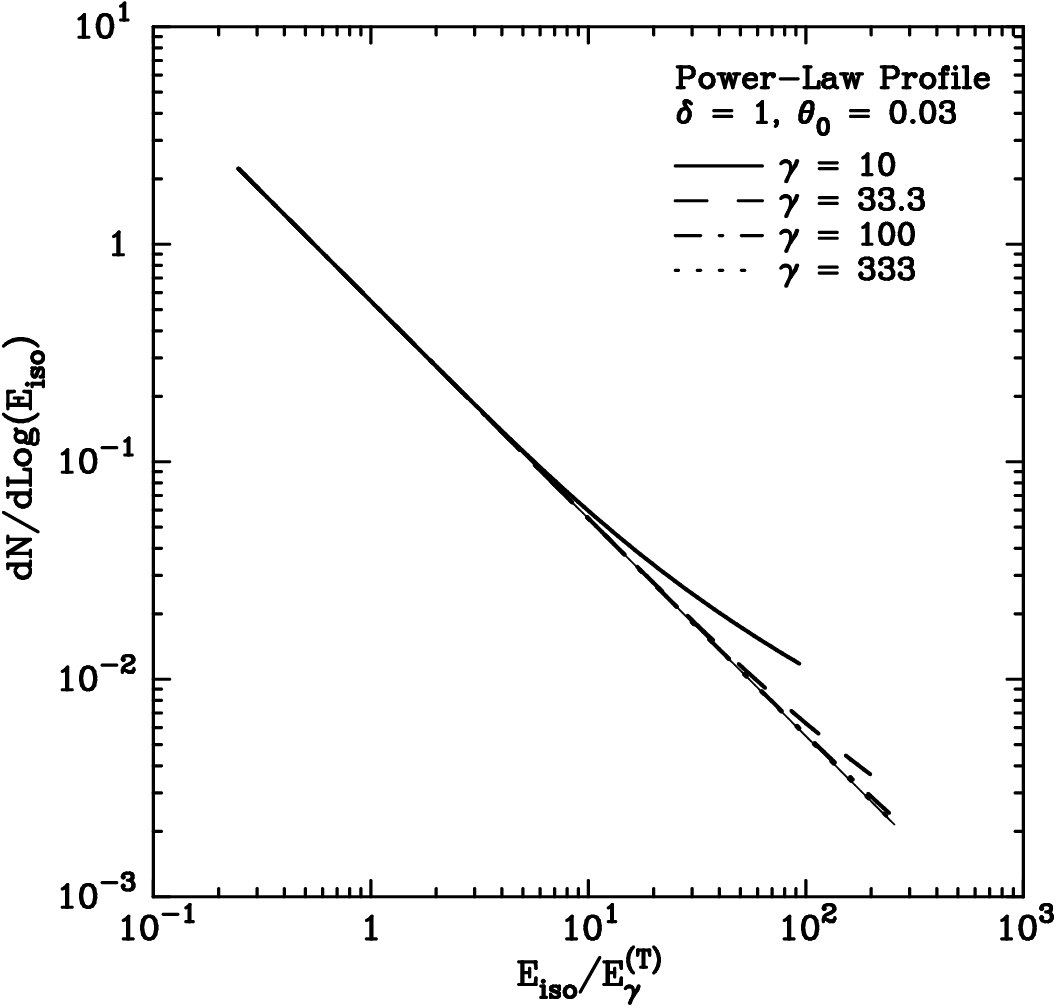}\\
\includegraphics[width=3.2in,clip=]{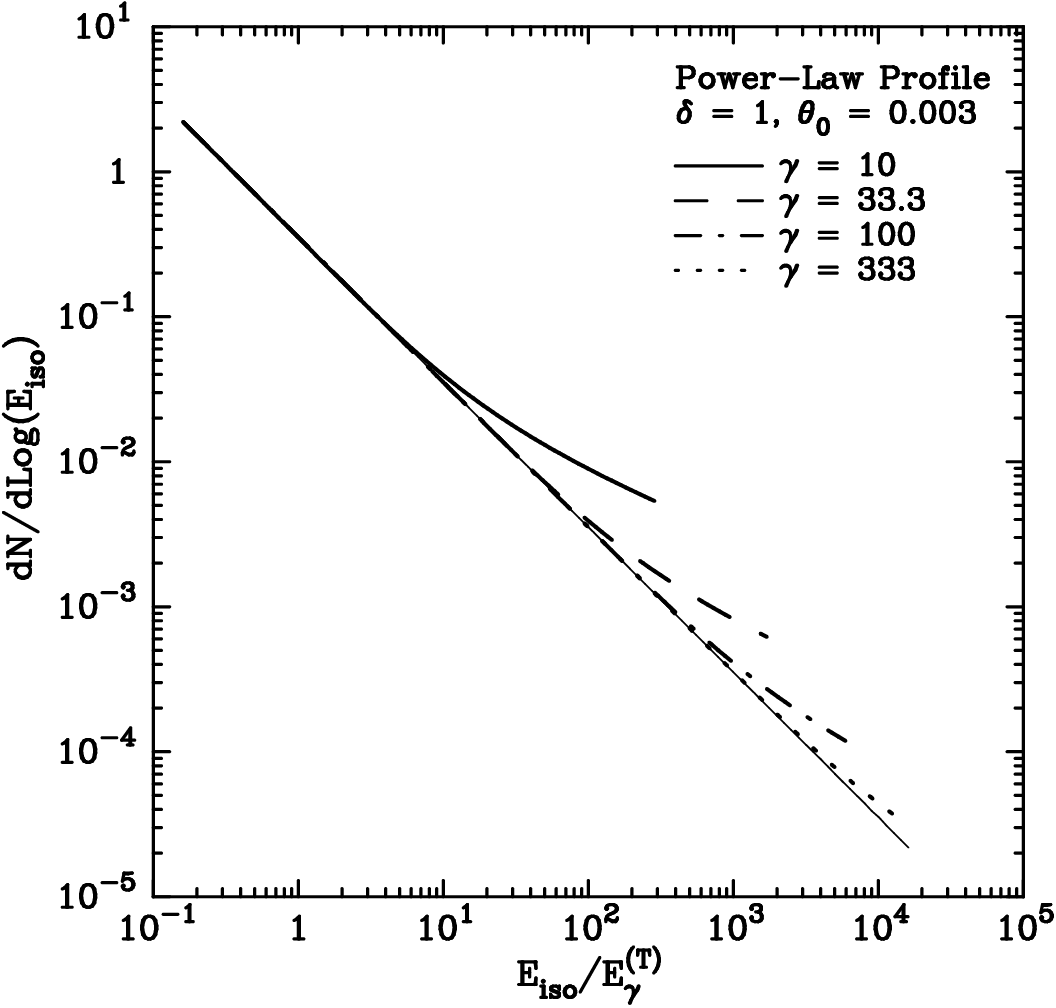}
\end{minipage}
\caption{$dN/d\log(\eiso)$ distributions for universal jet models based on
$\delta=1$ power-law profiles.  Each panel shows the result for $\gamma$ =
10, 33.3, 100, and 333.  Top left:  $\theta_0=0.1$.  Top right:
$\theta_0=0.03$.  Bottom left: $\theta_0=0.01$.  Bottom right:
$\theta_0=0.003$.  The vertical scale is arbitrary.  The light solid line
shows the ``bare'' distribution due to the underlying profile ---
essentially the $\gamma\rightarrow\infty$ limit.  Its slope is -1, as
expected from Eq.~\ref{plxiso}.}
\label{pl_1.0_dndxiso}
\end{figure}

The four panels of Fig.~\ref{pl_1.0_dndxiso} show the effect of
relativistic kinematics on the \eiso\ frequency distribution of $\delta=1$
power-law jets --- see Eq.~(\ref{plxiso}).  Again, the ``bare''
($\gamma\rightarrow\infty$) distribution is shown by a light solid line, whose
slope in this case is -1.

We see here that the distribution of $dN/d\eiso$ appears to be
irreducibly narrow, and relativistic kinematics are powerless to broaden
it.  From this it appears that this ``extremal'' power-law profile cannot
under any circumstances reproduce the broad --- and very nearly uniform in
$\log\eiso$ --- distribution that is observed empirically in GRBs.

\begin{figure}[t]
\begin{minipage}[t]{3.2in}
\includegraphics[width=3.2in,clip=]{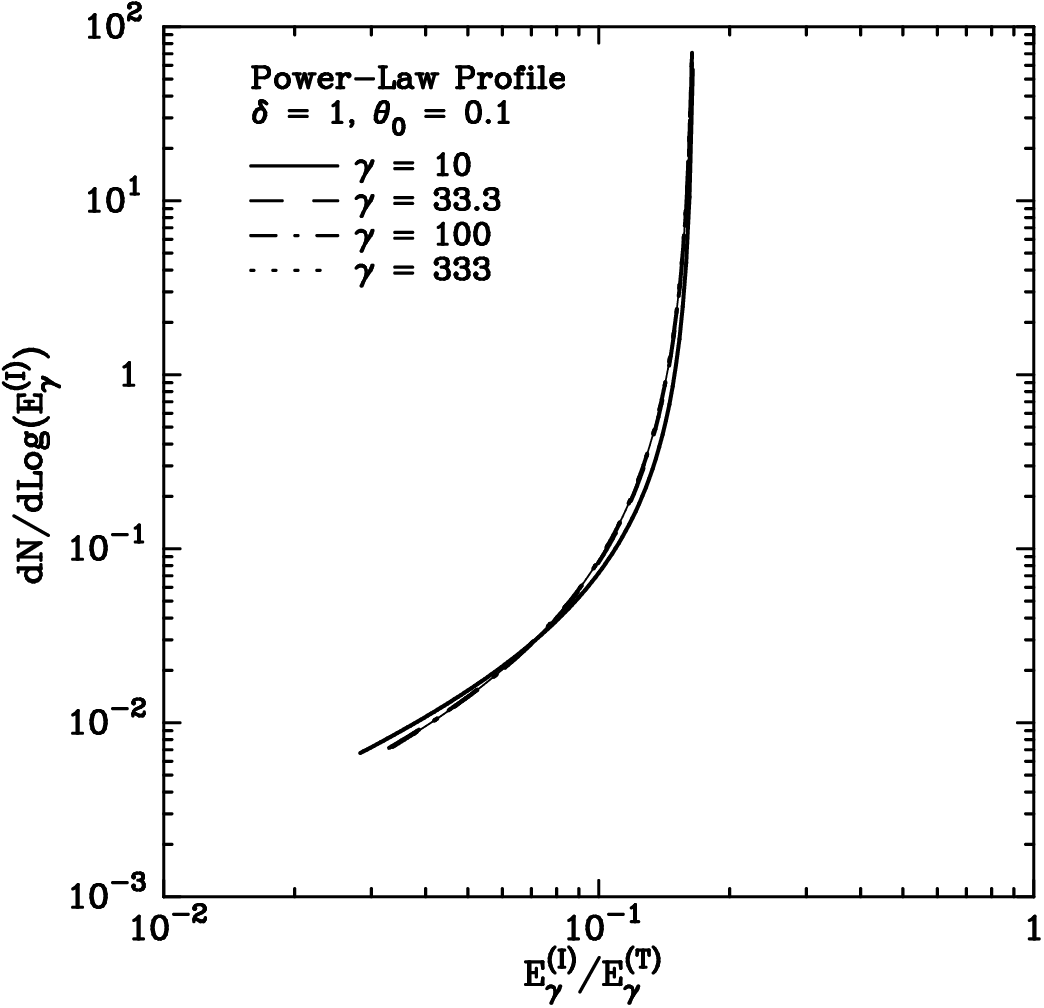}\\
\includegraphics[width=3.2in,clip=]{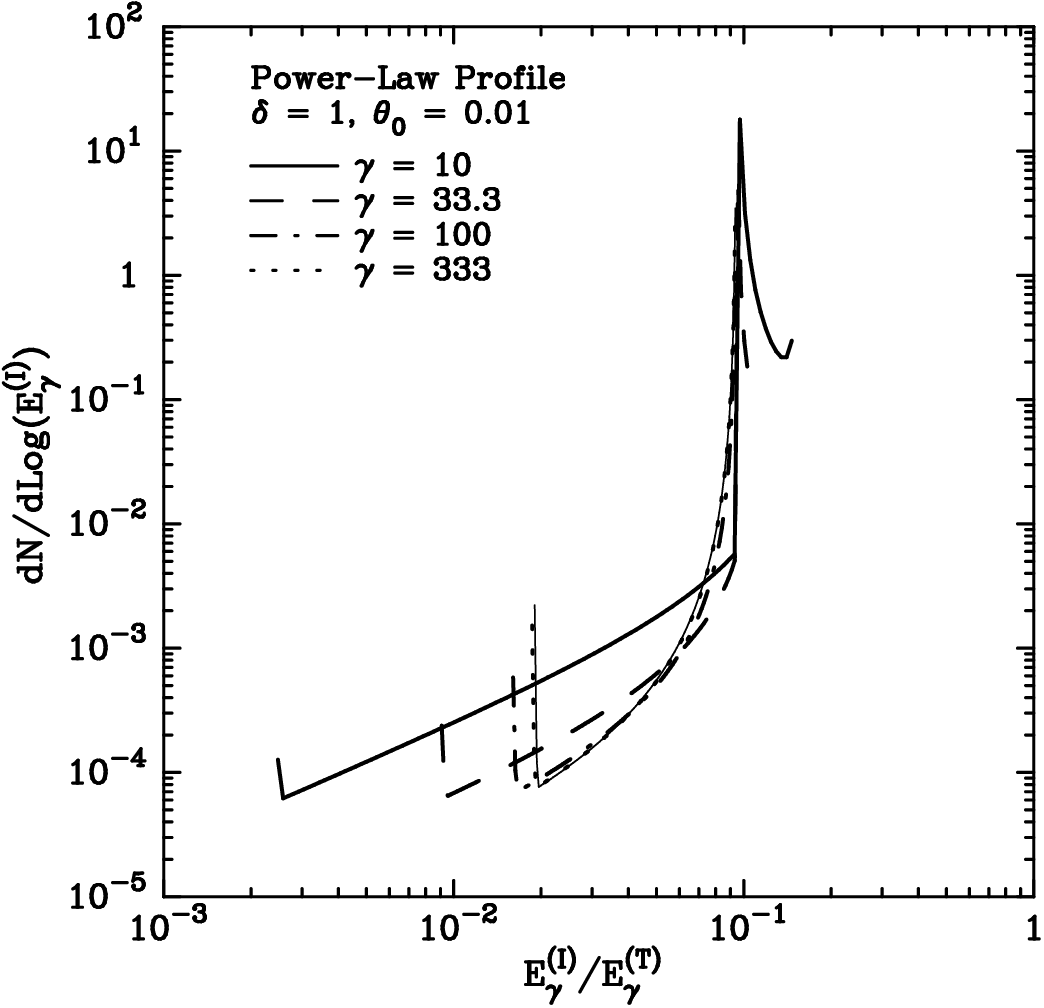}
\end{minipage}
\begin{minipage}[t]{3.2in}
\includegraphics[width=3.2in,clip=]{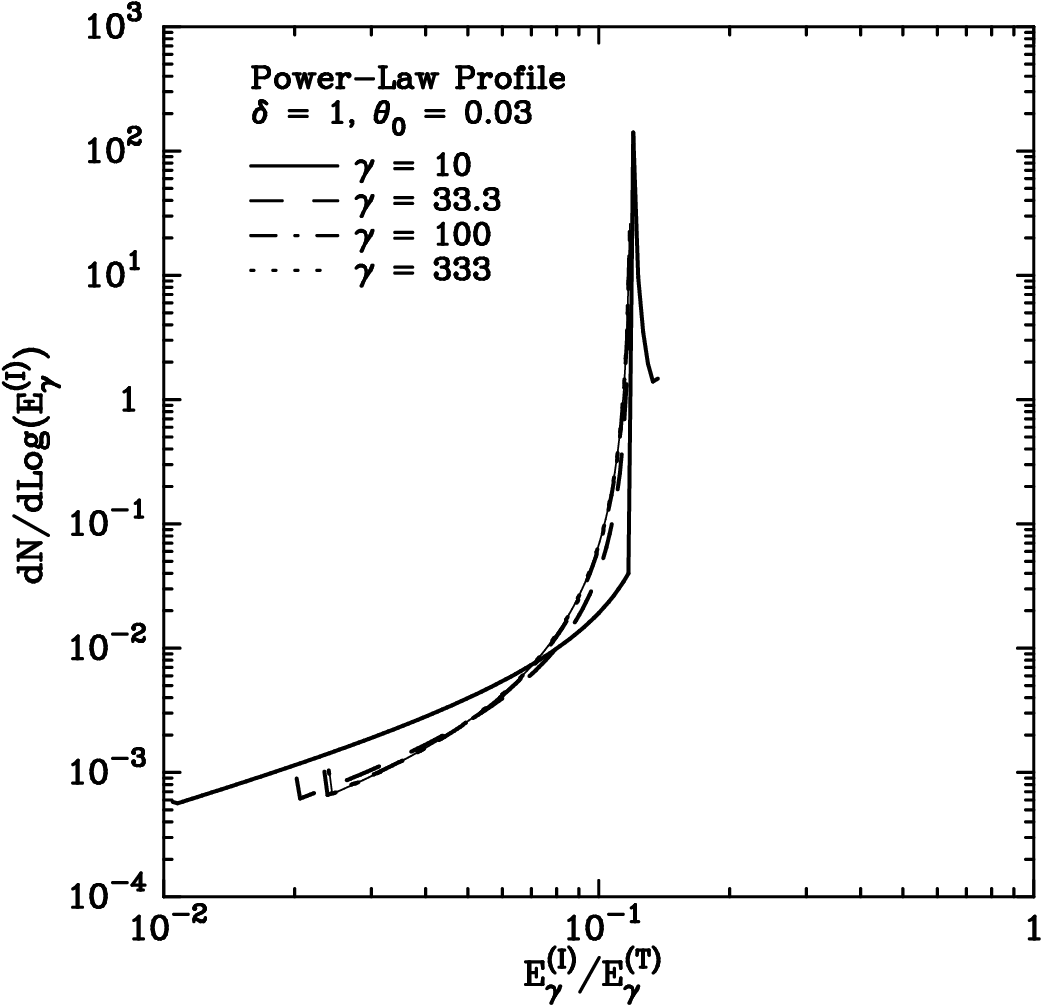}\\
\includegraphics[width=3.2in,clip=]{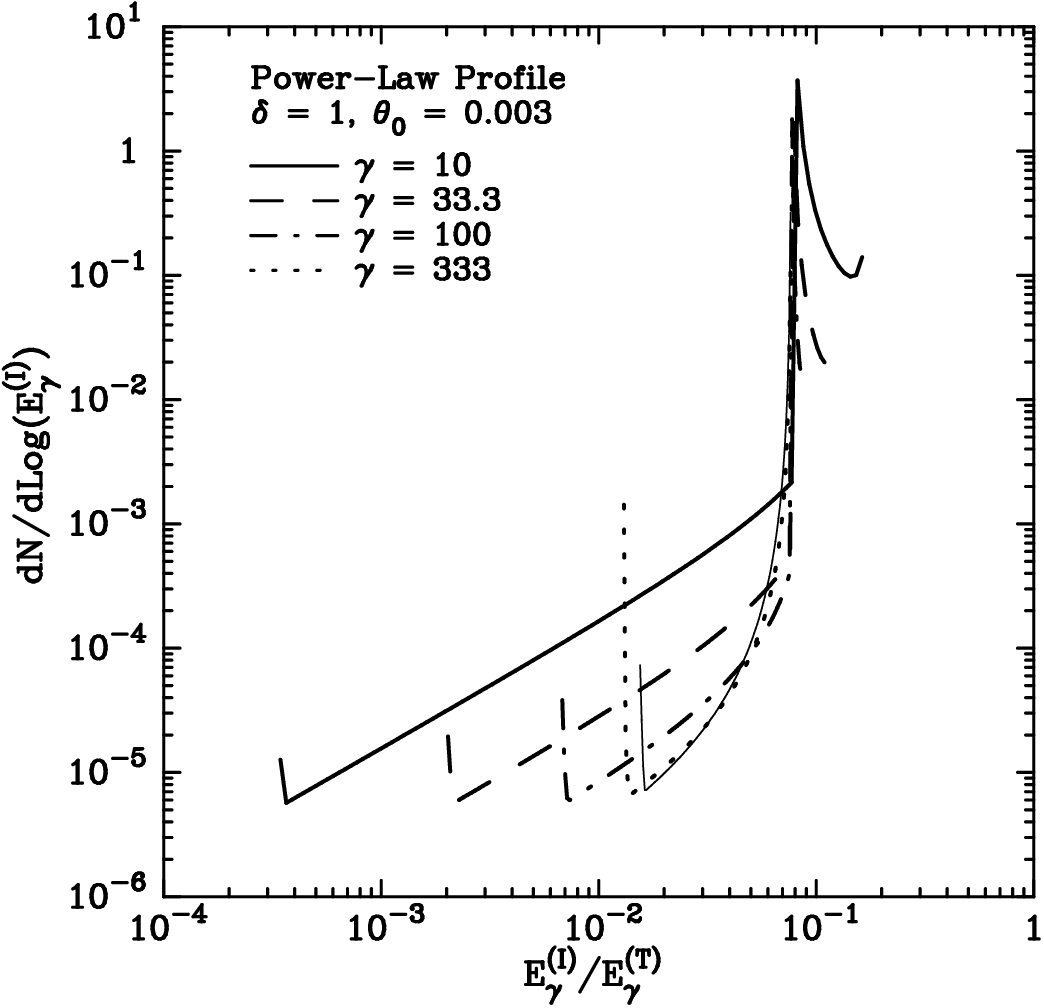}
\end{minipage}
\caption{$dN/d\log(\egi)$ distributions for universal jet models based on
$\delta=1$ power-law profiles.  Each panel shows the result for $\gamma$ =
10, 33.3, 100, and 333.  Top left:  $\theta_0=0.1$.  Top right:
$\theta_0=0.03$.  Bottom left: $\theta_0=0.01$.  Bottom right:
$\theta_0=0.003$.  The vertical scale is arbitrary.  The light solid line
shows the ``bare'' distribution due to the underlying profile ---
essentially the $\gamma\rightarrow\infty$ limit.  If it were not for the
$\theta_0$ cutoff and the relativistic kinematics, this distribution would
be a $\delta$-function, by construction.}
\label{pl_1.0_dndxgi}
\end{figure}

The four panels of Fig.~\ref{pl_1.0_dndxgi} show the relativistic
broadening of the $\delta$-function distribution designed into this
profile --- see Eq.~(\ref{plxgi}).  It clearly succeeds in preserving its
empirically-desirable narrowness.

%%%%%%%%%%%%%%%%%%%%%%%%%%%%%%%%%%%%%%%%%%%%%%%%%%%
%%%%%%%%%%%%%%%%%%%%%%%%%%%%%%%%%%%%%%%%%%%%%%%%%%%
\subsection{$\delta=2$ Power-Law Jet Profiles}
\label{sec_pl_2.0}

The results of \S\ref{sec_pl_1.0} are to some extent anomalous for
power-law jets, since $\delta=1$ is a singular case as far as \egi\ is
concerned.  The $\delta=2$ case comprises more typical behavior.

\subsubsection{Emission}

\begin{figure}[t]
\begin{minipage}[t]{3.2in}
\includegraphics[width=3.2in,clip=]{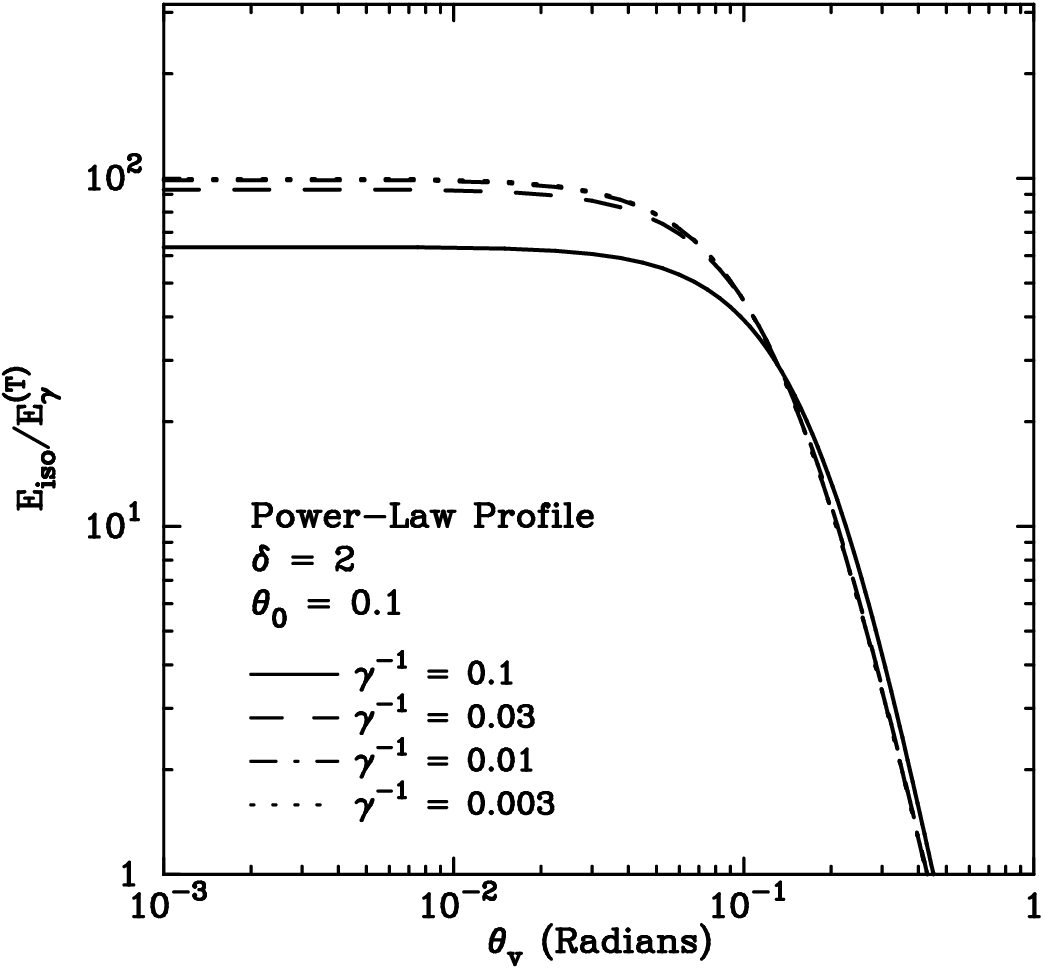}\\
\includegraphics[width=3.2in,clip=]{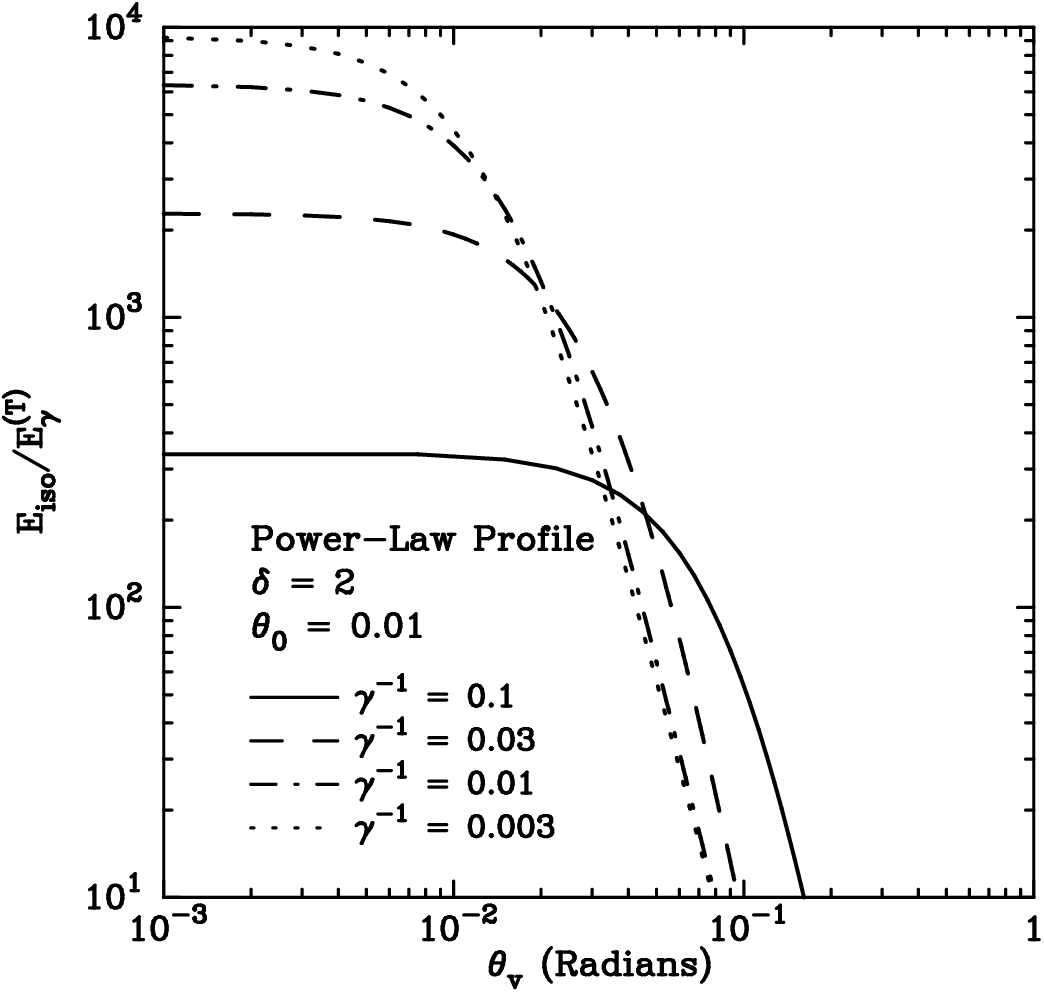}
\end{minipage}
\begin{minipage}[t]{3.2in}
\includegraphics[width=3.2in,clip=]{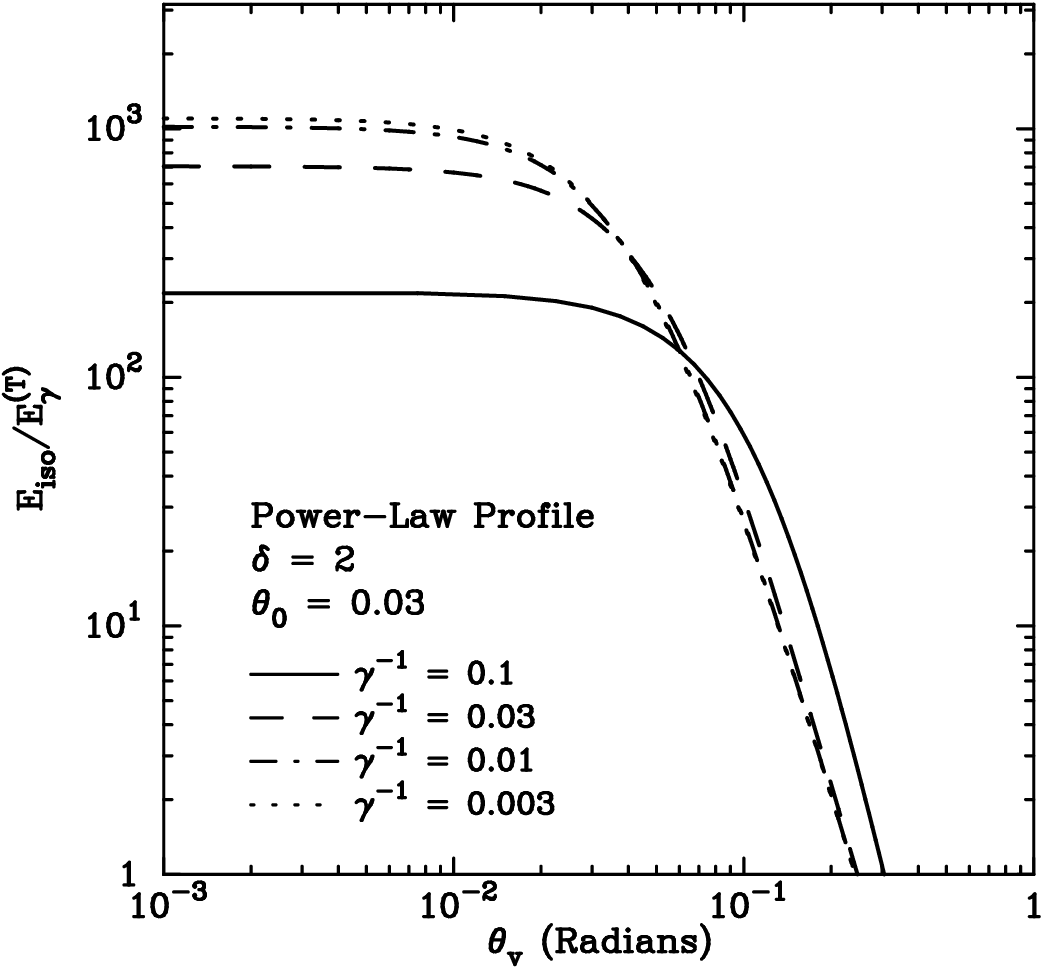}\\
\includegraphics[width=3.2in,clip=]{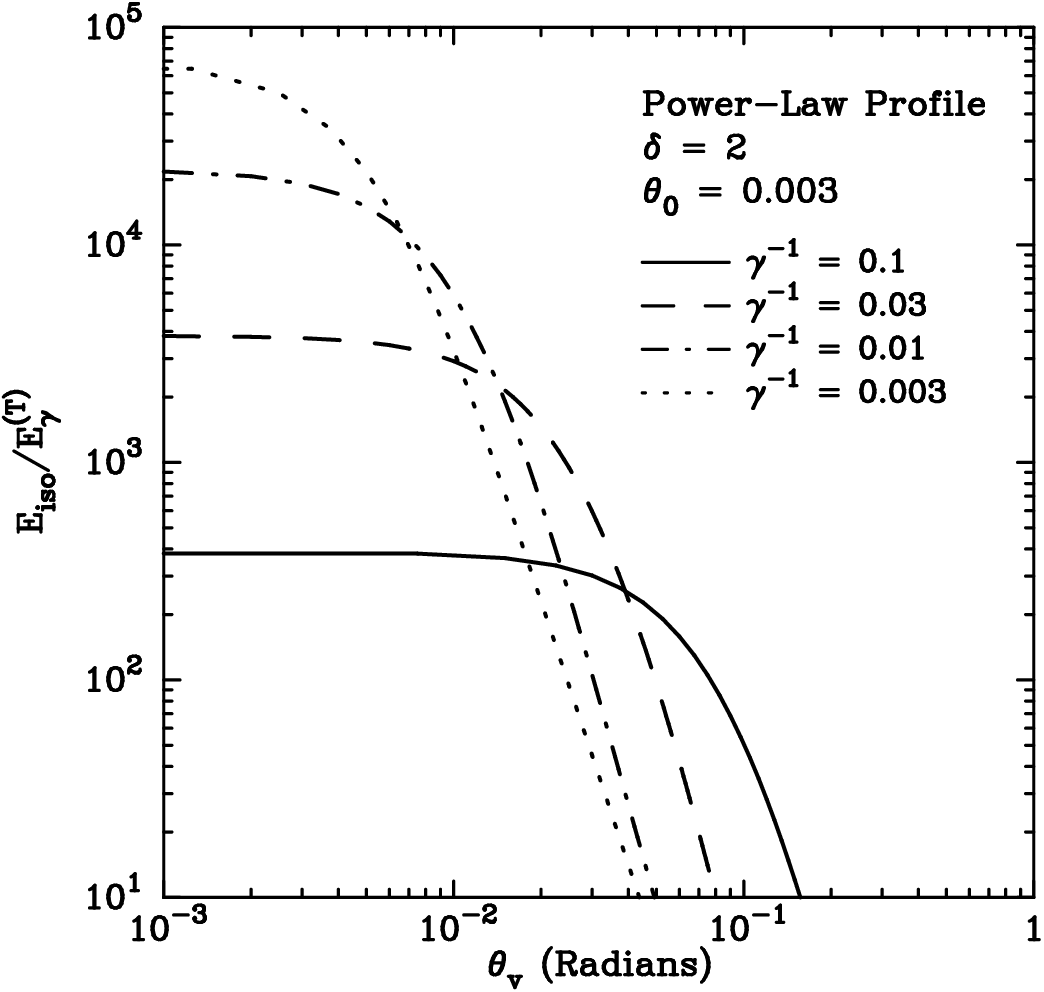}
\end{minipage}
\caption{\eiso\ as a function of viewing angle for emission from
relativistic jets with $\delta=2$ power-law emission profiles.  Each panel
shows the result for $\gamma^{-1}$ = 0.1, 0.03, 0.01, and 0.003.  Top
left:  $\theta_0=0.1$.  Top right: $\theta_0=0.03$.  Bottom left:
$\theta_0=0.01$.  Bottom right: $\theta_0=0.003$.}
\label{pl_2.0_emission}
\end{figure}

The four panels of Fig.~\ref{pl_2.0_emission} show the behavior of \eiso\ 
as a function of viewing angle $\theta_v$ for a power-law profile with
$\delta=2$.  We see jet broadening again, as expected.  The tails at high
$\theta_v$ are no longer independent of $\gamma$, as they were for
$\delta=1$, since now the profile function falls off as rapidly as the
Doppler photon distribution function.  The available dynamic range of
\eiso\ is thereby increased.

\begin{figure}[t]
\begin{minipage}[t]{3.2in}
\includegraphics[width=3.2in,clip=]{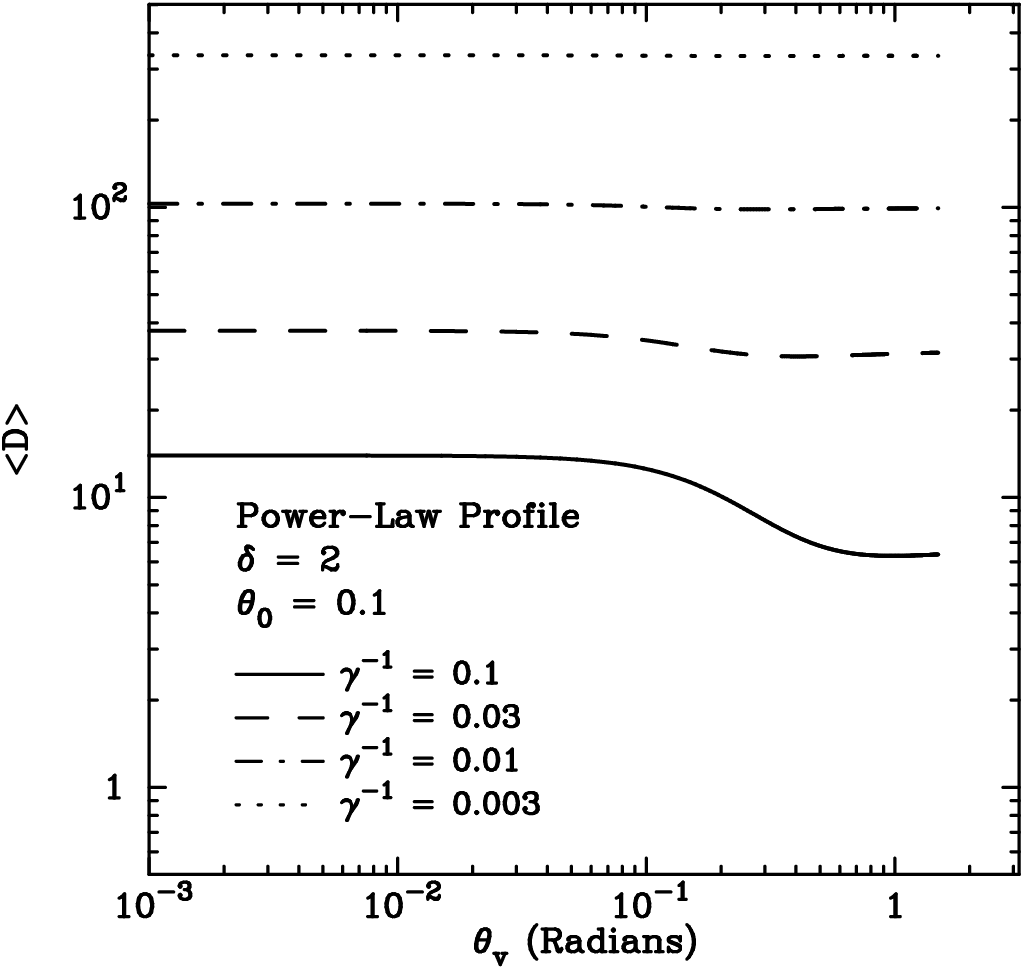}\\
\includegraphics[width=3.2in,clip=]{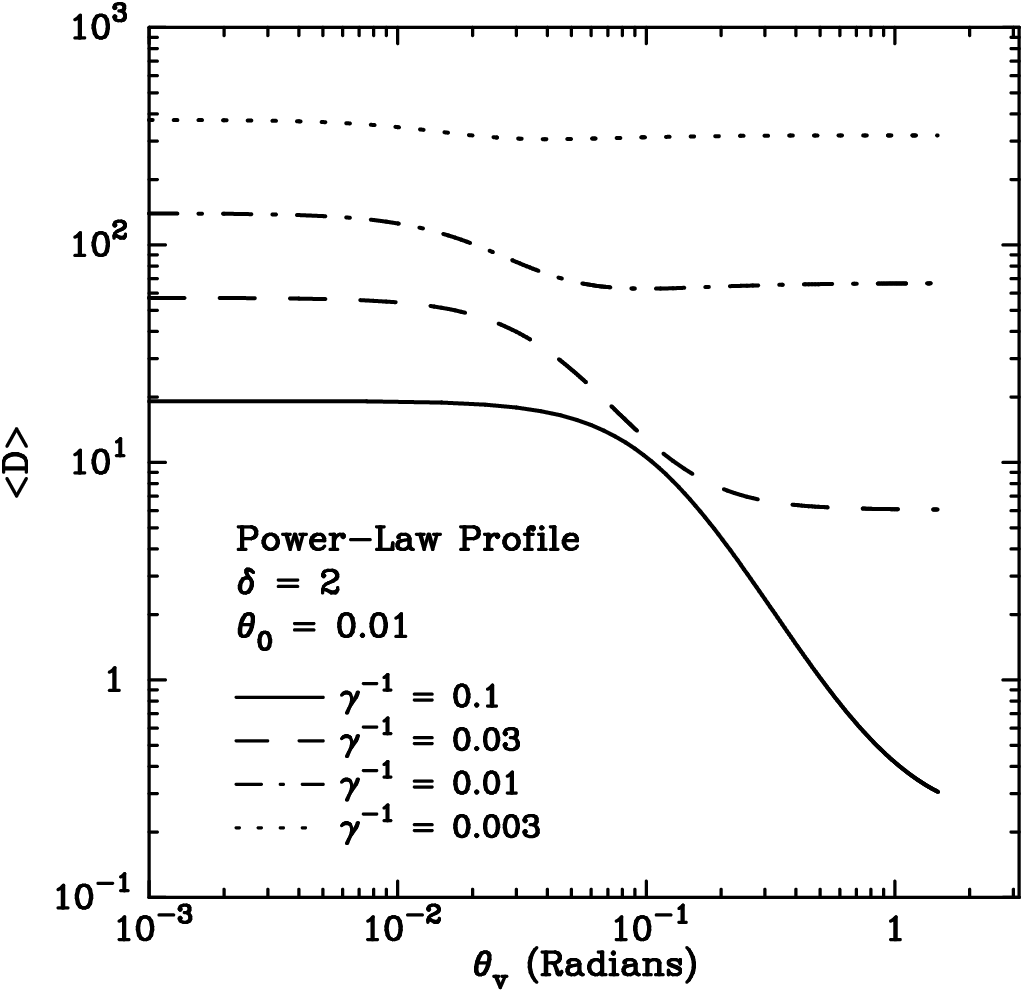}
\end{minipage}
\begin{minipage}[t]{3.2in}
\includegraphics[width=3.2in,clip=]{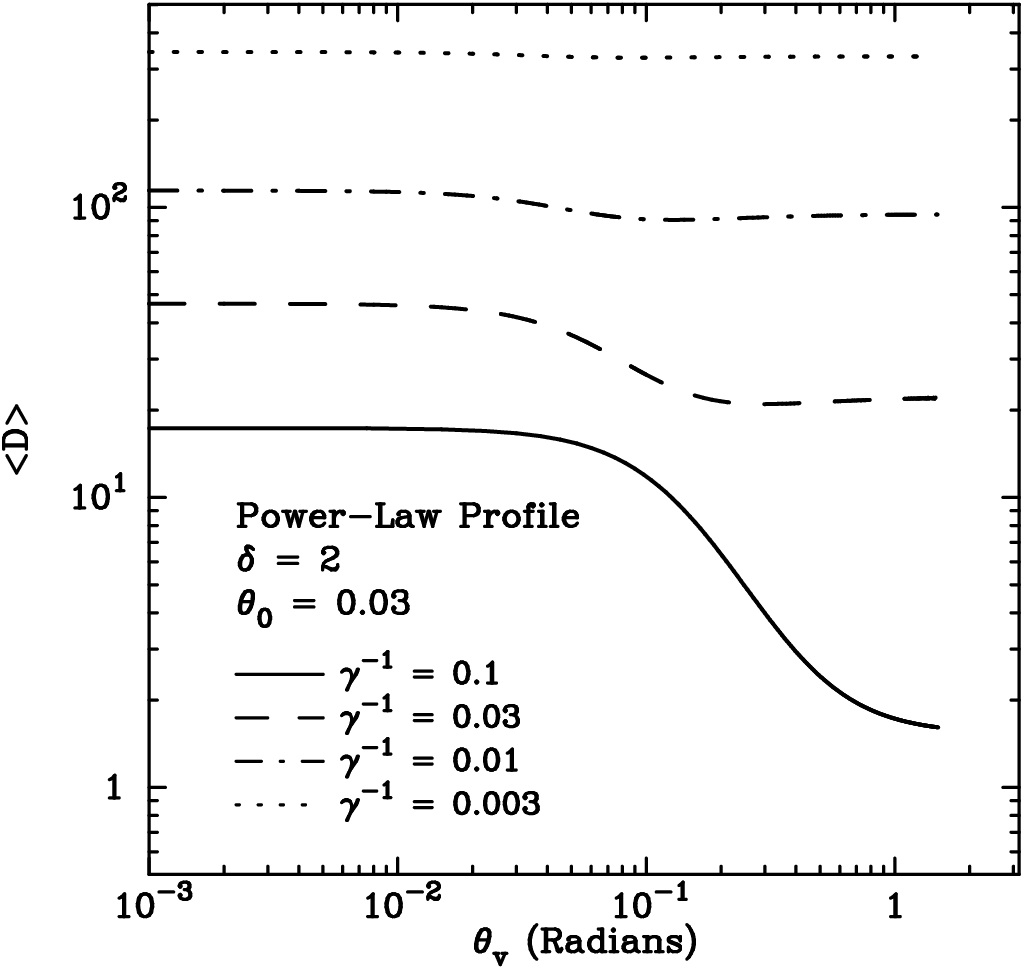}\\
\includegraphics[width=3.2in,clip=]{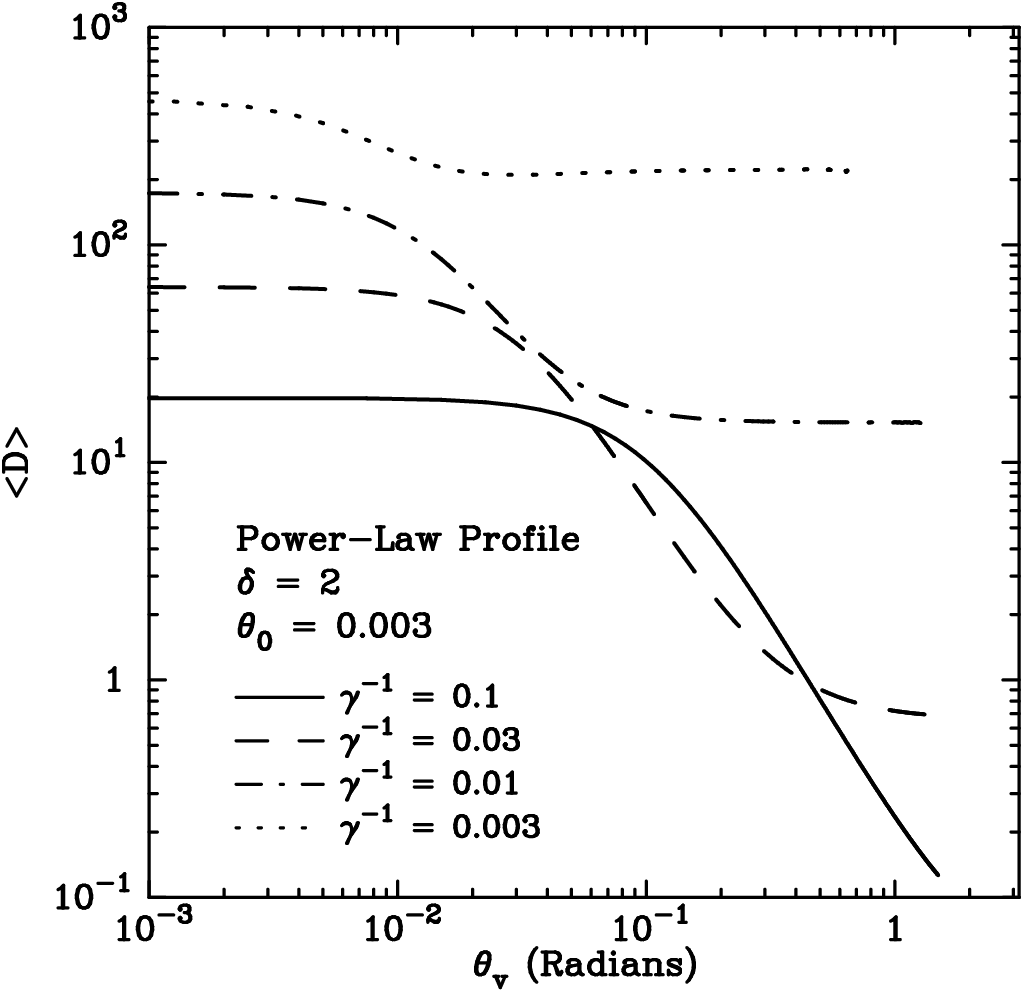}
\end{minipage}
\caption{Photon-number-averaged Doppler factor \davg\ as a function of
viewing angle for emission from relativistic jets with $\delta=2$
power-law emission profiles. Each panel shows the result for $\gamma^{-1}$ =
0.1, 0.03, 0.01, and 0.003.  Top left:  $\theta_0=0.1$.  Top right:
$\theta_0=0.03$.  Bottom left: $\theta_0=0.01$.  Bottom right:
$\theta_0=0.003$.  The dynamic range of \davg\ is still quite limited,
for almost all of the selected parameters.}
\label{pl_2.0_epk}
\end{figure}

The four panels of Fig.~\ref{pl_2.0_epk} show the behavior of \davg\ as a
function of viewing angle $\theta_v$.  We see that \davg\ can now just
about produce two orders of magnitude's worth of dynamic range, in the
``needle jet'' approximation ($\theta_0\ll\gamma^{-1}$), but the dynamic
range again becomes quite restricted in other regimes.  For
$\theta_0>\gamma^{-1}$, the large-$\theta_v$ limit of \davg\ is still
$\gamma$, as expected, and this limit still serves to constrain the
available range of \davg.

\subsubsection{\eiso\ -- \davg\ and \egi\ -- \davg\ Correlations}

\begin{figure}[t]
\begin{minipage}[t]{3.2in}
\includegraphics[width=3.2in,clip=]{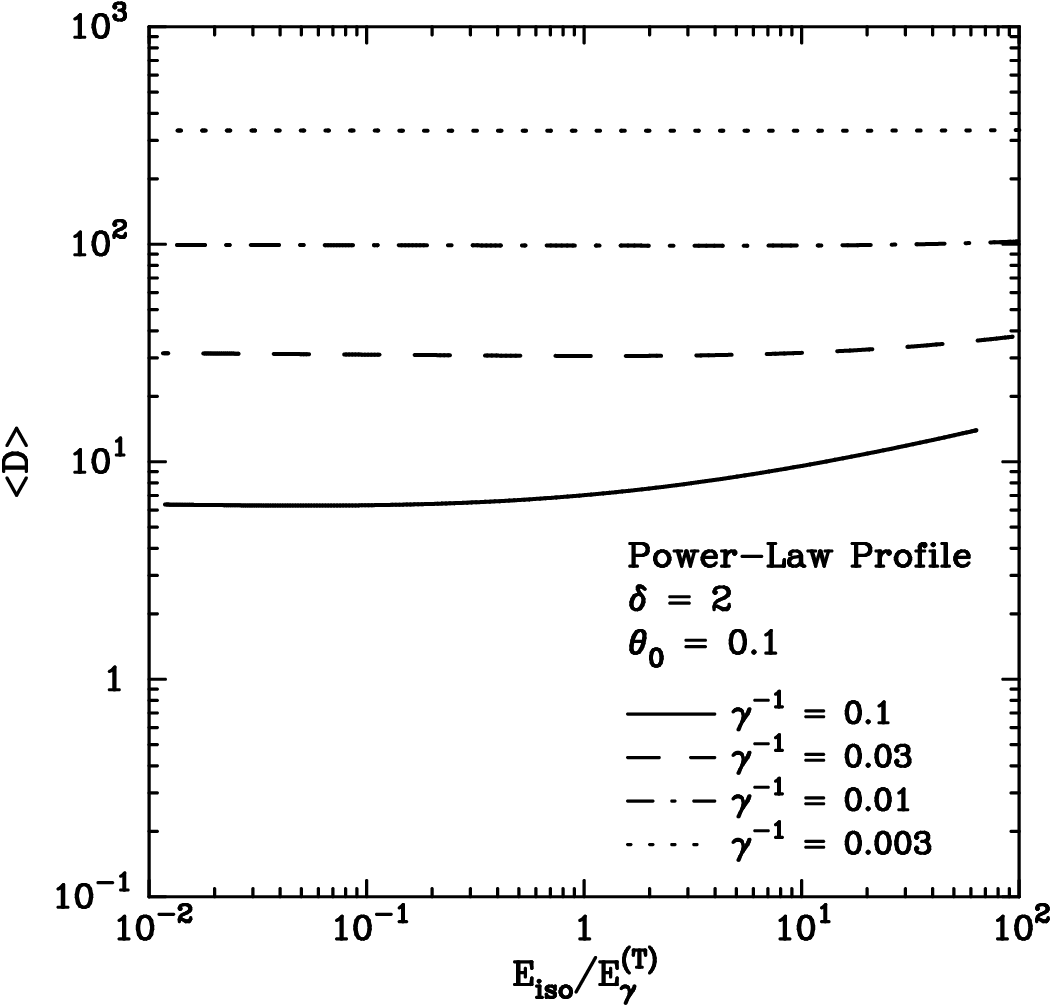}\\
\includegraphics[width=3.2in,clip=]{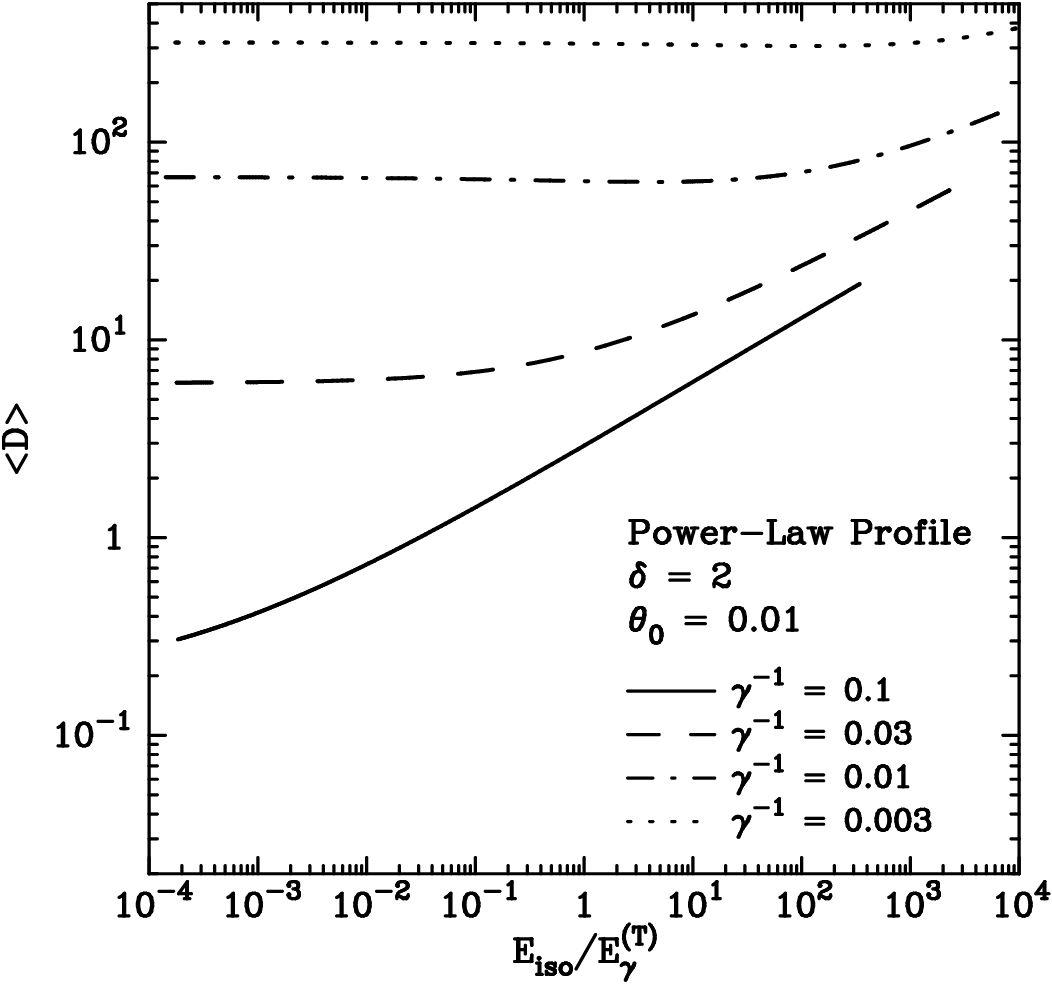}
\end{minipage}
\begin{minipage}[t]{3.2in}
\includegraphics[width=3.2in,clip=]{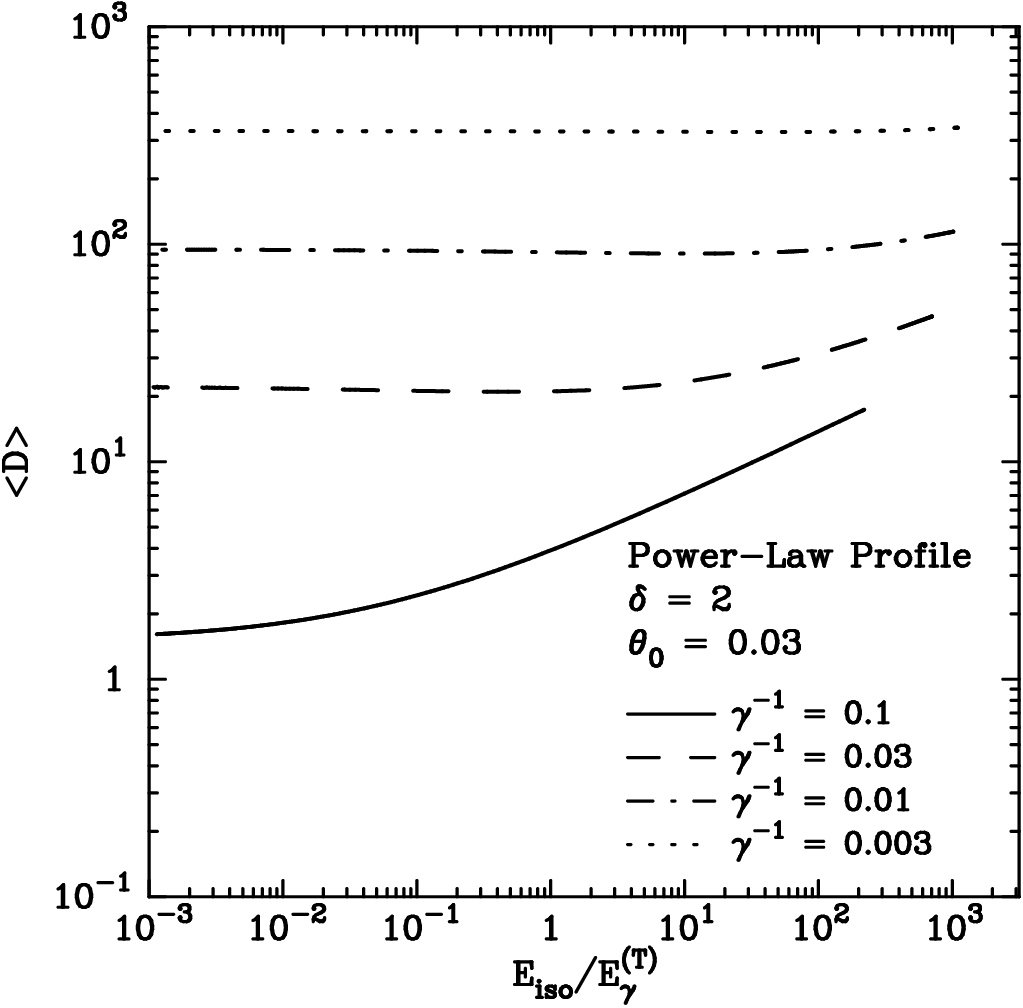}\\
\includegraphics[width=3.2in,clip=]{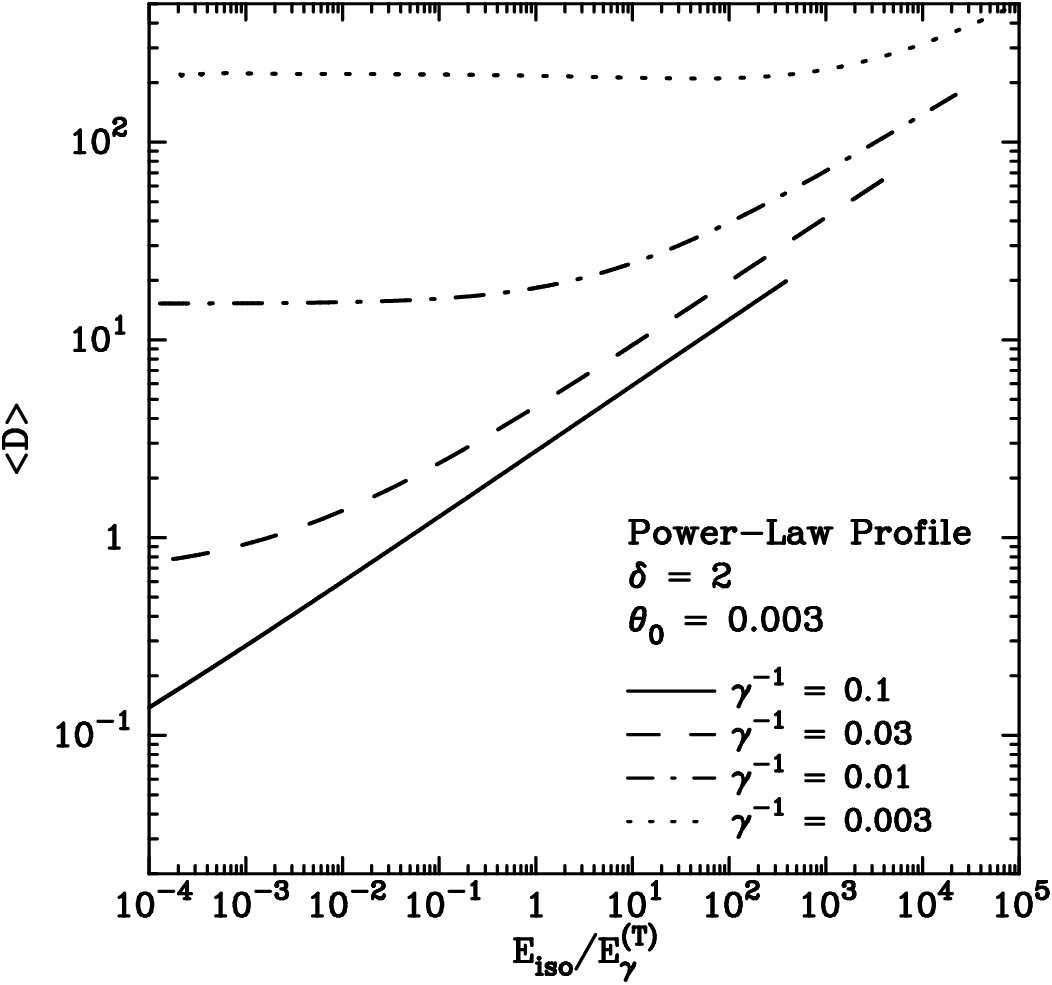}
\end{minipage}
\caption{\eiso--\davg\ (``Amati'') plots for universal jet models based on
$\delta=2$ power-law profiles.  Each panel shows the result for $\gamma^{-1}$ =
0.1, 0.03, 0.01, and 0.003.  Top left:  $\theta_0=0.1$.  Top right:
$\theta_0=0.03$.  Bottom left: $\theta_0=0.01$.  Bottom right:
$\theta_0=0.003$.}
\label{pl_2.0_amati}
\end{figure}

The four panels of Fig.~\ref{pl_2.0_amati} show the behavior of the
\eiso--\davg\ (``Amati'') correlation in universal $\delta=2$ power-law
models with our standard parameter choices.

Here it appears that this profile is trapped between a rock and a hard
place.  In order to make enough dynamic range in \davg, it must enter the
``needle jet'' regime ($\theta_0\ll\gamma^{-1}$).  But as soon as it does
so, the slope of the \eiso--\davg\ curve becomes 1/3, as expected for
off-axis emission from narrow jets.  It does not appear that this profile
can make an ``Amati'' relation spanning two decades of \epk.

\begin{figure}[t]
\begin{minipage}[t]{3.2in}
\includegraphics[width=3.2in,clip=]{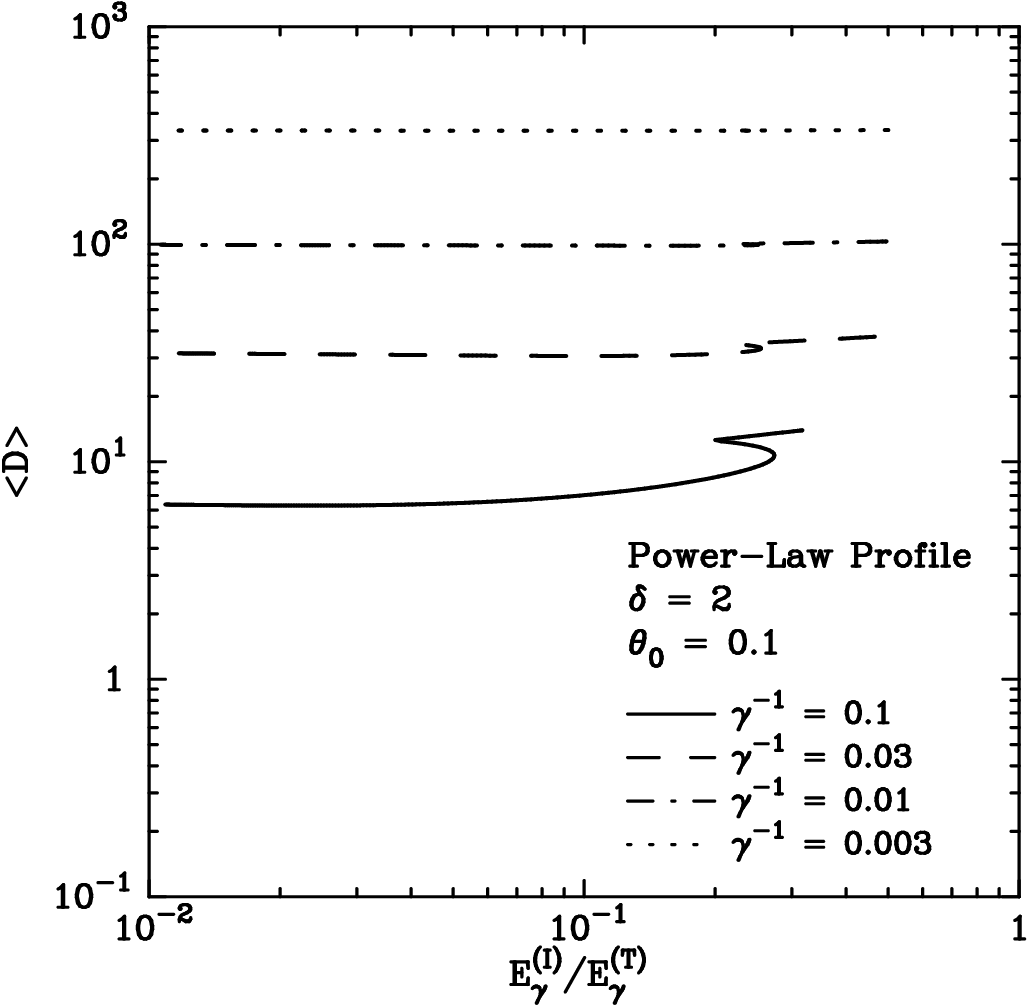}\\
\includegraphics[width=3.2in,clip=]{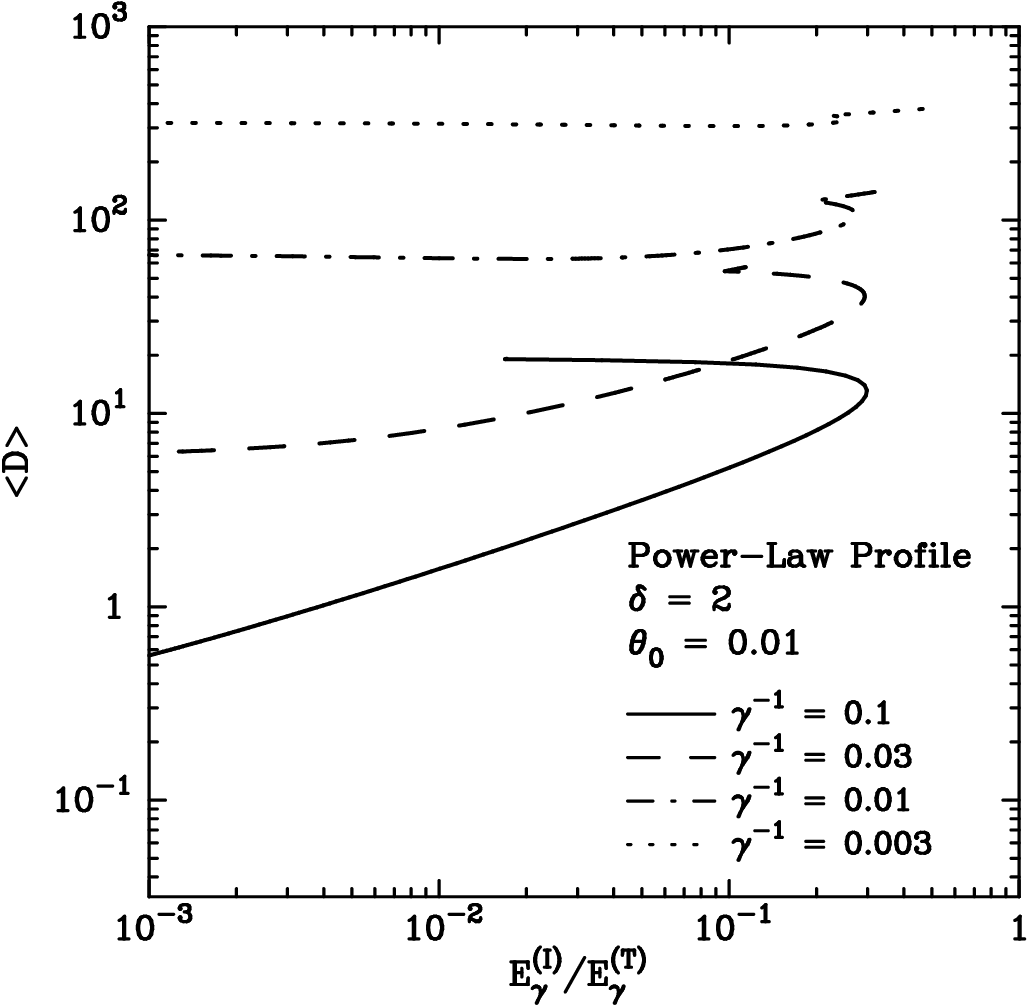}
\end{minipage}
\begin{minipage}[t]{3.2in}
\includegraphics[width=3.2in,clip=]{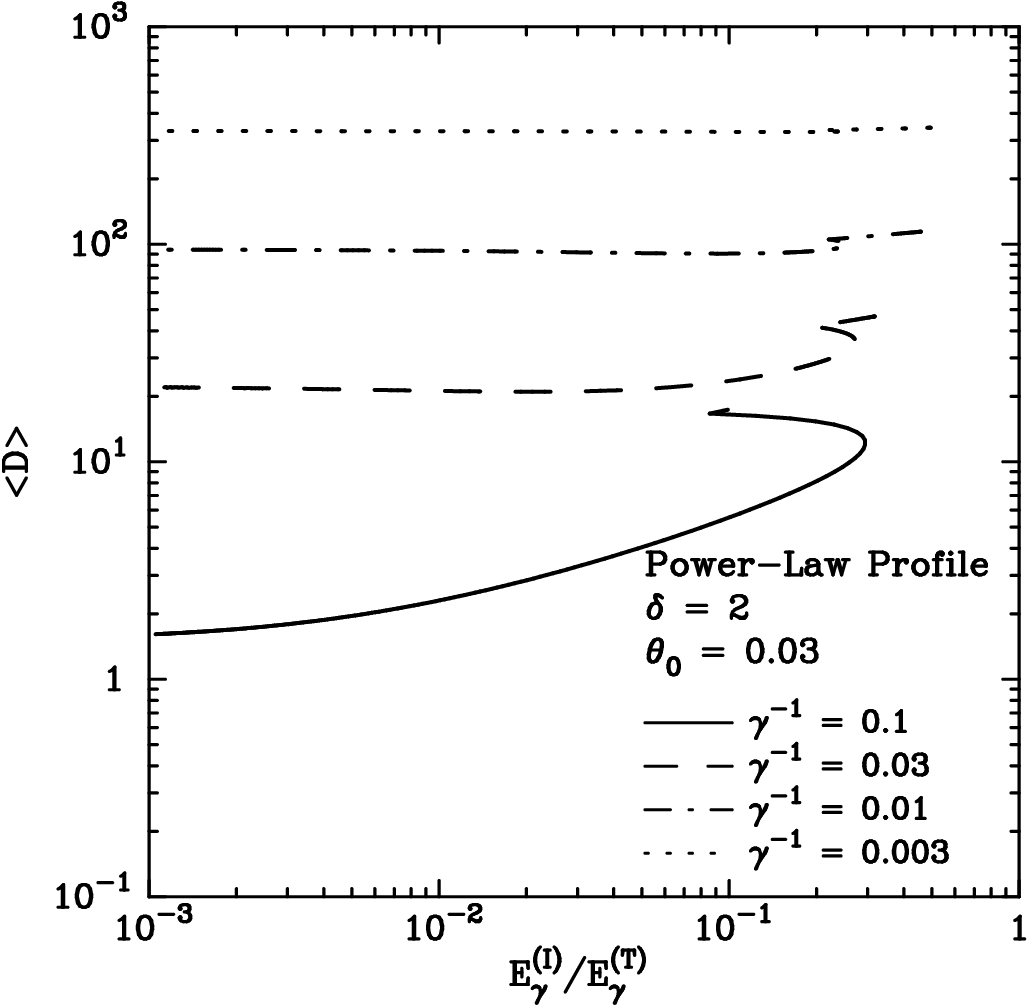}\\
\includegraphics[width=3.2in,clip=]{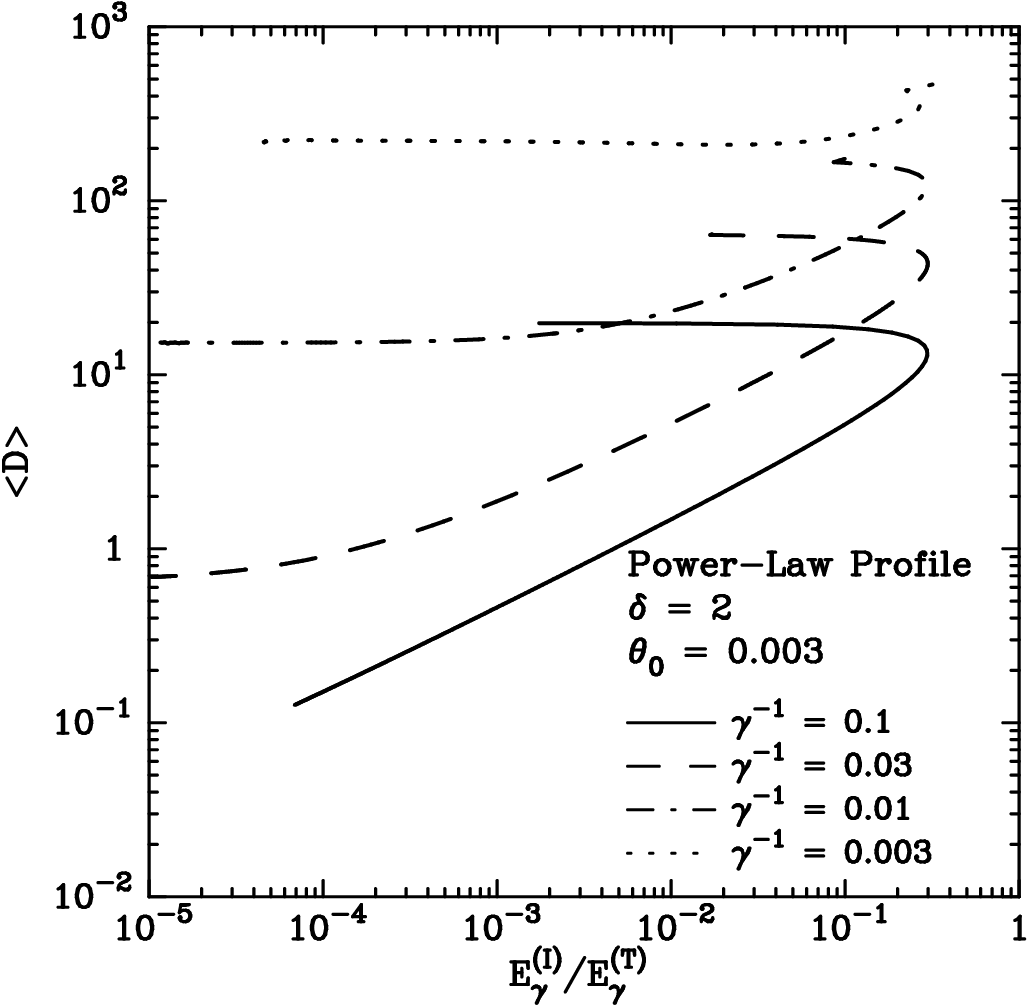}
\end{minipage}
\caption{\egi--\davg\ (``Ghirlanda'') plots for universal jet models based
on $\delta=2$ power-law profiles.  Each panel shows the result for $\gamma^{-1}$ =
0.1, 0.03, 0.01, and 0.003.  Top left:  $\theta_0=0.1$.  Top right:
$\theta_0=0.03$.  Bottom left: $\theta_0=0.01$.  Bottom right:
$\theta_0=0.003$.}
\label{pl_2.0_ghirlanda}
\end{figure}

The four panels of Fig.~\ref{pl_2.0_ghirlanda} show the behavior of the
\egi--\davg\ (``Ghirlanda'') correlation in universal $\delta=2$ power-law
models with our standard parameter choices.

We  can see from these curves that this profile's ability to make a
respectable ``Ghirlanda'' relation is no better than it's ability to make
``Amati'' relations, and for the same reasons.

\subsubsection{\eiso\ and \egi\ Frequency Distributions}

\begin{figure}[t]
\begin{minipage}[t]{3.2in}
\includegraphics[width=3.2in,clip=]{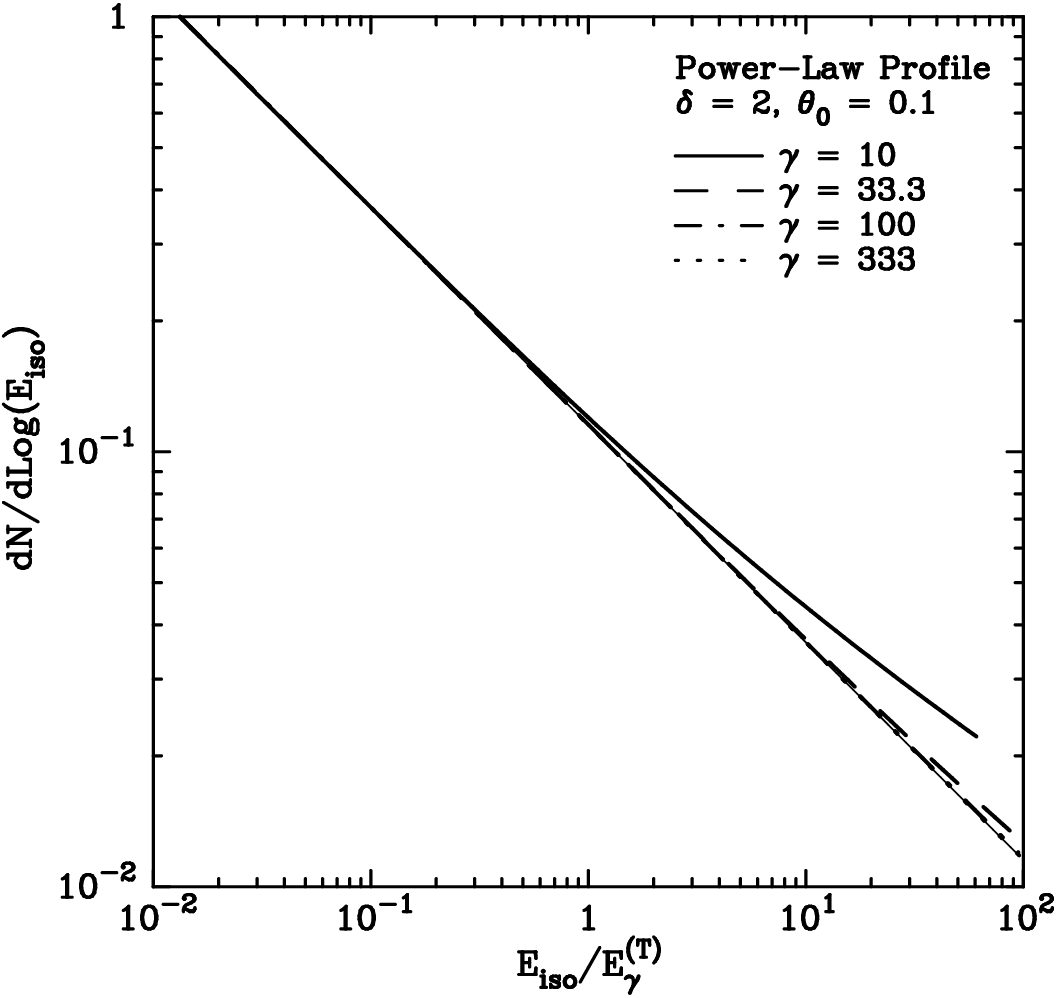}\\
\includegraphics[width=3.2in,clip=]{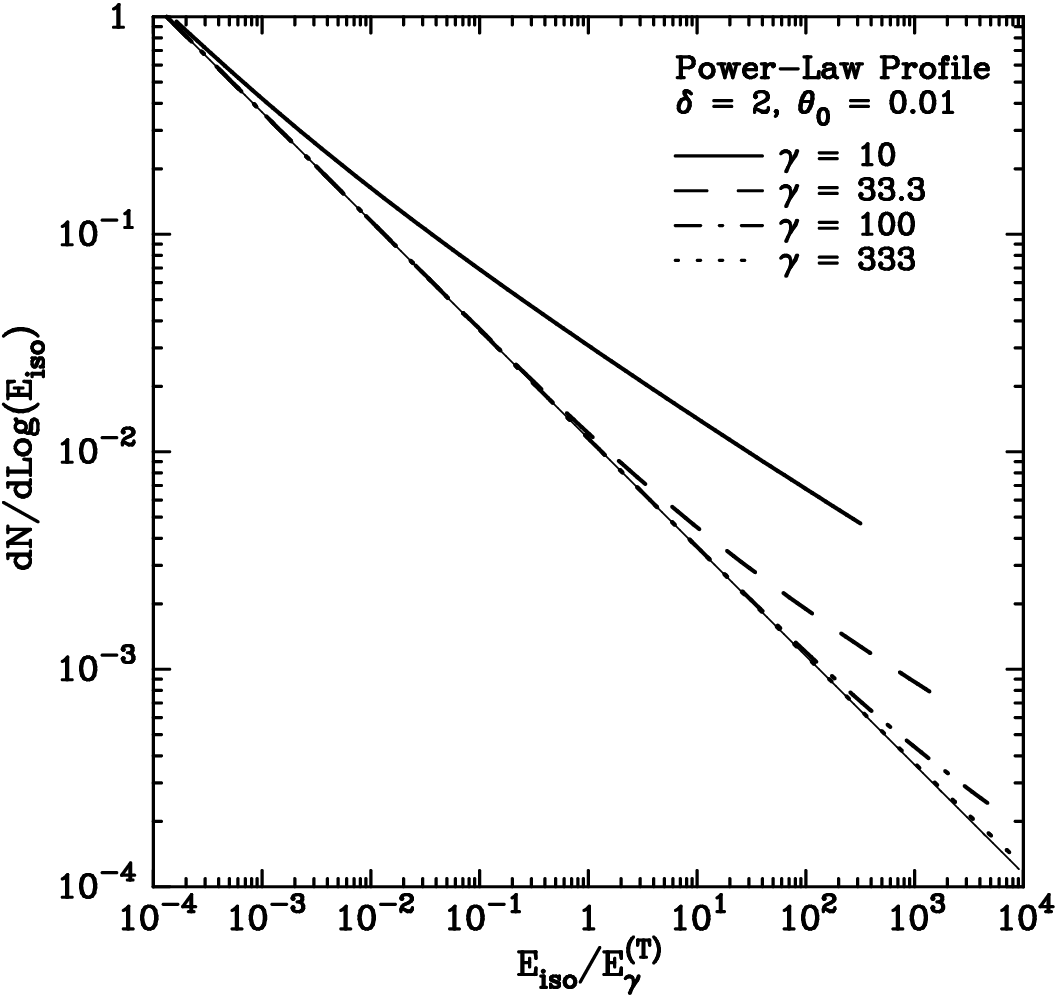}
\end{minipage}
\begin{minipage}[t]{3.2in}
\includegraphics[width=3.2in,clip=]{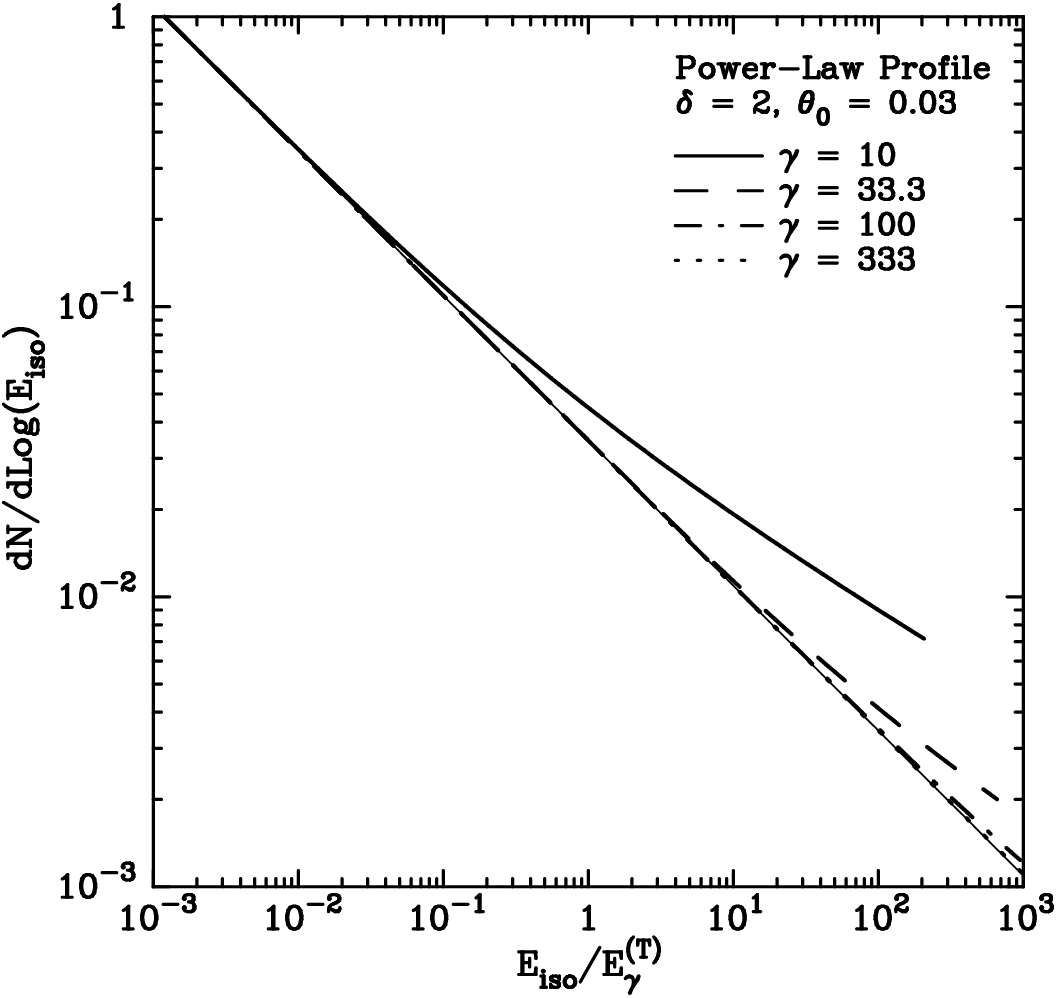}\\
\includegraphics[width=3.2in,clip=]{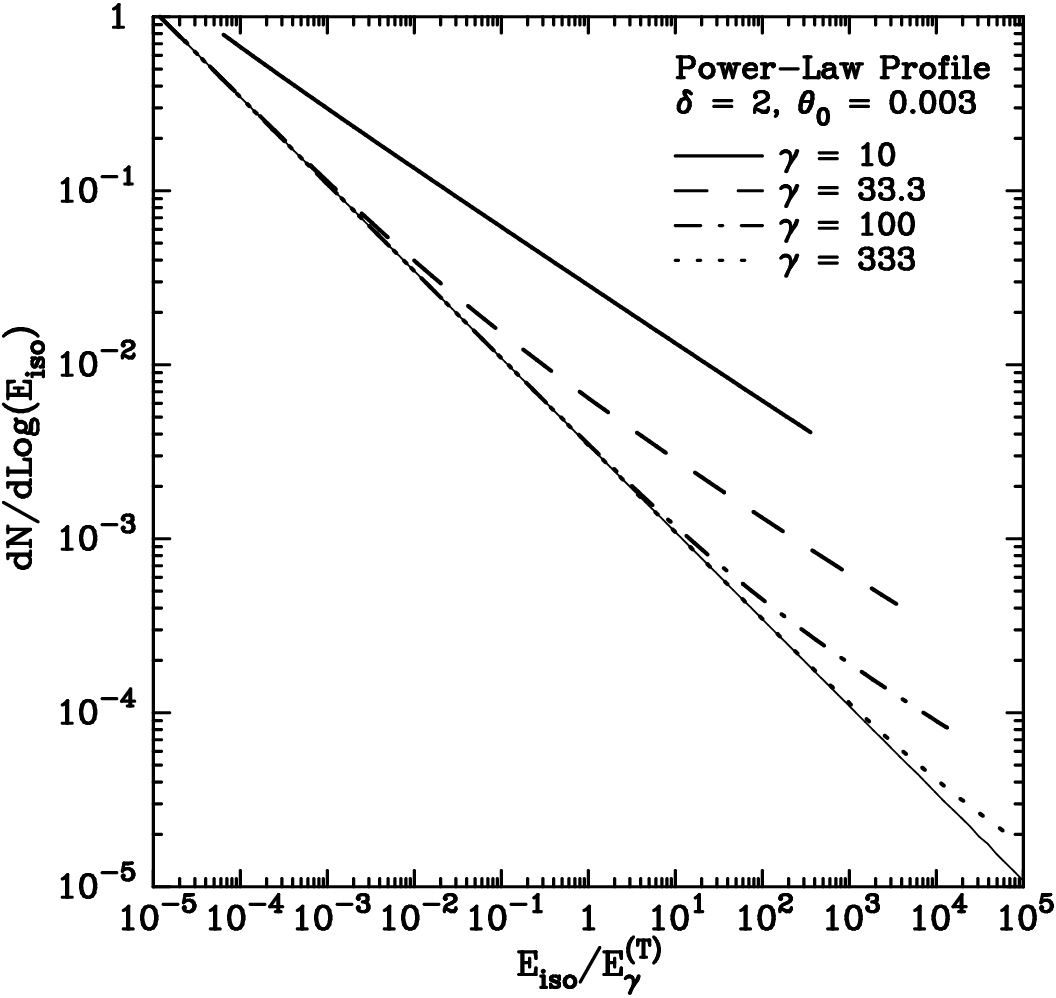}
\end{minipage}
\caption{$dN/d\log(\eiso)$ distributions for universal jet models based on
$\delta=2$ power-law profiles.  Each panel shows the result for $\gamma$ =
10, 33.3, 100, and 333.  Top left:  $\theta_0=0.1$.  Top right:
$\theta_0=0.03$.  Bottom left: $\theta_0=0.01$.  Bottom right:
$\theta_0=0.003$.  The vertical scale is arbitrary.  The light
solid line shows the ``bare'' distribution due to the underlying profile
--- essentially the $\gamma\rightarrow\infty$ limit.  Its slope is -1/2, as
expected from Eq.~\ref{plxiso}.}
\label{pl_2.0_dndxiso}
\end{figure}

The four panels of Fig.~\ref{pl_2.0_dndxiso} show the effect of
relativistic kinematics on the \eiso\ frequency distribution of $\delta=2$
power-law jets --- see Eq.~(\ref{plxiso}).  Again, the ``bare''
($\gamma\rightarrow\infty$) distribution is shown by a light solid line,
whose slope in this case is -1/2.

The distribution of $dN/d\eiso$ is still narrower than the data seem to
indicate, although in the ``needle-jet'' limit $\theta_0\ll\gamma^{-1}$,
the slope of the distribution is softened by relativistic kinematics to
the -1/3  index power-law expected from Eq.~(\ref{eiso_limit}).

\begin{figure}[t]
\begin{minipage}[t]{3.2in}
\includegraphics[width=3.2in,clip=]{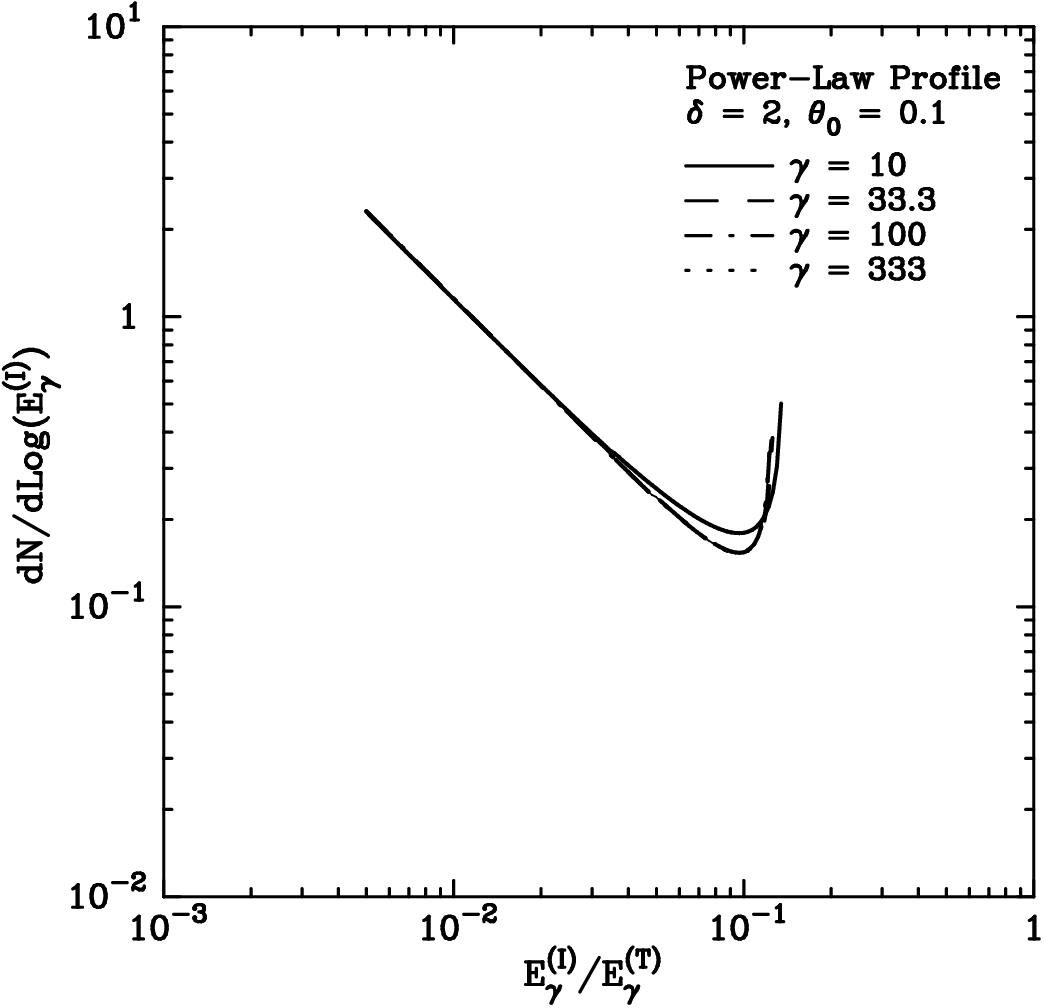}\\
\includegraphics[width=3.2in,clip=]{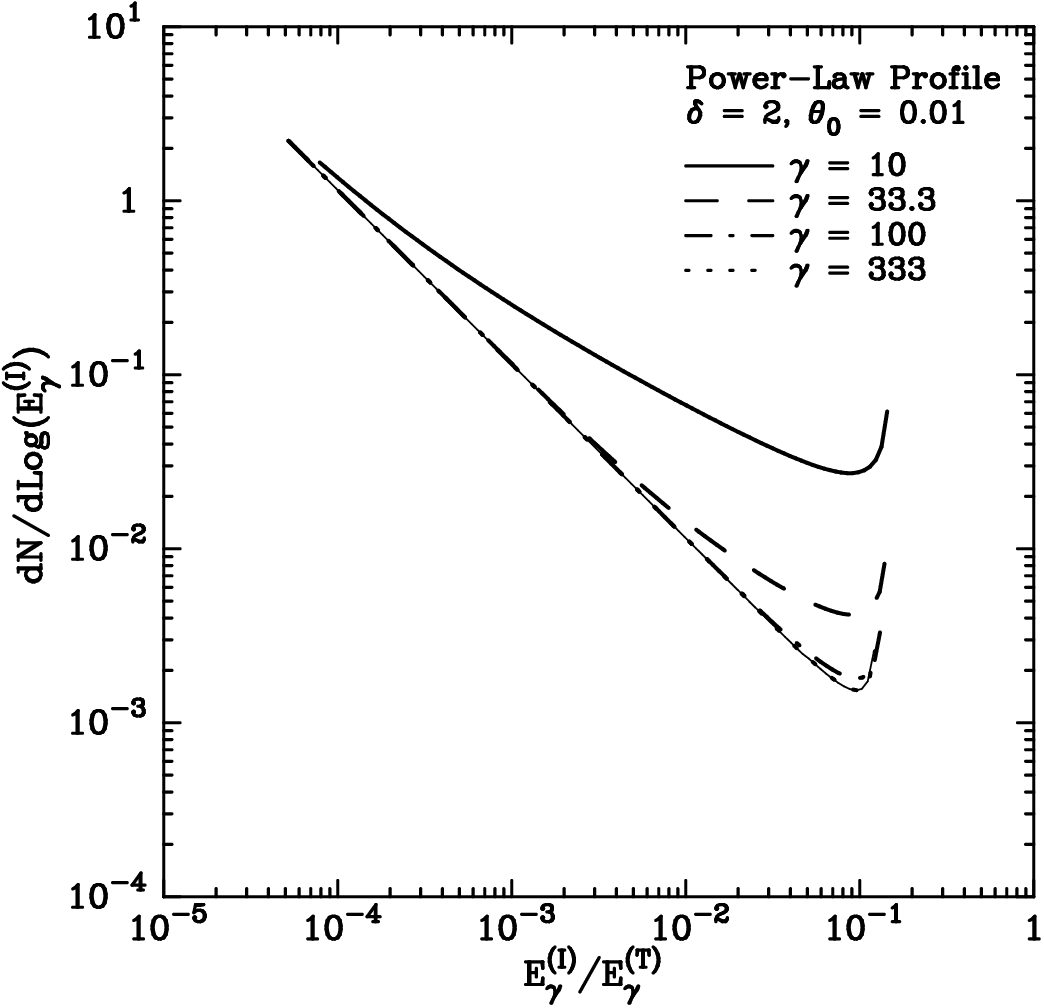}
\end{minipage}
\begin{minipage}[t]{3.2in}
\includegraphics[width=3.2in,clip=]{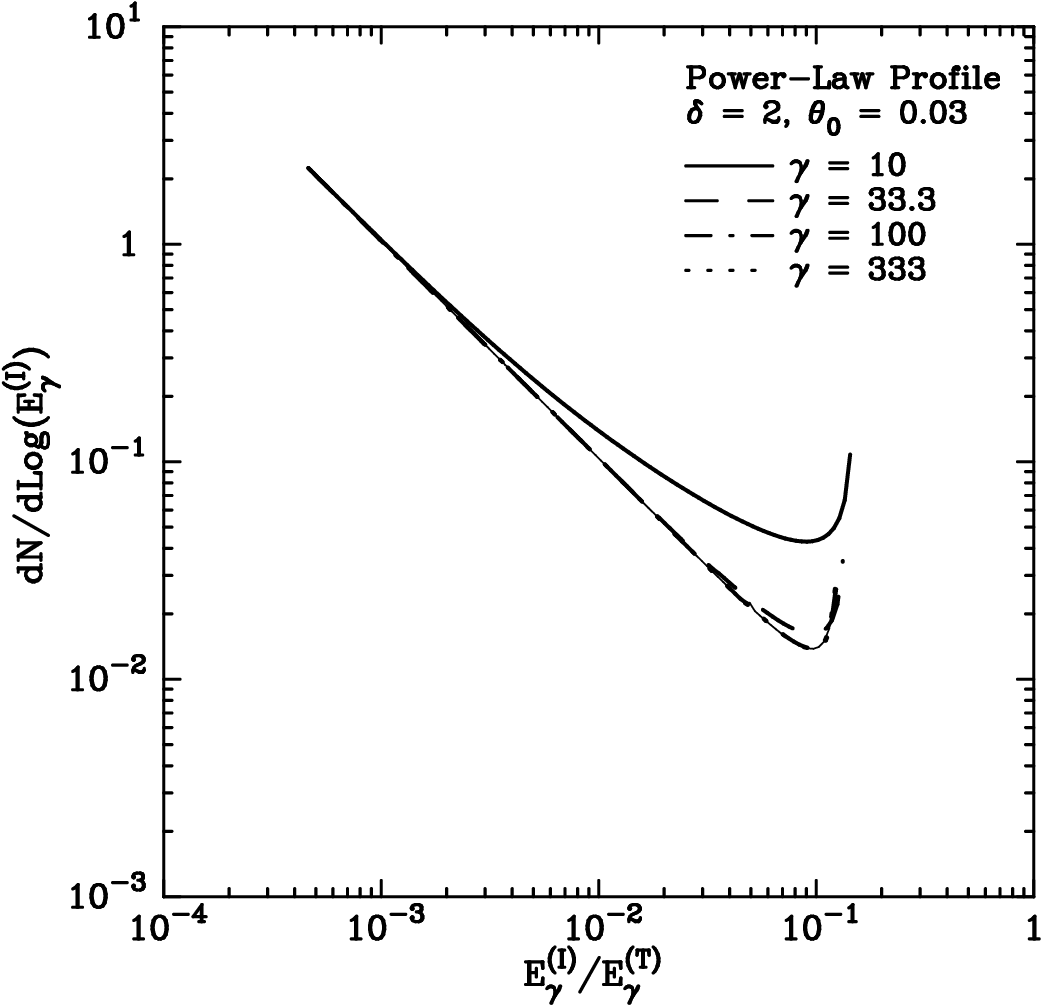}\\
\includegraphics[width=3.2in,clip=]{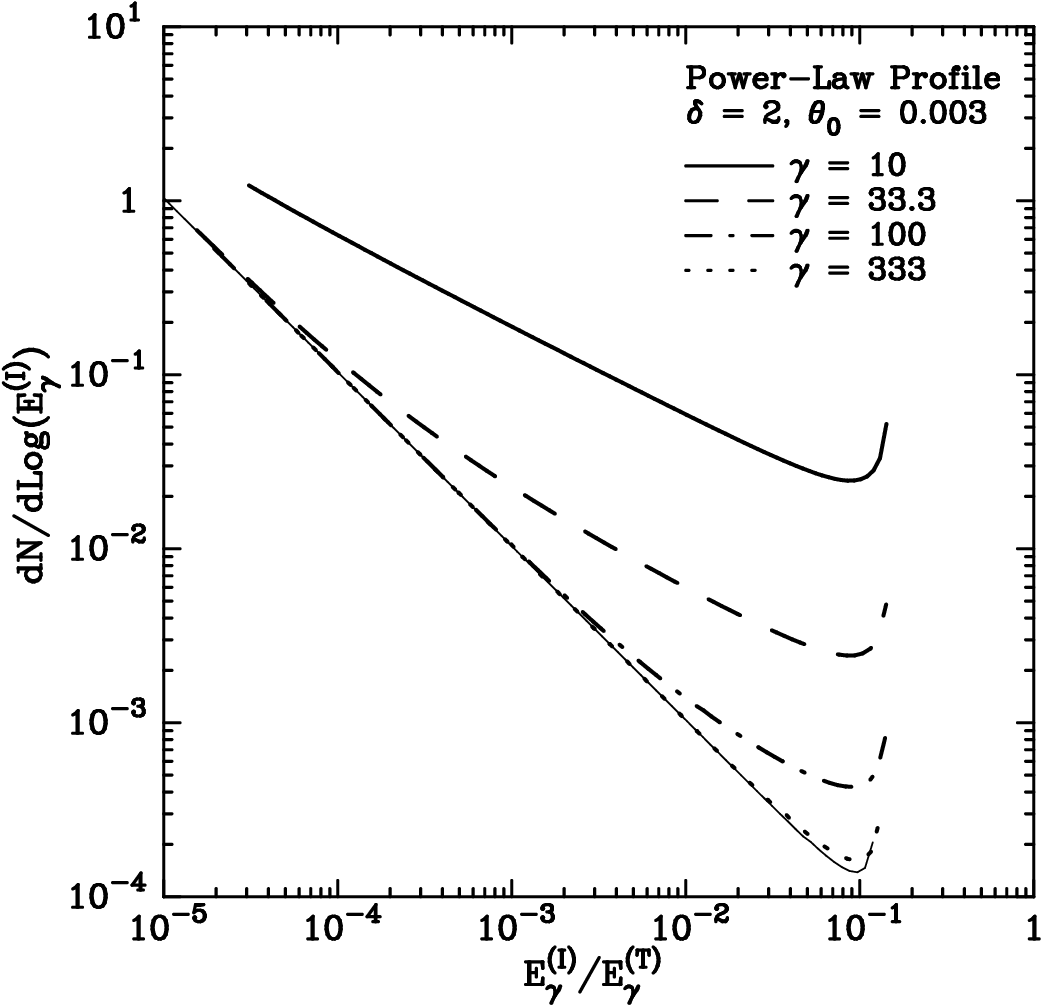}
\end{minipage}
\caption{$dN/d\log(\egi)$ distributions for universal jet models based on
$\delta=2$ power-law profiles.  Each panel shows the result for $\gamma$ =
10, 33.3, 100, and 333.  Top left:  $\theta_0=0.1$.  Top right:
$\theta_0=0.03$.  Bottom left: $\theta_0=0.01$.  Bottom right:
$\theta_0=0.003$.  The vertical scale is arbitrary.  The light solid line
shows the ``bare'' distribution due to the underlying profile ---
essentially the $\gamma\rightarrow\infty$ limit.  Its slope is -1, as
expected from Eq.~\ref{plxgi}.}
\label{pl_2.0_dndxgi}
\end{figure}

The four panels of Fig.~\ref{pl_2.0_dndxgi} show the effect of relativistic
kinematics on the \egi\ frequency distribution of power-law jets ---
see Eq.~(\ref{plxgi}).  The distribution can easily be made acceptably narrow,
although in the ``needle-jet'' limit $\theta_0\ll\gamma^{-1}$, the -1/2
slope expected from Eq.~(\ref{egi_limit}) seems somewhat less steep
than is consistent with the data.  This means that a fit of such a
universal model could potentially provide a (model-dependent) lower limit
on $\gamma$.

%%%%%%%%%%%%%%%%%%%%%%%%%%%%%%%%%%%%%%%%%%%%%%%%%%%
%%%%%%%%%%%%%%%%%%%%%%%%%%%%%%%%%%%%%%%%%%%%%%%%%%%
\subsection{$\delta=8$ Power-Law Jet Profiles}
\label{sec_pl_8.0}

The chief difficulty encountered by the relatively shallow $\delta=2$
index power-law of \S\ref{sec_pl_2.0} is a lack of dynamic range,
particularly for \davg.  It is therefore interesting to examine whether
this weakness can be addressed by adopting a steeper power-law index. 
Here we examine $\delta=8$.  As a reminder, our standard parameter values
are altered here, to $\theta_0$ = 0.4, 0.12, 0.04, and 0.012, so as to
make the effective width estimator $\theta_w$ of Eq.~(\ref{pl_width}) take
the values $\theta_w$ =0.1, 0.3, 0.01, and 0.003.

\subsubsection{Emission}

\begin{figure}[t]
\begin{minipage}[t]{3.2in}
\includegraphics[width=3.2in,clip=]{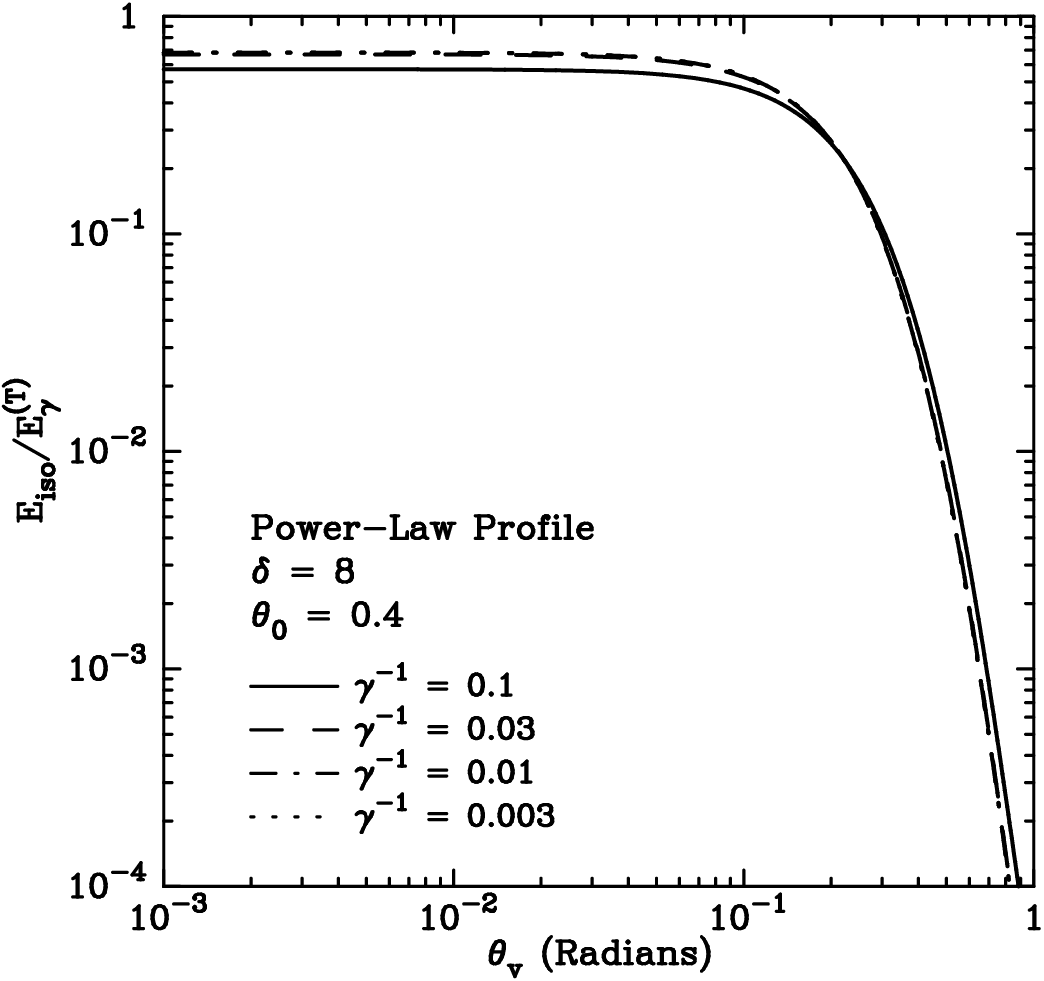}\\
\includegraphics[width=3.2in,clip=]{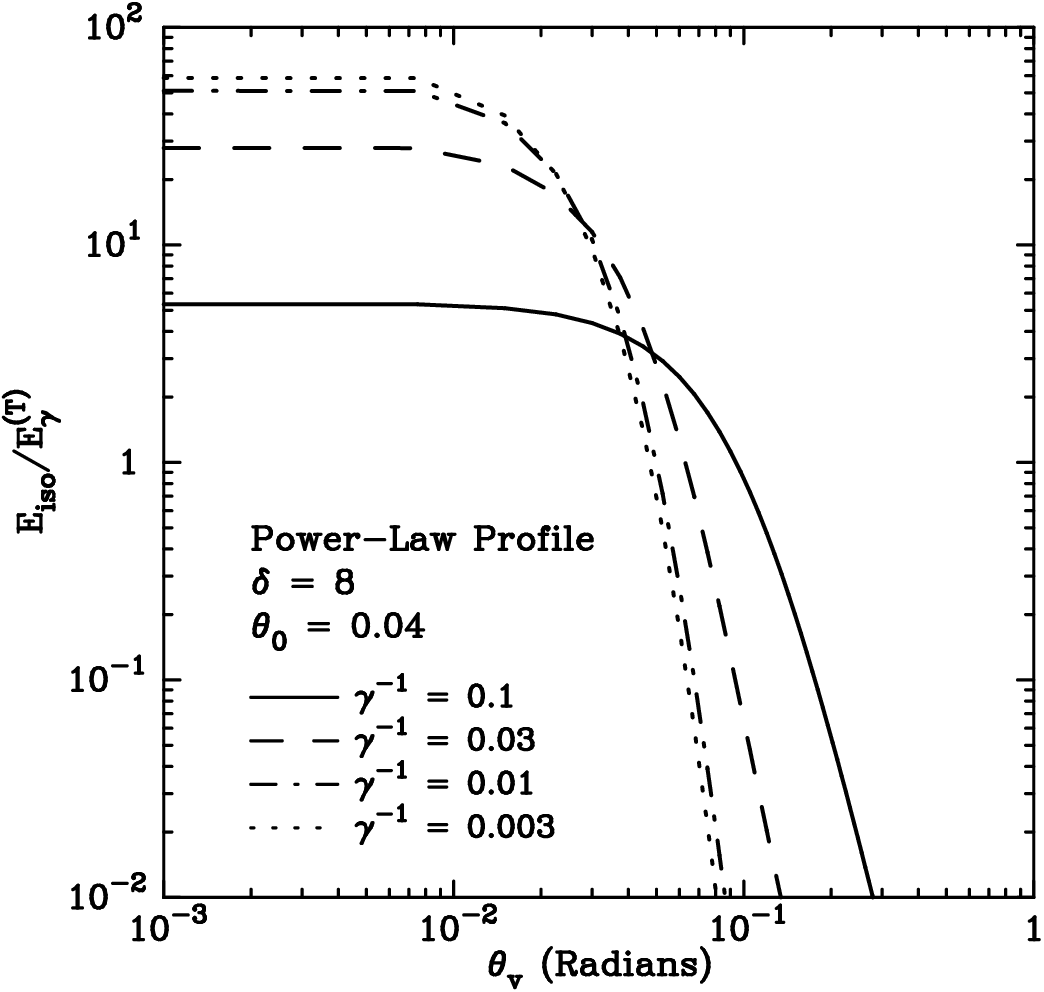}
\end{minipage}
\begin{minipage}[t]{3.2in}
\includegraphics[width=3.2in,clip=]{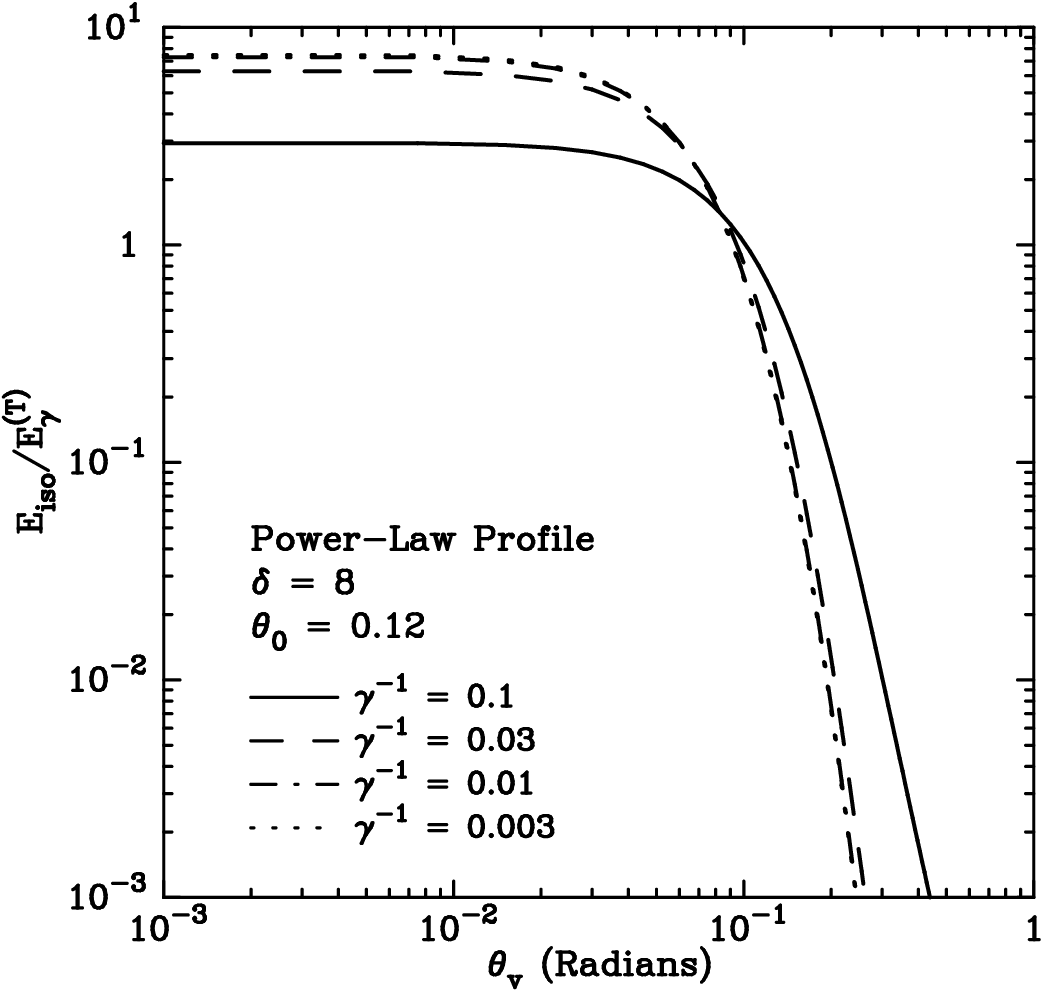}\\
\includegraphics[width=3.2in,clip=]{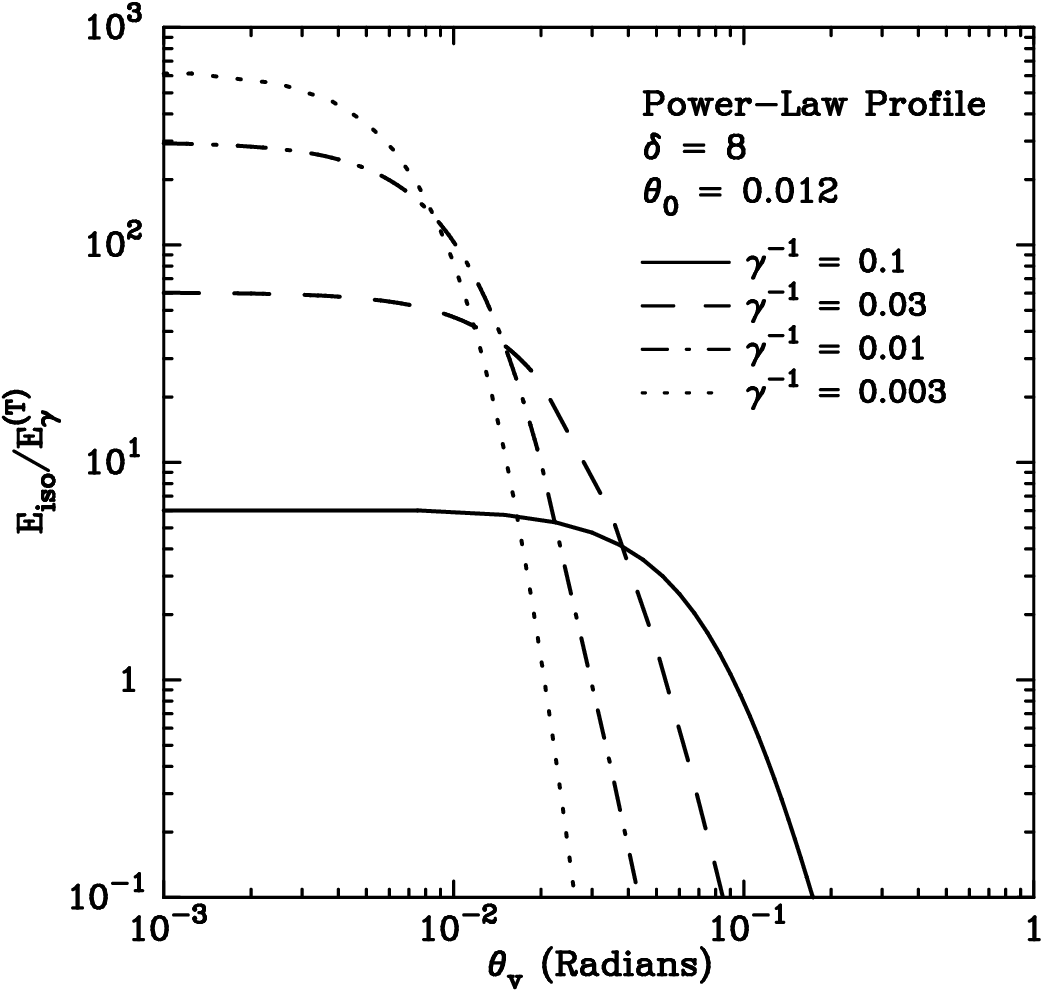}
\end{minipage}
\caption{\eiso\ as a function of viewing angle for emission from
relativistic jets with $\delta=8$ power-law emission profiles.  Each panel
shows the result for $\gamma$ = 10, 33.3, 100, and 333.  Top left: 
$\theta_0=0.4$.  Top right: $\theta_0=0.12$.  Bottom left:
$\theta_0=0.04$.  Bottom right: $\theta_0=0.012$.  These values of
$\theta_0$ are chosen so as to make the effective profile widths given by
Eq.~\ref{pl_width} be 0.1, 0.03, 0.01, and 0.003, respectively.}
\label{pl_8.0_emission}
\end{figure}

The four panels of Fig.~\ref{pl_8.0_emission} show the behavior of \eiso\ 
as a function of viewing angle $\theta_v$ for a power-law profile with
$\delta=8$.  We see jet broadening again, as expected.  For larger values
of $\gamma$, this profile appears to be able to produce a fairly wide
variety of radiation patterns, and great dynamic range of \eiso.

\begin{figure}[t]
\begin{minipage}[t]{3.2in}
\includegraphics[width=3.2in,clip=]{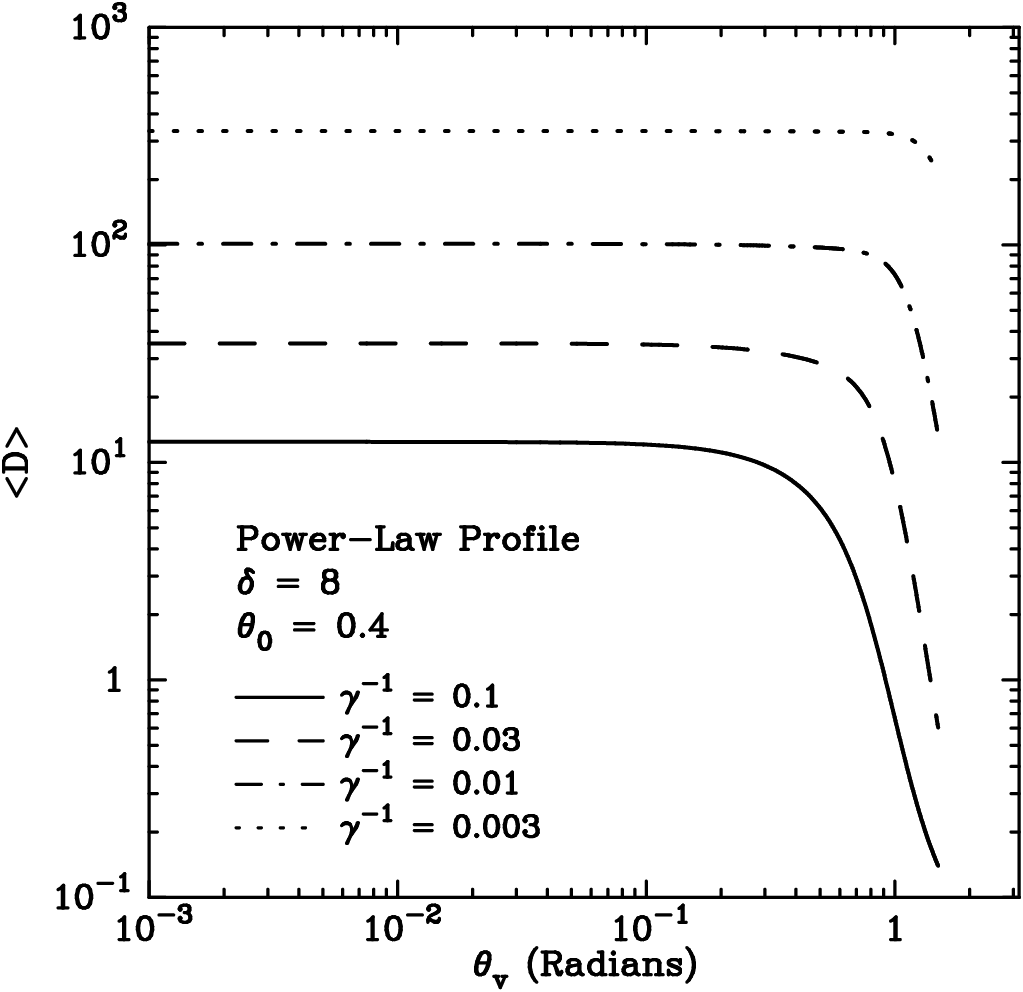}\\
\includegraphics[width=3.2in,clip=]{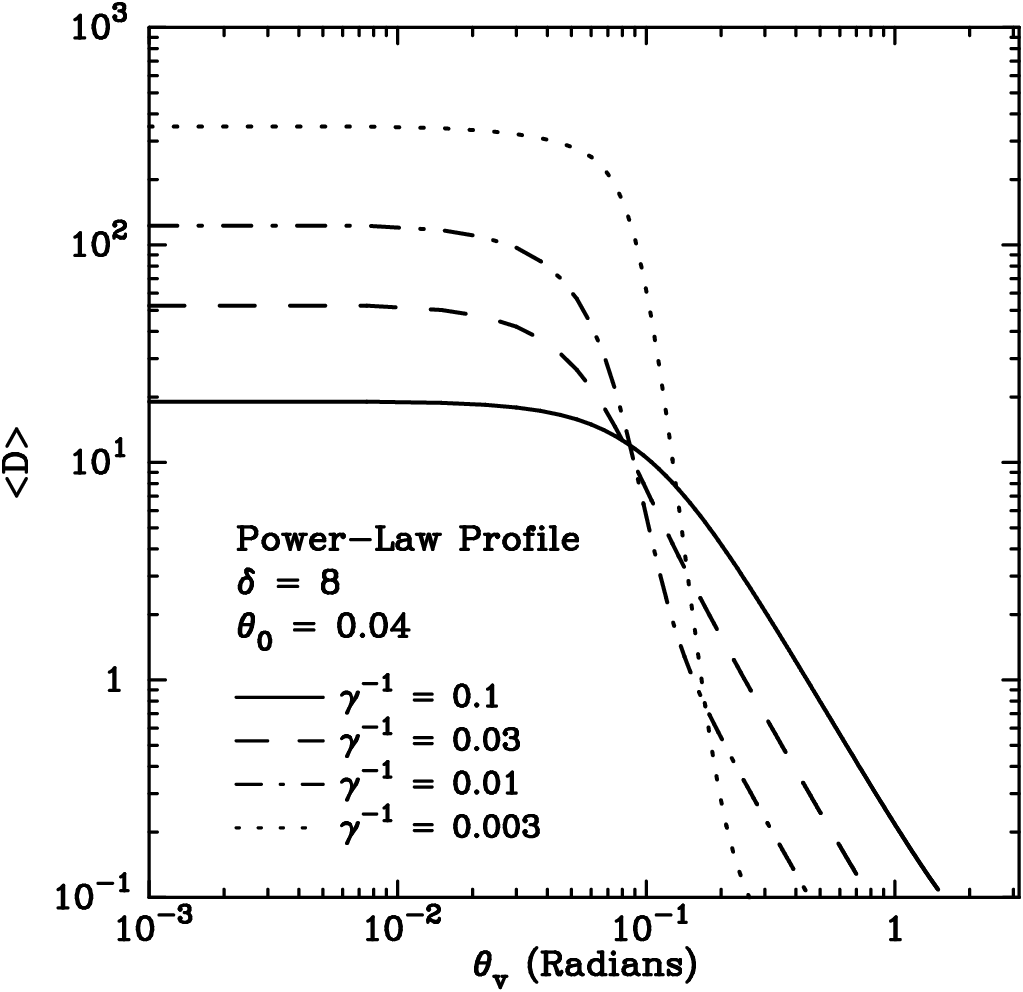}
\end{minipage}
\begin{minipage}[t]{3.2in}
\includegraphics[width=3.2in,clip=]{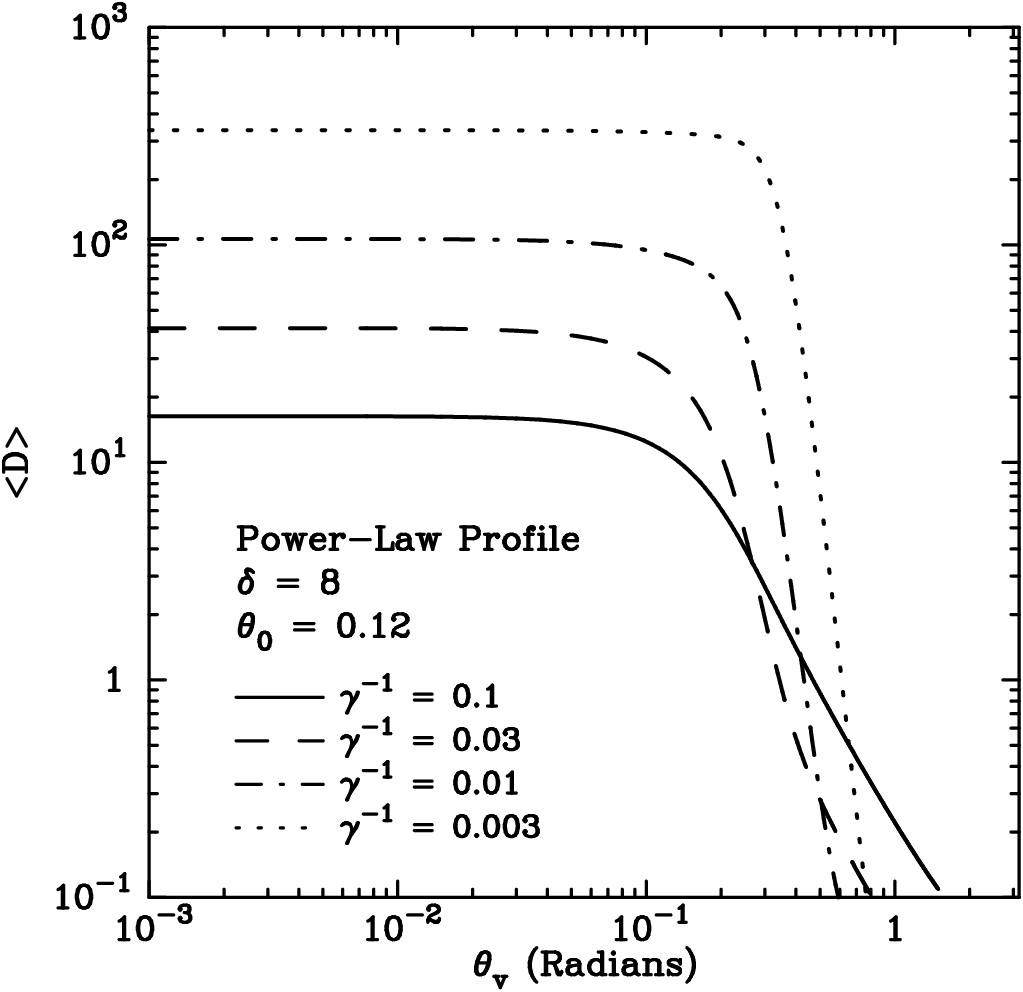}\\
\includegraphics[width=3.2in,clip=]{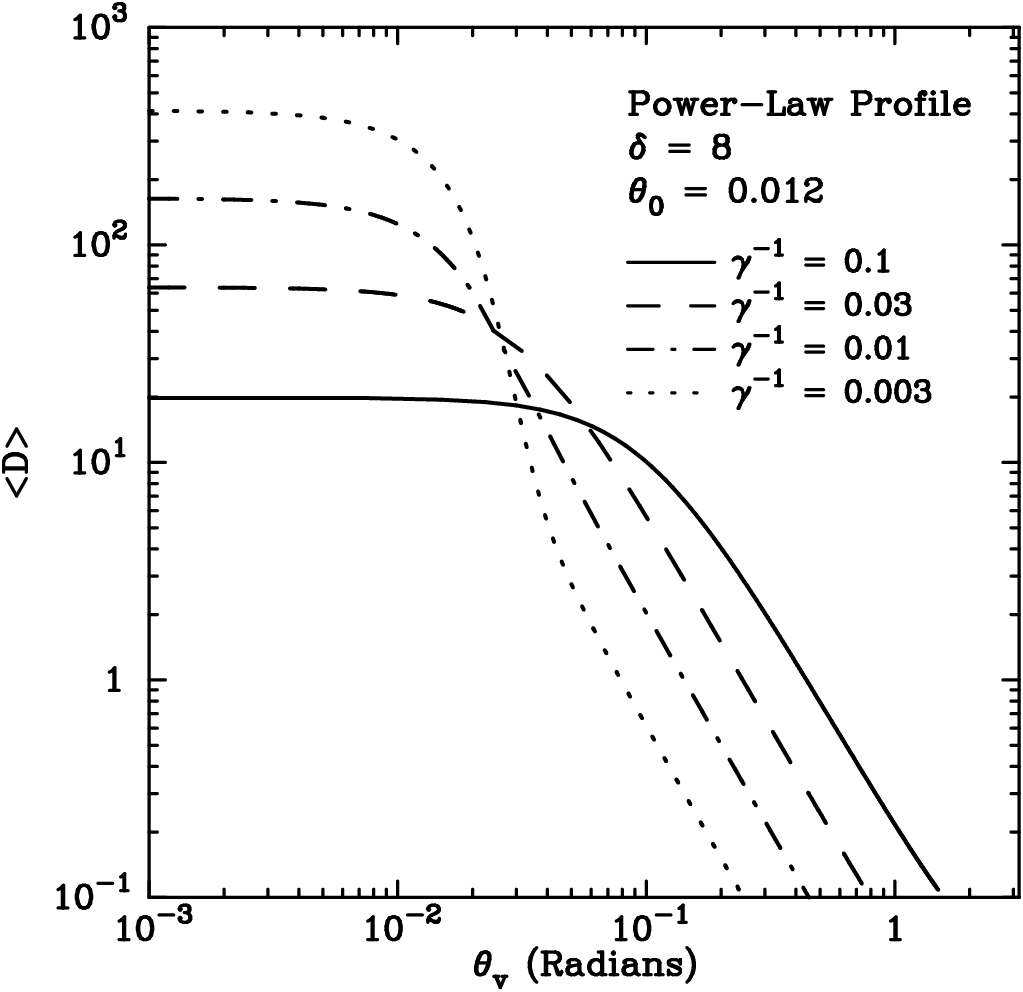}
\end{minipage}
\caption{Photon-number-averaged Doppler factor \davg\ as a function of
viewing angle for emission from relativistic jets with $\delta=8$
power-law emission profiles. Each panel shows the result for $\gamma$ =
10, 33.3, 100, and 333.  Top left:  $\theta_0=0.4$.  Top right:
$\theta_0=0.12$.  Bottom left: $\theta_0=0.04$.  Bottom right:
$\theta_0=0.012$.}
\label{pl_8.0_epk}
\end{figure}

The four panels of Fig.~\ref{pl_8.0_epk} show the behavior of \davg\ as a
function of viewing angle $\theta_v$.  Here we can finally see strong
dynamic range of \davg\ in a power-law model, at least for narrow jets.  We
also see the return of the ``shelf'' feature that was noted in
\S\ref{sec_fisher_emission}, which appears to be a generic feature of steep
profiles.

\subsubsection{\eiso\ -- \davg\ and \egi\ -- \davg\ Correlations}

\begin{figure}[t]
\begin{minipage}[t]{3.2in}
\includegraphics[width=3.2in,clip=]{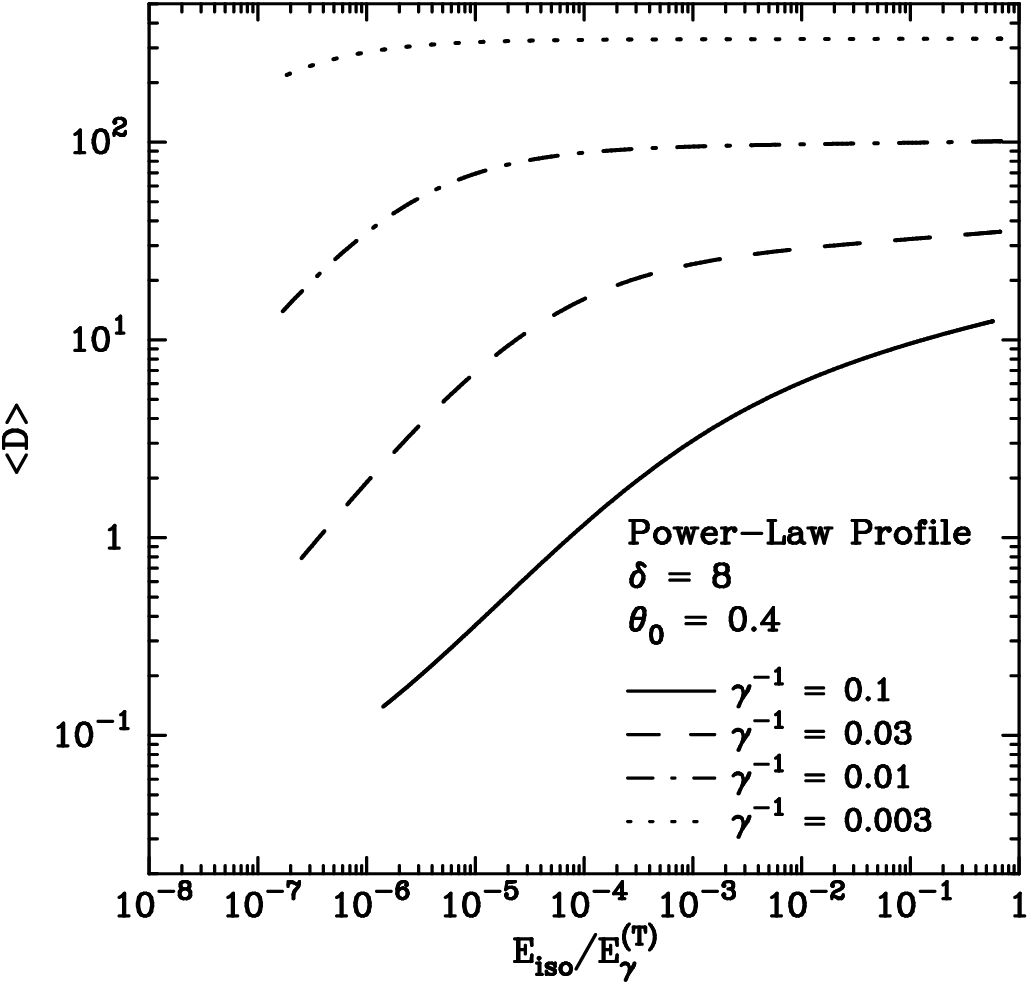}\\
\includegraphics[width=3.2in,clip=]{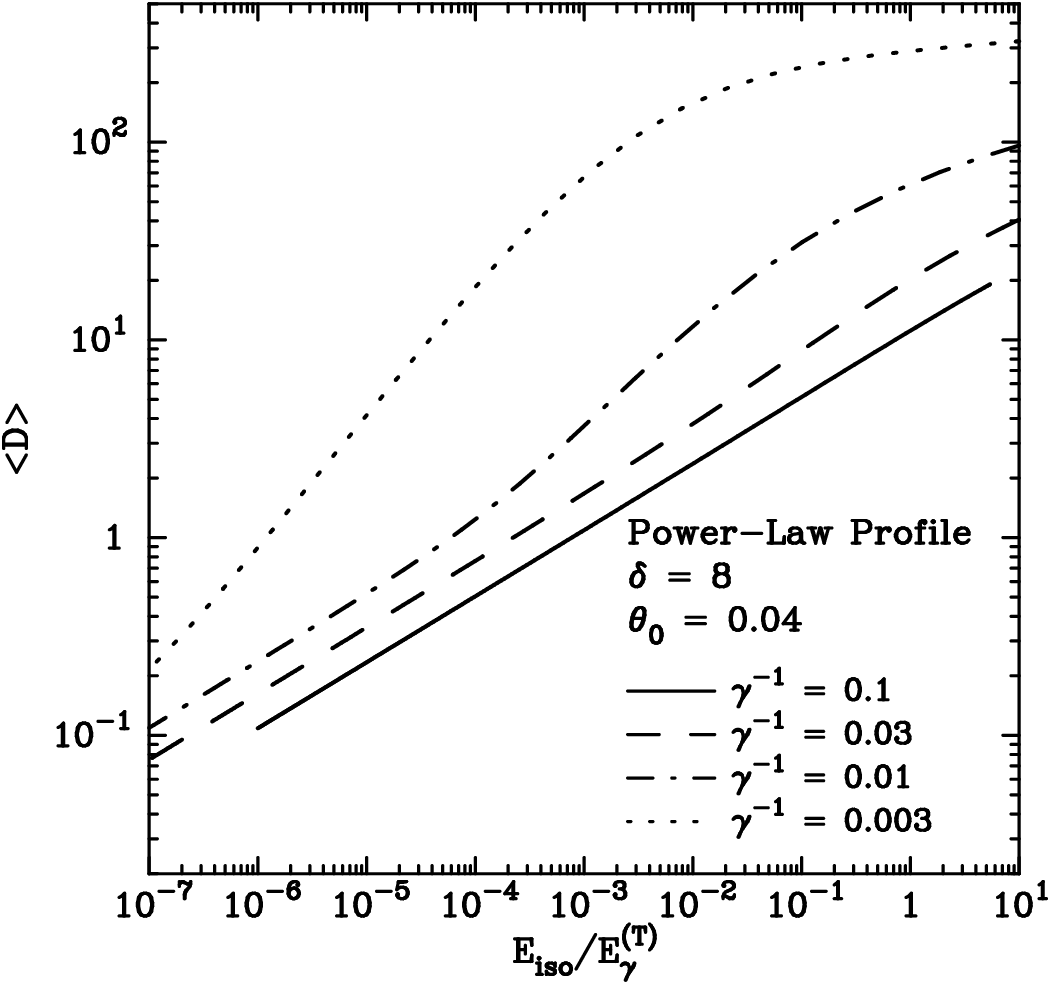}
\end{minipage}
\begin{minipage}[t]{3.2in}
\includegraphics[width=3.2in,clip=]{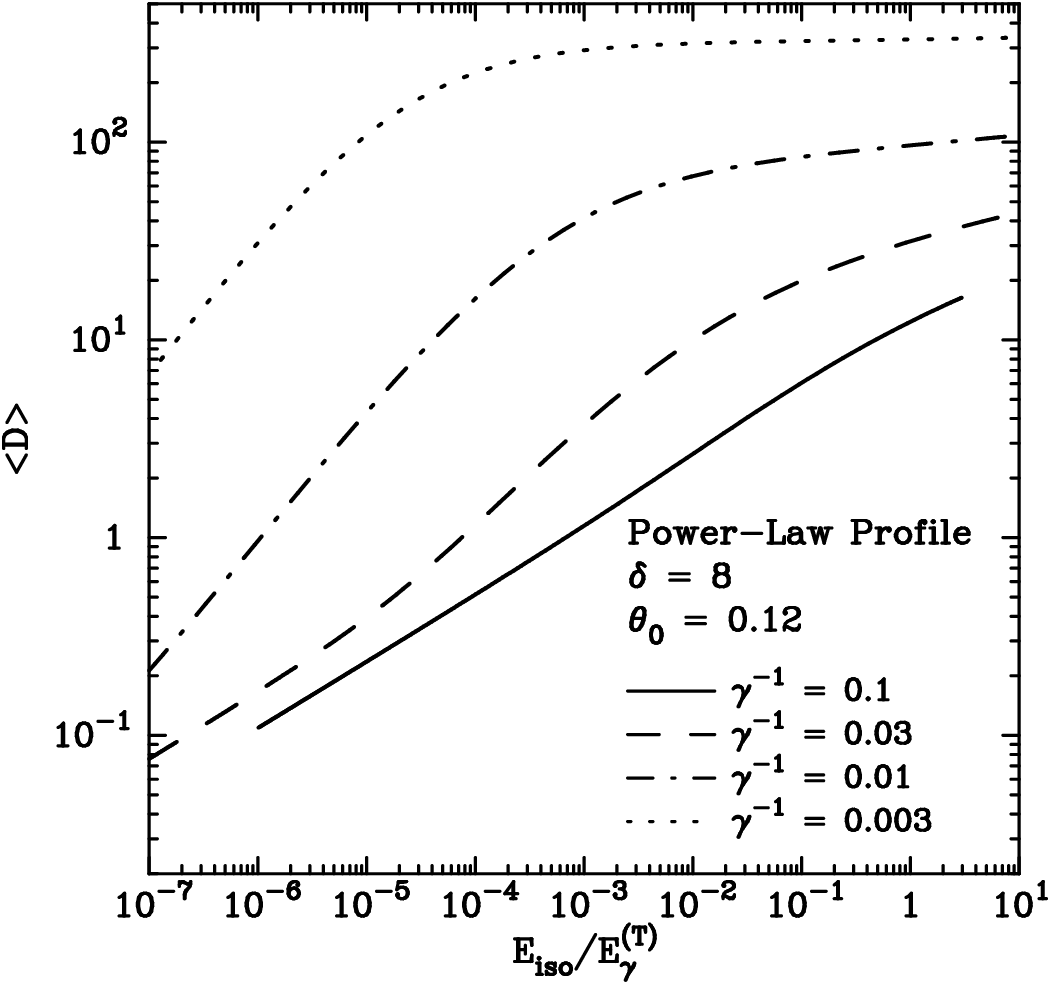}\\
\includegraphics[width=3.2in,clip=]{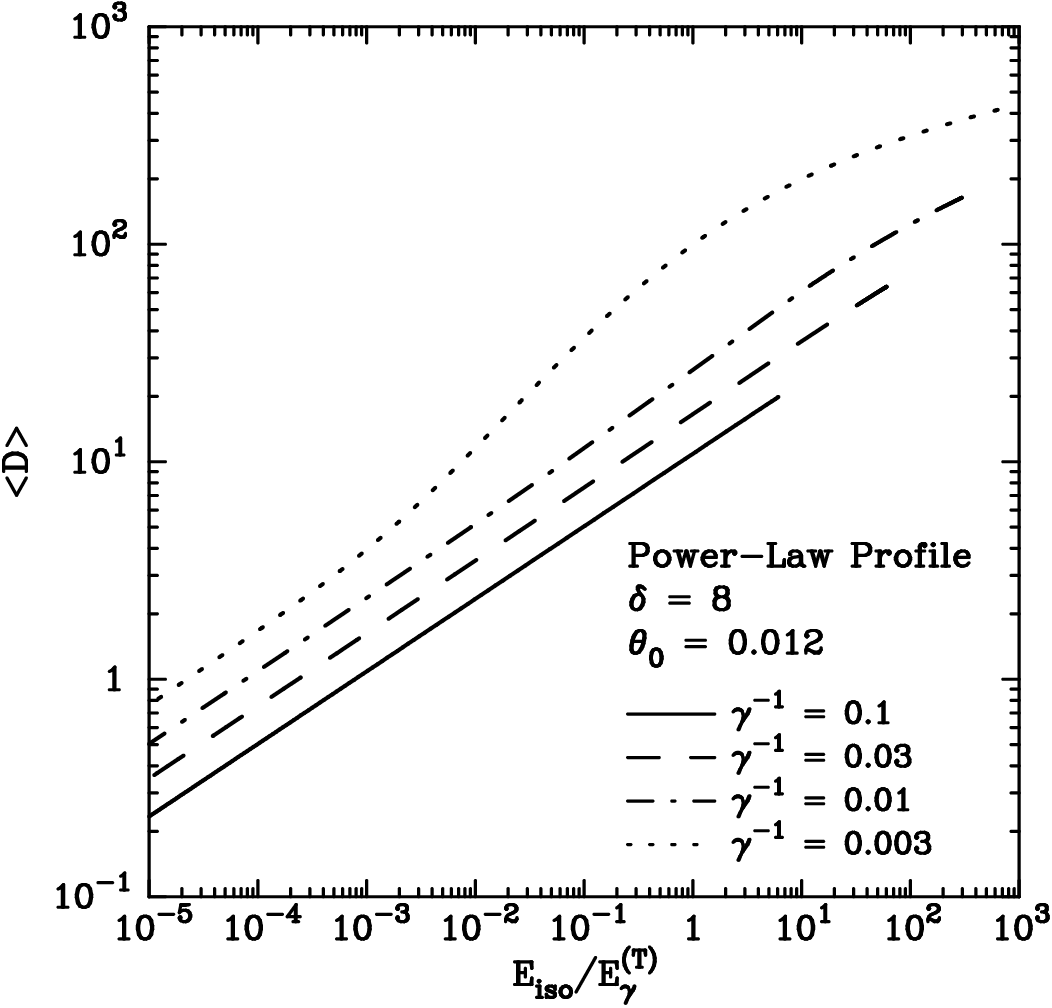}
\end{minipage}
\caption{\eiso--\davg\ (``Amati'') plots for universal jet models based on
$\delta=8$ power-law profiles.  Each panel shows the result for $\gamma$ =
10, 33.3, 100, and 333.  Top left:  $\theta_0=0.4$.  Top right:
$\theta_0=0.12$.  Bottom left: $\theta_0=0.04$.  Bottom right:
$\theta_0=0.012$.}
\label{pl_8.0_amati}
\end{figure}

The four panels of Fig.~\ref{pl_8.0_amati} show the behavior of the
\eiso--\davg\ (``Amati'') correlation in universal $\delta=8$ power-law
models.  The axis scales are chosen so that a slope of
0.5 appears as a 45$^\circ$ line, in order to ease the search for an
``Amati-like'' slope.

Here it appears that slopes steeper than 1/3 can be produced, possibly
even over a sufficiently large dynamic range of \eiso\ and \davg, at the
cost of some flattening out near the bright end of the distribution. The
slope of the $\theta_0=0.04$, $\gamma^{-1}=0.003$ curve (bottom left
panel), for example, is about 0.68 at intermediate \eiso, and by
comparison with Fig.~\ref{pl_8.0_epk} (bottom right panel), the flattening
out at high values of \davg\ only sets in at about $\theta_v=0.1$,
subtending only $5\times 10^{-3}$ of a hemisphere.  While the slope is
high compared to the standard Amati value of 0.5, it appears possible that
there could be a value of $\delta$ bracketed between $\delta=2$ and
$\delta=8$ that can reproduce the Amati slope.

\begin{figure}[t]
\begin{minipage}[t]{3.2in}
\includegraphics[width=3.2in,clip=]{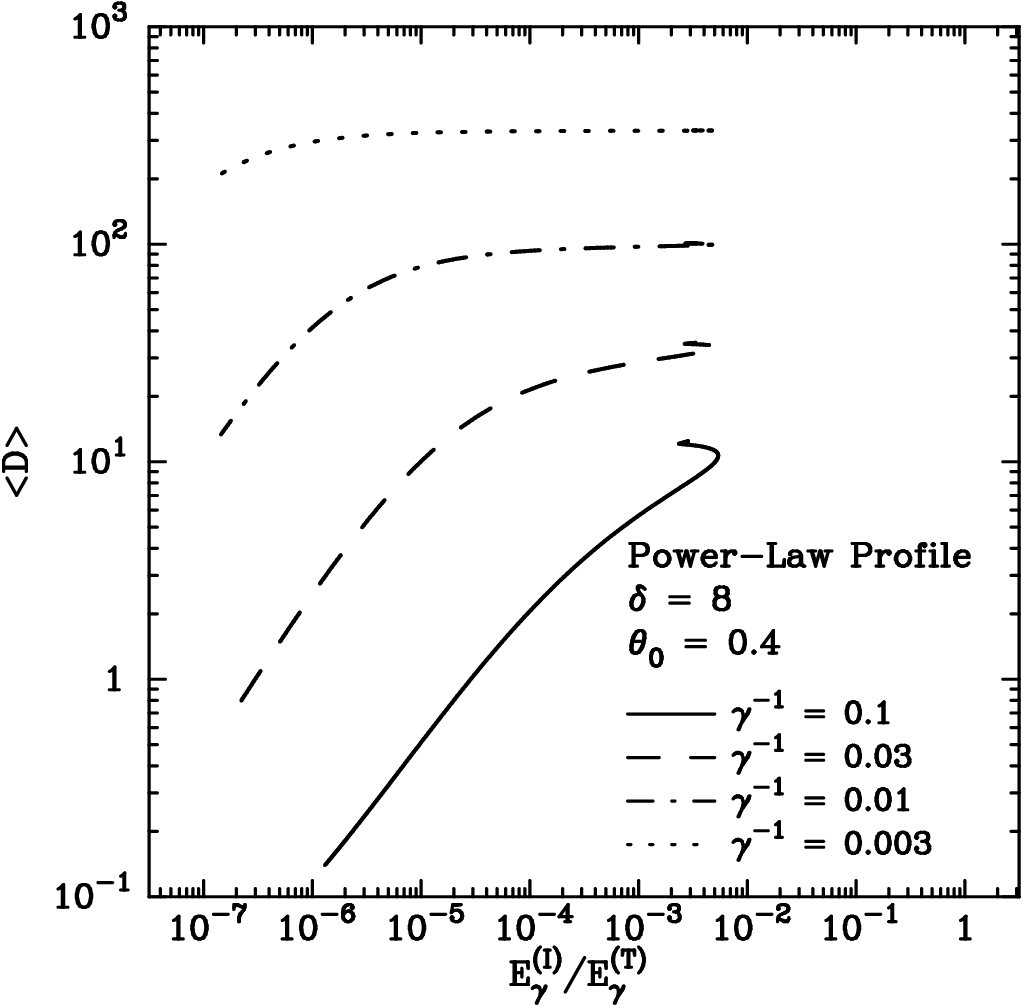}\\
\includegraphics[width=3.2in,clip=]{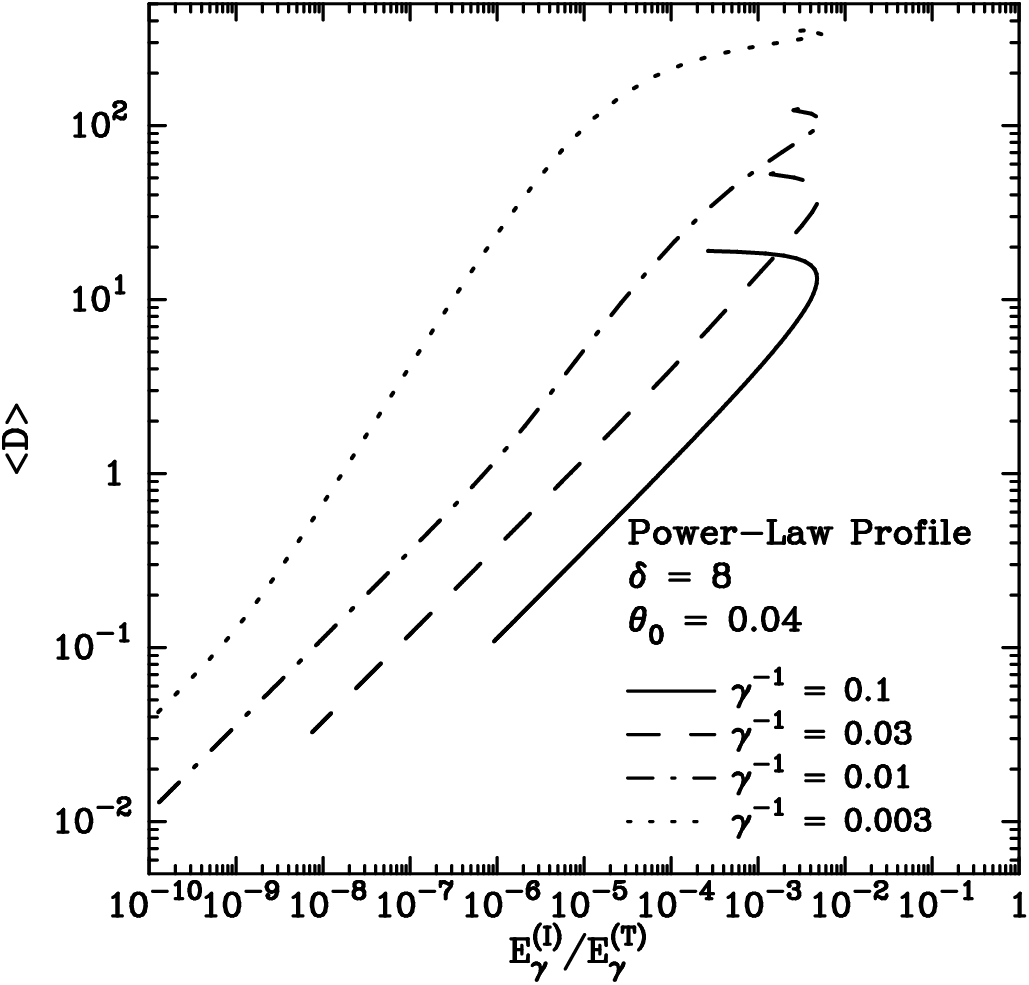}
\end{minipage}
\begin{minipage}[t]{3.2in}
\includegraphics[width=3.2in,clip=]{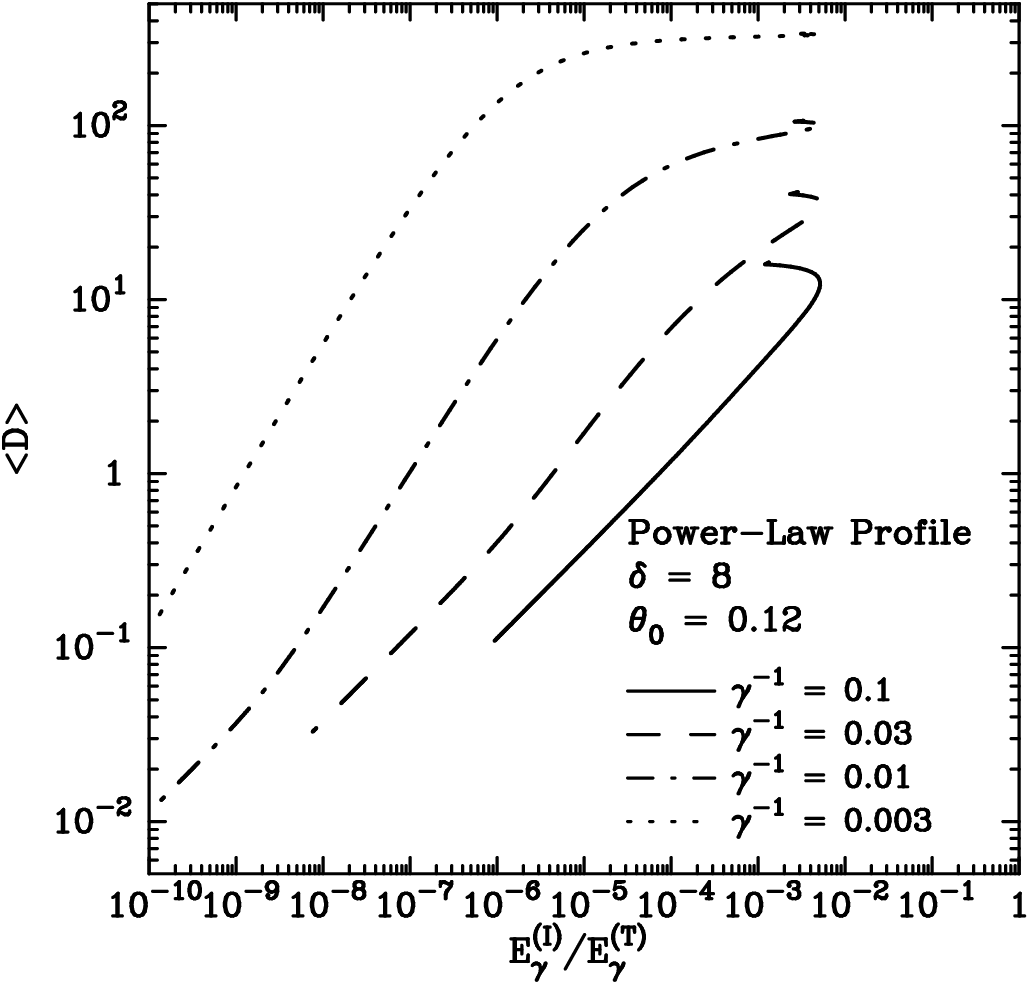}\\
\includegraphics[width=3.2in,clip=]{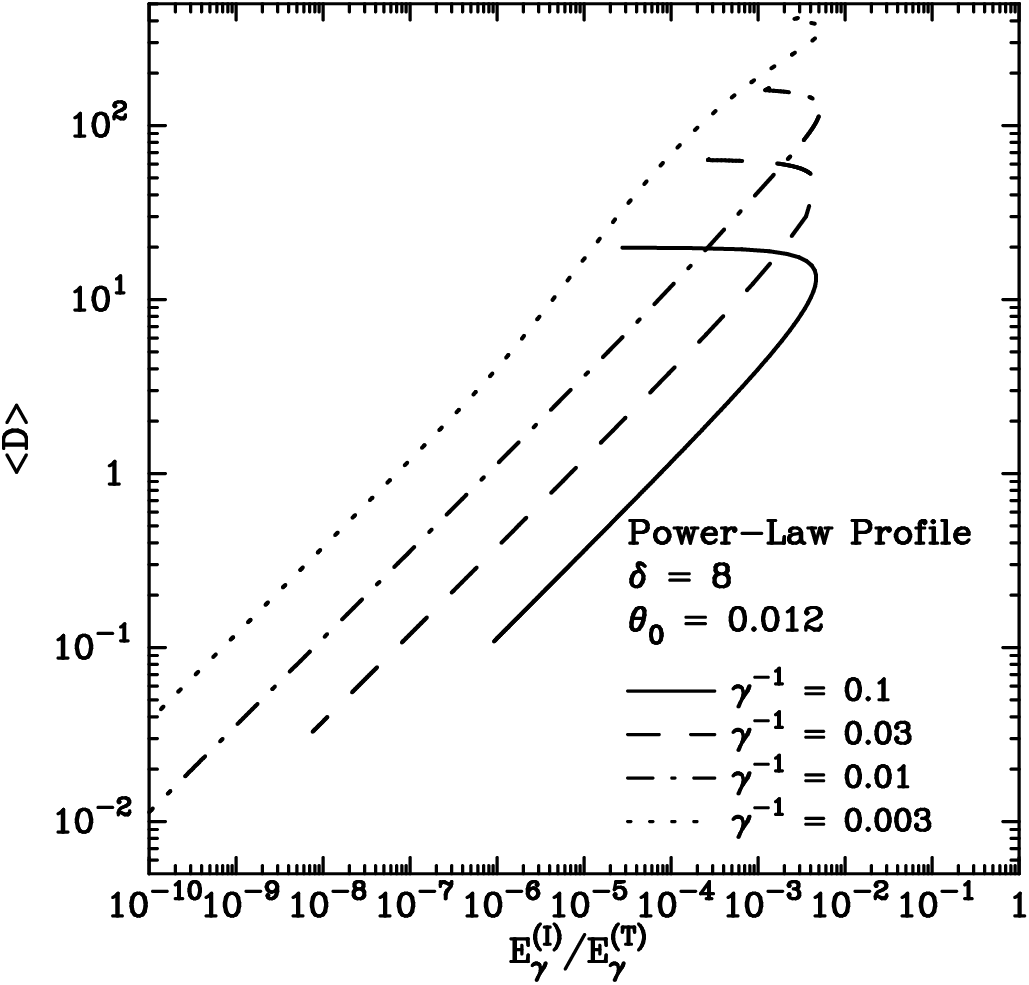}
\end{minipage}
\caption{\egi--\davg\ (``Ghirlanda'') plots for universal jet models based
on $\delta=2$ power-law profiles.  Each panel shows the result for
$\gamma$ = 10, 33.3, 100, and 333.  Top left:  $\theta_0=0.4$.  Top right:
$\theta_0=0.12$.  Bottom left: $\theta_0=0.04$.  Bottom right:
$\theta_0=0.012$.}
\label{pl_8.0_ghirlanda}
\end{figure}

The four panels of Fig.~\ref{pl_8.0_ghirlanda} show the behavior of the
\egi--\davg\ (``Ghirlanda'') correlation in universal $\delta=8$ power-law
models.

Just as for the ``Amati'' correlation plots, there appears to
be a range of parameter space with relatively steep slope and large
dynamic range of \davg.  The slope of the $\theta_0=0.04$,
$\gamma^{-1}=0.003$ curve (bottom left panel), for example, is about 0.83
at intermediate \eiso, and again the flattening at high energy subtends a
tiny solid angle.  It thus appears possible that there could be a value of
$\delta$ bracketed between $\delta=2$ and $\delta=8$ that can reproduce
the Ghirlanda slope.

\subsubsection{\eiso\ and \egi\ Frequency Distributions}

\begin{figure}[t]
\begin{minipage}[t]{3.2in}
\includegraphics[width=3.2in,clip=]{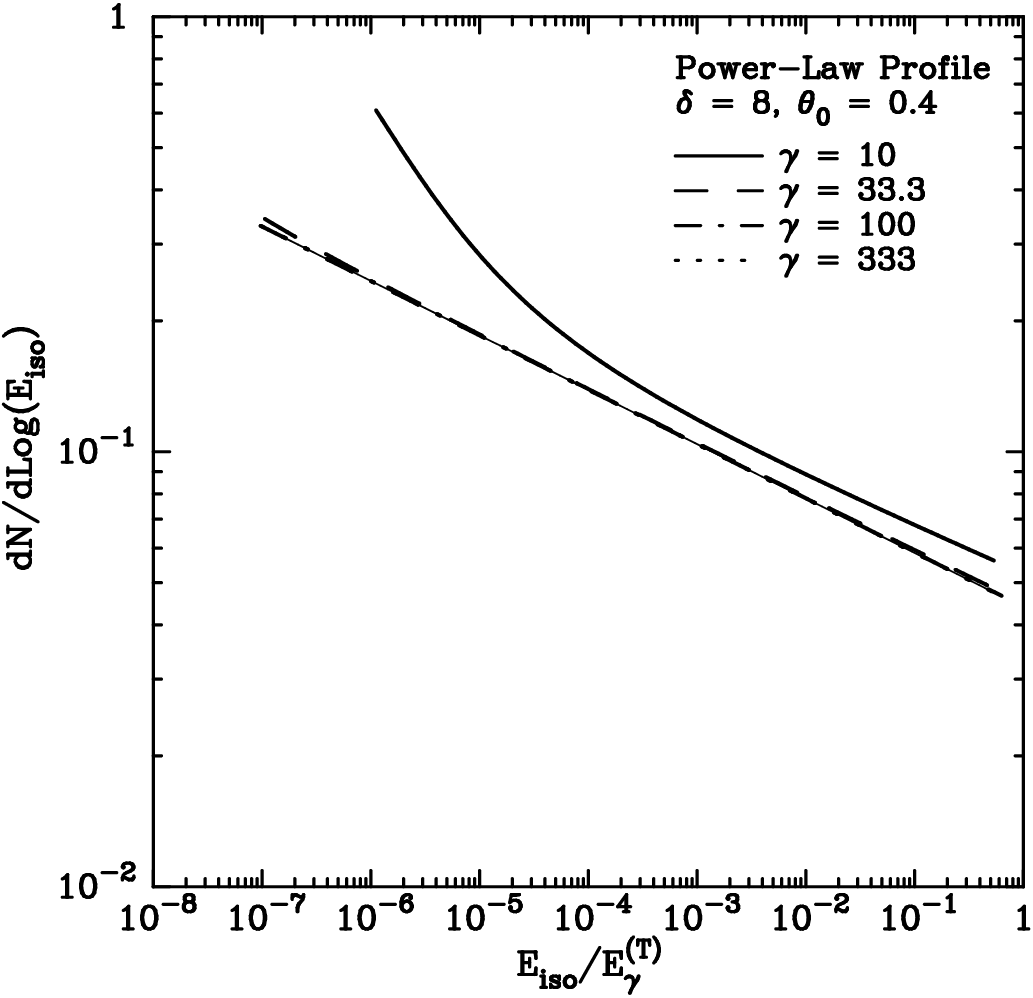}\\
\includegraphics[width=3.2in,clip=]{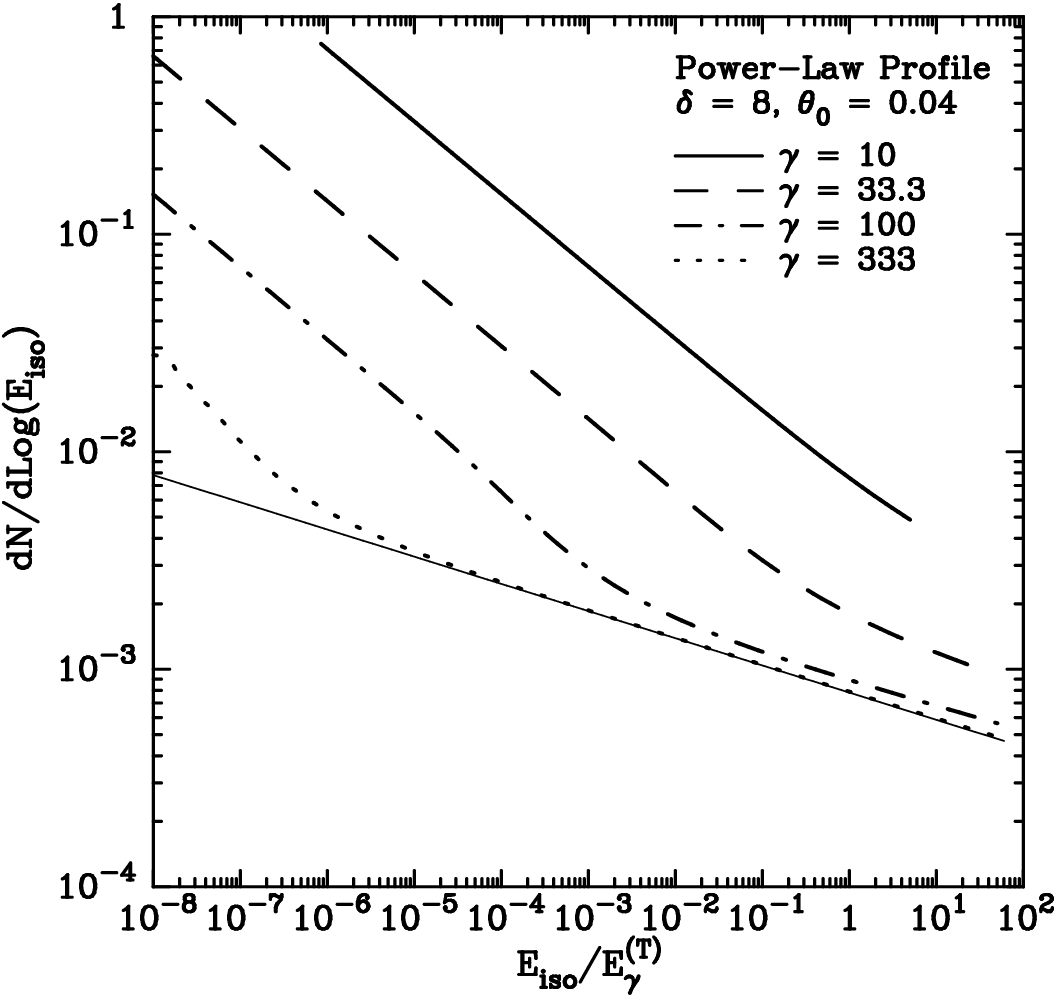}
\end{minipage}
\begin{minipage}[t]{3.2in}
\includegraphics[width=3.2in,clip=]{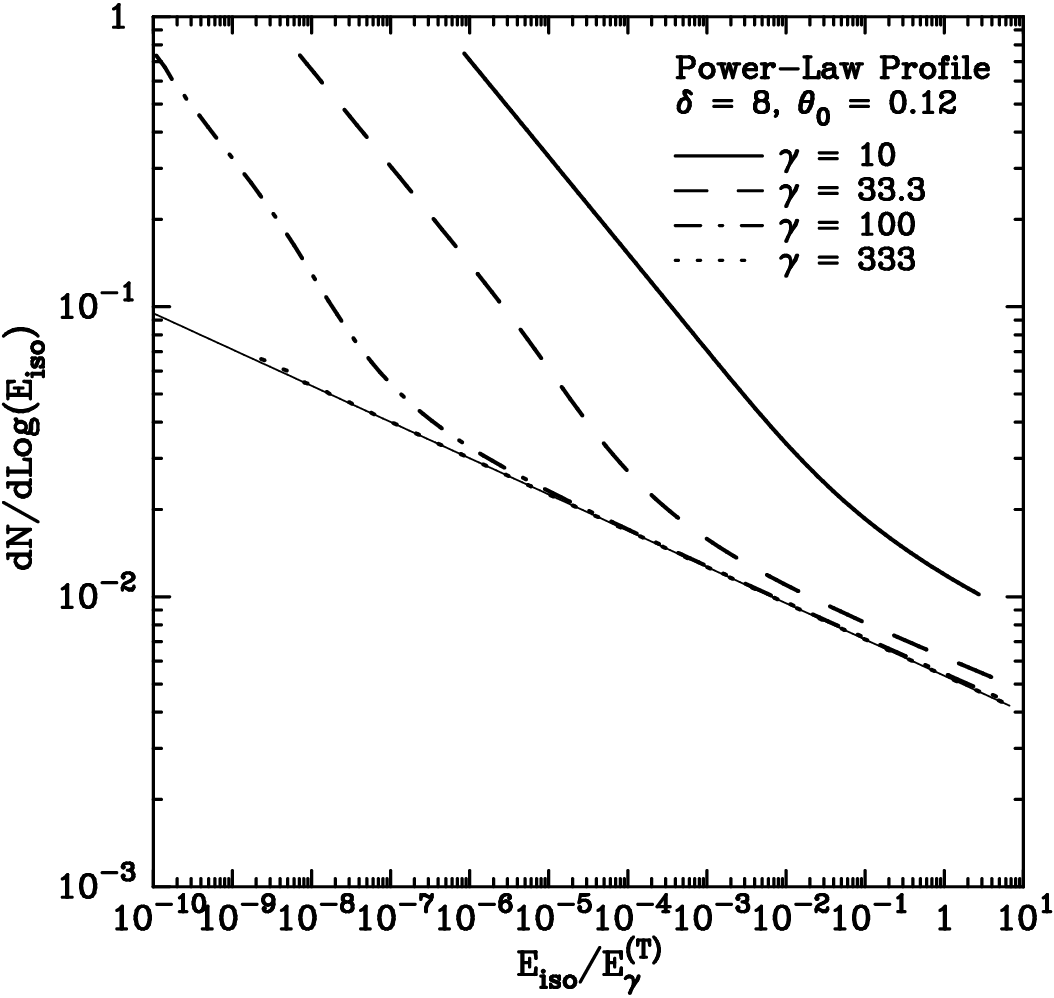}\\
\includegraphics[width=3.2in,clip=]{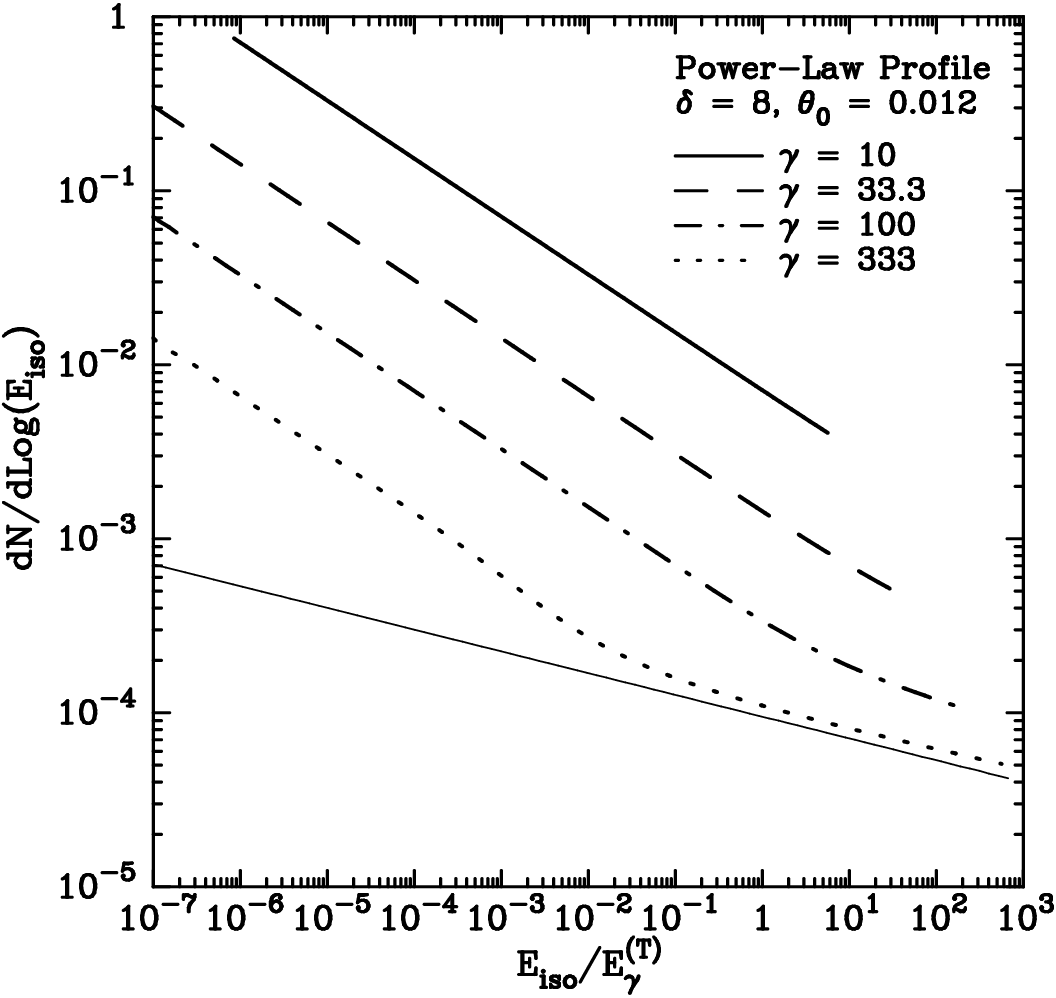}
\end{minipage}
\caption{$dN/d\log(\eiso)$ distributions for universal jet models based on
$\delta=8$ power-law profiles.  Each panel shows the result for $\gamma$ =
10, 33.3, 100, and 333.  Top left:  $\theta_0=0.4$.  Top right:
$\theta_0=0.12$.  Bottom left: $\theta_0=0.04$.  Bottom right:
$\theta_0=0.012$.  The vertical scale is arbitrary.  The light
solid line shows the ``bare'' distribution due to the underlying profile
--- essentially the $\gamma\rightarrow\infty$ limit.  Its slope is -1/8, as
expected from Eq.~\ref{plxiso}.}
\label{pl_8.0_dndxiso}
\end{figure}

The four panels of Fig.~\ref{pl_8.0_dndxiso} show the effect of
relativistic kinematics on the \eiso\ frequency distribution of $\delta=8$
power-law jets --- see Eq.~(\ref{plxiso}).  Again, the ``bare''
($\gamma\rightarrow\infty$) distribution is shown by a light solid line, whose
slope in this case is -1/8.

The large value of $\delta$ has finally made it possible for this
distribution to become relatively shallow and broad, possibly doing less
violence to the observed, nearly-uniform distribution than was possible
using shallower power-law indexes.  It remains the case that the ``needle
jet'' regime $\theta_0\ll\gamma^{-1}$ must be avoided to avoid introducing
too great a non-uniformity (see Eq.~(\ref{eiso_limit}). This is likely to
place a lower bound on $\gamma\theta_0$ in fits to the \eiso\ distribution
data.

\begin{figure}[t]
\begin{minipage}[t]{3.2in}
\includegraphics[width=3.2in,clip=]{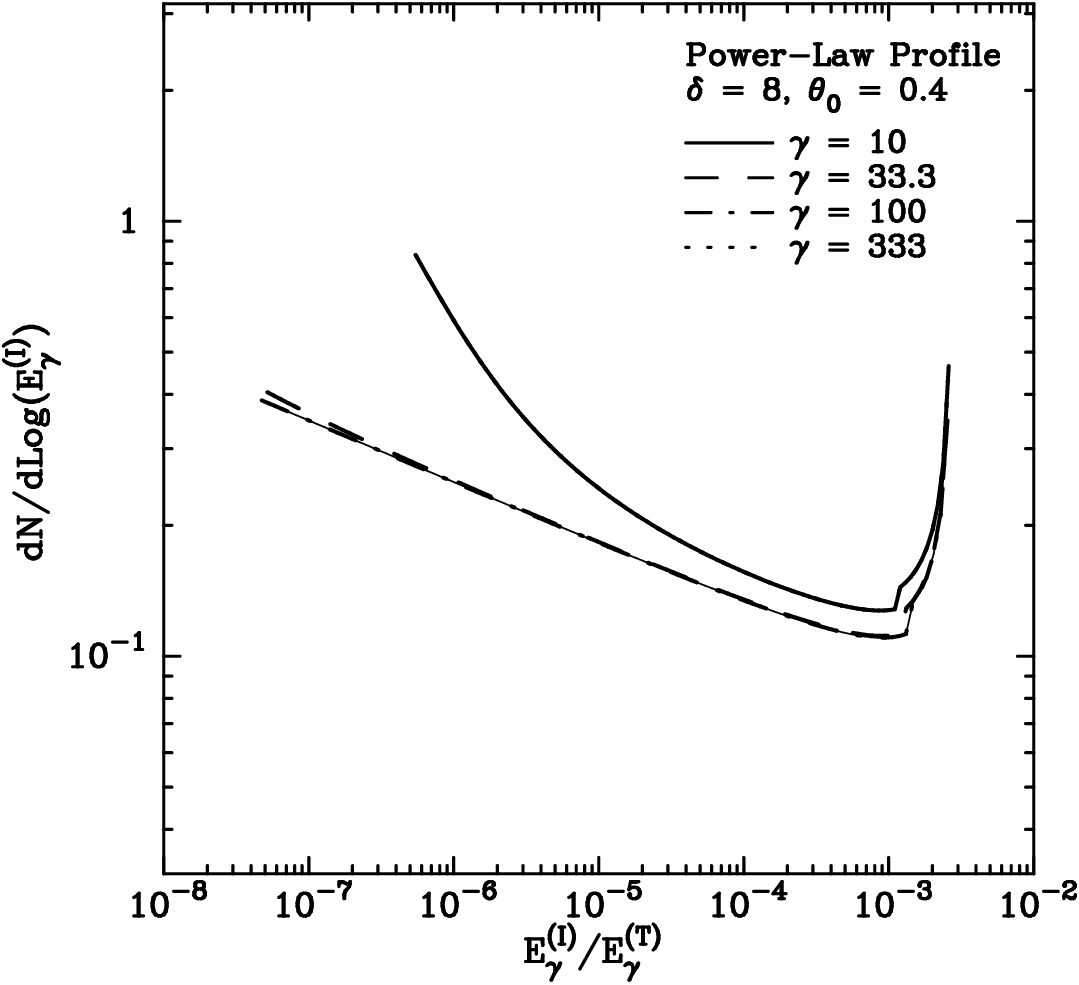}\\
\includegraphics[width=3.2in,clip=]{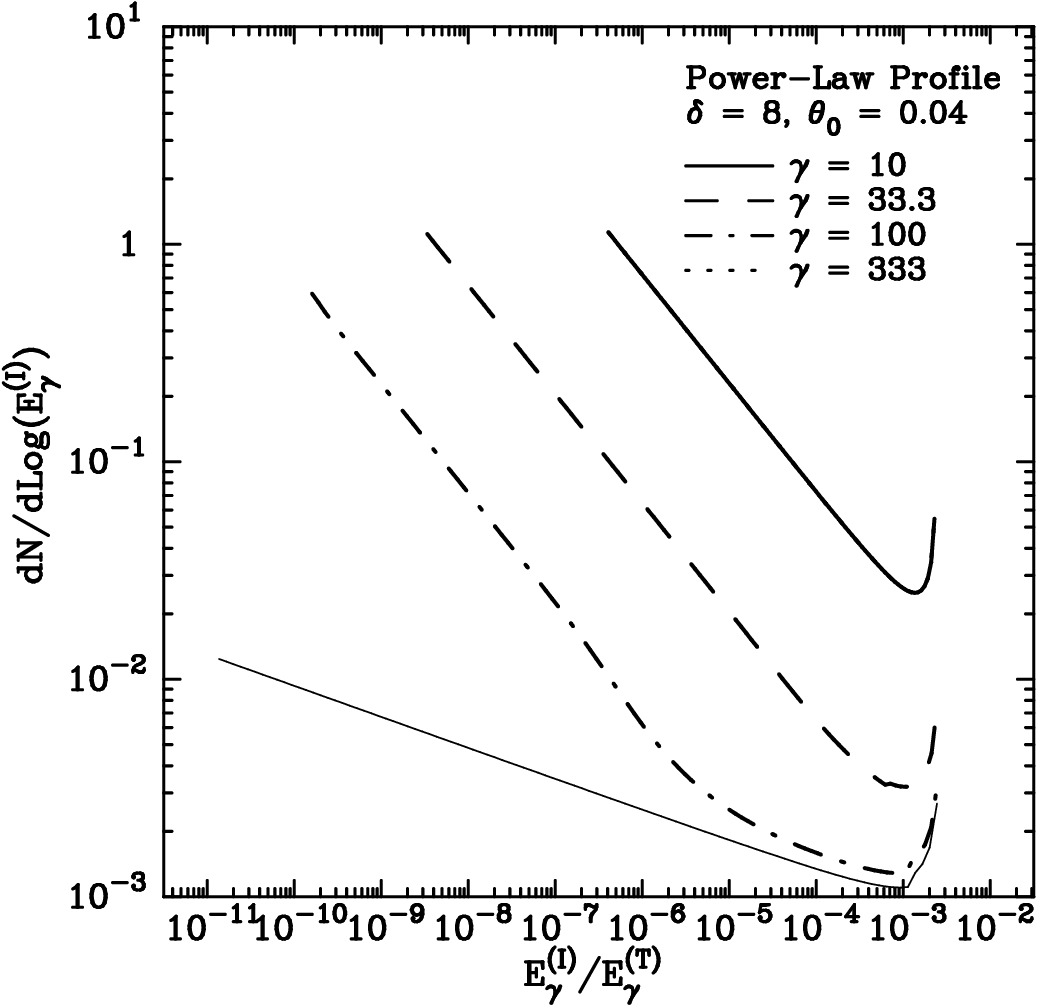}
\end{minipage}
\begin{minipage}[t]{3.2in}
\includegraphics[width=3.2in,clip=]{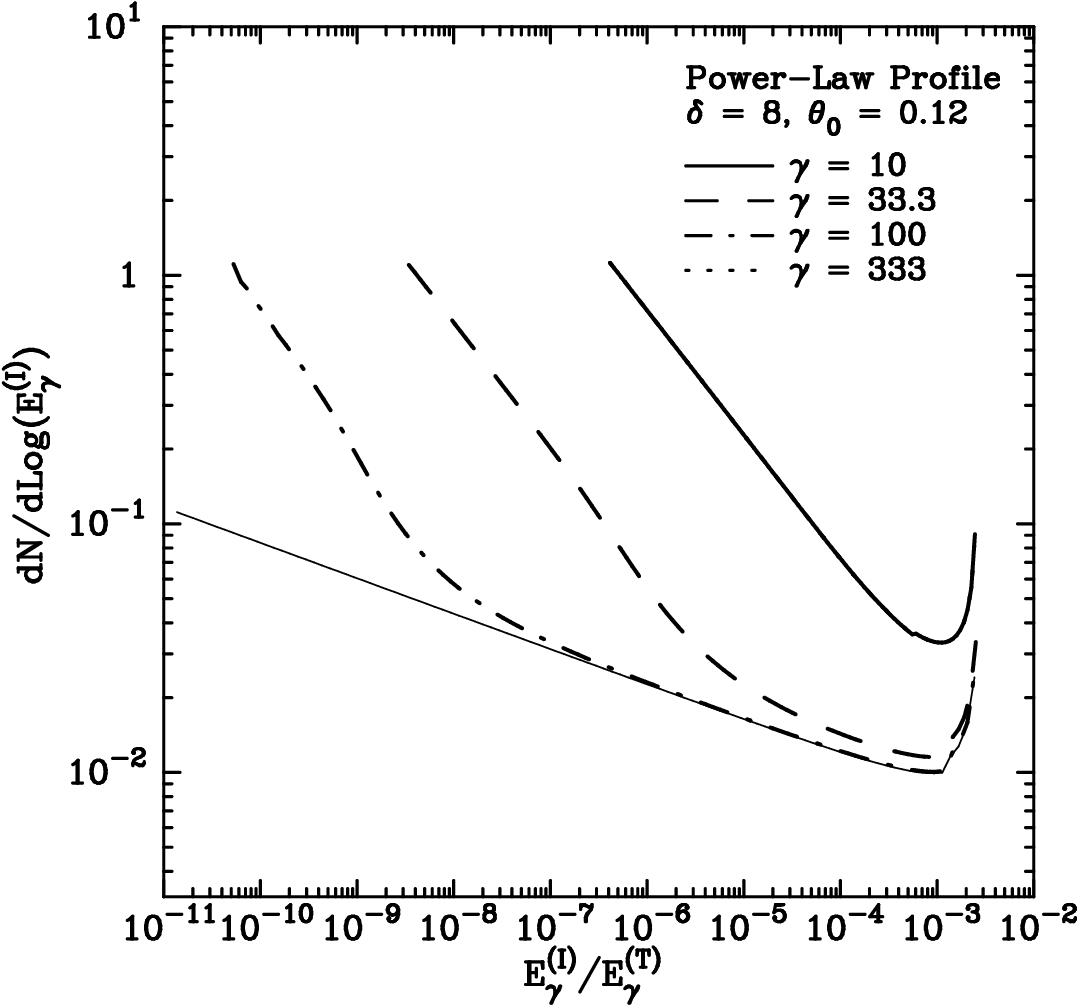}\\
\includegraphics[width=3.2in,clip=]{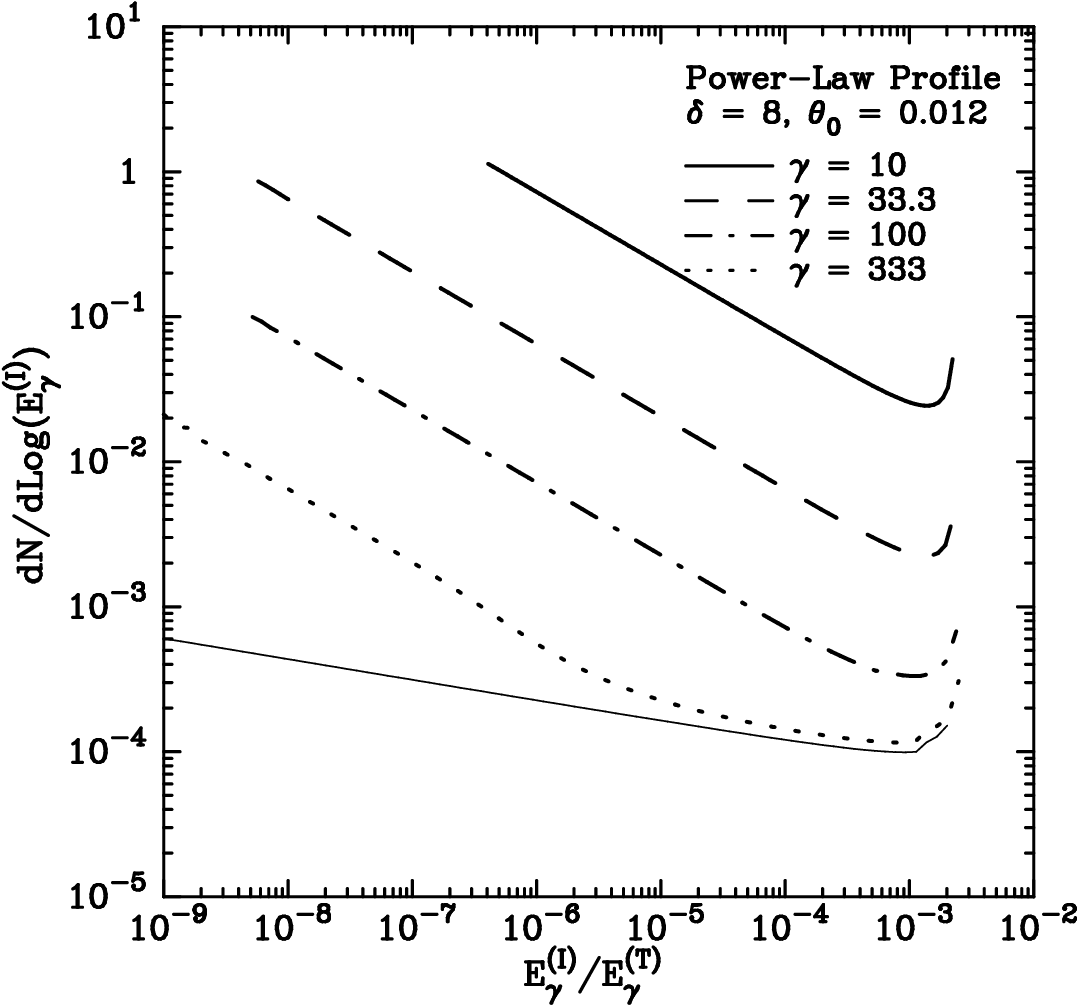}
\end{minipage}
\caption{$dN/d\log(\egi)$ distributions for universal jet models based on
$\delta=8$ power-law profiles.  Each panel shows the result for $\gamma$ =
10, 33.3, 100, and 333.  Top left:  $\theta_0=0.4$.  Top right:
$\theta_0=0.12$.  Bottom left: $\theta_0=0.04$.  Bottom right:
$\theta_0=0.012$.  The vertical scale is arbitrary.  The light solid line
shows the ``bare'' distribution due to the underlying profile ---
essentially the $\gamma\rightarrow\infty$ limit.  Its slope is -1/7, as
expected from Eq.~\ref{plxgi}.}
\label{pl_8.0_dndxgi}
\end{figure}

The four panels of Fig.~\ref{pl_8.0_dndxgi} show the effect of
relativistic kinematics on the \egi\ frequency distribution of $\delta=8$
power-law jets --- Eq.~(\ref{plxgi}).  Again, the ``bare''
($\gamma\rightarrow\infty$) distribution is shown by a light solid line, whose
slope in this case is -1/7.

Here the case is reversed.  The ``bare'' slope is far too shallow,
compared to the strongly-peaked distribution of the real data.  It seems
unlikely that relativistic kinematics in the ``needle jet'' regime can
save the day, since even ignoring the just-discussed deleterious effect of
this regime on the \eiso\ frequency distribution, the sharpest slope that
can be produced in this regime is -1/2, by Eq.~(\ref{egi_limit}), which
seems somewhat less steep than is consistent with the data.

%%%%%%%%%%%%%%%%%%%%%%%%%%%%%%%%%%%%%%%%%%%%%%%%%%%
%%%%%%%%%%%%%%%%%%%%%%%%%%%%%%%%%%%%%%%%%%%%%%%%%%%
\subsection{Emission From Top-Hat Jet Profiles}
\subsubsection{Emission}

\begin{figure}[t]
\begin{minipage}[t]{3.2in}
\includegraphics[width=3.2in,clip=]{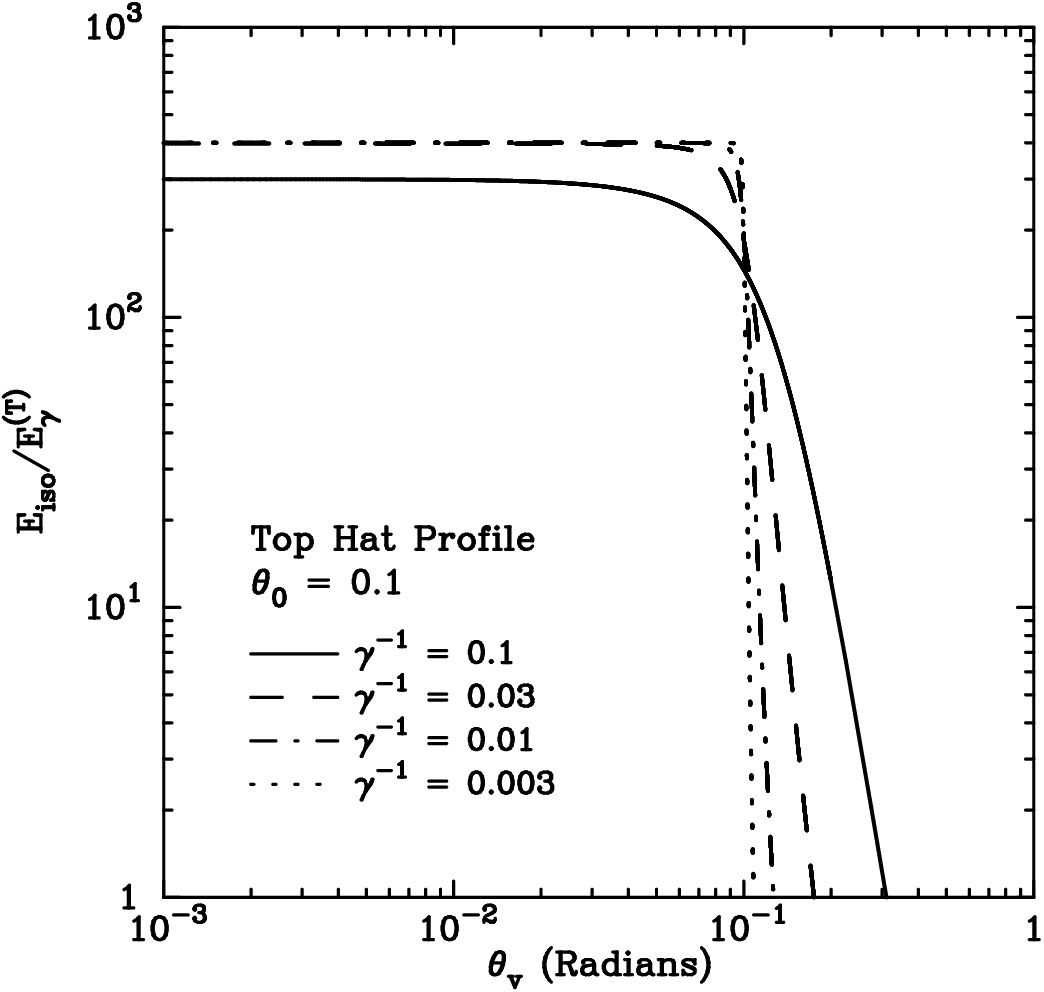}\\
\includegraphics[width=3.2in,clip=]{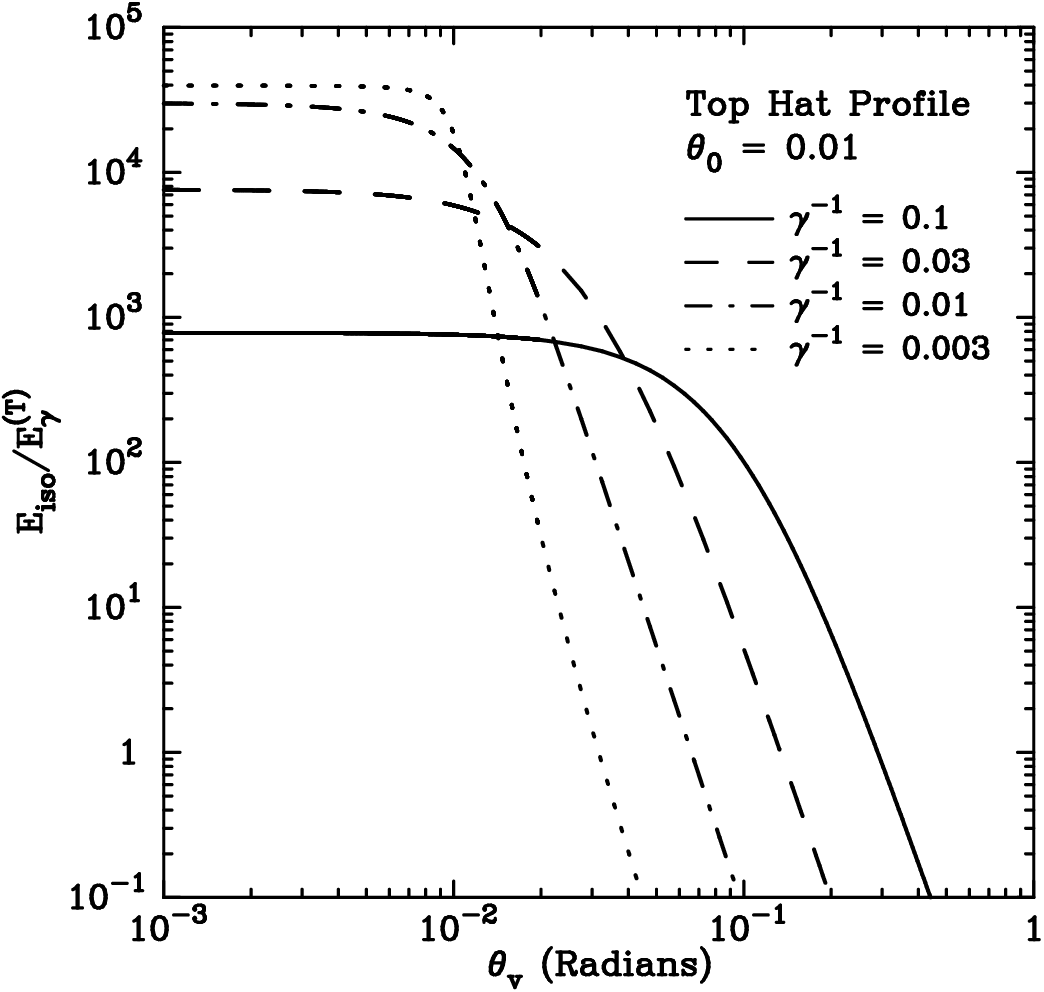}
\end{minipage}
\begin{minipage}[t]{3.2in}
\includegraphics[width=3.2in,clip=]{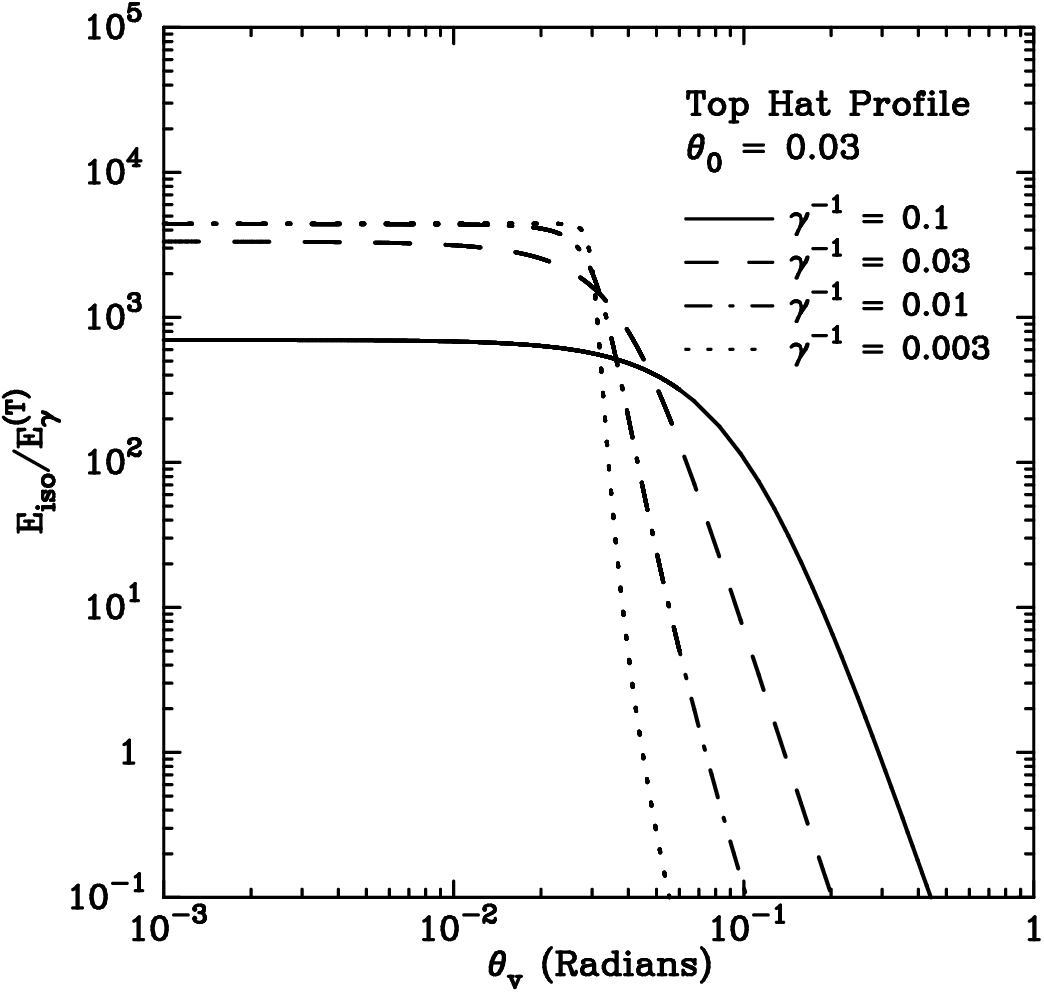}\\
\includegraphics[width=3.2in,clip=]{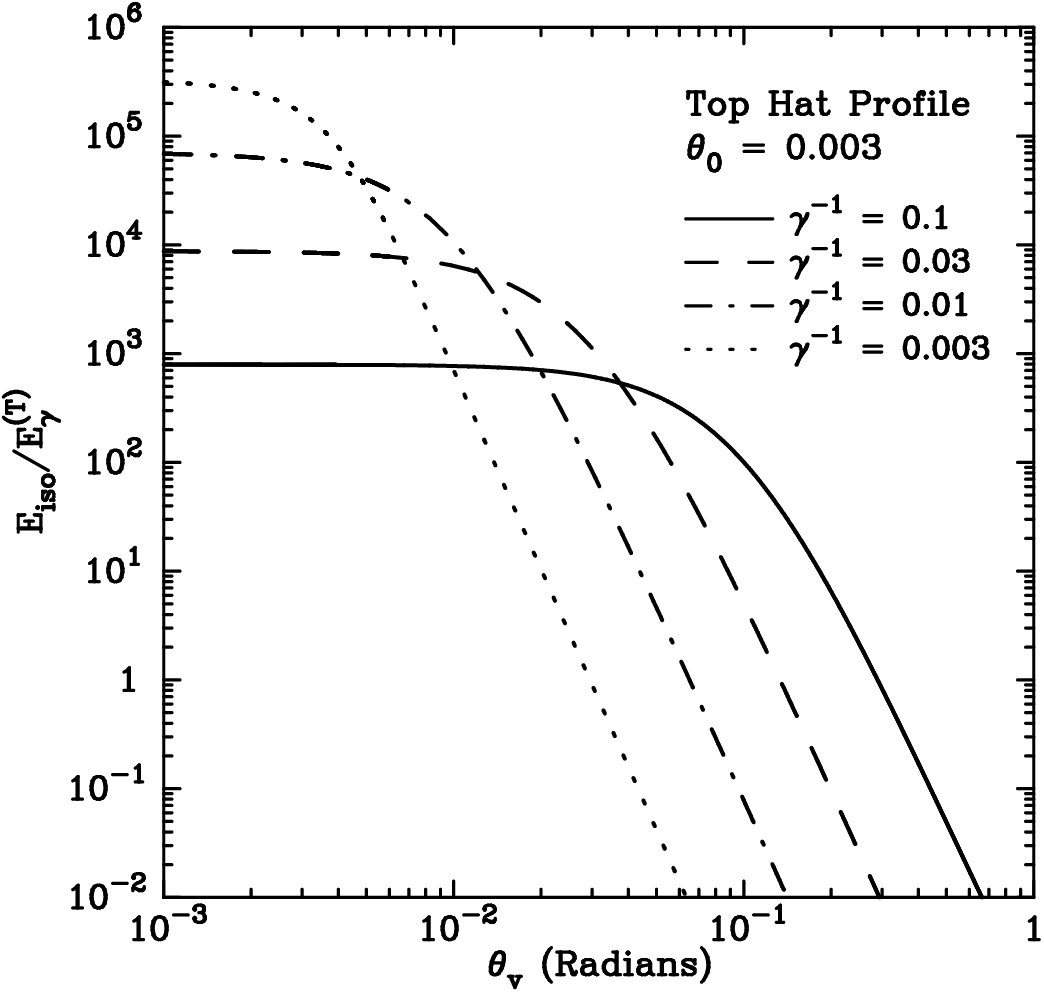}
\end{minipage}
\caption{\eiso\ as a function of viewing angle for emission from
relativistic jets with top-hat emission profiles.  These figures help
illustrate the fact that $\gamma^{-1}$ sets a lower limit on the effective
angular size of the jet. Each panel shows the result for $\gamma^{-1}$ =
0.1, 0.03, 0.01, and 0.003.  Top left:  $\theta_0=0.1$.  Top right:
$\theta_0=0.03$.  Bottom left: $\theta_0=0.01$.  Bottom right:
$\theta_0=0.003$.}
\label{uniform_emission}
\end{figure}

The four panels of Fig.~\ref{uniform_emission} show the behavior of
\eiso\  as a function of viewing angle $\theta_v$ for a top-hat profile. 
We see jet broadening again, as expected.

\begin{figure}[t]
\begin{minipage}[t]{3.2in}
\includegraphics[width=3.2in,clip=]{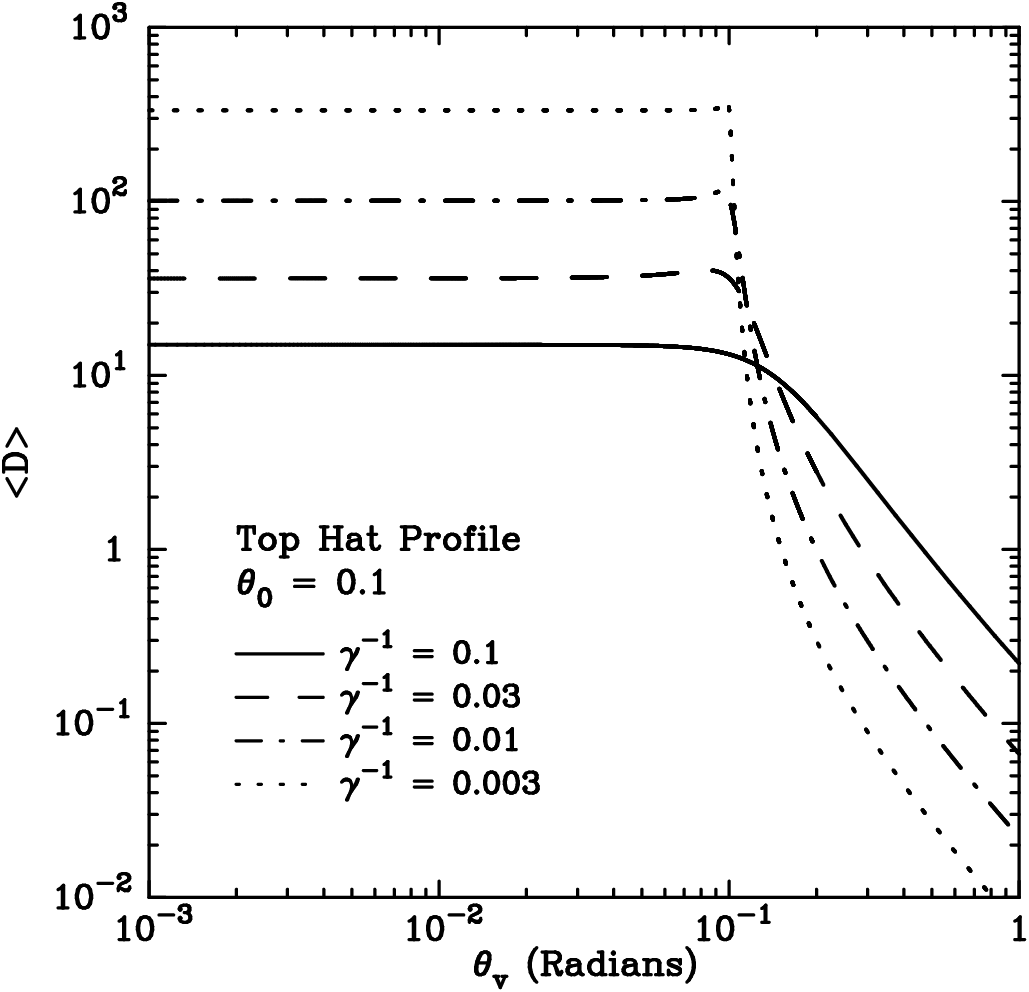}\\
\includegraphics[width=3.2in,clip=]{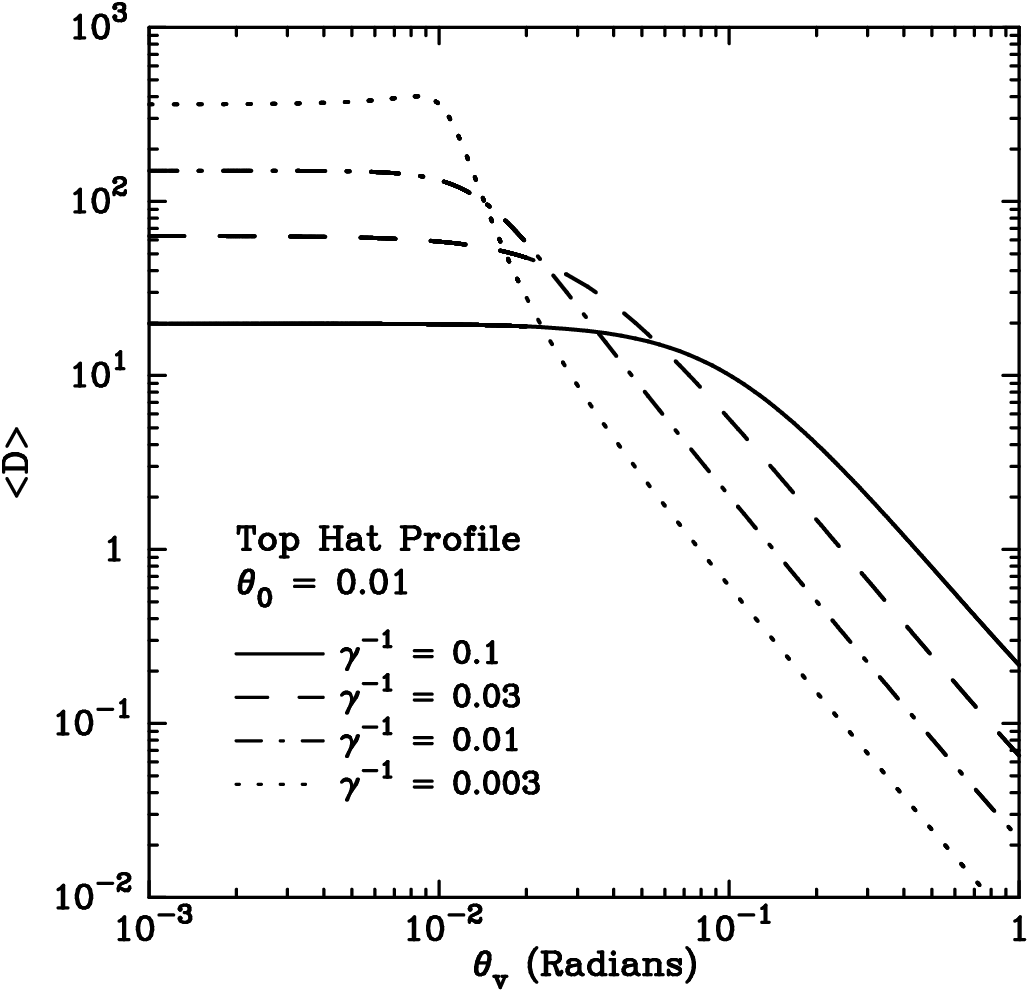}
\end{minipage}
\begin{minipage}[t]{3.2in}
\includegraphics[width=3.2in,clip=]{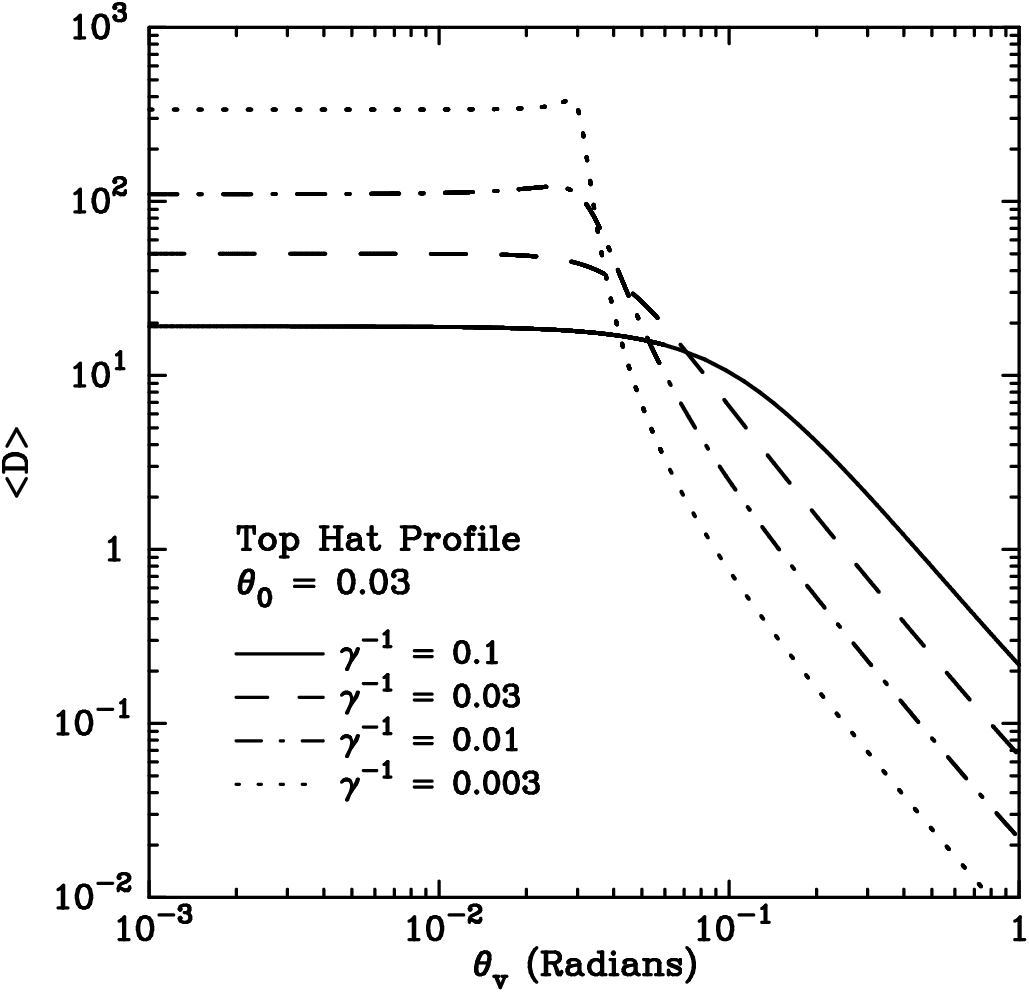}\\
\includegraphics[width=3.2in,clip=]{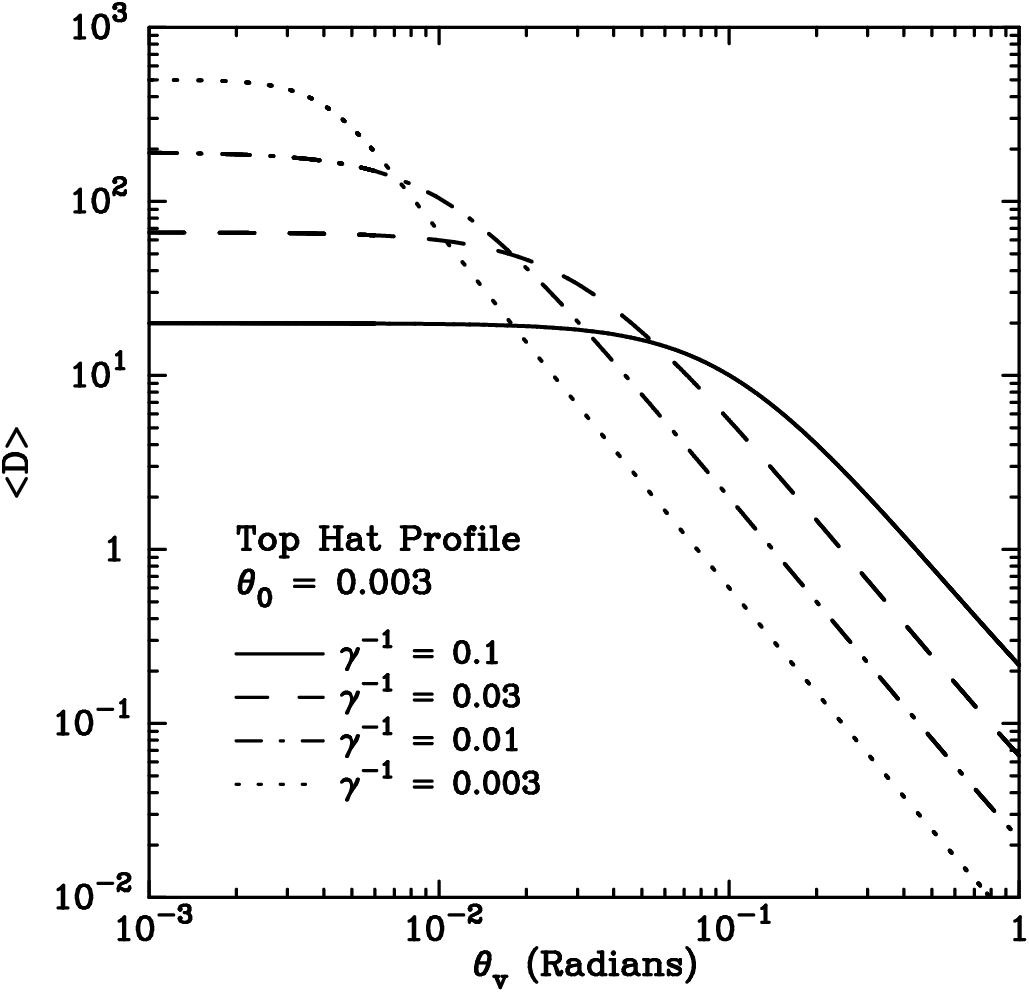}
\end{minipage}
\caption{Photon-number-averaged Doppler factor \davg\ as a function of
viewing angle for emission from relativistic jets with top-hat emission
profiles. Each panel shows the result for $\gamma^{-1}$ = 0.1, 0.03, 0.01,
and 0.003.  Top left:  $\theta_0=0.1$.  Top right: $\theta_0=0.03$. 
Bottom left: $\theta_0=0.01$.  Bottom right: $\theta_0=0.003$.}
\label{uniform_epk}
\end{figure}

The four panels of Fig.~\ref{uniform_epk} show the behavior of \davg\ as a
function of viewing angle $\theta_v$.  There is no ``shelf'' effect here,
in contrast to the Fisher and the $\delta=8$ power-law cases --- the sharp
edge of the profile is the feature that dominates the emission for
$\gamma^{-1}<\theta_0$.  One curiosity is the small ``horn'' structure
that appears at the edge of the profile when $\gamma^{-1}<\theta_0$.  For
example, the curve for $\theta_0=0.03$, $\gamma^{-1}=0.003$ (top right
panel) shows a brief upswing of \davg\ just before it plummets off the
edge.  This feature is easily explained: \davg\ represents an average of
the Doppler factor weighted by the Doppler photon distribution function. 
Both the Doppler factor and the Doppler photon distribution function peak
along the line-of-sight.  Over most of the face of the profile, the value
of the average Doppler factor is dragged down from the value at the peak
by the lower values corresponding to photons arriving from off the
line-of-sight.  However, within $\gamma^{-1}$ of the edge of the jet, some
of the photons that perform this ``dragging down'' office are simply
missing, so the average briefly surges up.

\subsubsection{\eiso\ -- \davg\ and \egi\ -- \davg\ Correlations}

\begin{figure}[t]
\begin{minipage}[t]{3.2in}
\includegraphics[width=3.2in,clip=]{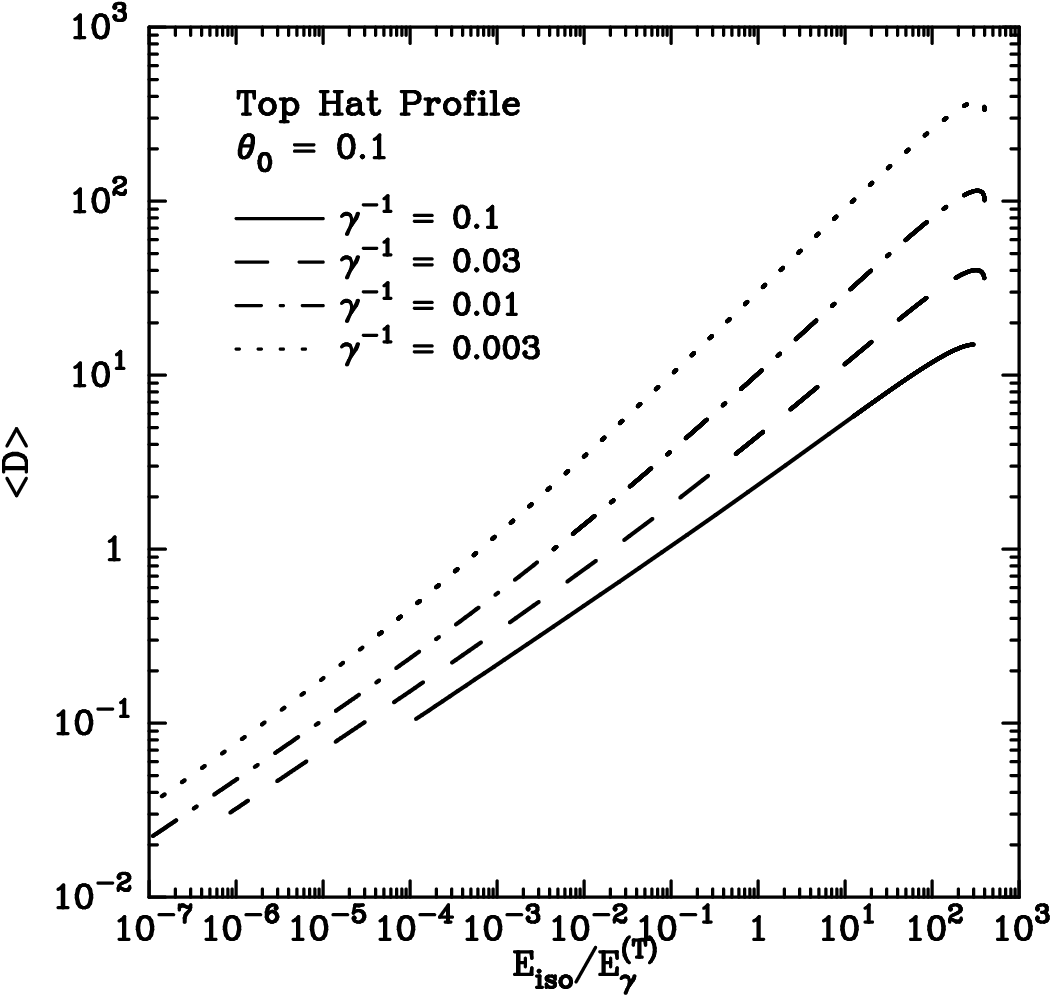}\\
\includegraphics[width=3.2in,clip=]{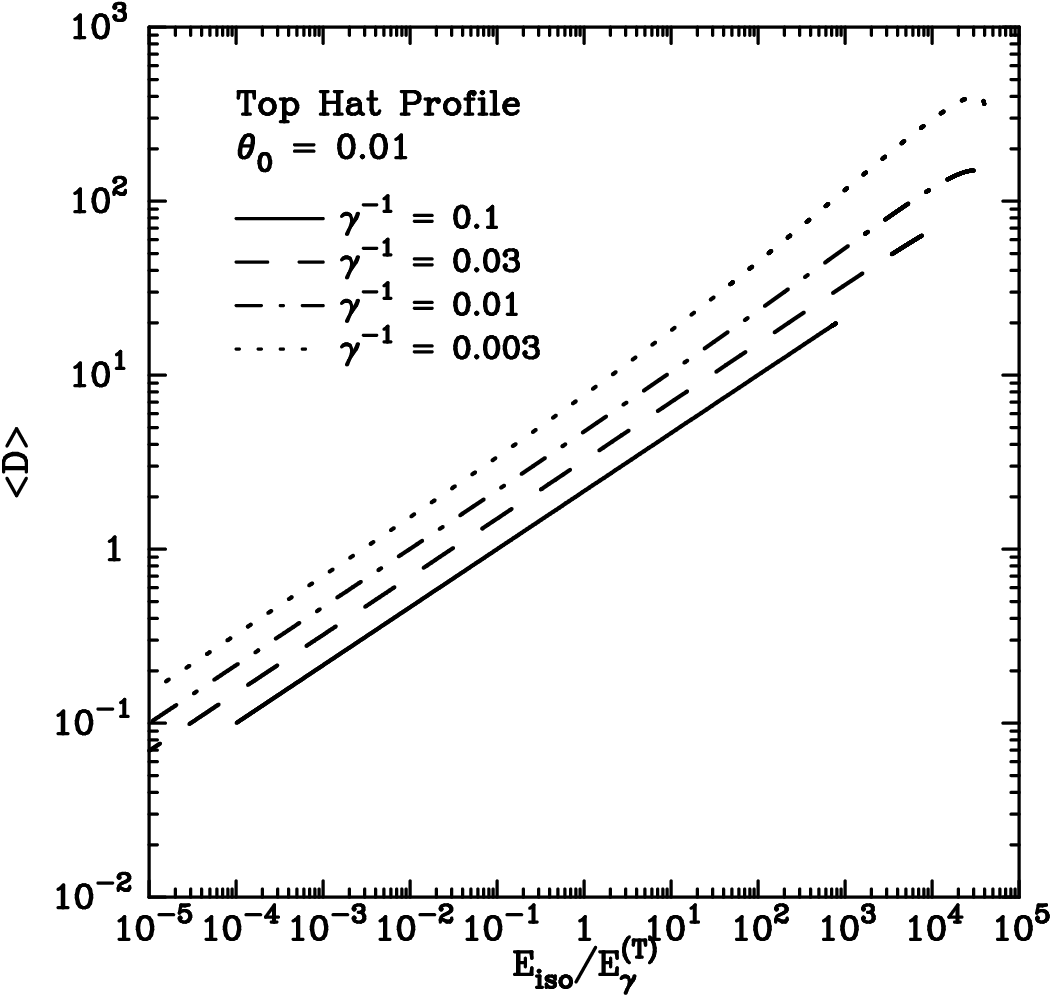}
\end{minipage}
\begin{minipage}[t]{3.2in}
\includegraphics[width=3.2in,clip=]{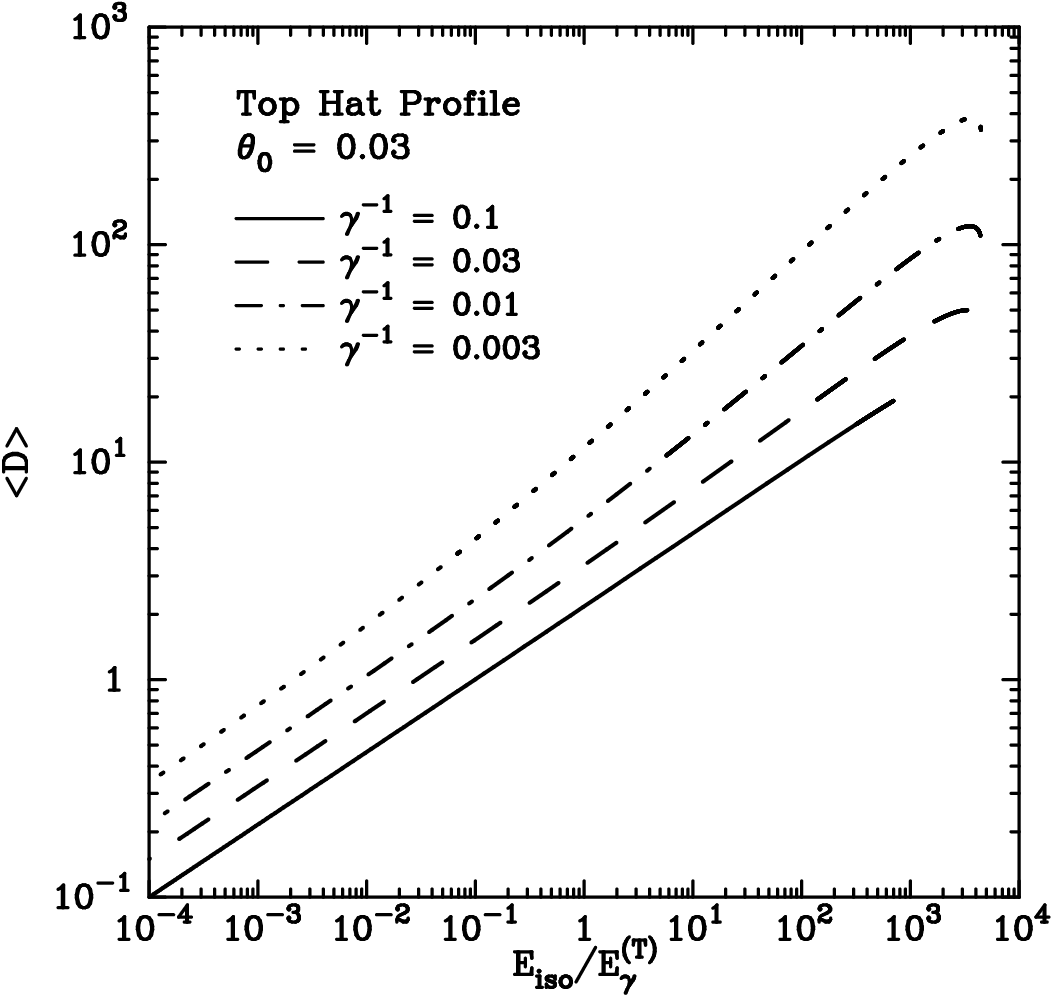}\\
\includegraphics[width=3.2in,clip=]{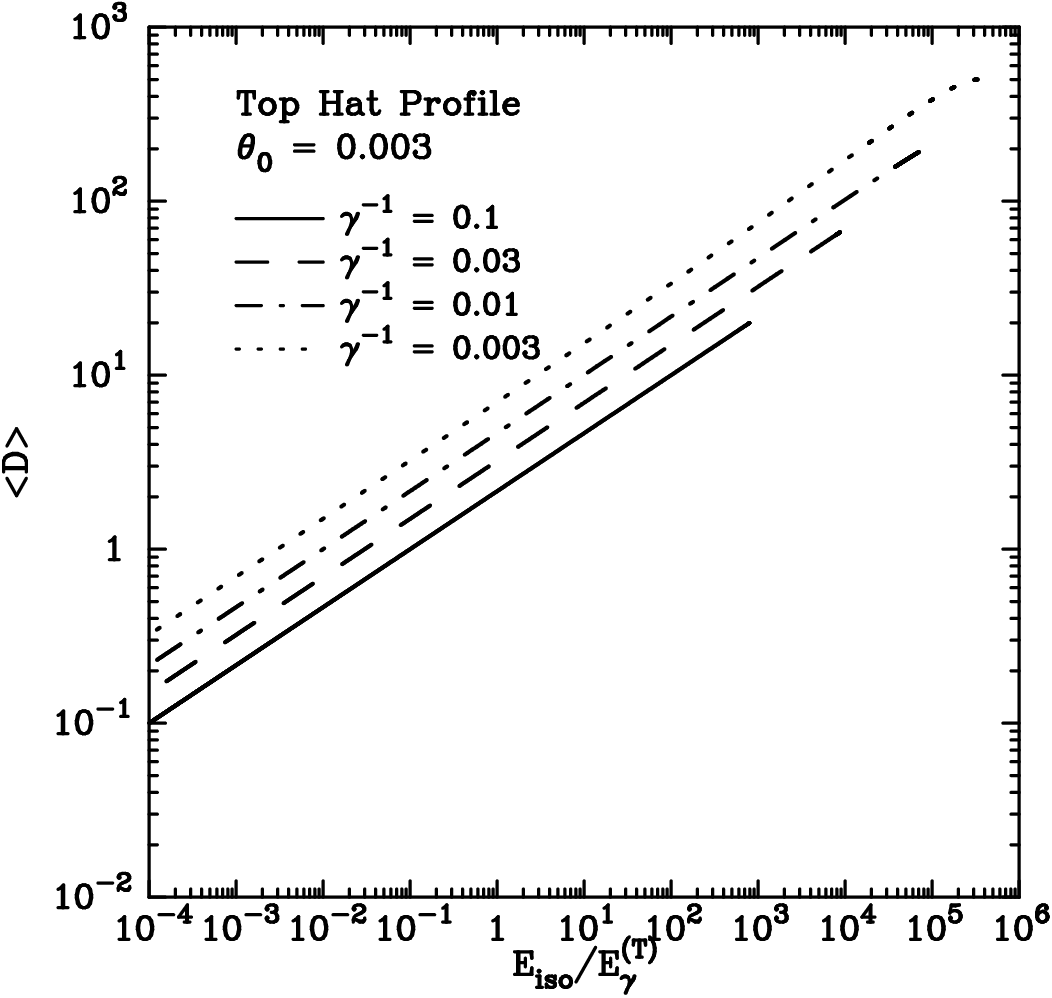}
\end{minipage}
\caption{\eiso--\davg\ (``Amati'') plots for universal jet models based on
top-hat profiles.  Each panel shows the result for $\gamma^{-1}$ = 0.1,
0.03, 0.01, and 0.003.  Top left:  $\theta_0=0.1$.  Top right:
$\theta_0=0.03$.  Bottom left: $\theta_0=0.01$.  Bottom right:
$\theta_0=0.003$.}
\label{uniform_amati}
\end{figure}

The four panels of Fig.~\ref{uniform_amati} show the behavior of the
\eiso--\davg\ (``Amati'') correlation in universal top-hat models.  The
axis scales are chosen so that a slope of 0.5 appears as a 45$^\circ$
line, in order to ease the search for an ``Amati-like'' slope.

As expected, the ``needle jet'' cases $\theta_0\ll\gamma^{-1}$ exhibit the
1/3 slope characteristic of off-axis emission.  These cases are therefore
unpromising.  The opposite case, $\theta_0>\gamma^{-1}$ is more
interesting.  For example, in the case $\theta_0=0.1$, $\gamma^{-1}=0.003$
(top left panel), the slope of the curve near the top is 0.46, and the
gentler slope only sets in for fainter events.  The similarity of the case
$\theta_0=0.03$, $\gamma=333$ (top right panel) to Figure 1 of
\citet{toma2005} (which assumed $\theta_0=0.02$, $\gamma^{-1}=300$ is
quite striking, particularly in view of of the contrast between the rather
detailed nature of the \citet{toma2005} model and the rather simple model
presented here.

Potentially troubling is the fact that all the events seen on-axis are
located at the tip of the hook structure at the end of the curve (this
structure is a side-effect of the ``horn'' structure remarked upon in the
previous section).  Therefore, fits of this model to ``Amati''-type data
must constrain $\theta_0$ to be small enough as not to overload the top
end of the Amati relation with an excessive concentration of events.  Note
also that in order to take this profile seriously in a universal model, it
is necessary to abandon the ``GRBs are on-axis, XRFs are off-axis''
paradigm of \citet{yamazaki2002,yamazaki2003}, since otherwise only XRFs
would obey the Amati relation, while all GRBs would cluster very tightly
in the \eiso-\epk\ plane.

\begin{figure}[t]
\begin{minipage}[t]{3.2in}
\includegraphics[width=3.2in,clip=]{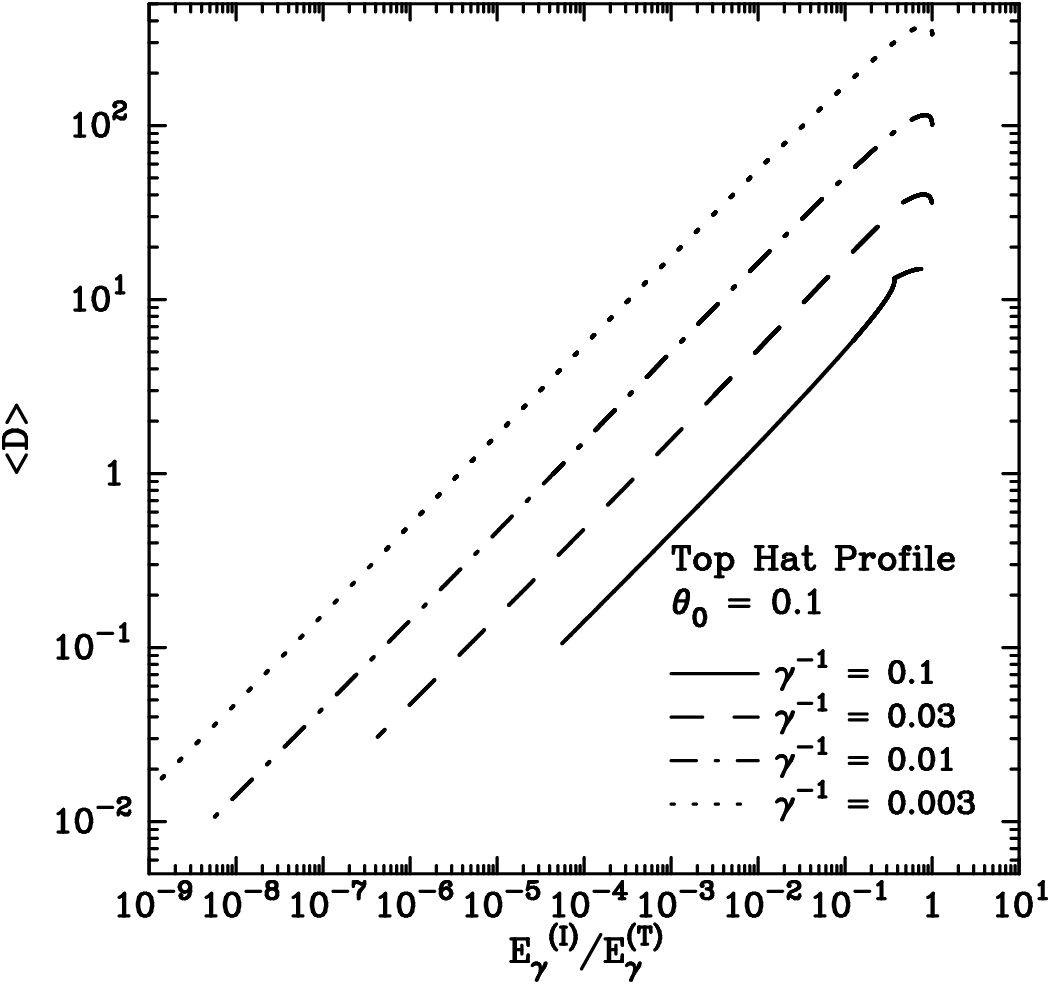}\\
\includegraphics[width=3.2in,clip=]{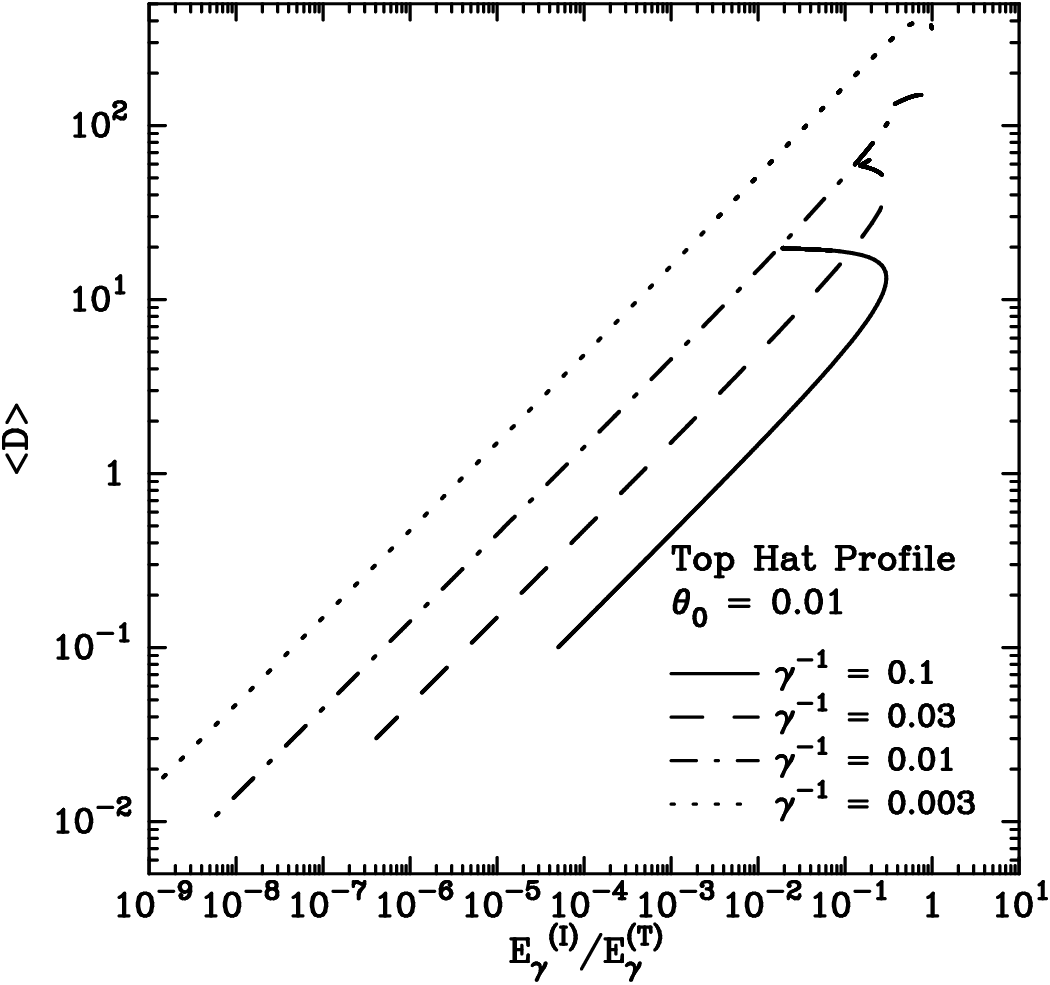}
\end{minipage}
\begin{minipage}[t]{3.2in}
\includegraphics[width=3.2in,clip=]{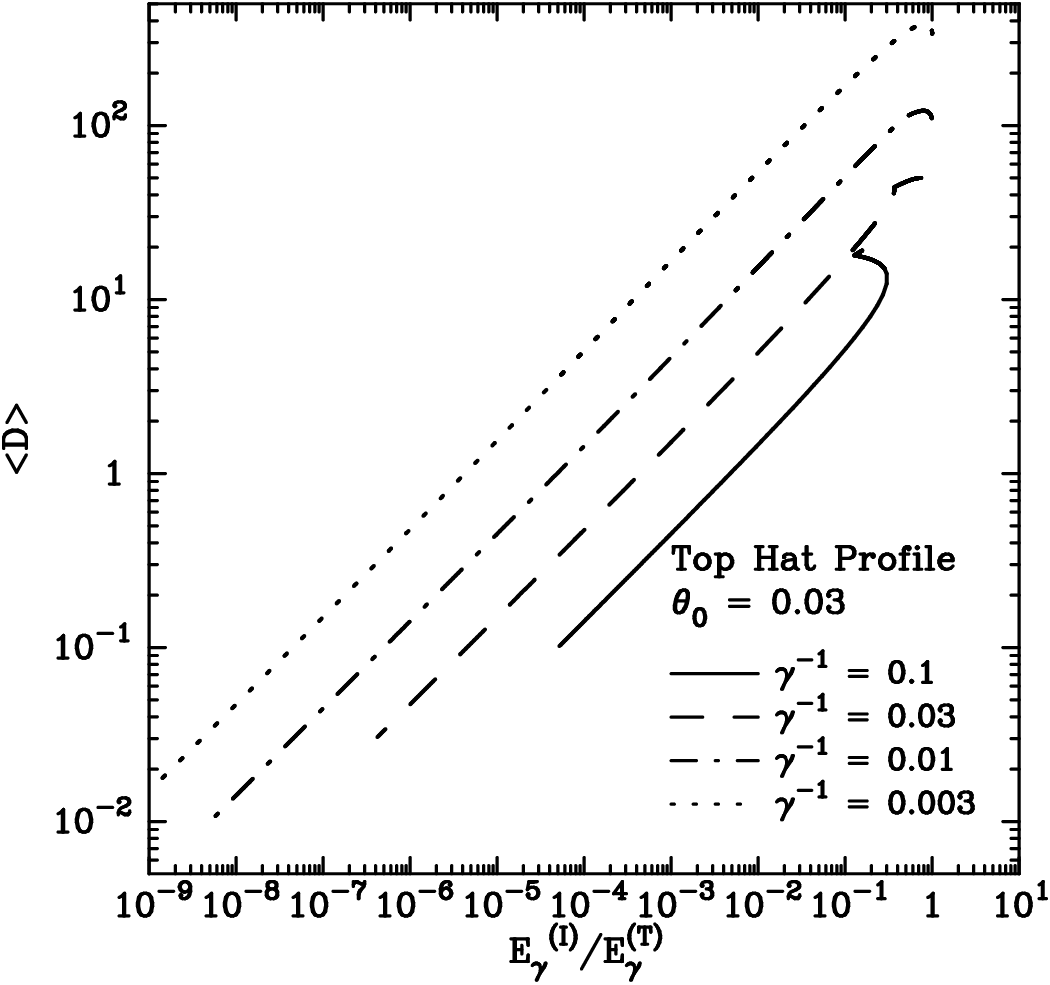}\\
\includegraphics[width=3.2in,clip=]{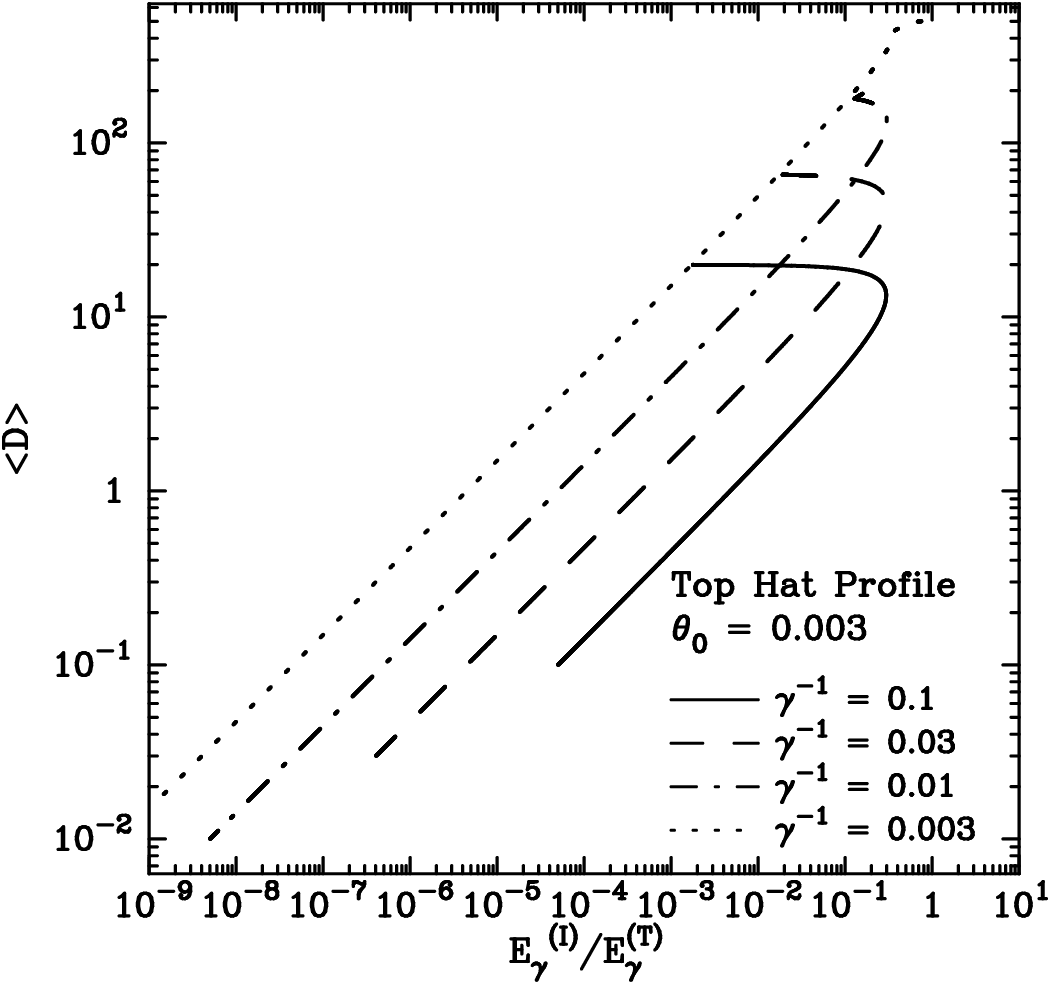}
\end{minipage}
\caption{\egi--\davg\ (``Ghirlanda'') plots for universal jet models based on
top-hat profiles.  Each panel shows the result for $\gamma^{-1}$ =
0.1, 0.03, 0.01, and 0.003.  Top left:  $\theta_0=0.1$.  Top right:
$\theta_0=0.03$.  Bottom left: $\theta_0=0.01$.  Bottom right:
$\theta_0=0.003$.}
\label{uniform_ghirlanda}
\end{figure}

The four panels of Fig.~\ref{uniform_ghirlanda} show the behavior of the
\egi--\davg\ (``Ghirlanda'') correlation in universal top-hat models.

On the rising slope part of these curves, it appears that
``Ghirlanda''-like slopes can indeed be made for $\gamma^{-1}<\theta_0$.
As was the case for the Amati relation, the delicacy required to prevent a
conspicuous blob of on-axis events from distorting the plot must
necessarily constrain $\theta_0$.

\subsubsection{\eiso\ and \egi\ Frequency Distributions}

\begin{figure}[t]
\begin{minipage}[t]{3.2in}
\includegraphics[width=3.2in,clip=]{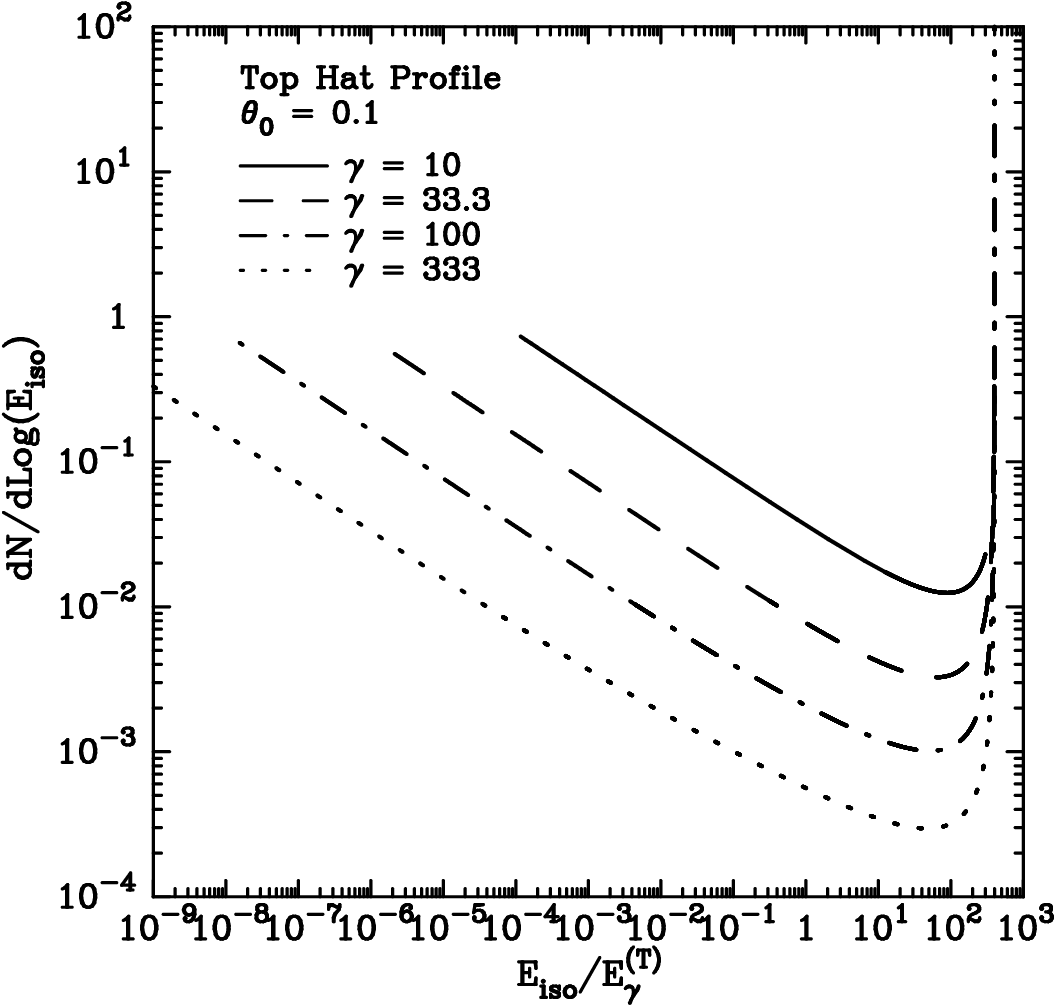}\\
\includegraphics[width=3.2in,clip=]{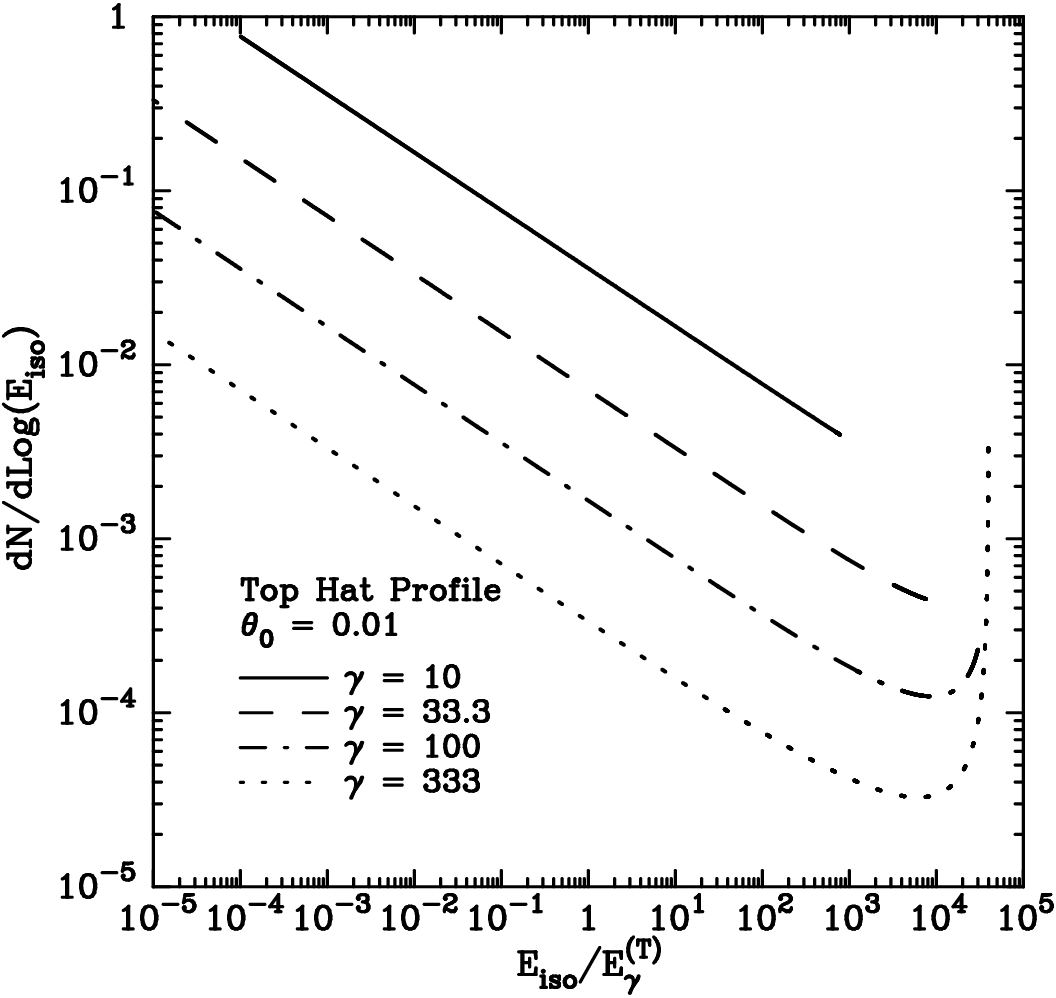}
\end{minipage}
\begin{minipage}[t]{3.2in}
\includegraphics[width=3.2in,clip=]{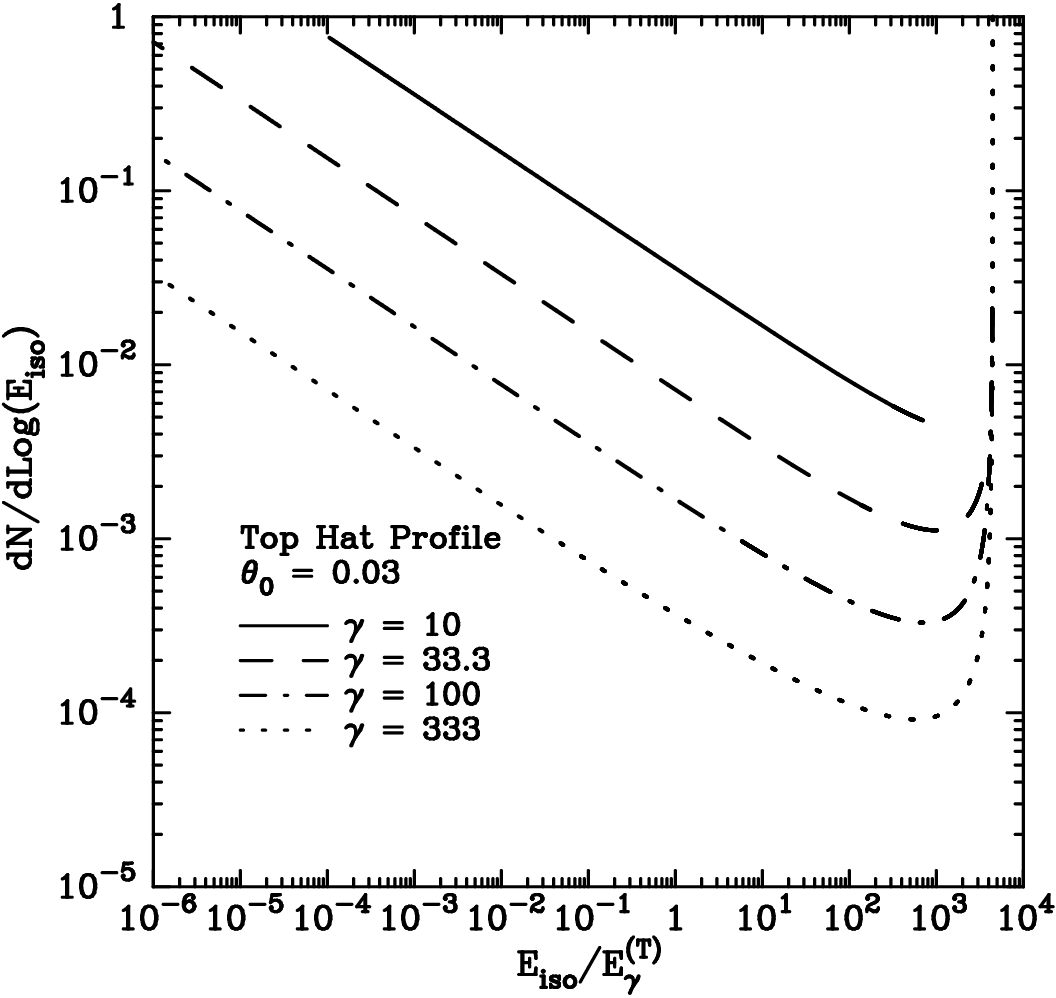}\\
\includegraphics[width=3.2in,clip=]{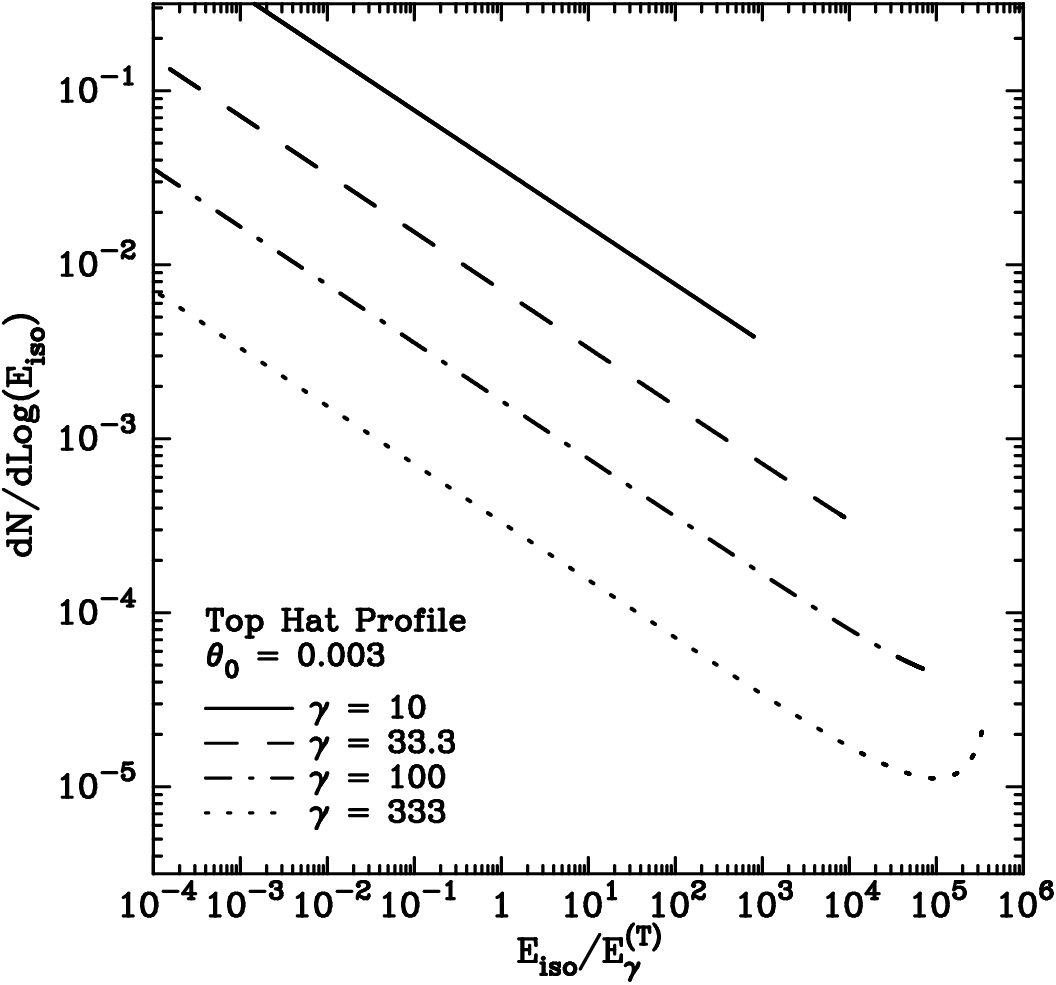}
\end{minipage}
\caption{$dN/d\log(\eiso)$ distributions for universal jet models based on
top-hat profiles.  Each panel shows the result for $\gamma$ =
10, 33.3, 100, and 333.  Top left:  $\theta_0=0.1$.  Top right:
$\theta_0=0.03$.  Bottom left: $\theta_0=0.01$.  Bottom right:
$\theta_0=0.003$.  The vertical scale is arbitrary.}
\label{uniform_dndxiso}
\end{figure}

The four panels of Fig.~\ref{uniform_dndxiso} show the effect of
relativistic kinematics on the \eiso\ frequency distribution of top-hat
jets.

Here is evident the main difficulty that inheres in these models:  the
distribution of $dN/d\log\eiso$ simply can not be made even approximately
uniform, as required by the data.  Rather, for most of the  dynamic range
$dN/d\log\eiso\sim\eiso^{-1/3}$, as expected for off-axis emission
(Eq.~\ref{eiso_limit}).  The distribution then piles up at the top of the
\eiso\ range, where all the events that are viewed on-axis reside.  It is
apparent that no choice of parameters can make this model even remotely
resemble the \eiso\ frequency data.

\begin{figure}[t]
\begin{minipage}[t]{3.2in}
\includegraphics[width=3.2in,clip=]{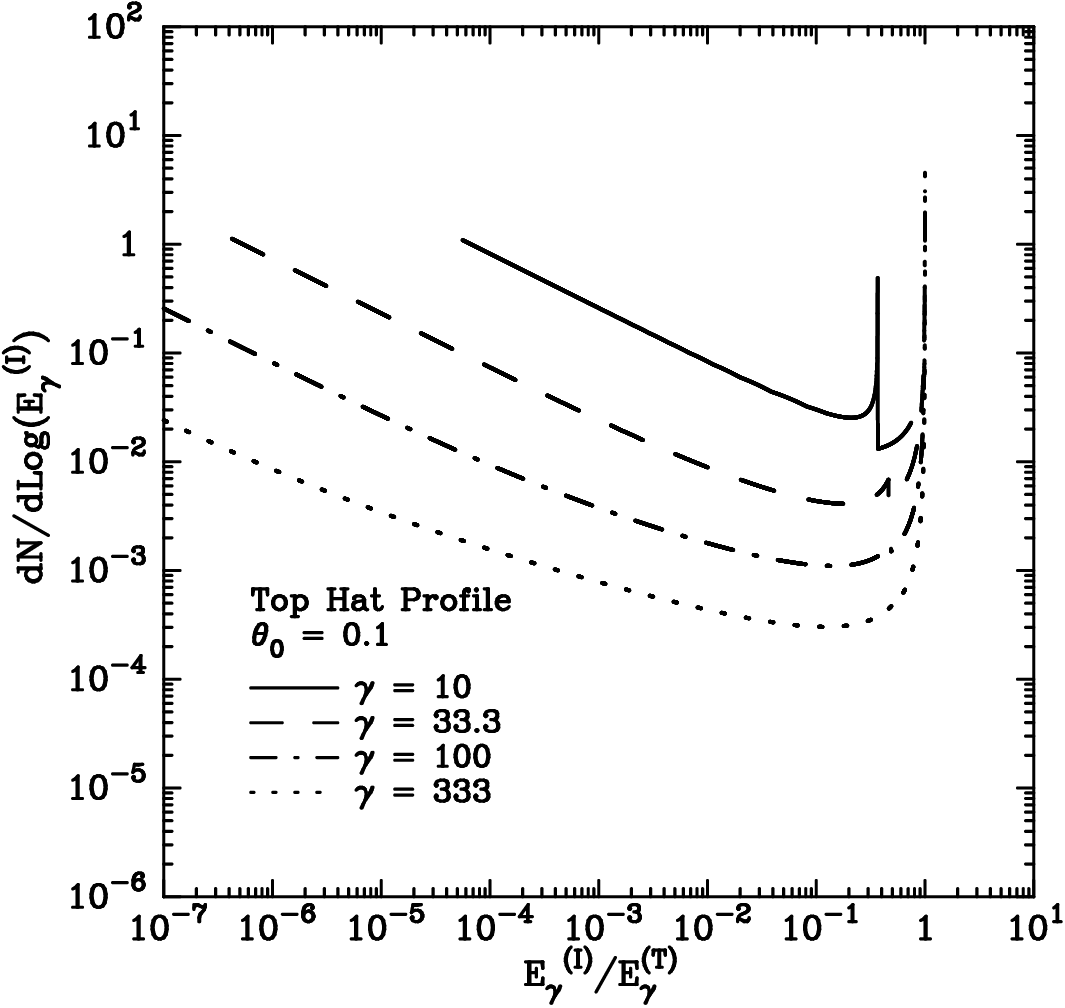}\\
\includegraphics[width=3.2in,clip=]{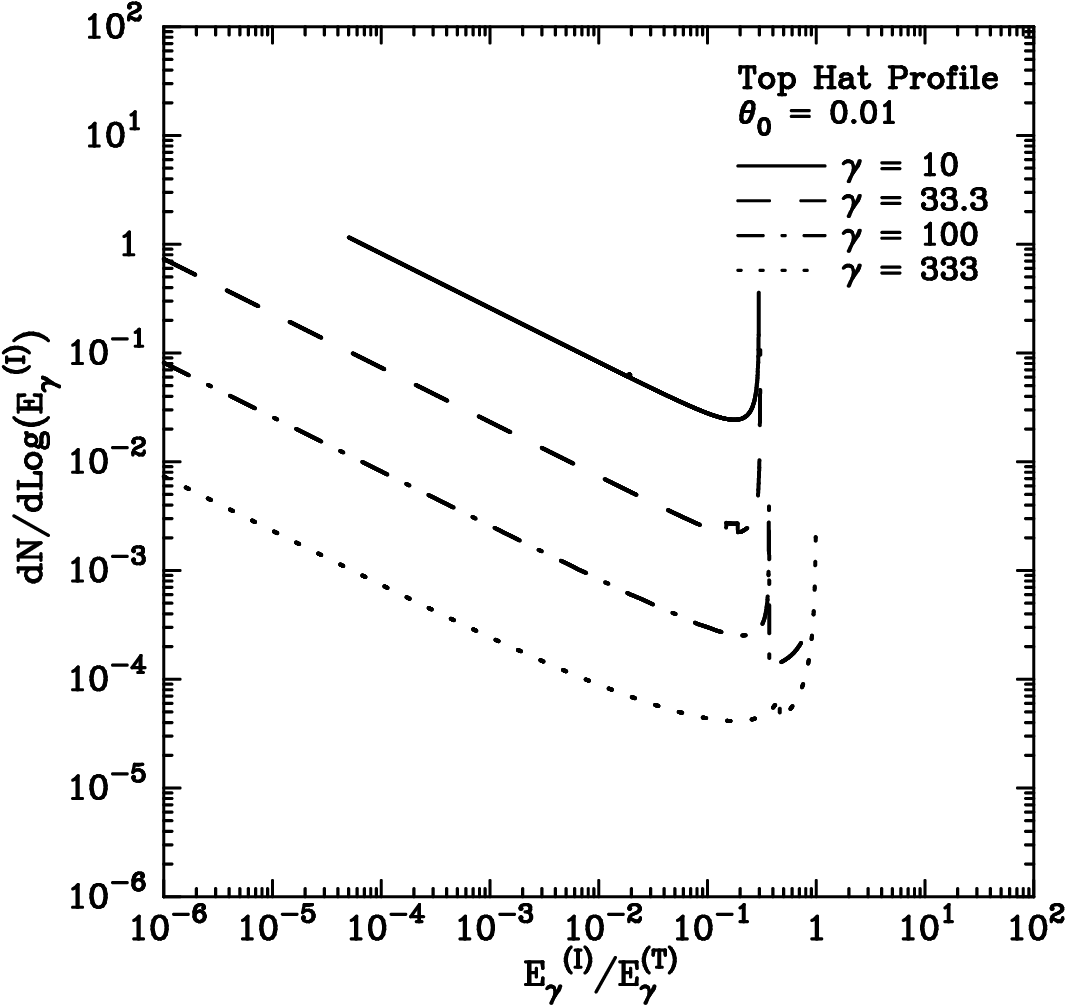}
\end{minipage}
\begin{minipage}[t]{3.2in}
\includegraphics[width=3.2in,clip=]{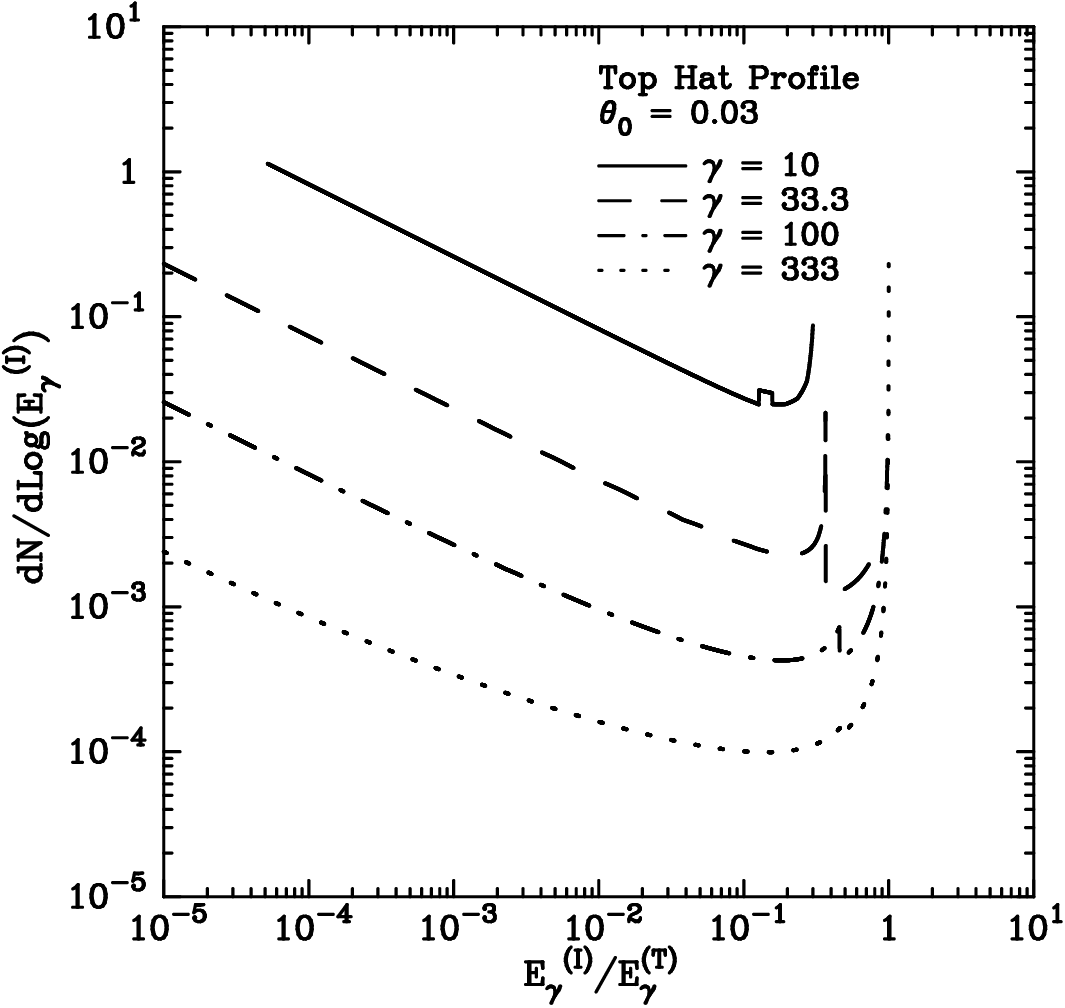}\\
\includegraphics[width=3.2in,clip=]{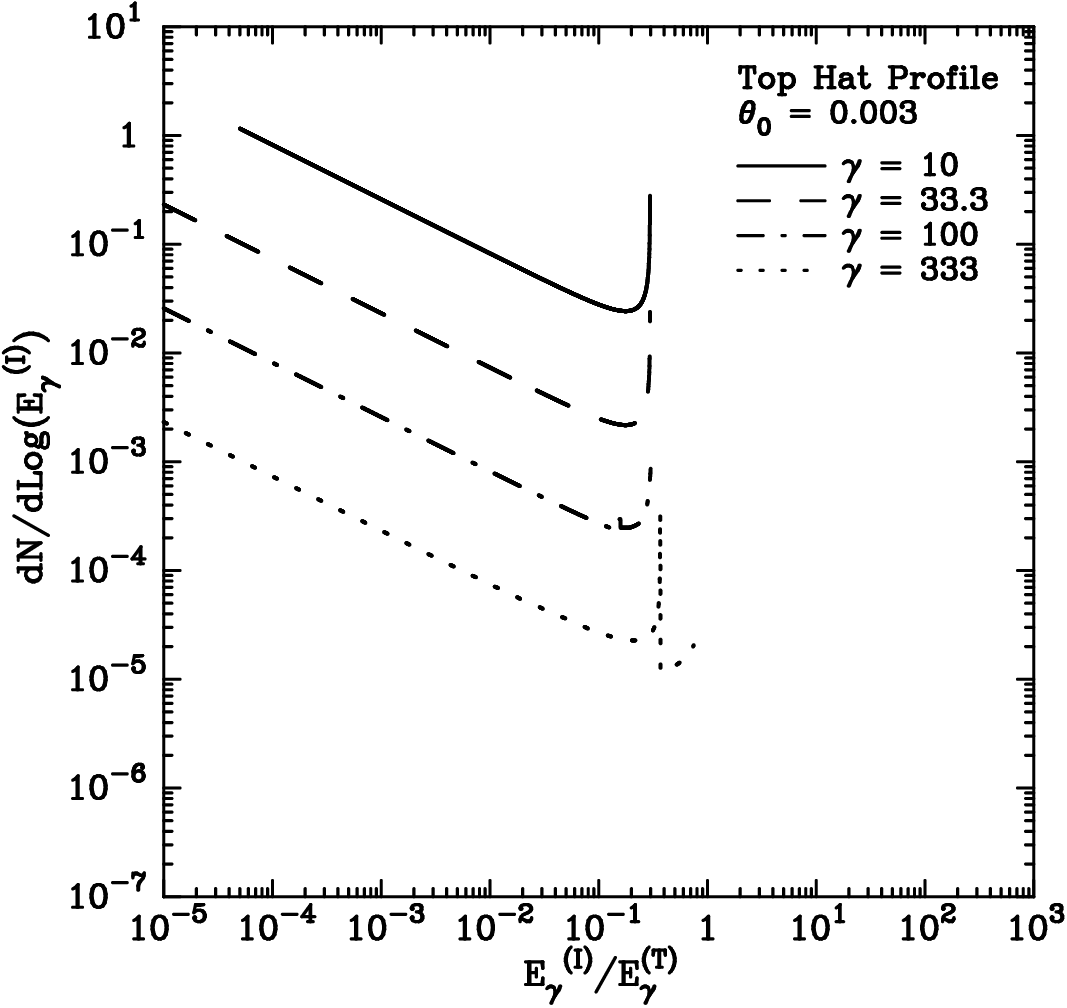}
\end{minipage}
\caption{$dN/d\log(\egi)$ distributions for universal jet models based on
top-hat profiles.  Each panel shows the result for $\gamma$ =
10, 33.3, 100, and 333.  Top left:  $\theta_0=0.1$.  Top right:
$\theta_0=0.03$.  Bottom left: $\theta_0=0.01$.  Bottom right:
$\theta_0=0.003$.  The vertical scale is arbitrary.}
\label{uniform_dndxgi}
\end{figure}

The four panels of Fig.~\ref{uniform_dndxgi} show the effect of
relativistic kinematics on the \egi\ frequency distribution of top-hat
jets.

The curious, two-peak structure observed in some of these curves is 
produced by the application of the \citet{f01} prescription to top-hat
profiles, in conjunction with the discontinuous nature of the break angle,
$\theta_{br}=\max(\theta_v,\theta_0)$.  A spike can appear near the energy
corresponding to $\theta_0$, since the functional dependence of \egi\ on
$\theta_v$ can produce a maximum there, which produces a spike in the
distribution function by Eq.~(\ref{dndegi}).

Events viewed near- or on-axis (the ``spiky'' components) appear capable of
producing relatively narrow \egi\ frequency distributions, especially for
$\gamma^{-1}<\theta_0$.  The case of $\theta_0=0.1$, $\gamma^{-1}=0.003$
(top left panel) is an example.  In this case, however, one must write
off the off-axis component, which is undeniably broad, perhaps under cover
of the fact that these would be XRFs, and cannot at present be placed in a
``Ghirlanda''-type plot, since no break times have yet been observed for
XRFs with redshifts.

\section{Discussion}
\label{discussion}

From the foregoing exploration, it appears that relativistic kinematics
cannot alleviate the weakness of universal jet models that we identified
in \S\ref{broad_vs_narrow}: it does not appear possible for a universal
jet model to produce a narrow distribution of \egi simultaneously with a
broad distribution of \eiso. This is a difficulty for universal jets,
although not necessarily a fatal one.  The currently-available sample of
\egi\ values is certainly more strongly affected by systematic selection
effects than the sample of \eiso\ values, since there is some fraction of
bursts for which no jet break is known.  In particular, there are no jet
breaks available for X-Ray Flash events, so that the frequency
distribution of \egi\ may in fact extend to XRFs, despite being truncated
by observational selection effects.  Nonetheless, even concentrating only
on GRBs, it is striking that the empirical width of the \egi distribution
of GRBs is is not much larger than a single decade, while the width of the
\eiso distribution is somewhere between three and four decades.  It is
hard to understand how such a discrepancy could be produced by selection
effects alone.

There is a generic effect of relativistic kinematics that is worth
highlighting:  The relativistic beaming angle $\gamma^{-1}$ imposes a
lower limit on the effective angular size of the jet.  Underlying jet
emission profiles that are narrower than $\gamma^{-1}$ will be broadened
to an angular size of $\gamma^{-1}$ by the relativistic Doppler effect. 
This effect has been noted in the case of afterglow emission
\citep{woods1999,rossi2002}, but its influence on prompt emission has not
been widely discussed. The net effect is to impose an upper bound on the
extent to which the prompt flux can be amplified by collimation of the
profile.  A concomitant effect is to curb the ability of very steep
profiles to produce a large dynamic range of emission intensity.

It seems worth commenting further upon the ``shelf'' feature of the
$\davg-\theta_v$ plots that was examined carefully in the case of Fisher
jet profiles, and remarked upon in the case of $\delta=8$ power-law
profiles.  This feature appears to be a generic property of steep
profiles, as might be expected from the explanation provided in
\S\ref{sec_fisher_emission}.  It illustrates the fact that relativistic
kinematics can not only serve to soften a source spectrum (as is the case
for the spectrum emitted by a narrow jet observed far off-axis), but can
also {\it harden} a GRB spectrum with respect to what might be expected
from a naive application of ``needle-jet'' kinematics.  For example, in
the case of a Fisher jet with $\theta_0=0.1$, one might naively expect
that for $\theta_v>\theta_0$, the jet profile presents the appearance of a
needle jet, so that \davg\ must drop away as $(1-\beta\muv)^{-1}$.  It is
therefore surprising to find that \davg\ can in fact remain relatively
constant and considerably higher than this naive argument would lead us
to believe, even quite far from the jet axis.

It is encouraging to note that Amati-like and Ghirlanda-like correlations
may emerge from ``steep'' profiles such as the Fisher profile, or the
$\delta=8$ power-law profile, under suitable choices of $\theta_0$ and
$\gamma$.  It is somewhat striking, even surprising, to find such
correlations emerging of their own accord from top-hat profiles.  Such
profiles are not usually considered promising candidates for universal
models, as they are not ``structured''.  The main exception to this view
has been put forward by Yamazaki and collaborators
\citep{yamazaki2002,yamazaki2003,yamazaki2004,toma2005}, as well as other
papers by the same authors.  It is certainly proven to be something of a 
surprise that top-hat models can produce Amati slopes much steeper than the
1/3 value expected from a relativistic point source (the ``needle jet''
approximation).

It is noteworthy that in the case of the jet profiles that have a decent
chance of making realistic-looking Amati/Ghirlanda relations (the Fisher
profile, the $\delta=8$ power-law profile, and the top-hat profile), it is
invariably the case that the Ghirlanda relation has a steeper slope than
the Amati relation.  This steepening is evidently a purely geometrical
effect.  It is clearly unrelated to the steepening effect noticed by
\citet{levinson2005}, since we have certainly not included any dependency
of the break angle on \eiso, the source of the steepening noted in that
paper.  Apparently, with a break angle that is conceived purely in terms of
geometry, it is still the case that Ghirlanda relations of universal models
are expected to be steeper than their corresponding Amati relations.

The inability of shallow power-law profiles ($\delta=1$ and $\delta=2$) to
produce any substantial dynamic range in \davg\ spells major trouble for
such models, above and beyond their expected inability to produce broad
\eiso\ distributions.  This aspect of the situation might be correctable,
if one is willing to incur the cost of introducing an angular dependence of
the spectrum, in conjunction with the angular dependence of the emission
profile.  In particular, one could consider the (purely empirical)
introduction of a relation like $\epk\propto\epsilon(\vec{n}\cdot\vec{b})^{1/2}$
as a microphysical relation satisfied on the surface of the shock.

There have been some investigations that appear at first glance to do
something like this.  \citet{yamazaki2004} introduced an intrinsic Amati
relation between \epk\ and \eiso\ in their universal off-axis model, to
examine how the correlation was modified by relativistic kinematics for
off-axis observers.  \citet{zhang2004} similarly assumed an intrinsic
Amati relation in their study of Gaussian jet profiles.

\citet{kobayashi2002} pointed out that ${\epk}^2$ is proportional to the
internal energy of the jet material.  \citet{zm02b} and \citet{rrlr2002}
observed that internal synchrotron models can certainly produce a
correlation between \epk\ and the luminosity, although the exact shape of
the correlation varies with the dependence of $L$ on $\gamma$.  It should
be pointed out, however, that these dependencies of luminosity on \epk\
appear to be ``global'' statements about average conditions prevailing in
the jet.  The question of whether an internal shock model radiating by
synchrotron emission might be expected to exhibit a {\it local} Amati-like
correlation between \epk\ and emissivity has not, to our knowledge, been
addressed.

We note in passing that it would be straightforward to modify our method
to calculate \epk\ instead of \davg\ in such a model, since all that is
necessary is to introduce an additional multiplicative function
$\epk(\vec{n}\cdot\vec{b})$ in the integral in the numerator of
Eq.~(\ref{dbar}).

\acknowledgments
We gratefully acknowledge Wayne Hu, Hiranya Peiris, Rosalba Perna, Davide
Lazzati, and Ryo Yamazaki for detailed discussions and comments. This
research was supported in part by NASA Contract NAGW-4690 and NASA Grant
NAG5-10759.

\clearpage

% Eiso/Egamma distributions

\clearpage

% Graphical analysis of dN/dEgamma

\clearpage

% Emission from fisher profile

\clearpage

% Epk function from Fisher profile

\clearpage

% Explanation of "step-function" behavior in <D> function of Fisher
% profile.

\clearpage

% ``Amati'' relations for the universal Fisher model

\clearpage

% ``Ghirlanda'' relations for the universal Fisher model

% dN/dLog(Eiso) for the universal Fisher model

% dN/dLog(Egamma)  for the universal Fisher model

\clearpage

% Emission from a power-law, delta=1.0 (Zhang-M\'esz\'aros) profile

\clearpage

% Epk function from a power-law, delta=1.0 (Zhang-M\'esz\'aros) profile

\clearpage

% ``Amati'' relations for the universal delta=1.0 power-law model

\clearpage

% ``Ghirlanda'' relations for the universal delta=1.0 power-law model

\clearpage

% dN/dLog(Eiso) for the universal delta=1.0 power-law model

\clearpage

% dN/dLog(Egamma)  for the universal delta=1.0 power-law model

\clearpage

% Emission from a power-law, delta=2 profile

\clearpage

% Epk function from a power-law, delta=2 profile

\clearpage

% ``Amati'' relations for the universal delta=2 power-law model

\clearpage

% ``Ghirlanda'' relations for the universal delta=2 power-law model

\clearpage

% dN/dLog(Eiso) for the universal delta=2  power-law model

\clearpage

% dN/dLog(Egamma)  for the universal delta=2.0 power-law model

\clearpage

% Emission from a power-law, delta=8 profile

\clearpage

% Epk function from a power-law, delta=8 profile

\clearpage

% ``Amati'' relations for the universal delta=8 power-law model

\clearpage

% ``Ghirlanda'' relations for the universal delta=8 power-law model

\clearpage

% dN/dLog(Eiso) for the universal delta=8  power-law model

\clearpage

% dN/dLog(Egamma)  for the universal delta=8 power-law model

\clearpage

% Emission from uniform profile

\clearpage

% Epk function from uniform profile

\clearpage

% ``Amati'' relations for the universal uniform model

\clearpage

% ``Ghirlanda'' relations for the universal uniform model

\clearpage

% dN/dLog(Eiso) for the universal uniform model

\clearpage

% dN/dLog(Egamma)  for the universal uniform model


\begin{thebibliography}{999}

\bibitem[Amati et al.(2002)]{amati2002} 
        Amati, L., et al. 2002, \aap, 390, 81

\bibitem[Band et al.(1993)]{band1993} Band, D., et al.\ 1993, 
\apj, 413, 281 

\bibitem[Bloom et al.(2003)]{bloom2003} Bloom, J.~S., Frail, 
D.~A., \& Kulkarni, S.~R.\ 2003, \apj, 594, 674 

\bibitem[Cooley \& Tukey(1965)]{ct65}
    Cooley,~J.~W., and Tukey,~J.~W. 1965, Math. Comput. 19, 297

\bibitem[Gradshteyn \& Rhyzik(1965)]{gr65}
    Gradshteyn,~I.~S., and Rhyzik,~I.~M. 1965, Tables of Integrals,
    Series, and Products (New York: Academic)

\bibitem[Dai \& Zhang(2005)]{dz05}
    Dai,~X., and Zhang,~B. 2005, ApJ 621, 875

\bibitem[Driscoll \& Healy(1994)]{dh94}
    Driscoll,~J.~R, and Healy,~D.~M. 1994, Adv. Appl. Math. 15, 202

\bibitem[Frail et al.(2001)]{f01}
    Frail,~D.~A. et al 2001, ApJ 562, L55

\bibitem[Ghirlanda et al.(2004)]{ggl2004} Ghirlanda, G., 
Ghisellini, G., \& Lazzati, D.\ 2004, \apj, 616, 331 

\bibitem[Heise et al.(2001)]{heise2001} Heise, J., in't Zand, J., 
Kippen, R.~M., \& Woods, P.~M.\ 2001, Gamma-ray Bursts in the Afterglow 
Era, 16 
 
\bibitem[Kippen et al.(2001)]{kippen2001} Kippen, R.~M., Woods, 
P.~M., Heise, J., in't Zand, J., Preece, R.~D., \& Briggs, M.~S.\ 2001, 
Gamma-ray Bursts in the Afterglow Era, 22

\bibitem[Granot et al.(1999)]{granot1999} Granot, J., Piran, T., 
\& Sari, R.\ 1999, \apj, 513, 679

\bibitem[Granot \& Kumar(2003)]{granot2003} Granot, J., \& Kumar, 
P.\ 2003, \apj, 591, 1086 

\bibitem[Kobayashi et al.(2002)]{kobayashi2002} Kobayashi, S., Ryde, 
F., \& MacFadyen, A.\ 2002, \apj, 577, 302 

\bibitem[Kumar \& Granot(2003)]{kumar2003} Kumar, P., \& Granot, 
J.\ 2003, \apj, 591, 1075

\bibitem[Guetta et al.(2005)]{guetta2005} Guetta, D., Granot, J., 
\& Begelman, M.~C.\ 2005, \apj, 622, 482 

\bibitem[Lapenta \& Kronberg(2005)]{lapenta2005} Lapenta, G., \& 
Kronberg, P.~P.\ 2005, \apj, 625, 37

\bibitem[Lamb et al.(2004)]{lamb03} Lamb, D.~Q., et al.\ 2004, 
New Astronomy Review, 48, 423

\bibitem[Lamb, Donaghy, \& Graziani(2005)]{lamb04} Lamb,
D.~Q., Donaghy, T.~Q., Graziani, C.\ 2005, \apj 620, 355-378

\bibitem[Levinson \& Eichler(2005)]{levinson2005}
Levinson,~A., and Eichler,~D.\ 2005, astro-ph/0504125

%\bibitem[Lloyd-Ronning \& Ramirez-Ruiz(2002)]{lrrr2002} 
%Lloyd-Ronning, N.~M., \& Ramirez-Ruiz, E.\ 2002, \apj, 576, 101 

\bibitem[Lloyd-Ronning et al.(2004)]{lr2004} Lloyd-Ronning, 
N.~M., Dai, X., \& Zhang, B.\ 2004, \apj, 601, 371 

\bibitem[Mardia(1972)]{m72}
    Mardia,~K.~V. 1972, Statistics of Directional Data (New York: Academic)

\bibitem[Oh, Spergel, \& Hinshaw(1998)]{osh98}
    Oh,~S.~P., Spergel,~D.~N., and Hinshaw,~G. 1998, ApJ 510, 551

\bibitem[Piran(1999)]{piran1999} Piran, T.\ 1999, \physrep, 314, 
575 

\bibitem[Ramirez-Ruiz \& Lloyd-Ronning(2002)]{rrlr2002} 
Ramirez-Ruiz, E., \& Lloyd-Ronning, N.~M.\ 2002, New Astronomy, 7, 197 

\bibitem[Rees \& M\'esz\'aros(1994)]{rees1994} Rees, M.~J., \& 
M\'esz\'aros, P.\ 1994, \apjl, 430, L93 

\bibitem[Rhoads(1999)]{rhoads1999} Rhoads, J.~E.\ 1999, \apj, 525, 
737 

\bibitem[Rossi et al.(2002)]{rossi2002} Rossi, E., Lazzati, D., 
\& Rees, M.~J.\ 2002, \mnras, 332, 945 

\bibitem[Sari et al.(1999)]{sari1999} Sari, R., Piran, T., \& 
Halpern, J.~P.\ 1999, \apjl, 519, L17 

\bibitem[Stern(2003)]{stern2003} Stern, B.~E.\ 2003, \mnras, 345, 
590 

\bibitem[Toma et al.(2005)]{toma2005}
Toma,~K., Yamazaki,~R., and Nakamura,~T.\ 2005, astro-ph/0504624

\bibitem[Wandelt \& Gorski(2001)]{wg01}
    Wandelt,~B.~D., and G\'{o}rski,~K.~M 2001, Phys. Rev. D, 63, 123002

\bibitem[Woods \& Loeb(1999)]{woods1999} Woods, E., \& Loeb, A.\ 
1999, \apj, 523, 187
 
\bibitem[Yamazaki et al.(2002)]{yamazaki2002} Yamazaki, R., Ioka, 
K., \& Nakamura, T.\ 2002, \apjl, 571, L31 
 
\bibitem[Yamazaki et al.(2003)]{yamazaki2003} Yamazaki, R., Ioka, 
K., \& Nakamura, T.\ 2003, \apj, 593, 941 

\bibitem[Yamazaki et al.(2004)]{yamazaki2004}
    Yamazaki,~R., Ioka,~K., and Nakamura,~T. 2004, ApJ 606, L33

\bibitem[Zhang et al.(2004)]{zhang2004} Zhang, B., Dai, X., 
Lloyd-Ronning, N.~M., \& M{\' e}sz{\' a}ros, P.\ 2004, \apjl, 601, L119 

\bibitem[Zhang \& M\'esz\'aros(2002)]{zm02}
    Zhang,~B., and M\'esz\'aros,~P. 2002, ApJ 571, 876

\bibitem[Zhang \& M{\' e}sz{\' a}ros(2002b)]{zm02b} Zhang, B., 
\& M{\' e}sz{\' a}ros, P.\ 2002, \apj, 581, 1236 

\end{thebibliography}
\end{document}